\documentstyle[epsfig]{elsart}

\newcommand{\bk}{{\bf k}}
\newcommand{\bp}{{\bf p}}
\newcommand{\bq}{{\bf q}}
\begin{document}

\begin{frontmatter}
\title{Singular Fermi Liquids}
\author{C. M. Varma$^{a,b}$, Z. Nussinov$^b$,
  and Wim van Saarloos$^b$}

\address{$^a$Bell Laboratories, Lucent Technologies, Murray Hill, NJ
  07974,
U.S.A.\thanksref{bell} }
\thanks[bell]{Present and permanent address}
\address{
$^b$Instituut--Lorentz, Universiteit Leiden, Postbus 9506,
  2300 RA Leiden,\\ The Netherlands }

\begin{abstract}

An introductory survey of the theoretical ideas and calculations and the
experimental results which depart from  Landau Fermi-liquids  is
presented. Common themes and possible routes to the singularities
leading to the breakdown of Landau Fermi liquids are categorized following an
elementary discussion of the theory. Soluble examples of Singular Fermi
liquids  include models of impurities in metals
with special symmetries and one-dimensional interacting
fermions. A review  of these is
followed by a discussion of Singular Fermi
liquids in a wide variety of experimental
 situations and theoretical models. These include the effects of
 low-energy collective fluctuations,
 gauge fields due either to symmetries in the hamiltonian or
 possible dynamically generated symmetries,
fluctuations around quantum
critical points, the normal state of high temperature superconductors
and the two-dimensional metallic state. For the last three systems, the
principal experimental results are summarized and the outstanding
theoretical issues highlighted.

\end{abstract}
\end{frontmatter}
\tableofcontents


\section{Introduction}

\subsection{Aim and scope of this paper}
In the last two decades a variety of
metals have been discovered which display thermodynamic
and transport properties at low
temperatures which are fundamentally different
from those of the usual metallic systems which are well described by the Landau
Fermi-liquid theory. They have often been referred to as
Non-Fermi-liquids. A fundamental characteristic of such systems is
that the low-energy properties in a wide range of their phase diagram
are dominated by singularities as a function of energy and temperature. Since
these problems still relate to a liquid state of fermions and since it
is not a good practice to name things after what they are not, we prefer
to call them Singular Fermi-liquids (SFL).

The basic notions of Fermi-liquid theory have actually been with us at
an intuitive level since the time of Sommerfeld: He showed that the
linear low temperature specific heat behavior of metals as well as
their asymptotic low temperature resisitivity and optical conductivity
could be understood by assuming that the electrons in a metal could be
thought of as a gas of non-interacting fermions, i.e., in terms of
quantum mechanical particles which do not have any direct interaction
but which do obey Fermi statistics. Meanwhile Pauli calculated that
the paramagnetic susceptibility of non-interacting electrons is
independent of temperature, also in accord with experiments in metals.
At the same time it was understood, at least since the work of Bloch
and Wigner, that the interaction energies of the electrons in the
metallic range of densities are not small compared to the kinetic
energy. The rationalization for the qualitative success of the
non-interacting model was provided in a masterly pair of papers by
Landau \cite{landau1,landau2} who  initially was concerned with the
properties of liquid
$^3He$. This work introduced a new way of thinking about the
properties of interacting systems which is a cornerstone of our
understanding of condensed matter physics. The notion of
quasiparticles and elementary excitations and the methodology of
asking useful questions about the low-energy excitations of the system
based on concepts of symmetry, without worrying about the myriad
unnecessary details, is epitomized in Landau's phenomenlogical theory
of Fermi-liquids. The microscopic derivation of the theory was also
soon developed.

Our perspective on Fermi-liquids has changed significantly in the last
two decades or so. This  is due both to changes in our
theoretical perspective, and due to the experimental developments: on
the experimental side, new materials have been found which exhibit
Fermi-liquid behavior in the temperature dependence of their low
temperature properties with the coefficients often a factor of order
$10^3$ different from the non-interacting electron values. These observations
dramatically illustrate the power and range of validity of the
Fermi-liquid ideas.
On the other hand, new materials have been discovered whose properties
are qualitatively different from the predictions of Fermi-liquid
theory (FLT). The most prominently discussed of these materials are the
normal phase of high-temperature superconducting materials for a range
of compositions near their highest $T_c$.  Almost every idea discussed
in this review has been used to understand the high-$T_c$ problem, but
there is no consensus yet on the solution.

\begin{figure}
\begin{center}   
  \epsfig{figure=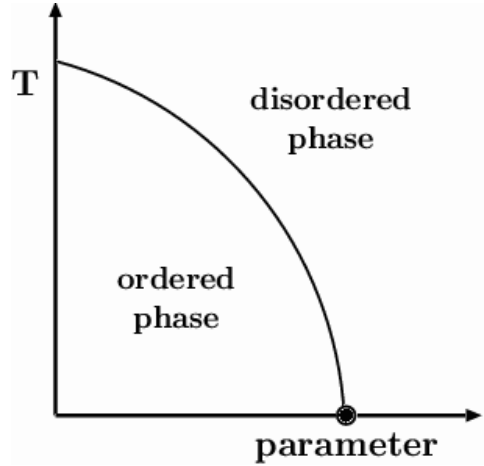,width=0.45\linewidth}
\end{center}
\caption[]{Schematic phase diagram near a Quantum Critical Point.
  The parameter along the $x$-axis can be quite general, like the pressure or a ratio of coupling constants.  Whenever the
  critical temperature vanishes, a QCP, indicated with a dot
  in the  figure, is encountered. In the vicinity of such a point quantum mechanical, zero-point fluctuations become
  very important.  However when $T_{c}$ is finite, critical slowing
down implies that the relevant frequency scale goes as $\omega \sim
  |T_{c}-T|^{\nu z}$, dwarfing quantum effects; the standard classical
  critical methodology then applies. An example of a phase diagram of
this type for $MnSi$ is shown in Fig. \ref{datamnsi1} below.}
\label{figphasediagram}
\end{figure}

It has of course been known for a long time that FLT breaks down in
the fluctuation regime of classical phase transitions. This breakdown
happens in a more substantial region of the phase diagram around the
quantum critical point (QCP) where the transition temperature tends to
zero as a function of some parameter, see
Fig. \ref{figphasediagram}. This phenomenon has been
extensively investigated for a wide variety of magnetic transitions in
metals where the transition temperature can be tuned through
application of pressure or by varying the electronic density through
alloying. Heavy fermion with their close competition between
states of magnetic order with localized moments and itinerant states
due to Kondo-effects appear particularly prone to such QCP's. Equally
interesting are
 questions having to do with
the change in properties due to impurities in systems which are near
a QCP in the pure limit.

The density-density correlations of itinerant disordered electrons
at long wavelengths and low energies must have a diffusive form.
In two-dimensions this leads to logarithmic singularities in the
effective interactions
when the interactions are
treated perturbatively. The problem of finding the ground state
and low-lying excitations in this situation is unsolved.
 On the
experimental side the discovery of the metal-insulator transition in
two dimensions and the unusual properties observed in the metallic
state makes this an important problem to resolve.

The one-dimensional electron gas reveals logarithmic singularities
in the effective interactions even in a second-order perturbation
calculation. A variety of mathematical techniques have been used to
solve a whole class of interacting one-dimensional problems and
one now knows the essentials of the correlation functions even in the
most general case. An important issue is whether and how this knowledge
can be used in higher dimensions.

The solution of the Kondo problem and the realization that its low
temperature properties may be discussed in the language of FLT has led
in turn to the formulation and solution of impurity models with singular
low energy properties. Such models have a QCP for a
particular relation between the coupling constants; in
some examples they exhibit a quantum critical line. The thermodynamic
and transport
properties around such critical points or lines are those of local
singular Fermi-liquids. Although the direct experimental relevance of
such models (as of one-dimensional models) to experiments is often
 questionable, these models, being
soluble, can be quite instructive in helping to understand the
conditions necessary for the breakdown of FLT and associated
quasiparticle concepts. The knowledge from zero-dimensional
 and one-dimensional
problems must nevertheless be used with care.

A problem which we do not discuss but which belongs in the study
of SFL's is the Quantum Hall Effect problem. The massive degeneracies of
 two-dimen\-si\-o\-nal electrons in a magnetic-field leads to spectacular
 new properties and involves new fractional quantum numbers. The
essentials of this problem were solved following
 Laughlin's inspired variational calculation. The principal reason
 for the omission is firstly that excellent papers reviewing the developments
 are available \cite{prangebook,dassarmabook,girvin}
  and secondly that the methodology used in this problem is
 in general distinct from those for discussing the other SFL's which have
a certain unity. We will however have occasions to refer to aspects
  of the quantum Hall effect problem often. Especially interesting from our point of view is
  the weakly singular Fermi-liquid behavior predicted in the
  $\nu=1/2$ Quantum Hall Effect \cite{halperinleeread}.
 
One of the aspects that we want to bring to the foreground in this
review is the fact that SFL's all have in common some fundamental
features which can be stated usefully in several different ways.  {\em
  (i)} they have degenerate ground states to within an energy of order
$k_B T$. This degeneracy is not due to static external potentials or
constraints as in, for example the spin-glass problem, but
degeneracies which are dynamically generated.{\em (ii)} Such
degeneracies inevitably lead to a breakdown of perturbative
calculations because they generate infra-red singularities in the
correlation functions. {\em (iii)} If a bare particle or hole is added
to the system, it is attended by a divergent number of low energy
particle-hole pairs, so that the one-to-one correspondence between the
one-particle excitation of the interacting problem and those of the
non-interacting problem, which is the basis for FLT, breaks down.

 On the theoretical side, one may now view Fermi-liquid theory as a
forerunner of the Renormalisation Group ideas. The renormalisation
group has led to a sophisticated understanding of singularities in the
collective behavior of many-particle systems. Therefore these methods
have an important role to play in understanding the breakdown of FLT.

The aim of this paper is to provide a pedagogical introduction to
SFL's, focused on the essential conceptual ideas and on issues which
are settled and which can be expected to survive future
developments. Therefore, we will
not attempt to give an exhaustive review of the literature on this
problem or of all the experimental systems which show hints of SFL
behavior. The experimental examples we discuss have been
selected to illustrate both what is essentially understood and what is
not
understood even in principle.
On the theoretical side, we will shy away from presenting in depth the
sophisticated methods necessary for a detailed evaluation of
correlation functions near QCP --- for this we refer to the book by
Sachdev  \cite{sachdev} --- or for an exact solution of local impurity
models (see, e.g.,  \cite{hewson,multichannelreview,tsvelik}). Likewise, for a
discussion of the application of quantum critical scaling ideas to
Josephson arrays or quantum Hall effects, we refer to the nice
introduction by Sondhi {\em et al.}  \cite{sondhi}.

\subsection{Outline of the paper} The outline of this paper is as
follows. We start by summarizing in
section \ref{landausfermiliquid} some of the key features of Landau's
FLT --- in doing so, we will not attempt to retrace all of the
ingredients which can be found in many
of the classical textbooks  \cite{pn,bp}; instead our
discussion will be focused on those elements of the theory and the
relation with its microscopic derivation that allow us to understand
the possible routes in which the FLT can break down. This is followed
in section \ref{localfermil} by the Fermi-liquid formulation of the
Kondo problem and of the SFL variants of the Kondo problem and of
two-interacting Kondo impurities. The intention here is
to reinforce the concepts of FLT in a different context as well as to provide
examples of SFL behavior which offer important insights because they
are both simple and solvable.
We then discuss the problem of one spatial dimension ($d=1$), presenting the
principal features of the solutions obtained. We discuss
 why $d=1$ is special, and the problems encountered in
extending the methods and the physics to $d>1$. We move then from
the comforts of
solvable models to the reality of the
discussion of possible mechanisms for
SFL behavior in higher dimensions. First we analyze in section
\ref{sflfromgauge} the paradigmatic case of long
range interactions. Coulomb interactions will not do in this regard,
since they are always screened in a metal, but transverse
electromagnetic fields do give rise to long-range interactions. The
fact that as a result no metal is a Fermi-liquid for sufficiently low
temperatures was already realized long ago  \cite{holstein} --- from a
practical point of view this mechanism is not very relevant, since the
temperatures where these effects  become important are of order
$10^{-16} $ Kelvin; nevertheless, conceptually this is important since it is a
simple example of a gauge theory giving rise to SFL behavior. Gauge theories
on lattices
have been introduced to discuss problems of fermions moving with the constraint of
only zero or single occupation per site.  We then
discuss in section \ref{qcpsection} the properties near a quantum
critical point, taking first an
example in which the ferromagnetic transition temperature
goes to zero as a function of some externally chosen suitable parameter.
We refer in this section to several experiments in heavy fermion
compounds which are only partially understood or not understood even
in principle. We then turn to a discussion of the marginal Fermi-liquid
phenemenology for the SFL state of
copper-oxide High - $T_c$ materials and discuss the requirements
on a microscopic theory that the phenemenology imposes. A sketch
of a microscopic derivation of the phenemenology is also given.
We close the paper in section \ref{mitrans}  with
a discussion  of the metallic state in $d=2$ and the state of the
theory
 treating the diffusive
singularities in $d=2$ and its relation to the metal-insulator
transition.

\section{Landau's Fermi-liquid}\label{landausfermiliquid}

\subsection{Essentials  of Landau Fermi-liquids}

The basic idea underlying Landau's Fermi-liquid theory
 \cite{landau1,landau2,pn,bp} is that of {\em
  analyticity}, i.e. that states with the same symmetry can be
adiabatically connected. Simply put this means that whether or not we
can actually do the calculation we know that the eigenstates of the
full Hamiltonian of the same symmetry can be obtained perturbatively
from those of a simpler Hamiltonian. At the same time states of
different symmetry can not be obtained by ``continuation'' from the
same state. This suggests that given a hard problem which is
impossible to solve, we may guess a right simple problem. The low
energy and long wavelength excitations, as well as the correlation and the
response functions of the impossible problem bear a one-to-one
correspondence with the simpler problem in their analytic properties.
This leaves fixing only numerical values. These are to be determined
by parameters, the minimum number of which is fixed by the symmetries.
Experiments often provide intuition as to what the right simple
problem may be: for the interacting electrons, in the metallic range
of densities, it is the problem of kinetic energy of particles with
Fermi statistics (If one had started with the opposite limit, just
the potential energy alone, the starting ground state is the Wigner
crystal --- a bad place to start thinking about a metal!). If we start
with non-interacting fermions, and then turn on the interactions, the
qualitative behavior of the system does not change as long as the
system does not go through (or is close to) a phase transition.
 Because of the analyticity, we can even consider
strongly interacting systems --- the low energy excitations in these
 have strongly renormalized values of their parameters compared to
the non-interacting problem, but their qualitative behavior is the
same of that of the simpler problem.

\begin{figure}
\begin{center}   
\epsfig{figure=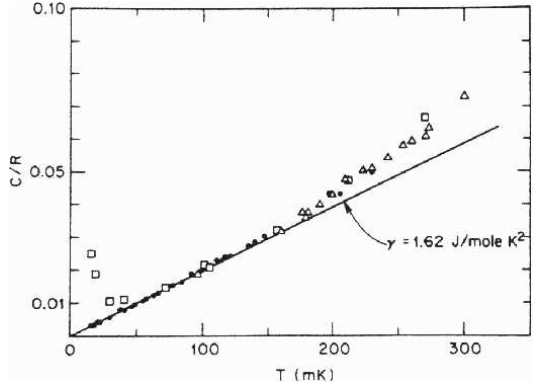,width=0.5\linewidth}
\end{center}
\caption[]{Specific heat of $CeAl_3$ at low temperatures from
Andres {\em et al.} \cite{andres}. The slope of the linear specific heat is
about 3000 times that of the linear specific heat of say $Cu$.  But
the high-temperature cut-off of this linear term is smaller than that
of $Cu$ by a similar amount. The rise of the specific heat in a
magnetic field at low temperatures is the nuclear contribution,
irrelevant to our discussion.} \label{fig1}
\end{figure}

\begin{figure}
\begin{center}   
\epsfig{figure=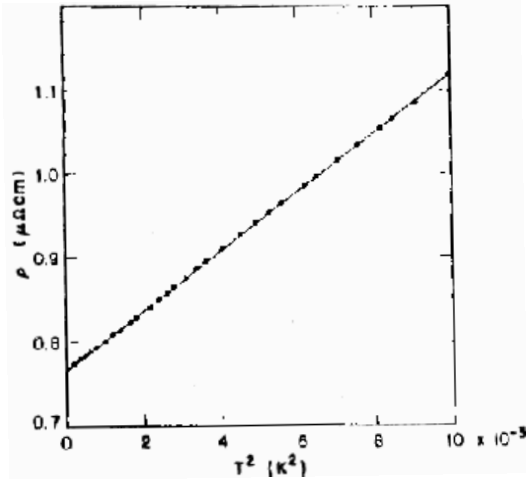,angle=0.8,width=0.5\linewidth}
\end{center}
\caption[]{Electrical resistivity of $CeAl_3$ below 100 $mK$, plotted
against $T^2$. From Andres {\em et al.} \cite{andres}.} \label{fig2}
\end{figure}

The heavy fermion problem provides an extreme example of the domain of
validity of the Landau approach. This is illustrated in Fig.
\ref{fig1}, which shows the specific heat of the heavy fermion
compound $CeAl_3$. As in the Sommerfeld model, the specific heat is
linear in the temperature at low $T$, but if we write $C_v \approx
\gamma T$ at low temperatures, the value of $\gamma$ is about a
thousand times as large as one would estimate from the density of
states of a typical metal, using the free electron mass. For
a Fermi gas the density of states $N(0)$ at the Fermi energy is
proportional to an effective mass $m^*$:
\begin{equation}
N(0)= {{m^* k_F}\over{\pi^2 \hbar^2}}~,
\label{DOSeqn.}
\end{equation}
with $k_F$ the Fermi wavenumber.  Then the fact that
the density of states at the chemical potential is a thousand times
 larger than for normal
metals can be expressed by the statement that
 the effective mass $m^*$ of the quasiparticles is a
thousand times larger than the free electron mass $m$.
Likewise, as
Fig. \ref{fig2} shows, the resistivity of $CeAl_3$ at low temperatures
increases as $T^2$. This also is a characteristic sign of a
Fermi-liquid, in which the quasiparticle lifetime
$\tau$ at the Fermi surface, determined by electron-electron interactions,
 behaves as $\tau \sim 1/T^2$.\footnote{In heavy fermions, at least in
the observed range of temperatures, the transport lifetime determining the
temperature dependence of resistivity is proportional to the single-particle
lifetime.} However, just as the prefactor $\gamma$ of the specific
heat is a factor thousand times larger than usual, the prefactor of
the $T^2$ term in the resistivity is a factor $10^6$ larger --- while
$\gamma$ scales linearly with the effective mass ratio $m^*/m$, the
prefactor of the $T^2$ term in the resistivity increases for this class of
Fermi-liquids as
$(m^*/m)^2$.

It should be remarked that the right simple problem is not always easy
to guess. The right simple problem for liquid $^4He$ is not the
non-interacting Bose gas but the weakly interacting Bose gas (i.e. the
Bogoliubov problem \cite{bogolubov,landaulifshitz}). The right simple
problem for the
Kondo problem (a low-temperature local Fermi liquid) was
 guessed  \cite{nozieres}
only after the numerical renormalization group solution was obtained by
Wilson  \cite{wilson}. The right simple
problem for two-dimensional interacting disordered electrons in the
''`metallic'' range of densities (section 8 in this paper) is at
present unknown.

For SFL's the problem is
different: usually one is in a regime of parameters where no simple
problem is a starting point --- in some cases the fluctuations between
solutions to different simple problems determines the physical
properties, while in others even this dubious anchor is lacking.

\subsection{Landau Fermi-liquid and the wave function renormalization $Z$}\label{landauwavefren}
Landau
theory is the forerunner of our modern way of thinking about
low-energy effective
 Hamiltonians in complicated problems and of the renormalisation group.
The formal statements of Landau theory in their original form are
often somewhat cryptic and mysterious --- this both reflects
Landau's style and his ingenuity. We shall take a more
pedestrian approach.
\begin{figure}
  \vspace{0.5cm}
\begin{center}   
  \epsfig{figure=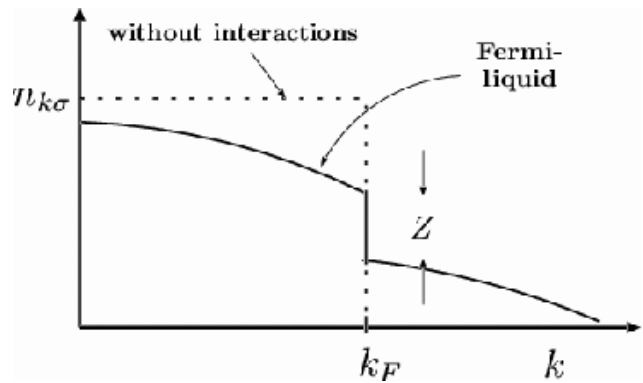,width=0.6\linewidth}
\end{center}
\vspace*{2mm}
\caption[]{Bare-particle distribution at $T=0$ for a given
spin-direction in a translationally invariant Fermi-system with
interactions (full line) and without interactions (dashed line). Note
that the position of the discontinuity, i.e. the Fermi wavenumber
$k_F$, is not renormalized by interactions. }\label{figzerotempocc}
\end{figure}

Let us consider the
essential difference between non-interacting fermions and an
interacting Fermi-liquid from a simple microscopic perspective.  For
free fermions, the momentum states $|{\bf k} \rangle $ are also
eigenstates with energy eigenvalue
\begin{equation} \label{1.2a}
\epsilon_{\bf k} = {{\hbar^2 k^2}\over{2m}}
\end{equation}
of the Hamiltonian. Moreover, the distribution of particles is given
by the Fermi-Dirac function for the thermal occupation $n^0_{{\bf
    k}\sigma}$, where $\sigma$ denotes the spin label.  At $T=0$ the
distribution jumps from 1 (all states occupied within the Fermi
sphere) to zero (no states occupied within the Fermi sphere) at$
|{\bf k}| =k_F$ and energy equal to the chemical potential $\mu$. This
is illustrated in Fig. \ref{figzerotempocc}.

\begin{figure}
\begin{center}   
  \epsfig{figure=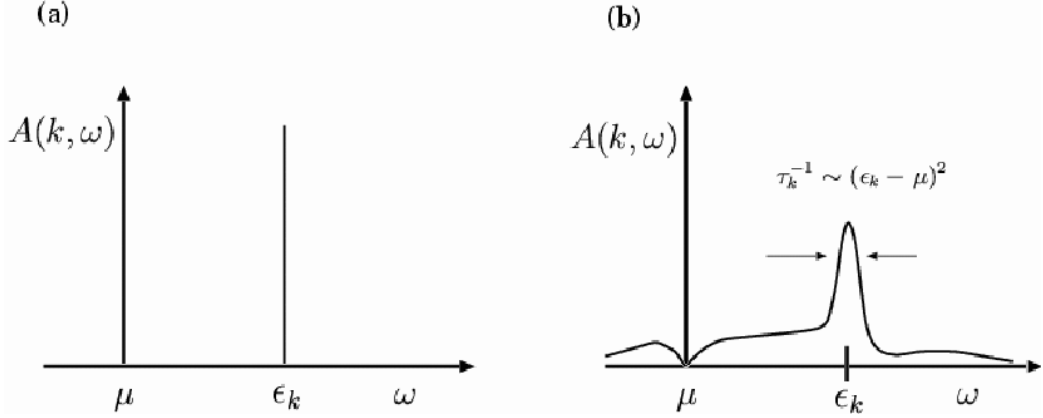,width=0.99\linewidth}
\end{center}
\caption[]{{\em (a)} The non-interacting spectral function
$A(k,\omega)$ at fixed
$k$ as a function of $\omega$. {\em (b)} The spectral function of
single-electron excitations in a Fermi-liquid at fixed $k$ as a
function of $\omega$. If $\frac{1}{\pi} A(k,\omega)$ is normalized to
1, signifying one bare particle, the weight under the Lorentzian,
i.e. the quasi-particle part, is $Z$. As explained in the text, at the
same time $Z$ is  the discontinuity in
Fig. \ref{figzerotempocc}.}  \label{fig4}\label{figspectraldensob}
\end{figure}

A good way to probe a system is to investigate the {\em spectral
  function}; the spectral function $A(k,\omega )$ gives the
distribution of energies $\omega$ in the system when a particle with
momentum ${\bf k}$ is added or removed from it (Remember that removing a particle
excitation below the Fermi energy means that we add a hole
excitation).  As
sketched in Fig. \ref{fig4}{\em (a)}, for the non-interacting system,
$A_0(\bk,\omega)$ is simply a delta-function peak at the energy
$\epsilon_{\bf k}$, because all momentum states are also energy
eigenstates:
\begin{eqnarray}
\label{1.3a}
A_0({\bf k},\omega) & = &  \delta ( \omega -(\epsilon_{\bf k} -\mu)) ~, ~~~\mbox{for}~ \omega>\mu~,\\
 & = & - {{1}\over{\pi}} \mbox{Im} {{1}\over{\omega -(\epsilon_{\bf k} -\mu) +
     i\delta}} = -{{1}\over{\pi}} \mbox{Im} G^0({\bf k},\omega) \label{1.4b}~.
\end{eqnarray}
Here $\delta$ is small and positive; it reflects that particles or holes
are introduced adiabatically, and is taken to zero at the end of the calculation
for the pure non-interacting problem.  The
first step of the second line is just a simple mathematical rewriting
of the delta function; in the second line the Green's function $G^0$ for
non-interacting electrons is introduced. More generally the
single-particle Green's function $G(\bk,\omega)$ is defined in terms of
the correlation function of particle creation and annihilation
operators in standard textbooks
\cite{nozieresbook,agd,rickayzen,mahan}. For our present
purposes, it is sufficient to note that it is related to the spectral
function $A(\bk,\omega)$, which has a clear physical meaning and which can be deduced
 through Angle Resolved Photoemission Experiments :
\begin{equation}
\label{1.4a}
G(\bk ,\omega)=\int_{-\infty}^{\infty} dx~
{{A(\bk,x)}\over{\omega-\mu-x+i\delta \: \mbox{sgn}(\omega-\mu)}}.
\label{speceqn.}
\end{equation}
$A(\bk,\omega)$ thus is the spectral representation of the complex
function $G(k,\omega)$. Here we have defined the so-called ${\it retarded}$
Green's function which is especially useful since its real and imaginary parts
obey the Kramers-Kronig relations. In the problem with interactions
$G(\bk,\omega)$ will differ from $G^0(\bk,\omega)$. This difference
can be quite generally defined through the single-particle self-energy
function $\Sigma(\bk,\omega)$:
\begin{equation}
\left( G(\bk,\omega)\right)^{-1} =  \left( G^0(\bk,\omega)\right)^{-1}
- \Sigma(\bk, \omega)~.
\label{sigmaeq}
\end{equation}
Eq. (\ref{speceqn.}) ensures the relation between $ G(\bk,\omega)$ and
$A(\bk,\omega)$
\begin{equation}
A(\bk, \omega) = - {{1}\over{\pi}} \mbox{Im} \; G(\bk,\omega)~.
\end{equation}

With these preliminaries out of the way, let us consider the form of
$A(\bk,\omega)$ when we add a particle to an interacting system of
fermions.

\begin{figure}
\begin{center}   
  \epsfig{figure=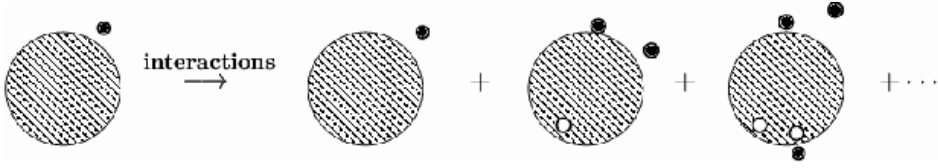,width=0.9\linewidth}
\end{center}
\caption[]{Schematic illustration of the perturbative expansion
(\ref{1.4}) of the change of wave function as a result of the
addition of an electron to the
Fermi sea due to   interactions with the particles in the Fermi sea.}\label{fig6}\label{figperturbcalc}
\end{figure}

Due to the interaction (assumed repulsive) between the added particle
and those already in the Fermi-sea, the added particle will kick
particles from below the Fermi-surface to above.  The possible terms
in a perturbative description of this process are constrained by the
conservation laws of charge, particle number, momentum and spin.
Those which are allowed by the aforementioned conservation laws are
indicated pictorially in Fig. \ref{figperturbcalc}, and lead to an expression of
the type
\begin{eqnarray}
|\psi^{N+1}_{\bk \sigma} \rangle & = &  Z_{\bk}^{1/2} c_{\bk \sigma}^\dagger |
\psi^N\rangle  \nonumber
  + {{1}\over{V^{3/2}} }
\sum_{ {\bk_1,\bk_2,\bk_3} } \sum_{\sigma_1, \sigma_2,
      \sigma_3} \alpha_{\bk_1\sigma_1 \bk_2\sigma_2
      \bk_3\sigma_3}
c^\dagger_{\bk_3}c_{\bk_2}c^\dagger_{\bk_1}~ \times \\
& & \hspace*{3cm} \times~ \delta_{\bk,\bk_1-\bk_2+\bk_3}\delta(\sigma; \sigma_1,\sigma_2,\sigma_3)
|\psi^N\rangle + \cdots~.  \label{1.4}
\end{eqnarray}
Here the $c^\dagger_\bk$'s and $c_\bk$'s are the {\em bare} particle
creation and annihilation operators, and the dots indicate higher
order terms, for which two or more particle-hole pairs are created and
$\delta(\sigma; \sigma_1,\sigma_2,\sigma_3)$ expresses conservation of
spin under vector addition. The multiple-particle-hole pairs for a
fixed total momentum can be created with a continuum of momentums of the
individual bare particles and holes. Therefore an added particle with fixed
total momentum has a wide distribution of energies. However if $Z_\bk$
defined by Eqn. (\ref{1.4}) is finite, there will be a well-defined
feature in this distribution at some energy which is in general
different from the non-interacting value $\hbar^2 k^2/(2m)$. The
spectral function in such a case will then be as illustrated in Fig.
\ref{fig4}. It is useful to separate the well-defined feature from
the broad continuum by writing the spectral function as the sum of two
terms, $A(\bk,\omega) = A_{coh} (\bk,\omega) + A_{incoh}(\bk,\omega)$.
The single-particle Green's function can be expressed as
a sum of two corresponding terms, $G(\bk,\omega)
= G_{coh} (\bk,\omega) + G_{incoh}(\bk,\omega)$. Then
\begin{equation}
\label{1.8}
G_{coh}(\bk,\omega) = {{Z_\bk}\over{\omega- \tilde{\epsilon}_\bk + i/
    \tau_\bk}}~,
\end{equation}
which for large lifetimes $\tau_\bk$ gives a Lorentzian peak in
the spectral density at the quasiparticle energy
$\tilde{\epsilon}_\bk\equiv \epsilon_\bk -\mu$.  The incoherent
Green's function is smooth and hence for large $\tau_\bk$ corresponds to
the smooth background in the spectral density.

The condition for the occurrence of the well-defined feature
can be expressed as the condition on the self-energy $\Sigma({\bf k},\omega)$
that it
has an analytic expansion about $\omega=0$ and $\bk=\bk_F$ and that its
real part be much larger than its imaginary part. One can easily see that were it
not so the expression (\ref{1.8}) for $G_{coh}$ could not be obtained. These conditions are
necessary for a Landau Fermi-liquid.  Upon expanding
$\Sigma(\bk,\omega)$ in (\ref{1.6}) for small $\omega$ and small
deviations of $\bk$ from $\bk_F$ and writing it in
the form (\ref{1.8}), we make the identifications
\begin{equation} \label{1.10}
\tilde{\epsilon}_\bk = \epsilon_\bk Z_\bk \hat{Z}_\bk ~,~~~~~~~~
{{1}\over{\tau_\bk}} = Z_\bk\:\mbox{Im}  \Sigma (\bk_F,\omega=0)~,
\end{equation}
where
\begin{equation} \label{1.11}
Z_\bk = \left( 1- {{\partial \Sigma}\over{\partial
      \omega}}\right)_{\omega=0,k=k_F}^{-1}~,~~~~~~~~~~
\hat{Z} = \left( 1+ {{1}\over{v_F}} {{\partial \Sigma}\over{\partial
      k}}\right)_{\omega=0,k=k_F}~.
\end{equation}

From Eq. (\ref{1.4}) we have a more physical definition of $Z_\bk$:
$Z_\bk$ is the projection amplitude of $|\psi_{\bk}^{N+1}\rangle$ onto the
state with one bare particle added to the ground state, since all
other terms in the expansion vanish in the thermodynamic limit in the
perturbative expression embodied by (\ref{1.4}),
 \begin{equation}
 Z_\bk^{1/2} = \langle \psi_\bk^{N+1}| c^\dagger_\bk
  |\psi^N\rangle~.\label{1.6}
\end{equation}
In other words, $Z_\bk$ is the overlap of the ground state wavefunction
of a system of
interacting $N\pm 1$ fermions of total momentum $\bk$ with the wave
function of $N$ interacting particles and a bare-particle of momentum
$\bk$. $Z_\bk$ is called the quasiparticle amplitude.

The Landau theory tacitly assumes that $Z_\bk$ is finite. Furthermore it
asserts that for small $\omega$ and $\bk$ close to $\bk_F$, the physical
properties can be calculated from quasiparticles which carry the same
quantum numbers as the particles, i.e. charge, spin and momentum and
which may be defined simply by the creation operator
$\gamma^\dagger_{\bk,\sigma}$:
\begin{equation}
\label{1.5}
|\psi^{N+1}_\bk \rangle =  \gamma^\dagger_\bk |\psi^N \rangle  ~.
\end{equation}
Close to $\bk_F$, and for $T$ small compared to the Fermi-energy, the
distribution of the quasiparticles {\em is assumed to be the
  Fermi-Dirac distribution} in terms of the renormalized quasiparticle
energies. The bare particle distribution is quite
different. As is illustrated in Fig.\ \ref{figzerotempocc}, it is
depleted below $\bk_F$ and augmented above $\bk_F$, with a discontinuity
at $T=0$ whose value is shown in microscopic theory to be $Z_\bk$.
A central result of
Fermi-liquid theory is that close to the Fermi energy at zero
temperature, the width $1/\tau_\bk$ of the coherent quasiparticle peak is
proportional to $(\epsilon_\bk -\mu)^2$ so that near the Fermi
energy the lifetime is long and quasiparticles are well-defined.
Likewise, at the Fermi energy  $1/\tau_\bk $ varies with temperature as
$T^2$. From the microscopic derivation of this result, it follows that
the weight in this peak, $Z_\bk$, becomes equal to the jump $Z$ in
$n_{\bk\sigma}$ when we approach the Fermi surface: $Z_\bk
{\rightarrow Z}$ for ${\bk\rightarrow \bk_F}$.  For heavy fermions,
as we already mentioned,
$Z$ can be of the order of $10^{-3}$. But
as long as $Z$ is nonzero, one has Fermi-liquid properties for
temperatures lower than about $ZE_F$. Degeneracy is effectively
lost for temperatures much higher than
$ZE_F$ and classical statistical  mechanics prevails. \footnote{It
is an unfortunate common mistake to think of the properties in
this regime as SFL behavior.}
 
An additional result from microscopic theory is the so-called
Luttinger theorem, which states that the volume enclosed by the
Fermi-surface does not change due to interactions  \cite{nozieresbook,agd}.
The mathematics behind this theorem is that with the assumptions of FLT,
the number of poles in the interacting Green's function below the chemical
potential is the same as that for the non-interacting Green's function.
Recall that the latter is just the number of particles in the system.

Landau actually started his discussion of the Fermi-Liquid by writing the
equation for the deviation of the (Gibbs) free-energy from its
ground state value as a functional of the deviation of the
quasiparticle distribution function $n(\bk,\sigma)$ from the
equilibrium distribution function $n_0(\bk,\sigma)$
\begin{equation}
\delta n(\bk,\sigma)= n(\bk,\sigma)-n_0(\bk,\sigma),
\end{equation}
as follows:

\begin{equation}
G= G_0 + {{1}\over{V}} \sum_{\bk,\sigma} (\epsilon_\bk-\mu) \delta n_{\bk\sigma} +
{{1}\over{2V^2}} \sum_{\bk \bk', \sigma \sigma'} f_{\bk\bk',\sigma \sigma'}
\delta n_{\bk\sigma} \delta n_{\bk' \sigma'}+\cdots \label{energy}
\end{equation}
 Note that $(\epsilon_\bk-\mu)$
is itself a function of $\delta n$; so the first term contains at least
a contribution of order $(\delta n)^2$ which makes the second term quite necessary.
 In principle, the unknown
function $f_{\bk \bk', \sigma \sigma'}$ depends on spin and momenta.
However, spin rotation invariance allows one to write the spin part in
terms of two quantities, the symmetric and antisymmetric parts $f^s$
and $f^a$. Moreover, for low-energy and long-wavelength
phenomena only momenta with $\bk\approx \bk_F$
play a role; if we consider the simple case of $^3He$ where the
Fermi surface is spherical, rotation invariance implies that for
momenta near the Fermi momentum $f$ can only depend on the relative
angle between $\bk$ and $\bk'$; this allows one to expand in Legendre
polynomials $P_l(x)$ by writing
\begin{equation}
N(0) f^{s,a}_{\bk \bk',\sigma \sigma'} \stackrel{\bk\approx \bk'\approx
  \bk_F}{\longrightarrow} \sum_{l=0}^\infty F_l^{s,a} P_l (\hat{\bk}\cdot
\hat{\bk}')~.
\end{equation}
From the expression (\ref{energy}) one can then relate the
lowest order so-called Landau coefficients $F_0$ and $F_1^s$ and the
effective mass $m^*$ 
to thermodynamic quantities like the specific
heat $C_v$, the compressibility $\kappa   $, and the susceptibility $\chi$:
\begin{equation}
{{C_v}\over{C_{v0}}} = {{m^*}\over{m}}~, ~~~~~
{{\kappa}\over{\kappa_0}} = (1+F^s_0) {{m^*}\over{m}}~,~~~~~
{{\chi}\over{\chi_0}} =  (1+F^a_0) {{m^*}\over{m}}~.
\end{equation}
Here subscripts 0 refer to the quantities of the non-interacting
reference system, and $m$ is the mass of the fermions.  For a Galilean
invariant system (like $^3He$), there is a  a simple relation between
the mass enhancement and the Landau parameter $F^s_1$, and there is no
renormalization of the particle current ${\bf j}$; however,
there {\em is} a renormalization of the velocity: one has
\begin{equation}
{\bf j}= \bk/m ~, ~~~~~~~{\bf v}=\bk/m^*~,~~~~~~~{{m^*}\over{m}} =  \left( 1+
  {{F^s_1}\over{3}}\right)~.\label{jvm*}
\end{equation}
 The transport properties are calculated by defining a distribution
function  $\delta n
(\bk\sigma;r,t)$ which is slowly varying in space and time
and writing a Boltzmann equation for it  \cite{pn,bp}.

It is a delightful conceit of the Landau theory that the expressions
of the low-energy properties in terms of the quasiparticles  in no
place involve the quasiparticle amplitude $Z_\bk$. In fact in a
translationally invariant problem as liquid $^3He$, $Z_\bk$ cannot be
measured by any thermodynamic or transport measurements . A masterly use of
conservation laws ensures that $Z$'s
cancel out in all physical properties (One can
extract $Z$ from measurement of the momentum distribution.
By neutron scattering measurements,
it is found that $Z \approx 1/4$
\cite{glyde} for $He^3$ near the melting line). This is no
longer true on a lattice, in the electron-phonon interaction
problem \cite{prange} or in heavy fermions \cite{varma3}  or even
 more generally in any situation where the interacting
problem contains more than one type of particle with different
characteristic frequency scales.

\subsection{Understanding microscopically why Fermi-liquid Theory works}
\label{whyworks}

Let us try to understand from a
more microscopic approach why the Landau theory works so well. We present a
qualitative discussion in this subsection and outline the principal features
of the formal derivation in the next subsection.

As we already
remarked, a crucial element in the approach is to choose the proper
non-interacting reference system. That this is possible at all is due to
the fact that the number of states to which an added particle can scatter
due to interactions is severely limited due to the Pauli principle. As a result
non-interacting
fermions are a good stable system to perturb about: they have
a finite compressibility and susceptibility in the ground state, and
so collective modes and thermodynamic quantities change smoothly when
the interactions are turned on. This is not true for non-interacting bosons
which do not support collective modes like sound waves. So one cannot
 perturb about the  non-interacting
bosons as a  reference system.

Landau also laid the foundations for the formal justification of Fermi
Liquid theory in two and three dimensions.  The flurry of activity in
this field following the discovery of high-$T_c$ phenomena has led to
new ways of justifying Fermi-liquid theory (and understanding why
the one-dimensional problem is different). But the principal physical
reason, which we now discuss, remains
 the phase space restrictions due to kinematical constraints.
 
We learned in section \ref{landauwavefren} that to be able to define quasiparticles,
it was necessary to have a finite $Z_{\bk_F}$ and that this in turn
needed a self-energy function $\Sigma({\bk_F,\omega})$ which is smooth near
the chemical potential, i.e. at $\omega=0$. Let us first see
why a Fermi gas has such properties when interactions are calculated
perturbatively.

\begin{figure}
\begin{center}   
  \epsfig{figure=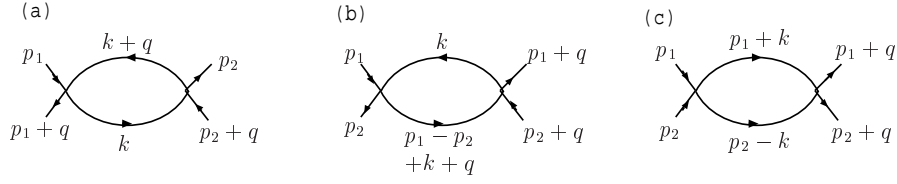,width=0.85\linewidth}
\end{center}
\caption[]{The three second-order processes in a perturbative
calculation of the correction to the bare interaction in a
Fermi-liquid}\label{2ndorderpt}
\end{figure}

In
Fig. \ref{2ndorderpt} we show the three possible processes that arise
in second order perturbation theory for the scattering of two
particles with fixed initial energy $\omega$ and momentum $q$.
Note that in two of the diagrams, Fig. \ref{2ndorderpt}{\em (a)} and \ref{2ndorderpt}{\em (b)} the
 intermediate state has a
particle and a hole while the
intermediate state
in diagram \ref{2ndorderpt}{\em (c)} has a pair of particles.

\begin{figure}
\begin{center}   
  \epsfig{figure=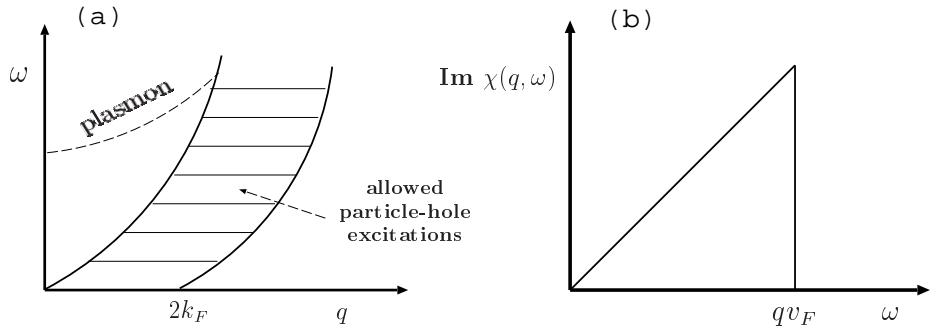,width=0.9\linewidth}
\end{center}
\caption[]{{\em (a)} Restriction on allowed particle-hole excitations
in a Fermi-sea due to kinematics. The plasmon mode has been drawn for
the case $d=3$. {\em (b)} The absorptive part of the
particle-hole susceptibility (in the charge, current and spin
channels) for $\omega <q v_F$ in the Fermi gas.}\label{spectrum}
\end{figure}

We will find that, for our present purpose, the contribution of
diagram \ref{2ndorderpt}{\em (a)}
 is more important than the other two. It gives a contribution
\begin{equation}
\label{diagram(a)}
g^2 \sum_\bk {{f_{\bf k+q}-f_\bk}\over{\omega-(E_{\bf k+q}-E_\bk)+i \delta }}~.
\end{equation}
 Here $g $ is a measure of the strength of the
scattering potential (the vertex in the diagram) in the limit of small
$\bq  $. The denominator
ensures that the largest contribution to the scattering comes from
small scattering momenta $\bq$: for these the energy difference is
linear in $\bq$, $ E_{\bk+\bq} -E_\bk \approx {\bf q}\cdot {\bf v_\bk}$, where
${\bf v_\bk}$ is a vector of length $v_F$ in the direction
 of ${\bf
k}$. Moreover, the term in the numerator is nonzero only
in the area contained between two circles (for $d=2$) or spheres (for $d=3$)
with their centers
displaced by $\bq$ --- here is the phase space restriction
due to the Pauli principle!
This area is also proportional to ${\bf q}\cdot {\bf v_k}$, and
so in the small $\bq$ approximation we get
from diagram \ref{2ndorderpt}{\em (a) } a term proportional to
\begin{equation}
g^2 {{{\bf q} \cdot {\bf v_\bk}}\over{\omega - {\bf q}\cdot {\bf v_\bk} +
i \delta}} ~{{df}\over{d\epsilon_{\bf k}}}~.
\end{equation}
Now we see why we diagram \ref{2ndorderpt}{\em (a)} is
special. There is a singularity
at $\omega={\bf q \cdot v_k}$ and its value for small $\omega$ and
$\bq$ depends on which of the two is smaller. This singularity is
responsible for the low energy-long wavelength collective modes of the
Fermi liquid in Landau theory. At low temperatures,
$df/d \epsilon_{\bf k}= - \delta(\epsilon_{\bf k} - \mu) $, so the summation is {\em restricted} to the Fermi surface.
The real part of (\ref{diagram(a)}) therefore vanishes in the
limit $qv_F/\omega \rightarrow 0$, while
it
approaches a {\em finite} limit for $\omega \rightarrow 0$. The
imaginary part in this limit is proportional to\footnote{This behavior implies that this scattering contribution
is a marginal term in the renormalization group sense, which means
that it affects the numerical factors, but not the qualitative
behavior.} $\omega$,
\begin{equation}
 Im \chi (\bq,\omega) =  g^2 \pi N(0) {{\omega}\over{ q
v_F}}~~~~\mbox{for}~\omega < q v_F~,
\end{equation}
while $Im \chi(\bq,\omega)= 0 $ for $\omega>v_F q$.
This behavior is sketched in Fig. \ref{spectrum}{\em (b)}.
Explicit evaluation
yields for the real part
\begin{equation}
Re \chi(\bq,\omega) = g^2 N(0) \left[ 1+ \frac{\omega}{qv_F} \ln \left|
\frac{\omega -qv_F}{\omega+qv_F}\right|\: \right]~,  \label{chiformula}
\end{equation}
which gives a constant (leading to a finite compressibility and spin susceptibility)
at $\omega$ small compared to $qv_F$.
For diagram \ref{2ndorderpt}{\em (b)}, we get a term
 $\omega -(E_{\bf{p_1-p_2+k+q}} - E_\bk)$
 in the
denominator. This term is always finite for  general momenta ${\bf p_1}$ and
${\bf p}_2$, and hence the contribution from this diagram can always be
neglected relative to the one from \ref{2ndorderpt}{\em (a)}. Along
similar lines, one
finds that diagram \ref{2ndorderpt}{\em (c)}, which describes
scattering in the
particle-particle channel, is irrelevant except when
${\bf p}_1=-{\bf p}_2$, when it diverges as
$\ln \omega$.

Of course, this scattering process is the one which gives
superconductivity. Landau noticed this singularity but ignored
its implication\footnote{Attractive interactions in any angular
momentum channel
(leading to superconductivity) are therefore {\em relevant}
operators.}.
Indeed, as long as the effective interactions do not favor
superconductivity or as long as we are at temperatures much
higher than the superconducting transition
temperature, it is not important for
Fermi-liquid theory.

Let us now look further at the absorptive spectrum of particle-hole
excitations in two and three
dimensions, i.e., we examine the imaginary part of Eq.\ (\ref{diagram(a)}).
 When the total energy $\omega$ of the pair is small, both
the particle and the hole have to live close to the Fermi surface. In
this limit, we can make any excitation with momentum $q\le 2
k_F$. For fixed but small values of $q$, the maximum excitation
energy is $\omega \approx q v_F$; this occurs when ${\bf q}$ is in the
same direction as the main momentum ${\bf k}$ of each
quasiparticle. For $q$ near $2k_F$ the maximum possible energy is
 $\omega=v_F|q-2k_F|$. Combining these results, we get the sketch
in Fig. \ref{spectrum}{\em (a)}, in which the shaded area in the
$\omega$-$q$ space is the region of allowed
 particle-hole excitations\footnote{
In the presence of  long-range Coulomb interactions one gets in
addition to the particle-hole excitation spectrum a collective mode
with a finite
plasma frequency as $q \to 0$ in $d=3$ and a $\omega \sim \sqrt{q}$ behavior in
$d=2$.}.
From this spectrum one can calculate the polarizability,
or the magnetic
susceptibility.

The  behavior sketched above is valid generally in two and three
dimensions (but as we will see in section 4, not in one dimension).
The important point to remember is that the density of particle-hole
excitations decreases linearly with $\omega$ for $\omega$ small
compared to $qv_F$.
We shall see later that one way to undo Fermi-liquid theory is
 to have $\omega$ vary as $k^2$ in two dimensions or $\omega\sim
k^3$ in three dimensions.

\begin{figure}
\begin{center}
  \epsfig{figure=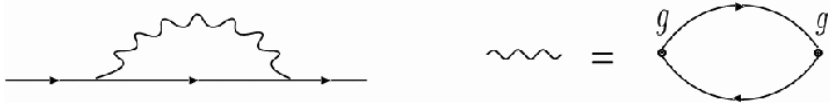,width=0.8\linewidth}
\end{center}
\caption[]{The single-particle self-energy diagram in second order.}\label{selfenergy2ndorder}
\end{figure}

We can now use $\mbox{Im}\;\chi({\bf q},\omega)$ to calculate the
single-particle self-energy to second order in the interactions.  This
is shown in Fig. \ref{selfenergy2ndorder} where the wiggly line
denotes $\chi ({\bf q},\nu)$ which in the present approximation is just
given by the diagram of Fig. \ref{2ndorderpt}{\em (a)}.

For the
perturbative evaluation of this process, the intermediate particle
with energy-momentum $(\omega + \nu)$, $(k+q)$ is a free
particle. Second order perturbation theory then yields an imaginary
part or decay rate
\begin{equation}
\mbox{Im}\, \Sigma (\bk,\omega) =  {{1}\over{\tau(\bk,\omega)}}  =  g^2 N(0)
\left( {{\omega}\over{E_F}}\right)^2 ~,\label{imsigma}
\end{equation}
in three dimensions for $\bk \approx \bk_F$.
In two dimensions, the same process yields $\mbox{Im}\, \Sigma(\bk_F,\omega) \sim \omega^2
\ln (E_F/\omega )$.

The $\omega^2$ decay rate is intimately related to the analytic result
(\ref{chiformula}) for $\mbox{Im}~\chi (q,\omega) $ exhibited in Fig.
(\ref{spectrum}). As may be found in text books, the same calculation
for electron-phonon interactions or for interaction with spin waves in an antiferromagnetic
metal gives  $\mbox{Im}\, \Sigma(\bk_F,\omega) \sim (\omega/\omega_c)^3$,
with $\omega_c $ is the phonon Debye
frequency in the former and  the characteristic zone-boundary
spin-wave frequency
in the latter.

The real part of the self-energy may be obtained directly or by
Kramers--Kronig transformation of (\ref{imsigma}). It is proportional
to $\omega$. Therefore, if the quasiparticle amplitude $Z_{k_F}$ is
evaluated perturbatively\footnote{This quantity has been precisely evaluated by Galitski \cite{galitski}
 for the model of a dilute Fermi gas characterized by a scattering length.}
\begin{equation}
Z_{k_F} \approx 1 - 2g^2 N(0)/E_F~.
\end{equation}
Thus in a perturbative calculation of the effect of interactions
the basic analytic structure of the Green's function is
 left the same as for non-interacting fermions. The general proof of
  the validity of Landau theory
consists in showing that what we have seen in second order in $g$
remains true to all orders in $g$. The original proofs  \cite{agd} are
self-consistency arguments --- we will consider them briefly in
section \ref{landauconsist}. They assume a finite $Z$ in exact
single-particle Green's functions and show effectively that to any
order of perturbation, the polarizability functions retain the analytic
structure of the non-interacting theory, which in turn ensures a finite
$Z$.

\begin{figure}
\begin{center}
  \epsfig{figure=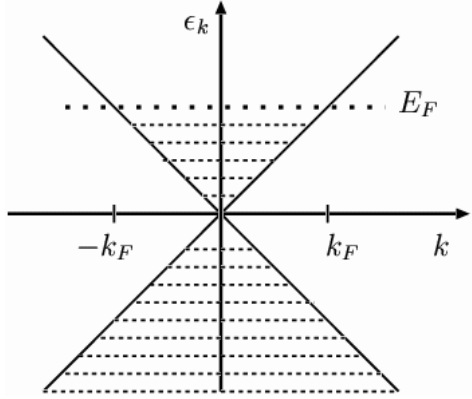,width=0.45\linewidth}
\end{center}
\caption[]{Single-particle energy $\epsilon_k$  in one dimension, in
the approximation that the dispersion relation is linearized about
$k_F$. Note the Fermi surface
consists of just two points. The spectrum of particle-hole excitations
is given by
  $\omega(q)= \epsilon(k+q) -\epsilon(k) = k_F q /m$. Low-energy
particle-hole excitations are only possible for  $q$ small or for $q$ near
$2k_F$.}\label{spectrum1d}
\end{figure}

\begin{figure}
\begin{center}
  \epsfig{figure=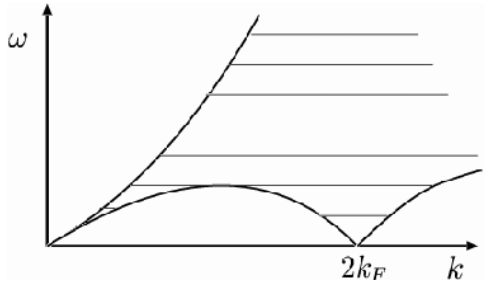,width=0.5\linewidth}
\end{center}
\caption[]{Phase space for particle-hole excitation spectrum in one
dimension. Compare with the same in higher dimensions,
Fig. \ref{spectrum}. For linearized single-particle kinetic energy
$\epsilon_k=\pm v_F
(k-k_F)$, particle-hole excitations are only possible  on lines going
through $k=0$ and $k=2k_F$.}\label{phasespace1d}
\end{figure}

In one dimension, phase space restrictions on the possible excitations
are crucially different\footnote{ It might appear surprising that they are not different
in any essential way between higher dimensions.}, since the Fermi surface consists of just
two points in the one-dimensional space of momenta --- see
Fig. \ref{spectrum1d}. As a result,
whereas in $d=2$ and $d=3$ a continuum of low-energy excitations with
finite $q$ is possible, at low energy only excitations with small $k$
or $k \approx 2 k_F$ are possible. The subsequent equivalence of
Fig. \ref{spectrum} for the one-dimensional case is the one shown in
Fig. \ref{phasespace1d}. Upon integrating over the momentum $k$ with
a cut-off of ${\cal{O}}(k_F)$ the
contribution from this particle-hole
scattering channel to $\mbox{Re}\, \chi(q,\omega)$ is
\begin{equation}
\label{1dphscattering}
\int dk {{1}\over{\omega + (2k+q) v_F }} \sim \ln [(\omega+qv_F)/E_F]~.
\end{equation}
(Note that (\ref{1dphscattering}) is true for both $q<<k_F$ and $|q-2k_F|<<k_F$.)
This in turn leads to a single particle self-energy calculated by
the process in Fig. \ref{selfenergy2ndorder} to be
$\mbox{Re}\, \Sigma(\bk_F,\omega)\sim \omega \ln \omega$
giving a hint of trouble.
Also the Cooper (particle-particle) channel has the same phase space
restrictions, and  gives a contribution also proportional to
$\ln \omega$. The fact that these sngular contributions are of the same order,
leads to a competition between charge/spin fluctuations
and Cooper pairing fluctuations, and to power law singularities.
Also, the fact that instead of a continuum
of low energy excitations as in higher dimensions, the width of the band of allowed
particle hole excitations vanishes as $\omega \rightarrow 0$, is the
reason that the properties of one-dimensional interacting metals can
be understood in terms of bosonic modes. We will present a brief
summary of the results for the single-particle Green's function and
correlation functions in section \ref{1dorthog}.

\begin{figure}
\begin{center}
 \epsfig{figure=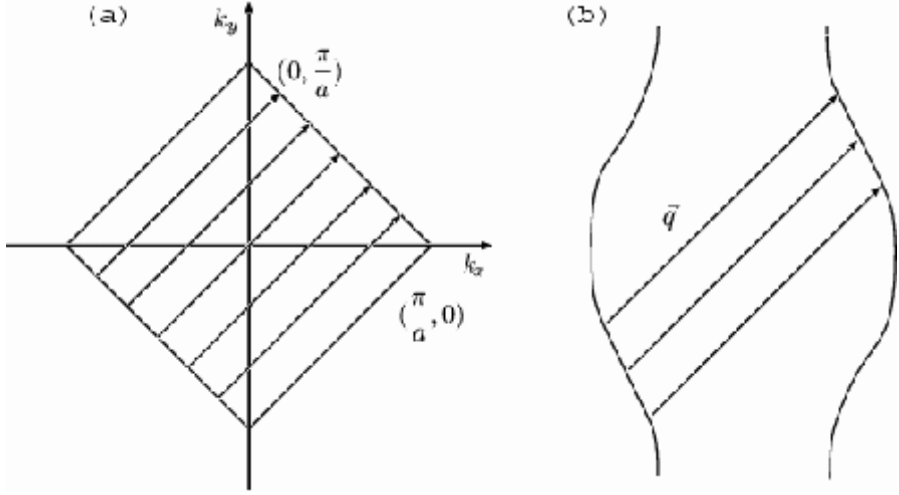,width=0.85\linewidth}
\end{center}
\caption[]{{\em (a)} The nested Fermi surface on obtained in a tight
binding model on a square lattice with nearest neighbor hopping; {\em
(b)} A partially nested Fermi surface which leads to charge density
wave or antiferromagnetic instabilities}\label{nesting}
\end{figure}

In special cases of nesting in two or three dimensions, one can have
situations that resemble the above one-dimensional
case. When the non-interacting Fermi
surface in a tight binding model has the square shape
sketched in Fig. \ref{nesting}{\em (a)} --- this happens for a
tight-binding model with nearest neighbor hopping on a square lattice
at half-filling
---  a continuous range of
momenta on opposite sides of the Fermi surface can be transformed into
each other by one
and the same wavenumber. This so-called nesting leads to $\log$ and
$\log^2$ singularities
for a continuous range of $\bk$ in the perturbation theory for the self
energy $\Sigma(\bk,\omega)$. Likewise, the partially nested Fermi
surface of Fig. \ref{nesting}{\em (b)} leads to charge density wave
and antiferromagnetic instabilities. We will come back to these
in sections  \ref{routesto} and \ref{qcpsection}.

\subsection{Principles of the Microscopic Derivation of
Landau Theory}\label{landauconsist}

In this section we will sketch how the conclusions
in the previous section based on second-order perturbation calculation
are generalized to all orders in perturbation theory. This section is slightly
more technical than the rest; the reader may choose to skip
to section \ref{routesto}.

We follow the microscopic approach whose foundations
were laid by Landau himself and which is discussed in detail in excellent textbooks
\cite{nozieres,pn,bp,agd}. For more recent methods with the same conclusions
see \cite{shankar,houghton}. Our emphasis will be on highlighting
the assumptions in the theory so that in the next section we can
summarize the routes by which the Fermi-liquid theory may break down.
These assumptions are usually not stated explicitly.

\begin{figure}
\begin{center}
 \epsfig{figure=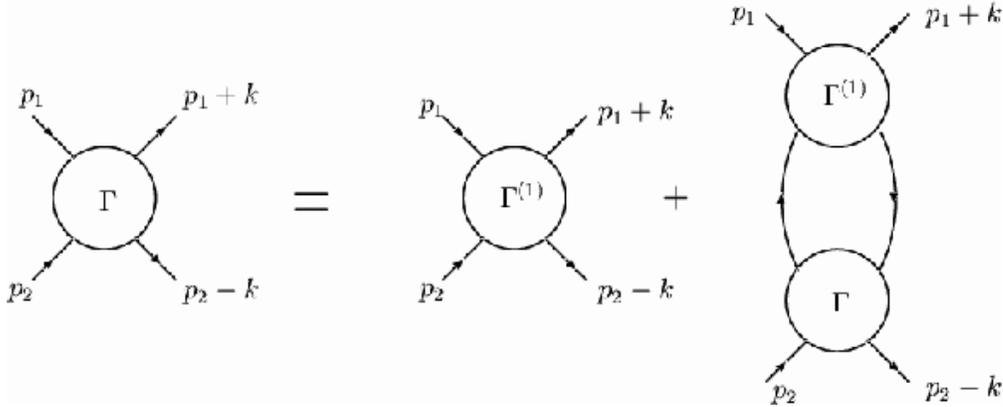,width=0.95\linewidth}
\end{center}
\caption[]{Diagrammatic representation of Eq. (\ref{Gamma1}). }\label{Gammagamma1diagram}
\end{figure}

The basic idea is that due to kinematic constraints, any perturbative
process with $n$ particle-hole pairs in the intermediate state gives
contributions to the polarizability proportional to
$(\omega/E_F)^n$. Therefore the low energy properties can be
calculated with processes with the same ``skeletal'' structure as those
in Fig. \ref{2ndorderpt}, which have only one particle-hole pair in the intermediate
state. So one may concentrate on the modification of the four-legged
vertices and the single-particle propagators due to interactions to
all orders. Accordingly the theory is formulated in terms of the single particle Green's
function $G (p)$  and the two-body scattering vertex,
\begin{equation}
\Gamma \left(p_{1},\  p_{2},\  p_{1}\  +\  k,\  p_{2}\  -k\right)\  =\Gamma
\left(p_{1},\  p_{2},\  k\right)
\end{equation}
Here and below we use, for sake of brevity, $p$, etc. to denote
the energy-momentum four-vector $({\bf p},\omega)$ and we suppress the
spin labels.
   The equation for $\Gamma $ is expanded in
{\em one of the two} particle-hole channels  as\footnote{To second order in the
interactions the correction to the vertex in the two possible particle-hole channels
has been exhibited in the first two parts of Fig. \ref{2ndorderpt}.}
\begin{eqnarray}
\Gamma (p_{1},p_{2},k)\ &  = &\  \Gamma ^{(1)}(p_{1},p_{2},k)\nonumber \\
&  &~~~- i\smallint \  \Gamma
^{(1)}(p_{1},q,k)\  G(q)G(q+k)\  \Gamma (q,p_{2},k){d^{4}q\over (2\pi )^{4}}
\   ,\label{Gamma1}
\end{eqnarray}
where $\Gamma^{(1)}$ is the irreducible part in the
particle hole channel in which Eq. (\ref{Gamma1}) is expressed.
In other words, $\Gamma^{(1)}$ can not be split up into two parts by
cutting two Green's function lines with total momentum $k$. So $\Gamma^{(1)}$
includes the complete vertex in the other (often called cross-)
 particle-hole channel. The
diagrammatic representation of Eq. (\ref{Gamma1}) is shown in
Fig. \ref{Gammagamma1diagram}. In the
simplest approximation $\Gamma^{(1)}$ is just the bare two-body
 interaction.
Landau theory {\em assumes} that $\Gamma^{(1)}$
has no singularities\footnote{The theory has been generalized for
Coulomb interactions \cite{pn,nozieres,agd}.The general
results remain unchanged because a screened short-range interaction
 takes the place of $\Gamma^{(1)}$.This is unlikely to be true in the critical
 region of a metal-insulator transition, because on the insulating side
 the Coulomb interaction is unscreened.}
An {\it assumption} is now further made that $G(p)$ does
have a coherent quasiparticle part at
$|{ \bf p}|\simeq p_{F}$ and $\omega \simeq 0$,
\begin{equation}
G(p)={{Z}\over{ \omega -\tilde{\epsilon}_{\bf p} +i\delta
\:\mbox{sgn}(\epsilon _{\bf p})}}
\  +\  G_{inc} ~,\label{Geq}
\end{equation}
where $\tilde{\epsilon}_{ \bf p}$ is to be identified as the excitation energy of the
quasiparticle, $Z$ its weight, and $G_{inc}$ the non-singular
part of $G$ (The latter provides the smooth background part of
the spectral function in Fig. \ref{figspectraldensob}{\em (b)} and the
former the sharp peak, which
is proportional to the $\delta$ function for $\tilde{\epsilon}_{\bf p}
= \epsilon_{\bf p} - \mu$.
  It follows \cite{nozieresbook,agd} from (\ref{Geq}) that
\begin{equation}
G(q)G(q+k)\  =\  {2i\pi z^{2}\over v_{F}}
\  {{{\bf v_{q}}\cdot \bk}\over {\omega -{\bf v_{q}}\cdot \bk}}
\delta (\nu )\delta (|{\bf q}|-p_{F})+\phi (q)~,\label{GGeq}
\end{equation}
for small $\bk$ and $\omega$, and where $\nu$ and $(\nu+\omega)$ are
frequencies of the two Green's functions.
Note the crucial role of kinematics in the form of the
first term which comes from the product of the quasiparticle
parts of $G$;  $\phi (q)$ comes from the scattering of the
incoherent part with itself and with the coherent part
and  is {\em assumed}
smooth and featureless (as it is indeed, given that $G_{inc}$ is smooth
and featureless and the scattering does not produce an infrared singularity
at least perturbatively in the interaction). The vertex $\Gamma$
in regions close to
$\bk \approx \bk_F$ and $\omega \approx 0$ is therefore
 dominated by the first term.  The derivation of
Fermi-liquid theory consists in proving that the equations (\ref{Gamma1})
for the vertex  and (\ref{Geq}) for the Green's function
 are mutually consistent.
 
The proof proceeds by defining a quantity
$\Gamma^{\omega} (p_1, p_2, k)$ through
\begin{eqnarray}
\Gamma ^{\omega }\left(p_{1},p_{2},k\right)\  & = & \  \Gamma
^{(1)}(p_{1},p_{2},k)  \nonumber \\
 & & ~~~-i\smallint \  \Gamma ^{(1)}(p_{1},q,k)  \phi
(q)\  \Gamma ^{\omega }(q,p_{2},k)\  {d^{4}q\over (2\pi )^{4}} \label{Gammaomega}~.
\end{eqnarray}
$\Gamma^{\omega}$ contains repeated scattering of the incoherent
part of the particle-hole pairs among itself and with the coherent part, but no scattering
of the coherent part with itself.  Then, provided the irreducible
part of $\Gamma^{(1)}$ is {\it smooth and not too large}, $\Gamma^{\omega}$ is
 smooth in $k$ because $\phi (q)$ is
by construction quite smooth.

Using the fact that the first part of (\ref{GGeq}) vanishes for
$v_F |k| / \omega \rightarrow 0$, and comparing (\ref{Gamma1}) and
(\ref{Gammaomega})
one can write the {\it forward scattering amplitude}
\begin{equation}
\lim_{\omega \rightarrow 0}\:\left[\lim_{k\rightarrow 0}\
\Gamma (p_{1},p_{2},k)\right]\  =\  \Gamma ^{\omega }(p_{1},p_{2})\  .
\end{equation}
This is now used to write the equation for the complete
vertex $\Gamma$ in terms of $\Gamma^{\omega}$:
\begin{eqnarray}
\Gamma (p_{1},p_{2},k) & = & \Gamma ^{\omega }(p_{1},p_{2})  \nonumber \\
& & ~~~+ {Z^{2}p_{F}^{2}\over (2\pi
)^{3}v_{F}}
\  \int \  \Gamma ^{\omega }(p_{1},q)\Gamma (q,p_{2},k) { { {\bf v_{q}}\cdot
\bk} \over { \omega -{\bf v_{q}} \cdot \bk}}
d\Omega _{q}~, \label{Gamma}
\end{eqnarray}
where in the above  $|q| =p_F$ and one integrates
only over the solid angle $\Omega_q$.

\begin{figure}
 \begin{center}
 \epsfig{figure=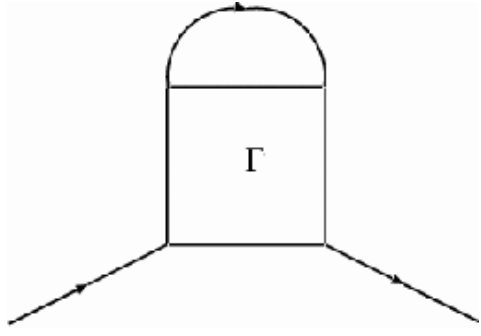,width=0.45\linewidth}
\end{center}
\caption[]{Diagram for the exact single-particle self-energy in terms
  of the exact vertex $\Gamma$ and the exact single-particle Green's
  function. }\label{Gamma1diagram}
\end{figure}

Given a non-singular $\Gamma^{\omega}$, a non-singular
$\Gamma$ is produced (unless the denominator in Eq. (\ref{Gamma})
produces singularities after the indicated integration-the Landau-Pomeranchuk
singularities discussed below).  The one-particle Green's function
$G$ can be expressed exactly in terms of $\Gamma$ --- see Fig. \ref{Gamma1diagram}.
  This
leads to Eq. (\ref{Geq}) proving the self-consistency of the ansatz
with a finite quasiparticle weight $Z$. The quantity
$Z^2 \Gamma^{\omega}$ is then a smooth function and goes
into the determination of the Landau parameters.

The Landau parameters can be written in terms of the forward
scattering amplitude.  In effect they parametrize the momentum
and frequency independent scattering of the incoherent parts
among themselves and with the coherent parts so that the end
result of the theory is that the physical properties can be expressed
purely in terms of the quasiparticle part of the single-particle
Green's function and the Landau parameters.  No reference to
the incoherent parts needs to be made for low energy
properties.  For single-component translational invariant
fermions (like liquid $^3$He) even the quasiparticle amplitude
$Z$ disappears from all physical properties.
This last is not true for renormalization due to
electron-phonon interactions and in multi-component systems
such as heavy fermions.  Special simplifications of the
Landau theory occur in such problems and in other
problems where the single-particle self-energy is nearly
momentum independent \cite{metzner,vollhardt,georgi,varma3,miyake}.
 
\begin{figure}
\begin{center}
 \epsfig{figure=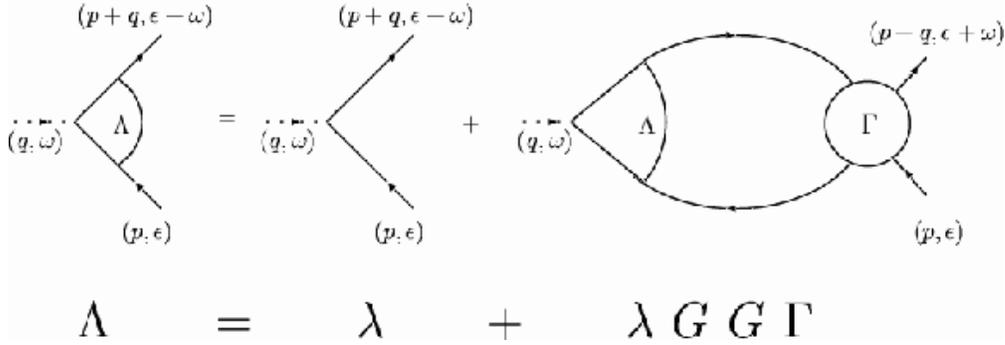,width=0.95\linewidth}
\end{center}
\caption[]{Vertex for coupling to external perturbations at
  energy-momentum $(\omega,q)$; $\lambda$ is the bare vertex.}\label{vertexcoupl}
\end{figure}

As we also mentioned, the single-particle self-energy $\Sigma$ can
be written exactly in terms of the vertex $\Gamma$: the relation
between the two is represented diagrammatically in
Fig. \ref{Gamma1diagram}.  The
relations between $\Sigma$ and $\Gamma$ are due to
conservation laws which Landau theory, of course,
obeys.  But the conservation laws are more general than
Landau theory.  It is often more convenient to express
these conservation laws as relations between the self-energy
and the three-point vertices,
$\Lambda_{\alpha} ( p \epsilon ; q \omega )$ which
couple external perturbations to either the density (the fourth
component, $\alpha = 4$) or  the current density in the
$\alpha=(1,2,3)$ direction The diagrammatic representation of the
equation for $\Lambda$ is shown in Fig. \ref{vertexcoupl}.
The following relations
(Ward identities) have been proven for
translationally invariant problems:
\begin{eqnarray}
\lim \left({\epsilon \over q}
\rightarrow 0,\  q\rightarrow 0\right)\Lambda _{\alpha }
({\bf p}\epsilon ;{\bf q}\omega) & = & {p_{\alpha }\over m}
-{\partial \over \partial p_{\alpha }}
\Sigma (\bp,\epsilon )~,~~~~(\alpha=1,2,3)\label{firstterm}
\\
\lim \left({\epsilon \over q}
\rightarrow 0,\  q\rightarrow 0\right)\Lambda _4
({\bf p}\epsilon ;{\bf q}\omega) & = & 1 +
{ {\partial \Sigma ({\bf p}, \epsilon)} \over {\partial \epsilon}}~,\label{secondterm}
\\
\lim \left({q\over \epsilon }
\rightarrow 0,\epsilon \rightarrow 0\right)\Lambda _{\alpha }
({\bf p}\epsilon ;\bq \omega ) & = & {p_{\alpha }\over m}
-{d\over dp_{\alpha }}
\Sigma ({\bf p},\epsilon )~,~~~~(\alpha=1,2,3) \label{thirdterm}
\\
\lim \left({q\over \epsilon }
\rightarrow 0,\epsilon \rightarrow 0\right)\Lambda _{4}
({\bf p}\epsilon ;\bq\omega
) & = & 1+{d \Sigma ({\bf p},\epsilon )\over d\mu }~. \label{fourthterm}
\end{eqnarray}

A relation analogous to (\ref{secondterm})  is derived for
fields coupling to spin for the case that interactions conserve spin.
The total derivative in (\ref{thirdterm}) and (\ref{fourthterm})
[rather than the partial
derivative in (\ref{firstterm}) and (\ref{secondterm})] represents that
$d \Sigma / d \mu$ is the variation in $\Sigma$
when $\epsilon$ is changed to $\epsilon + d \epsilon$
together with $\mu$ to $\mu + d \mu$, and
$d \Sigma / d p_{\alpha}$ represents the variation when the momentum
${\bf p}$ as well as the Fermi-surface is translated
by $d p_{\alpha}$.

Eq. (\ref{fourthterm}) is an expression of energy conservation,
and Eq. (\ref{secondterm}) of particle number conservation.
Eqs. (\ref{firstterm}) and (\ref{secondterm}) together signify the
continuity equation.
Eq. (\ref{thirdterm}) represents current conservation\footnote{
 The Ward identity Eq. (\ref{thirdterm}) does not hold for impure
system where the Fermi-surface cannot be defined in
momentum space. Since energy is conserved, a Fermi surface can still be
defined in energy space, and hence the other Ward identities continue
to hold.  This point is further discussed in
section \ref{hightcsection}.}.

In Landau theory, the right hand sides in
Eqs. (\ref{firstterm})-(\ref{fourthterm})
are expressible in
terms of the Landau
parameters. These relations are necessary
to derive the renormalization in the various thermodynamic quantities
that we quoted in Eqs. (17) and (18)
as well as the Landau transport equation. Needless to say, any theory of SFL
must also be consistent with the Ward identities.

\subsection{Modern derivations}

The modern derivations of Fermi-liquid theory start as
well by assuming the existence of a Fermi surface.  Kinematics
then inevitably leads to similar considerations as above.
  Instead of the division into coherent and
incoherent part made in Eq. (\ref{Geq}), the renormalization
group procedures are used to systematically generate a
successively lower energy and small momentum Hamiltonian with
excitations of particle ever closer to the Fermi surface.
The calculations are done either in terms of fermions \cite{shankar} or newly
developed bosonization methods in arbitrary dimensions \cite{houghton}.  The
end result is equivalent to Eqs. (\ref{Geq}), (\ref{Gammaomega})
and (\ref{Gamma}). These methods may well turn out to be very
important in finding the structure of SFL's and in
systematizing
them.

These derivations do the calculation in arbitrary
dimension  $d$ and conclude that the forward scattering
amplitude is
\begin{equation}
\Gamma ^{\omega }(p_{1},p_{2},k)\  \sim \  \left({k\over k_{F}}
\right)^{d-1}f(p_{1},p_{2},k)\  ,\
\end{equation}
where $f$ is a smooth function of all of its arguments.
In one-dimension the forward scattering amplitude has a
logarithmic singularity, as we noted earlier.

We can rephrase the conceptual framework of Landau
Fermi-liquid theory in the modern language of
Renormalization Group theory \cite{shankar}. As we discussed,
in Fermi-Liquid theory
one treats a complicated strongly interacting fermion problem by
writing the Hamiltonian ${\mathcal H}$ as
\begin{equation}
{\mathcal H} =   {\mathcal H}_{simple} +    {\mathcal H}_{rest}~. \label{2.1}
\end{equation}
In our discussion ${\mathcal H}_{simple}$ was the non-interacting Hamiltonian.
The non-inter\-ac\-ting  Hamiltonian is actually a member of a 'line' of fixed-point Hamiltonians
  ${\mathcal H}^*$
 all of which have the same symmetries but differ in their
 Landau parameters $F_{l}^{s,a}$ etc. The $F_{l}$'s, obtained from the forward scattering
 in Landau theory are associated with marginal operators and distinguish the
 properties of the various systems associated with the line of fixed points.
 Landau fermi-liquid theory is first of all the statement of
  the domain of attraction of this line of fixed points. The theory also establishes
  the universal low temperature properties due to the "irrelevant"
  operators generated by ${\mathcal H}_{rest}$ due to
   scattering in channels other than the forward channel.
  Landau theory does not establish (at least completely) the domain of attraction of the
  "critical surface" bounding the domain of attraction of the Fermi-liquid fixed line
  from those of other fixed points or lines. If ${\mathcal H}_{rest}$ were to generate
  a "relevant" operator --- i.e., effective interactions which diverge at
  low energies (temperatures) --- the scheme breaks down. For example, attractive interactions
  between fermions generate relevant operators--they presage a transition
  to superconductivity, a state of different symmetry.
  But if we stay sufficiently above $T_c$, we can usually continue using Landau theory\footnote{We note that in a Renormalization
Group terminology, all Landau parameters $f_{\bk \bk^{\prime},\sigma
  \sigma^{\prime}}$ originating from forward scattering (i.e. zero
momentum transfer), are ``marginal operators''
  \cite{krishna-murthy,shankar}.
All other operators that determine finite temperature observable
 properties are ``irrelevant''.
 Thus, in a
"universal" sense, condensed matter physics may be deemed to be an
``irrelevant'' field. So much for technical terminology!}.

\subsection{Routes to Breakdown of Landau Theory}\label{routesto}

From Landau's phenomenological theory, one can only say that the
theory breaks down when the physical properties --- specific heat divided
by temperature,\footnote{The specific heat of a system of
fermions can be written
in terms of integrals over the phase angle of the exact single-particle Green's
function (\cite{agd}).
Given any singularity in the self-energy, $C_v/T$ is never
more singular than $\ln T$. This accounts for the numerous experimental examples
of such behavior we will come across.} compressibility, or the magnetic
susceptibility --- diverge or when the collective modes
representing oscillations of the Fermi-surface in
any harmonic and singlet or triplet spin combinations become unstable.
The latter, called the Landau-Pomeranchuk singularities, are
indeed one route to the breakdown of Landau theory and occur when the
Landau parameters $F_l^{s,a}$ reach the critical value $-(2l+1)$. A
phase transition to a state of lower symmetry in then indicated. The new phase
can again be described in  Landau theory  by defining distribution
functions consistent with the symmetry of the new ground state.

The discussion following Eq. (\ref{1.4}) in section
\ref{landauwavefren} allows us to make
 a more general  statement.  Landau theory breaks
  down when the
quasiparticle amplitude $Z_\bk$ becomes zero; i.e. when the state
 $c_{\bf k}^{\dagger}|\psi^N \rangle$ and $|\psi_{\bf k}^{N+1} \rangle$
are orthogonal.
  This can happen
if the series expansion in Eq.\ (\ref{1.4}) in terms of the number
of particle-hole pairs is divergent. In other words, addition of a
particle or a hole to the system creates a divergent number of
particle-hole pairs in the system so that the leading term does not
have a finite weight in the thermodynamic limit. From Eq. (\ref{1.11})
connecting the $Z's$ to
$\Sigma's$, this requires that the single-particle self-energy be
singular as a function of $\omega$ at $k\simeq k_F$. This in turn
means that the Green's functions of SFL's contain branch cuts rather
than the poles unlike Landau Fermi-liquids.
The weakest singularity of this kind is encountered in the borderline
``marginal Fermi-liquids'' where\footnote{ To see why this is the borderline case, note
that a requisite for the definition of a quasiparticle
is that the quasiparticle peak width $\tau_{\bk}^{-1}= 2\Sigma^{''}$
should vanish faster than linear in $\omega$, the quasiparticle
energy. Thus $\Sigma^{''} \sim \omega$ is the   first power for which
this is not true. The $ \omega \ln (\omega_{c}/\omega)$ term in
Eqn. (\ref{marginal}) is then dictated by the Kramers-Kroning relation.}

\begin{equation}
\Sigma(k_{F},\omega) \simeq \lambda \left[ \omega \ln
  \frac{\omega_{c}}{\omega}+ i | \omega|   \right].
\label{marginal}
\end{equation}

If a divergent number of low-energy particle-hole pairs is created
upon addition of a bare particle, it means that the
low-energy response functions (which all involve creating
particle-hole pairs) of SFL's are also divergent. Actually
the single-particle self-energy can be written in terms of
integrals over the complete particle-hole interaction vertex as
in Fig. \ref{Gamma1diagram}.  The
implication is that the interaction vertices are actually more
divergent than the single-particle self-energy.

Yet another route to SFL's is the case in which the interactions
generate new quantum
numbers which are not descriptive of the non-interacting problem. This
happens most famously in the Quantum Hall problems and in
one-dimensional problems (section \ref{1dphysics}) as well as problems
of impurity
scattering with special symmetries (section \ref{localfermil}). In
such cases the
new quantum numbers
characterize new low-energy topological excitations. New quantum numbers
of course imply $Z=0$, but does $Z=0$ imply new quantum numbers.
One might wish to conjecture that this is so. But there
 is no proof of this\footnote{It would indeed be a significant step forward if such a
conjecture could be proven to be true or if the conditions in which it is true were
known.}.
 
In the final analysis all breakdowns of Landau theory are due
to degeneracies leading
to singular low-energy fluctuations. If the characteristic energy of
the fluctuations is lower than the temperature, a quasi-classical
statistical mechanical problem results. On the basis of our qualitative
discussion in section \ref{whyworks} and the sketch of the
microscopic derivation
in section \ref{landauconsist}, we may divide up the various routes to breakdown of Landau theory into the
 following (not necessarily orthogonal) classes:

{\em (i)}  {\em Landau-Pomeranchuk Singularities:}
 Landau
theory  points to the possibility of its breakdown through
the instability of the collective modes of the Fermi-surface which
 arise from the solution of the
homogeneous part of Eqn.(\ref{Gamma}).
These collective modes can be characterized by the
angular momentum $\ell$ of oscillation of the Fermi-surface
and whether the oscillation is symmetric ``$s$''
 or
anti-symmetric ``$a$'' in spin.  The condition for the
instability derived from the condition of zero frequency
of the collective modes are \cite{pn,bp}
\begin{equation}
\matrix{ F_{\ell }^{s} \leq-(2\ell +1)~,~~~~~~ F_{\ell }^{a} \leq -(2\ell +1)}~.
\end{equation}
The $\ell = 0$ conditions refer to the divergence in the
compressibility and the (uniform) spin-susceptibility.
The former would in general occur via a first-order transition,
 so is uninteresting to us. The latter describes the
  ferromagnetic instability. No other Landau-Pomeranchuk
   instabilities
  have been experimentally identified. But such new and
  exotic possibilities should be kept in mind. Thus, for
  example, an $F_1^s$-instability corresponds to  the Fermi-velocity
  $\rightarrow 0$, a
  $F_2^s$ instability to a ``$d$-$wave$-like'' instability of the
  particle-hole excitations on the Fermi-surface
  etc. Presumably these instabilities are resolved
  by reconstruction of the Fermi-surface with (patches) of
  energy gaps. Coupling of the damped transverse-excitations
  of charged-fermions to zero-point fluctuations of the electromagnetic-fields
  produces an SFL which we study in section \ref{gaugesec}.
  The microscopic interactions necessary for the Landau-Pomeranchuk
instabilities and the critical properties near such instabilities
have not been well investigated, especially for fermions
  with a lattice potential.

  It is also worth noting that some of the
  instabilities are disallowed in the limit of translational
  invariance. Thus, for example, time-reversal breaking states,
  such as the ``anyon-state'' \cite{laughlin,chen}
  cannot be realized
  because in a translationally invariant problem
  the current operator
  cannot be renormalized by the interactions, as we have learnt from
Eqs. (\ref{jvm*}), (\ref{firstterm}).

{\em (ii)} {\em Critical regions of Large Q-Singularities:}
Landau theory concerns itself only with long wavelength
response and correlations.  A Fermi-liquid may have
instabilities at a nonzero wave-vector, for example a charge-density
wave (CDW) or spin-density wave (SDW) instability.  Only a microscopic
calculation can provide the conditions for such instabilities
and therefore such conditions can only be approximately
derived.  An important point to note is that they arise perturbatively
from repeated scattering between the quasiparticle parts
of $G$ while the scattering vertices (irreducible interactions)are regular.   The
superconductive instability for any angular momentum is
also an instability of this kind. In general such instabilities are easily seen in
RPA and/or $t$-matrix calculations.

Singular Fermi-liquid behavior is generally expected
to occur only in the critical regime of such instabilities \cite{halperinandhohenberg,ma}.
If the transition temperature $T_c$ is finite then there is usually a
stable low temperature phase in which unstable modes are
condensed to an order parameter, translational symmetry is broken, and
gaps arise in part or all of the Fermi-surface. For excitations on
the surviving part of the Fermi-surface, Fermi-liquid theory is
usually again valid.  The fluctuations in the critical regime
are classical, i.e. with characteristics frequency
$\omega_{f \ell} \ll k_BT_c$.

If the transition is tuned by some external parameter
so that it occurs at zero temperature, one obtains, as illustrated already
in Fig. \ref{figphasediagram}, a  Quantum
Critical Point (QCP). If the transition is approached at $T=0$ as
a function of the external parameter, the fluctuations are
quantum-mechanical, while if it is approached as a function
of temperature
for the external parameter at its critical value,
 the fluctuations have
a characteristic energy proportional to the temperature.
 A large region
of the phase diagram near QCP's often carries SFL properties.
 We shall discuss
such phenomena in detail in section \ref{qcpsection}.

{\em (iii)} {\em Special Symmetries:}
The Cooper instability at $q = 0$, Fig. \ref{2ndorderpt}{\em (c)},
 is due to
the ``nesting'' of the Fermi-surface in the
particle-particle channel.  Usually indications of finite
$q$-CDW or SDW singularities are evident pertubatively
from Fig. \ref{2ndorderpt}{\em (a)}  or Fig. \ref{2ndorderpt}{\em (b)}
for special Fermi-surfaces,
nested in some $\bq$-direction in particle-hole channels.
One-dimensional fermions are perfectly nested in both
particle-hole channels and particle-particle channels
(Figs. \ref{2ndorderpt}{\em (a)-(c)}) and hence they are both
logarithmically singular.
Pure one-dimensional fermions also have the extra
conservation law that right going and left going momenta
are separately conserved.  These introduce special
features to the SFL of one-dimensional fermions such
as the introduction of extra quantum numbers.  These issues
are discussed in section \ref{1dorthog}. Several soluble
impurity problems with special symmetries have SFL properties.
Their study can be illuminating and we discuss them in
section \ref{localfermil}.

{\em (iv)} {\em Long-Range Interactions:}
Breakdown of Landau Fermi-liquid may come about through
long-range interactions, either in the bare Hamiltonian
through the irreducible interaction or through a generated
effective interaction.  The latter, of course, happens in the
critical regime of phase transitions such as discussed above.
 Coulomb interactions will not do for the former because of
screening of charge fluctuations. The fancy way of
saying this is that the
longitudinal electromagnetic mode acquires mass in a metal.
The latter is not true for current fluctuations or
transverse electromagnetic modes which due to gauge invariance
must remain massless.
This is discussed in section \ref{gaugesec}, where it is shown that
no metal at low enough temperature is a Fermi-liquid.
However, the cross-over temperature is too low to be of
experimental interest.

An off-shoot of an SFL through current fluctuations is the
search for extra (induced) conservation laws for some
quantities to keep their fluctuations massless. This line of investigation
may be referred to generically as gauge theories. Extra conservation laws imply
 extra quantum numbers and associated orthogonality. We
discuss these in section \ref{generalgaugesec}. The one-dimensional interacting
electron problem
and the Quantum Hall effect problems may be usefully thought of in these terms.

{\em (v)} {\em Singularities in the Irreducible Interactions:}
 In all the possibilities discussed in {\em (i)-(iii)}
 above
the irreducible interactions $\Gamma^{(1)}$
defined after Eq. (\ref{Gamma1})are regular and not too large. As noted after
Eq. (\ref{Gammaomega}) this is necessary to get a regular $\Gamma^{\omega}$.
When these conditions are satisfied the
conceivable singularities
 arise only from the repeated scattering of low-energy
particle-hole (or particle-particle) pairs) as in Eq. (\ref{Gamma}) or its
equivalent for large momentum transfers.

A singularity in the
irreducible interaction of course invalidates the basis
of Landau theory. Such singularities imply that the parts of the problem
 considered  harmless perturbatively because they involve the incoherent
 and high energy parts of the single-particle spectral weight
 as in Eq. (\ref{Gammaomega}) are,
  in fact, not so. This is also true if $\Gamma^{(1)}$ is large enough
that the solution of Eq. (\ref{Gammaomega}) is singular. Very few investigations of
 such processes exist.

How can an irreducible interaction be singular when the
bare interaction is perfectly regular? We know of two examples:

In disordered metals the density correlations are
diffusive with characteristic frequency $\omega$
scaling with $q^2$.  The irreducible interactions made
from the diffusive fluctuations and interactions
are singular in $d = 2$.  This gives rise to a new class of
SFL's which are discussed in section \ref{mitrans}. One finds that
in this case
the singularity in the cross-particle-hole channel (the channel
different from the one through which the irreducibility of $\Gamma^{(1)}$
is defined) feeds back into a singularity in $\Gamma^{(1)}$.
This is very special because the cross-channel is integrated over and
the singularity in it must be very strong for this to be possible.

The second case concerns  the particle singularities in the
irreducible interactions because of excitonic singularities.
Usually the excitonic singularities due to particle-hole
between different bands occur at a finite energy and do not
introduce low energy singularities.  But if the interactions
are strong enough these singularities occur near
zero frequency.  In effect eliminating high energy
degrees of freedom generates low energy irreducible singular vertices.

\begin{figure}
\begin{center}   
  \epsfig{figure=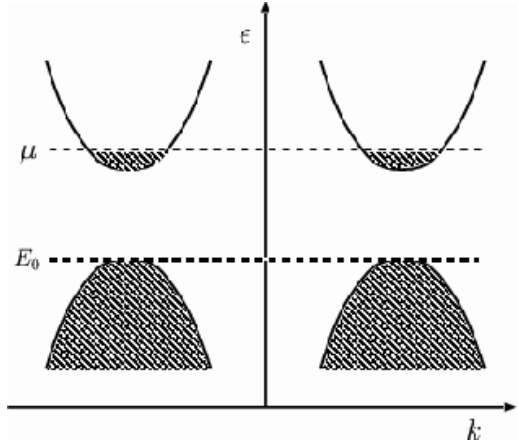,width=0.5\linewidth}
\end{center}
\caption[]{A model band structure for a solid with more than one atom
  per unit cell. Actually in $CaB_6$ where excitonic singularities have been invoked to
  produce a ferromagnetic state \cite{young} there are three equivalent points
  in the Brillouin zone where the conduction band minima and  the
  valence band maxima occur.}\label{bandstructureplot}
\end{figure}

Consider, for example, the band-structure of a solid
with more than one atom per unit cell with (degenerate)
valence band maxima and minima at the same points in the
Brillouin zone, as in Fig. (\ref{bandstructureplot}) .

Let the conduction band be partially filled, and the energy
difference between $\mu$ and the valence band marked
$E_0$ in Fig. \ref{bandstructureplot}  be much smaller than the attractive
particle-hole interactions $V$ between states in the valence ({\em v})
band and the conduction band ({\em c}).  For any finite $V$
excitonic resonances form
from scattering between $v$ and $c$ states, as in the $X$-ray edge
problem to be discussed in section \ref{xrayedge}.  For large
enough $V$ such resonances occur at asymptotically low
energy so that the Fermi-liquid description of states
near the chemical potential in terms of irreducible interaction
among the $c$-states is invalid.  The effective irreducible
interactions among the low-energy states integrate
over the excitonic resonance and will in general be
singular if the resonance is near zero-energy.

Such singularities require interactions above a critical
magnitude and are physically and mathematically of an
unfamiliar nature.  In a 2-band one-dimensional model, exact numerical
calculations have established the importance of such
singularities \cite{sudbo,stechel}.

Recently it has been found that $Ca B_6$ or
$Sr B_6$ with low density of trivalent $Eu$ or
quadrivalent $Ce$ ions substituting for ($Ca$, $Sr$) is a
ferromagnet \cite{young}.
  The most plausible explanation \cite{zhitomirsky,balents,barzykin} is that this
is a realization of the excitonic ferromagnetism predicted
by Volkov {\em et al.} \cite{volkov}. The instability to such a
state occurs because
the energy to create a hole in the valence band and a particle
in the conduction
band above the fermi-energy goes to zero if  the attractive
particle-hole (interband)interactions are large enough.
This problem has been investigated only in the mean-field approximation.
Fluctuations in
 the critical regime of
such a transition are well worth studying.

Excitonically induced singularities in the irreducible
interactions are also responsible for the Marginal Fermi-liquid
state of Cu$-$O metals in a theory to be discussed
in section \ref{hightcsection}.

\section{Local Fermi-Liquids \& Local Singular Fermi-Liquids}\label{localfermil}

In this section we discuss a particular simple form of Fermi-liquid
formed by electrons interacting with a dilute concentration of
magnetic impurity.
Many of the concepts of Fermi-liquid theory are revisited in this problem.
Variants of the problem provide an interesting array of soluble problems
of SFL behavior and  illustrate some of the principal themes of this article.

\subsection{The Kondo Problem}

The Kondo problem is at the same time one of the simplest and one of
the most subtle examples of the effects of strong correlation effects in
electronic systems. The experiments concern metals with a dilute
concentration  of  magnetic
impurities.  In the Kondo
model one considers only a single impurity; the
Hamiltonian then is
\begin{equation} \label{kondo}
{\mathcal H } = t \sum_{\langle ij \rangle} c_{i \alpha}^{\dagger} c_{j \alpha} + J {\bf S} \cdot c_{0}^{\dagger}
{\bf \sigma} c_{0}~,
\end{equation}
where $(c_{i \alpha}, c_{i \alpha}^{\dagger})$ denote the annihilation and
creation operators of a conduction electron at site $i$ with projection $\alpha$
 in the $z$-direction of spin
$\sigma$.  The second term is the exchange  interaction between a single
magnetic impurity at the origin (with  spin ${ S }=1/2$) and a
conduction electron spin.

When the exchange constant $J>0$ the system is a Fermi-liquid.
Although not often discussed, the ferromagnetic ($J<0$) variant of this
problem is one of the simplest examples of a singular Fermi-liquid.

\begin{figure}
\begin{center}   
  \epsfig{figure=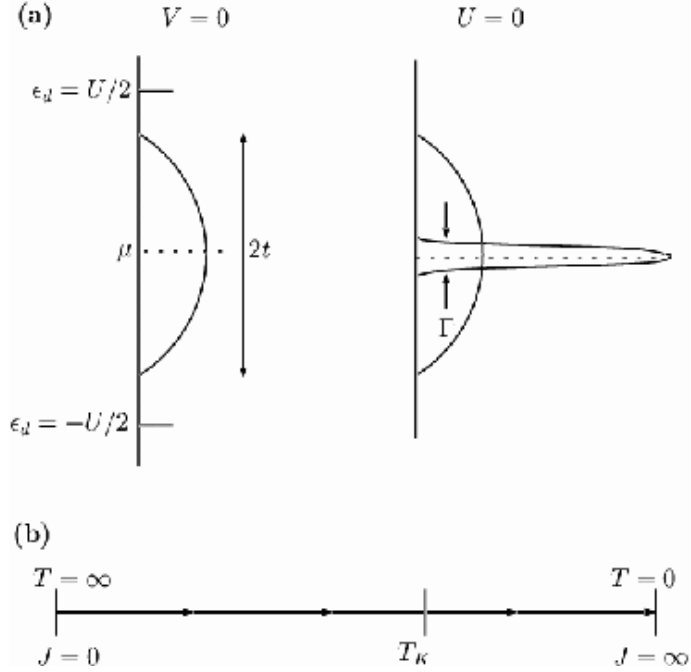,width=0.65\linewidth}
\end{center}
\caption[]{{\em (a)} Hartree-Fock excitation spectrum of the Anderson
  model in the two limits of zero hybridization, $V=0$ and zero
  interaction, $U=0$. {\em (b)} Renormalization group flow of the
  Kondo problem.} \label{limitfig}\label{kondolimits}
\end{figure}

There are two seemingly simple  starting points for the problem:
{\em (i)} $ J=0$: This turns out to describe
the unstable high temperature fixed point\footnote{ For the reader unfamiliar with reading a
renormalization group
diagram like that of Fig. \ref{kondolimits}{\em (b)} or
\ref{kondoflow}, the following explanation might be helpful.
The flow in a renormalization group diagram signifies the following.
The original problem, with bare parameters, corresponds to the starting
point in the parameter space in which we plot the flow. Then we imagine
"integrating out" the high energy scales (e.g. virtual excitations to high
energy states); effectively, this means that we consider the system
at lower energy (and temperature) scales by generating effective Hamiltonians with
new parameters so that the low energy properties remain invariant. The "length" along the flow
direction is essentially a measure of how many energy scales have been
integrated out --- typically, as in the Kondo problem, this decrease is
logarithmic along the trajectory. Thus, the regions towards which the
flow points signify the effective parameters of the model at lower and
lower temperatures. Fixed points towards which all
trajectories flow in a neighborhood describe the universal low temperature asymptotic
behavior of the  class of models to which the model under consideration belongs.
When a fixed point of the flow is unstable, it means that a model
whose bare parameters initially lie close to it flows away from this
point towards a stable fixed point; hence it has a
low-temperature behavior which does not correspond to the model
described by the unstable fixed point.
A fixed line usually corresponds with a class of models which have some
asymptotic behavior, e.g. an exponent, which varies continuously.}. The term proportional
to $J$ is
a marginal operator about the high temperature fixed point
because as discovered by Kondo\cite{kondo} in  a
third order perturbation calculation, the effective interaction acquires a
singularity $\sim J^{3}/t^{2} \ln(t/\omega )$.  {\em  (ii)} $ t=0$:
The perturbative expansion about
this point is well behaved. This turns out to describe the low temperature
Fermi-liquid fixed point. One might be surprised by this, considering that
typically the bare $t/J$ is of order $10^{+3}$. But such is the power of
singular renormalizations\footnote{A particularly lucid discussion
 of the renormalization procedure may be found in
\cite{krishna-murthy}. Briefly, the
 procedure consists in generating a sequence of Hamiltonians with
successively lower energy
 cut-offs that reproduce the low-energy spectrum. All terms allowed by symmetry
 besides those in the bare Hamiltonian are kept.
  The coeffcients of
 these terms {\it scale} with the cut-off. Those that decrease
 proportionately to the cut-off or change only logarithmically,
 are coefficients of marginal operators, those that grow/decrease
(algebraically)
 of relevant/irrelevant
 operators. Marginal operators are marginally relevant or marginally
irrelevant.
  Upon  renormalization the flow
 is to the strong
coupling $J=\infty$ fixed point, see Fig. \ref{kondolimits}. The terms
 generated from $t\ne 0$ serve as
irrelevant operators at this fixed point; this means that they do not
affect the ground state but determine
the measurable low-energy properties.}.

The interaction between conduction electrons and the
localized electronic level is not a direct spin interaction. It
 originates from quantum-mechanical charge fluctuations
that (through the Pauli principle) depend on the relative spin
orientation. To see this explicitly it is more instructive to
consider the Anderson
model \cite{anderson61}
in which
\begin{eqnarray}
{\mathcal H} &= & t \sum_{\langle ij \rangle} c_{i \alpha}^{\dagger} c_{j \alpha} + \epsilon_{d} \sum_{\sigma}
c_{0 \sigma}^{\dagger} c_{0 \sigma} + U c_{0,\uparrow}^{\dagger}
c_{0,\uparrow} c_{0,\downarrow}^{\dagger} c_{0,\downarrow} \nonumber +
\\& & \hspace*{3cm} + \sum_{k,\sigma} (V_{k} c_{k,\sigma}^{\dagger} c_{0,\sigma} + h.c.)
\end{eqnarray}

The last term in this Hamiltonian is the hybridization   between the
localized impurity state and the conduction electrons, in which spin
is conserved. In the particle-hole symmetric case,  $\epsilon_{d}= -U/2$ is
the one-hole state on the
impurity site in the Hartree-Fock approximation and  the one-particle state has
the energy $U/2$.

Following a perturbative treatment in the limit $\frac{t}{V},
\frac{U}{V} \gg 1$ the Anderson model reduces to the Kondo Hamiltonian
with an effective exchange constant $J_{eff} \sim (V^{2}/t)^{2}/U$.

The Anderson model has two simple limits, which are illustrated in
Fig. \ref{limitfig}:

{\em (i)} $V=0$: This describes a local moment with  Curie
susceptibility $\chi \sim \mu_B^2/T$.  This limit is the correct point of
departure for an investigation for the high temperature regime. As
noted one soon encounters the Kondo divergences.

(ii) $U=0$: In this limit the impurity forms a resonance of width
$\Gamma \sim V^{2}/t$  at the
chemical potential which in the particle-hole symmetric case is
half-occupied. The ground state is a spin singlet. This limit is the
correct starting point for an
examination of the low temperature properties ($T \ll T_{K}$).  A
temperature independent contribution to the susceptibility and a
linear contribution to the specific heat ($\sim N(0) T/\Gamma$) are contributed
by the resonant state. Hence the name {\em local Fermi-Liquid}.

The conceptually hard part of the problem was to realize that {\em (ii)}
is the correct stable low temperature fixed point and the technically
hard problem is to derive the passage from the
high-temperature regime to the low-temperature regime. This was first done
correctly by
Wilson \cite{wilson} through the invention of the
 Numerical Renormalization Group (and almost
correctly by Anderson and Yuval \cite{andersonyuval1,andersonyuval2}
by analytic methods).
The analysis  showed
that under Renormalization Group scaling transformations the ratio
$(J/t)$ increases monotonically as illustrated in
Fig. \ref{kondolimits}{\em (b)} --- continuous RG flows are observed from
the high temperature extreme  {\em (i)} to the low temperature extreme
{\em (ii)} and a smooth crossover between between the two regimes occurs
at the Kondo temperature
\begin{equation}
T_{K} \sim t \exp(-t/2J).
\end{equation}
Because all flow is towards the strong-coupling fixed point, universal
forms for the thermodynamic functions are found. For example, the
specific heat $C_{v}$ and the susceptibility $\chi$ scale as
\begin{equation}
C_{v} = T f_c(T/T_{K}),~~~~~~~~~~~ ~\chi = \mu_{B}^{2} f_\chi (T/T_{K})~,
\end{equation}
where the  $f$'s are universal scaling functions.

An important theoretical result is  that compared to a non-interacting
resonant level at the chemical potential, the ratio of the magnetic
susceptibility enhancement to the specific heat enhancement,
\begin{equation}
R_W = {{\delta \chi / \chi}\over{\delta C_v / C_v}}
\end{equation}
for spin 1/2 impurities at $T\ll T_K $ is precisely 2 \cite{wilson,nozieres}.
 In a noninteracting model, this ratio, nowadays
called the Wilson ratio, is equal to 1, since both $\chi$ and $C_v$
are proportional to the density of states $N(0)$. Thus the Wilson ratio
is a measure of the importance of correlation effects. It is in fact
the analogue of
the Landau parameter $F_0^a$ of Eq. (18).

\subsection{Fermi-liquid Phenemenology for the Kondo problem}\label{flkondo}

Following Wilson's solution \cite{wilson}, Nozi\`eres \cite{nozieres}
showed that the low-tempe\-ra\-tu\-re properties of the Kondo problem can be
understood simply through a (local) Fermi-liquid framework. This is a
beautiful example of the application of the concept of analyticity and
of symmetry principles about a fixed point. We present the key
arguments below. For the application of this line of approach to the
calculation of a variety of
properties we refer the reader to papers by Nozi\`eres and Blandin
\cite{nozieres,blandin}.

The properties of a local impurity can be characterized by the
energy-dependent $s$-wave phase shift $\delta_\sigma (\epsilon)$,
which in general also depends on the spin of the conduction electron
being scattered. In the spirit of Fermi-liquid theory the phase shift
may be written in terms of  the deviation of the distribution function $\delta
n(\epsilon) $ of conduction electrons from the equilibrium
distribution,
\begin{equation}
\delta_{\sigma}(\epsilon) = \delta_{0}(\epsilon) +
 \sum_{\epsilon^{\prime} \sigma^{\prime}}
 \delta_{\sigma \sigma^{\prime}}(\epsilon,\epsilon^{\prime})
\delta n_{\sigma^{\prime}}(\epsilon^{\prime}) +...~.\label{firstdeltaeq}
\end{equation}
About a stable fixed point the energy dependence is analytic near the
chemical potential $(\epsilon=0)$, so that we may expand
\begin{equation}
\delta_{0}(\epsilon)  =  \delta_{0} + \alpha \epsilon + ... ~,~~~~~~~~ \delta_{\sigma \sigma^{\prime}}(\epsilon,\epsilon^{\prime}) =  \phi_{\sigma \sigma^{\prime}}+...~.\label{seconddeltaeq}
\end{equation}
Just as  the Landau parameters are expressed
in terms of symmetric and antisymmetric parts, we can write

\begin{equation}
\phi_{\uparrow \uparrow} = \phi_{\downarrow\downarrow} =
\phi^{s}+\phi^{a} ~,~~~~~~~~~~~~~, ~ \phi_{\downarrow
    \uparrow}=\phi_{\uparrow \downarrow}= \phi^{s}-\phi^{a}~.
\end{equation}
Taken together, this leaves three parameters $\alpha$, $\phi^s$ and $\phi^a$ to
determine the low-energy properties. Nozi\`eres \cite{nozieres} showed
that in fact there is only one independent parameter (say $\alpha$ which
is of ${\mathcal O}(1/T_K)$, with a prefactor which can be obtained by
comparing with Wilson's detailed numerical solution).  To show this
note that by the Pauli principle same spin states do not interact, therefore
 \cite{nozieres}
\begin{equation}
\phi_{\uparrow \uparrow} = \phi^s + \phi^a = 0~.
\end{equation}
Secondly a shift of the chemical potential by $\delta \mu$ and a
simultaneous increase in $\delta n $ by $N(0) \delta \mu$ should have
no effect on the phase shift, since the Kondo-effect is tied to
the chemical potential. Therefore according to
\ref{firstdeltaeq}) and (\ref{seconddeltaeq})
\begin{equation}
[\alpha + N(0) \phi^s ] \delta \mu = 0~, ~~~~~\Longrightarrow \phi^s =
-\alpha/ N(0)~.
\end{equation}
Thirdly, one may borrow from Wilson's solution that the fixed point
has $\delta_0 = \pi/2$. This expresses that the tightly bound spin
singlet state formed of the impurity spin and conduction electron spin
completely blocks the impurity site to other conduction electrons;
this in turn implies maximal scattering and phase shift of $\pi/2$ for
the effective scattering potential \cite{nozieres}. In other words, it
is a strong-coupling
fixed-point where one conduction electron state is pushed below the
 chemical potential in the vicinity of the impurity to form a singlet
resonance with the impurity spin. One may now calculate all physical
properties in term of $\alpha$. In particular, one finds $\delta C_v /
C_v = 2 \alpha /(\pi V N(0)$ and a similar expression for the
enhancement of $\chi$,such that the Wilson ratio of 2.

\subsection{Ferromagnetic Kondo problem and the anisotropic Kondo problem}

 The  {\em ferromagnetic Kondo
  problem} provides us with the simplest example of SFL behavior. We will
 discuss this below after relating the problem to a
 general $X$-ray edge problem in which the connection to the
 so-called {\em orthogonality
  catastrophe} is clearer. As discussed in Sec.2, orthogonality plays
  an important role in SFLs generally.

 we start with the
anisotropic generalization of the Kondo Hamiltonian, which is the
proper starting model for a perturbative scaling analysis
\cite{poormans,fowler},
\begin{eqnarray} \nonumber
H & = & t \sum_{\langle i j \rangle, \sigma} c_{i \sigma}^{\dagger} c_{j
  \sigma} + \sum_{k,k^{\prime}} \left[J_{\pm}
\left( S^{+} c_{k \downarrow}^{\dagger} c_{k^{\prime}\uparrow}
+ S^{-} c_{k \uparrow}^{\dagger} c_{k^{\prime}\downarrow} \right)
\right. + \\
 & & \hspace*{3cm} +  J_{z}S^{z} \left. \left( c_{k,\uparrow}^{\dagger}c_{k^{\prime},\uparrow}
   -c_{k,\downarrow}^{\dagger}c_{k^{\prime},\downarrow} \right) \right]~ .
\end{eqnarray}
Long before the solution of the Kondo problem, perturbative
Renormalization Group for the effective vertex coupling constants
$J_\pm$ and $J_z$ as a function of temperature were obtained
\cite{poormans,fowler}. The scaling
relation between them is found to be exact to all orders in the $J's$:
\begin{figure}
\begin{center}
 \epsfig{figure=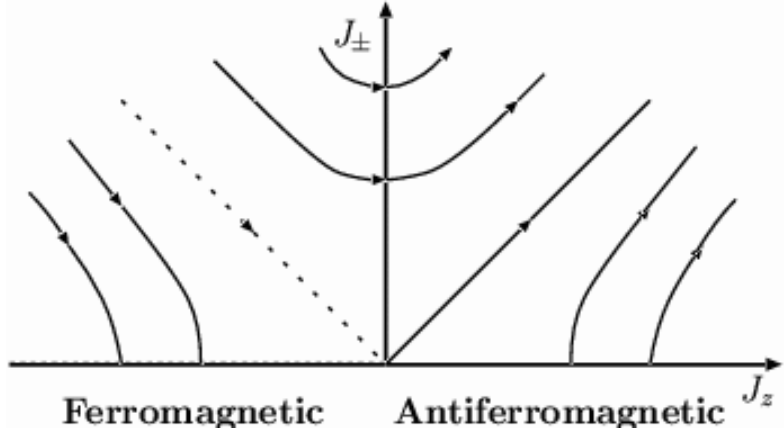,width=0.75\linewidth}
\end{center}
\caption[]{Renormalization group flows for the Kondo problem,
  displaying the line of critical points for the ``ferromagnetic''
  problem and the flow towards the fixed point $J^*_{\pm} = J^*_z \to
  \infty$ for the ``antiferromagnetic'' problem. } \label{kondoflow}
\end{figure}

 \begin{equation}
(J_{z}^{2}- J_{\pm}^{2}) = \mbox{const,} ~~~~~(J_{+}=J_{-})~.
\end{equation}

In the flow diagram  \ref{kondoflow} we show the scaling trajectories for the
anisotropic problem.
In the
antiferromagnetic regime the flows
continuously veer towards larger and larger
$(J_{\pm},J_{z})$ values; at the resulting
fixed point  $(J*_{\pm}, J*_{z}) =(\infty,\infty)$
singlets form between the
local moment and the
conduction
electrons.

The 'ferromagnetic' regime spans the
region satisfying the inequalities
$J_{z}<0$ and $|J_{z}| \ge J_{\pm}$.
Upon reducing the bandwidth the coupling parameters
flow towards negative $J_{z}$ values.
Observe the {\em line of fixed points}
on the negative $J_{z}$ axis.
Such a {\em continuous line} is also seen
in the Kosterlitz-Thouless transition \cite{kosterlitz}
of the two-dimensional XY model.
Moreover, in both problems
continuously varying exponents in physical properties
are obtained along these lines (in fact, the renormalization group
flow equations of the Kondo model for small coupling are
mathematically identical to those for the XY model).
This is an instance of {\em a zero temperature
Quantum Critical line}. The physics of the
Quantum Critical line has to do with an
"Orthogonality Catastrophe" which we describe next.
Such orthogonalities are an important part of
the physics of SFL's generally.

\subsection{Orthogonality Catastrophe}\label{orthogonalitysection}

As we saw in section \ref{landausfermiliquid}, a Fermi-liquid description
is appropriate so long as the
spectrum retains a coherent single particle
piece of finite weight $Z>0$. So if by some miracle the evaluation of
$Z$
reduces to an overlap
integral between two orthogonal
wave functions then
the system is a
SFL.

In the thermodynamic ($N \rightarrow \infty$)
limit, such a miracle is more generic than might
appear at first sight. In fact, 
such an {\em orthogonality catastrophe}
  arises if the injection of an
{\em infinitely massive particle in more than one dimension}
 produces an effective
finite range scattering potential for the
remaining $N$ electrons \cite{Anderson-1967}  (see Sec.(4.9). Such an
orthogonality is exact only
in the thermodynamic limit:
The single particle wave functions
are not orthogonal. It is only
the overlap between the ground state formed by their
Slater determinants\footnote{The results hold true also
for interacting fermions, at least when Fermi-liquid description
is valid for both of the states}which
vanishes as $N$ tends to
infinity.

More quantitatively, if the injection of
the  additional particle produces
an $s$-wave phase shift $\delta_{0}$
for the single particle
wave functions (all $N$ of them),
\begin{equation}
\phi(kr) = \frac{\sin kr}{kr} \rightarrow
\frac{\sin (kr+\delta_0)}{kr}
\end{equation}
then an explicit computation
of the Slater determinants
 reveals that their
overlap diminishes as
\begin{equation}
\langle \psi_{N} | \psi_{N}^{\prime} \rangle \sim  N^{-\delta_{0}^{2}/\pi^{2}}.
\end{equation}
Here  $|\psi_{N} \rangle$ is the determinant
Fermi sea wave function for $N$ particles
and $|\psi_{N}^{\prime} \rangle$ is the
wavefunction of the system
after undergoing a  phase shift
by the local perturbation
produced by the injected
electron\footnote{Through the Friedel sum
rule  $\delta_0/\pi$ has a physical meaning; it is  the charge
that needs to be transported
from infinity to the vicinity of the impurity in order to
screen the local
potential \cite{doniachsondheimer}.}.

Quite generally such an orthogonality ($Z=0$) arises also
if two $N$ particle
states of a system possess different
quantum numbers and almost the same energy. These
new quantum numbers might be
associated with novel topological
excitations.
This is indeed the case in
the Quantum Hall Liquid  where new quantum
numbers are associated with
fractional charge excitations. The SFL properties of the
interacting one-dimensional fermions (Section 4) may also be
looked on as
due to orthogonality. Often orthogonality
has the effect of making a quantum many-body problem
approach the behavior of a classical problem. This will be one of the
leitmotifs in this review. We turn first to a problem where
this orthogonality is well understood to lead to experimental consequences,
although not at low energies.

\subsection{X-ray Edge singularities}\label{xrayedge}

\begin{figure}
\begin{center}
\epsfig{figure=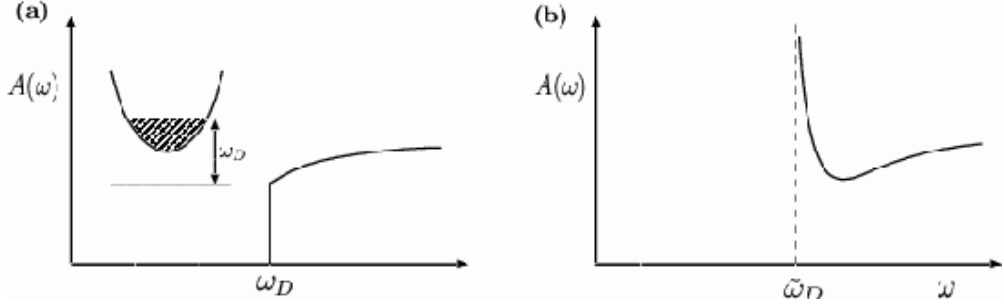,width=0.95\linewidth}
\end{center}
\caption[]{Absorption lineshape for transitions between a lower
  dispersionless level and a conduction band, {\em (a) } for zero
  interaction between the conduction electrons and local level, and
  {\em (b)} for small interaction.}\label{xrayfig}\label{fig0a}\label{fig0b}
\end{figure}

The term $X$-ray edge singularity is used for the line shape for
absorption in metals by creating a hole in an atomic core-level and a
particle in the conduction band above the chemical potential.
 In the non-interacting
particle description
of this process, the
absorption starts
at the threshold
frequency $\omega_{D}$, as sketched in Fig. \ref{xrayfig}.
In this case, a Fermi edge
reflecting the density
of unoccupied states in the conduction
band is expected to be visible the spectrum.

However, when a
hole is generated in the lower
level, the potential
that the conduction electrons
see is different. The relevant Hamiltonian
is now
\begin{equation}
{\mathcal H} = \epsilon_{d} \left( d^{\dagger} d - \frac{1}{2} \right)+ \sum_{k} \epsilon_{k} c_{k}^{\dagger} c_{k} +
\frac{1}{L} \sum_{k,k^{\prime}} V(k,k^{\prime}) \left( c_{k}^{\dagger}
c_{k^{\prime}} -\frac{1}{2} \right)
\left( d^{\dagger} d -\frac{1}{2} \right) ~, \label{xrayedgehamiltonian}
\end{equation}
where spin indices have been suppressed.
The operators $(d,d^{\dagger})$ annihilate or create holes in the core
level, which is taken to be dispersionless.
The first two terms in the Hamiltonian
represent the unperturbed energies
of the core hole and the free electrons.
The last term depicts the screened Coulomb interaction
between the conduction electrons and the  hole in the core level.

As a consequence of the interactions, the line shape is quite
different. This is actually an exactly solvable problem
\cite{nozieresdd}. There are two
kinds of effects, {\em (a)} excitonic ---
the particle and the hole
 attract, leading to a shift of the edge and a sharpening of the edge
singularity --- and {\em (b)} an orthogonality effect of the type just
discussed
above, which smoothens the edge irrespective of the sign of the
interaction. This changes the absorption spectrum to that of
Fig. \ref{xrayfig}{\em (b)} in the presence of interactions. The form
of the singularity is \cite{mahanart,nozieresdd}
\begin{equation} \label{Aequation}
A(\omega) \sim (\omega - \tilde{\omega}_{D})^{-2 \frac{\delta_{0}}{\pi} + \frac{\delta_{0}^{2}}{\pi^{2}}}.
\end{equation}
The exponent $\delta_{0}^{2}/\pi^{2}$ is a consequence
of the orthogonality catastrophe overlap integral;
the exponent $(- 2 \delta_{0}/\pi)$ is due
to the excitonic particle hole interactions.
 If the hole has finite mass we have a
problem with recoil which is not exactly solvable, but we know the
essential features of the solution. As we will discuss later in section
\ref{1dorthog}, recoil removes the singularity in two and three dimensions
and the absorption edge
acquires a characteristic width of the order of the dispersion of the hole
band. If the hole moves
only in one dimension, the singularity is not removed.

\subsection{A Spinless Model with Finite Range Interactions}

A model, which is a generalization of the Ferromagnetic Kondo
problem and
in which the low-energy physics is dominated by the orthogonality catastrophe,
is given by the following Hamiltonian :
\begin{eqnarray}
{\mathcal H} = \sum_{k,l} \epsilon_{k} \gamma_{k,l}^{\dagger} \gamma_{k,l}& + & \frac{t}{\sqrt{L}} \sum_{k} (\gamma_{k,0}^{\dagger} d+ h.c.) \nonumber
\\& & ~~~  +  \frac{1}{L} \sum_{k,k^{\prime}} V_{l}
\left(\gamma_{k,l}^{\dagger} \gamma_{k^{\prime},l} - \frac{1}{2}
\right) \left(d^{\dagger} d - \frac{1}{2} \right)~.
\end{eqnarray}
The operators $(\gamma, \gamma^{\dagger})$ are the annihilation and
creation operators of spinless conduction electrons with kinetic
energy $\epsilon_{k}$, while as before the  $d$ operators  create
or annihilate electrons in the localized level. The local chemical
potential has been set to
zero $(\epsilon_{d} =0)$ and the Hamiltonian is particle-hole symmetric.
The new index $l$ is an orbital angular
momentum index (or a channel index). Hybridization conserves
point-group symmetry, so the
localized orbital hybridizes with only one channel $(l=0)$.
By contrast, the impurity couples to all channels via the
interaction $V_{l}$. As we are summing over all
moments $(k,k^{\prime})$ this interaction
is local.

This problem
may be mapped onto the
anisotropic Kondo
model  \cite{Giamarchi-1993}. Indeed the transformation
\begin{eqnarray}
d^{\dagger} & \rightarrow & S^{\dagger}~, \nonumber
~~~~~~~~~~ d^{\dagger} d - \frac{1}{2} \rightarrow S_{z}~,
\\ t & \rightarrow & \frac{J_{\perp},0}{\sqrt{2 \pi a}}~,
~~~~~~~~~~~~ 2 V_{l} \rightarrow \sqrt{2} J_{z,l} - 2 \pi v_{F} (\sqrt{2}-1) \delta_{l,0}~,
\end{eqnarray}
produces
\begin{eqnarray}
{\mathcal H} = \sum_{k,\sigma,l} \epsilon_{k} c_{k,\sigma,l}^{\dagger} c_{k,\sigma,l} + \frac{1}{2} J_{\perp,0} (S^{\dagger}s_{l}^{-}+h.c.)
+\sum_{l} J_{z,l} S_{z} s_{z,l}.
\end{eqnarray}
Here $a$ is short distance cutoff.
In the resulting (anisotropic multi-channel) Kondo Hamiltonian
the spin operators $\bf{S}$ and ${\bf {s}}_{l}$ portray charge
excitations of the local orbital and conduction band;
the spin index in the resulting Kondo
Hamiltonian should now be
regarded as a charge label.
Physically, this mapping
is quite natural.
The impurity may or may not have
an electron,
this is akin to having spin up
or spin down. Similarly, the kinetic
hybridization term transforms into a
spin flip interaction
term of the form $(S^{\dagger} d^{-} + h.c.)$.
As $V_{l}$ couples to the occupancy of the
impurity site, we might anticipate
$J_{z}$ to scale with $V_{l}$. The
additional correction ($-2 \pi v_{F} (\sqrt{2}-1) \delta_{l,0})$
originates from the subtle transformation
taking the original fermionic system
into an effective spin model.

\begin{figure}
\begin{center}   
 \epsfig{figure=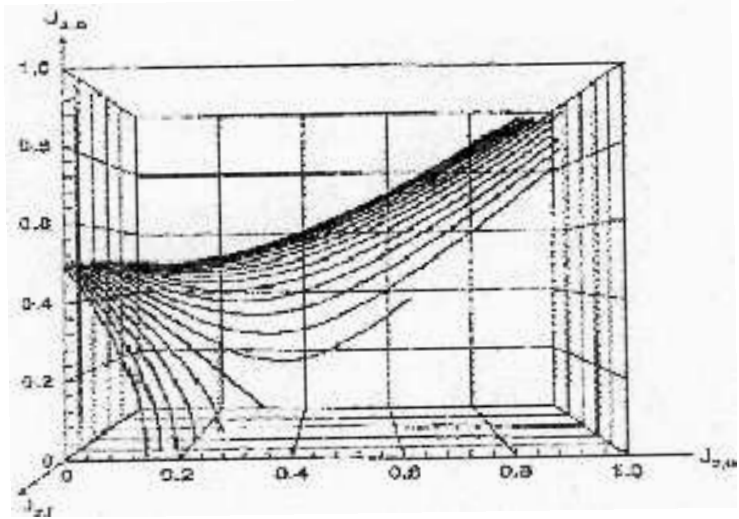,angle=0.8,width=0.7\linewidth}
\end{center}
\caption[]{The Renormalization flow diagram for the model with finite
  range interactions according to \cite{Giamarchi-1993}.  The initial
values are $J_{x,0}=0,J_{\perp,0}=0.5$, and $J_{x,l}$
varies from 0 to 1 in increments of 0.05. When $J_{z,l}$
becomes large enough (i.e. when $V$ is large enough)
the flow veers from the usual Kondo fixed
point to a zero hybridization ($J_{\perp,0}=0$)
singular Fermi liquid.}\label{prlgiam}
\end{figure}

This problem has been solved by renormalization group
methods, See Fig. \ref{prlgiam}. But simple arguments
 based on the $X$-ray edge singularity,
orthogonality and recoil give the correct qualitative physics.
When $t=0$, the problem is  that of the $X$-ray
edge Hamiltonian (with $\epsilon_{d}=0$).
When $t$ is finite, the charge at the
impurity orbital fluctuates
(the impurity site alternately
empties and fills). This generates, in turn,
a fluctuating potential. The $X$-ray absorption spectrum is the
Fourier transform of the particle-hole pair correlator
\begin{equation}
\Delta(\omega) \sim \langle \gamma^{\dagger}(t) d(t) d^{\dagger}(0) \gamma(0) \rangle_{\omega}.
\end{equation}
This quantity, which is none other than the hybridization correlation function,
should display the $X$-ray edge
characteristics for large frequencies ($\omega > \Delta_{eff}$)
where the effect of recoil is unimportant:
\begin{equation}
\Delta(\omega) \approx \Delta_0 (\omega/W)^{\gamma};~~ \gamma = -\frac{2 \delta_{0}}{\pi} + \sum_{l} \frac{\delta_{l}^{2}}{\pi^{2}}.
\end{equation}
The threshold frequency $\Delta_{eff}$
is determined by the recoil energy.
$W$ is the bandwidth.
The bare hybridization width $\Delta_0 \sim t^{2}/W$.
The exponent in the singularity contains
an exitonic shift ($-2 \delta_{0}/\pi$) as
well as an orthogonality contribution
($\sum_{l} \delta_{l}^{2}/\pi^{2}$).
 The recoil is cut off by $\Delta_{eff}$.
For $\omega < \Delta_{eff}$ the electron gas
becomes insensitive to the change in
the potential. As the $X$-ray edge
singularity is cut off at $\omega = O(\Delta_{eff})$,
self-consistency implies that
\begin{equation}
\Delta_{eff} = \Delta(\omega=\Delta_{eff})~.
\end{equation}
This leads to the identification
\begin{equation}
\Delta_{eff} = W (\Delta_0 /W)^{\beta}~,~~~~~\beta =
1/(1-\gamma)~,
\end{equation}
so that for
\begin{equation}
\gamma<1~,~~~ \Delta_{eff} \rightarrow 0~~~~~~\mbox{as}~ W \rightarrow \infty~.
\end{equation}
For $\gamma<1$, a
singular Fermi-liquid
emerges in which the hybridization of the localized
$d$-orbital with the electron gas
 scales to zero at zero frequency. The actual value of $\gamma$
  determines the singular properties at low energy or
temperature.
In the single channel
problem such a scenario
 occurs if the potential $V_{0}$ is
sufficiently attractive.
On mapping to the
spin problem we find that this region
corresponds to the singular Fermi-liquid
Ferromagnetic Kondo problem. The scaling
of the hybridization to zero corresponds,
in the  spin--model,
to $J_{\pm} \rightarrow 0$.
In a renormalization group
language, the flows will impinge on
the line of fixed points ($J_{\perp} =0$,
$J_{z}<0$). In this regime, we  recover,
once again, a continuous set of
exponents.

 If the number of
channels is large enough,
 the orthogonality
catastrophe associated
with the change in the number
of particles on the impurity site
is sufficiently strong to drive
the hybridization to zero even for the case of
repulsive interactions
$V$ or antiferromagnetic
$J_{z,l}$.

In the singular regime
various correlation
functions may be evaluated
 \cite{Giamarchi-1993}.
For instance, the Green's
function of the localized impurity
\begin{equation}
G_{d}(\tau) = -<T_{\tau} d(\tau) d^{\dagger}(0)>
\rightarrow_{\tau \rightarrow 0} e^{-\sum_{l} (V_{l}/\pi v_{F})^{2} \ln(|\tau|+a)}~.
\end{equation}
The orthogonality induced by the fluctuation in the
occupation number of the impurity site leads to the decay of
the correlation function with a nonuniversal
exponent. Because of this orthogonality
catastrophe the system behaves as a singular
Fermi-liquid. In the vicinity of the $t=0$ fixed point,
the self energy due to hybridization
\begin{equation}
\Sigma \sim \omega^{1-\gamma}.
\end{equation}
A line of critical points for $\gamma<1$ is found.
This bears a resemblance to
the Kosterlitz-Thouless phenomenon \cite{kosterlitz}.
The analogue to the emergence of
vortices in the Kosterlitz-Thouless
transition are instantons --- topological excitations
which are built of a succession
of spin flips
in time on the
impurity site.

\subsection{A model for Mixed-Valence Impurity}

We next consider a slightly more realistic model \cite{perakis,sikot,Sire-1993}
\begin{eqnarray}
{\mathcal H} = \sum_{k,\sigma,l} \epsilon_{kl} c^{\dagger}_{k \sigma
l} c_{k \sigma l} + \epsilon_{d} n_{d,\uparrow} n_{d,\downarrow} + t
\sum_{\sigma} \left( d_{\sigma}^{\dagger}c_{k \sigma 0} + h.c.\right) \nonumber
\\+ \sum_{k,k^{\prime},l} V_{k k^{\prime} l} \left( n_{d}-\frac{1}{2}
\right) \left[\sum_{\sigma} c_{k \sigma l}^{\dagger} c_{k^{\prime} \sigma l} - 1\right]~.
\end{eqnarray}
In this model both spin and charge
may be altered on the impurity site.
This enhanced number of degrees of freedom
implies that the states need to
be specified by more quantum numbers.
This also allows, a priori, for higher degeneracy.
In the following
$N$ screening channels, all of equal strength
$V$,
will be assumed. In the $U \rightarrow \infty$
limit the spectrum of the impurity site may be
diagonalized.
The two lowest states are
\begin{eqnarray}
\eta^{\dagger}\ 0 \rangle & = &  |0,1 \rangle,~~\mbox{with energy}~ E_{\eta} = -\frac{V \sqrt{N}}{4}-\frac{\epsilon_{d}}{2}, \nonumber
\\ \zeta_{\sigma}^{\dagger} |0 \rangle & = & | \sigma, 0
\rangle,~~\mbox{with energy}~ E_{\zeta} = -\left[
\left(\frac{\epsilon_{d}}{2} -\frac{V \sqrt{N}}{4}\right)^{2}+t^{2} \right]^{1/2},
\end{eqnarray}
where in the bras, the first entry is the charge and spin of the
impurity spin and the second one is  the compensating
charge in the screening channels. Other states are elevated by
energies  $ {\mathcal O} (V/\sqrt{N})$.
The states  satisfy the Friedel screening sum rule
by having a small phase shift  $\frac{\pi}{2 N}$ in each of the  $N$
channels. The parameter $V$ can be tuned to produce a degeneracy 
between the two states ($E_{\eta}=E_{\zeta}$) --- the mixed valence condition.
The enhanced degeneracy produces
Singular Fermi-liquid like behavior\footnote{There is a singularity only in the local
charge response at the impurity, not in the magnetic response. In this respect, the results of
(\cite{Sire-1993}) are not completely correct}.
A perturbative calculation
for small $\omega$ yields
a self-energy
\begin{equation}
\Sigma \sim [\omega \ln \omega + i \omega sgn(\omega)] + O[(\omega \ln \omega)^{2}]~.
\end{equation}
As a speculative note
we remark that this
physics might be of
relevance to quantum dot
problems. Quantum dots
are usually described in terms of the
Anderson model. However, there are certainly
other angular momentum channels whereby
the local charge on the dot and the external
environment can interact. As the external
potential in the leads is varied one is forced to
pass through a potential in which this mixed valence
condition must be satisfied. At this
potential the aforementioned singular
behavior should be observed.

\subsection{Multi-channel Kondo problem}\label{multichannelsection}

 Blandin and Nozi\`eres  \cite{blandin} invented the
multichannel  Kondo problem and  gave convincing
arguments for a local singular Fermi-liquid behavior of it. Since
then it has been solved by a multitude of sophisticated methods. For
an overview of these and of applications of the multichannel Kondo
problem, we refer to  \cite{multichannelreview}.

The multichannel Kondo problem is the generalization of the Kondo
problem to the case in which the  impurity spin has arbitrary spin $S$
and is coupled to  $n$ ``channels'' of conduction electrons. The
Hamiltonian  is
\begin{equation}
\label{mckondo}
{\mathcal H}_{mcK} = t \sum_{\ell=1}^n \sum_{i<j} c^{(\ell)\dagger}_{i \alpha} c^{(\ell)}_{j \alpha} + J {\bf S} \cdot c^{(\ell)\dagger}_{0}
{\bf \sigma} c_{0}^{(\ell)}~.
\end{equation}
Here $\ell$ is the channel index. Degeneracy, the key to SFL behavior,
is enforced
through equal antiferromagnetic coupling $J>0$ for all the channels.
 When the couplings to the various channels
are not all the same, at low enough temperatures a crossover to local
Fermi-liquid behavior in the channel with the largest $J_{\ell}$
always occurs \cite{affleck}.
 This crossover temperature is in general quite large
compared to $T_K$ because channel asymmetry is a {\em relevant}
perturbation about the symmetric fixed point.  Therefore
in comparing this theory with SFL
behavior in experiments, one should ensure that one is above the
crossover temperature.

The simple Kondo problem is the case $2 S= 1 =n$. In this case at low
temperatures  a singlet
state is formed of the impurity state and the conduction electron
electrons in the appropriate channel. In the general multi-channel
case in which $2S=n$,
 the physics is essentially the same, since there are exactly the right
number of conduction electron channels to compensate the impurity spin
at low temperatures. Thus, at low temperatures an effective spin 0 state is
formed again, and the properties of the {\em compensated
Kondo problem} $2S=n$ at finite temperature is that of crossover from
a weakly interacting
problem above the Kondo temperature $T_K$ to a strongly interacting
problem below $T_K$.

\begin{figure}
\begin{center}
\epsfig{figure=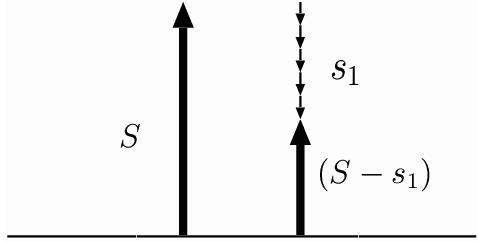,width=0.45\linewidth}
\end{center}
\caption[]{In a multichannel problem, a (Hund's rule coupled) spin $S$
  is compensated by total spin $S_1$ ($=n/2$) in $n$ different
  conduction electron channels leaving a net uncompensated spin as
  shown, or exact compensation or over-compensation when special
  properties may arise as discussed in the text. }\label{underscreenedfig}
\end{figure}

The physics of the {\em underscreened Kondo problem} $2S>n$ is
different\footnote{Since it is hard to imagine that
the angular momentum states of the impurity are larger than of the
conduction electron states about the impurity, such models may be
 regarded  of purely theoretical interest.
 See however Sec (6.5)}. In this case,
 there are not enough conduction electron
spins to compensate the impurity spin. As a result, when the
temperature is lowered and the effective coupling increases, only a
partial compensation of the impurity spin occurs by conduction electrons with
spin opposite to it. As depicted in Fig. \ref{underscreenedfig}, a net
spin in the same direction as the impurity spin then remains at the
impurity site. Since  the conduction electrons with
their spin in the same directions as the impurity spin can then still
make virtual excitations by hopping on that site while the site is
completely blocked for conduction electrons with opposite spin, a net
{\em ferromagnetic} coupling remains between the remaining effective
spin and the conduction electrons. As a result, the low temperature
physics of the
underscreened Kondo problem is
that of the ordinary {\em ferromagnetic Kondo problem}. To be more
precise, the approach to the fixed point is analogous to that in the
ferromagnetic Kondo problem along the boundary $J_z = -J_\pm$,
because the impurity must decouple (become pseudo-classical) at the
fixed point.

\begin{figure}
\begin{center}   
  \epsfig{figure=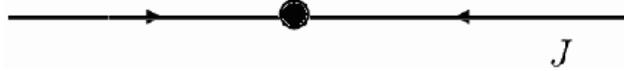,width=0.6\linewidth}
\end{center}
\caption[]{Flow diagram for the degenerate over-screened Kondo problem
  exhibiting a critical point.} \label{overscreenedfig}
\end{figure}

In the {\em overscreened Kondo problem}  $2S<n$,  there
are more channels than necessary to compensate the impurity spin. At
low temperatures, all $n$
channels tend to compensate the impurity spin due to the Kondo effect.
Channel democracy now causes an interesting problem.
As the effective interaction $J$ scales to stronger values,
 a local effective spin with direction {\em
opposite} to the impurity spin results. This effective spin must
 have an effective {\em antiferromagnetic } interaction with the
conduction electrons, since now the virtual excitations of conduction
electrons with spins {\em opposite to the effective local spin}  lower their
energy. This  then gives a new Kondo problem with a new effective
interaction, and so on. Of course, in reality one does not get a succession
of antiferromagnetic Kondo problems
 --- the net effect is that a new stable finite-$J$ fixed
point appears. As sketched in Fig. \ref{overscreenedfig}, the
renormalized effective
interaction flows to this fixed point both from the  strong-coupling  as well
as from the weak-coupling side. One can understand this intuitively
from the above picture: if one starts with a large initial value of
$J$, then in the next order of perturbation theory about it, the
interaction is   smaller, since in
perturbation theory the effective interaction due to virtual
excitations decreases with increasing $J$. This means that $J$ scales
to smaller values. Likewise, if we start from small $J$, then
initially $J$ increases due to the Kondo  scaling, but once $J$
becomes sufficiently large, the first effect which tends to decrease
$J$ become more and more important.
The finite-$J$ fixed point leads to nontrivial exponents for the
low-temperature   behavior of
quantities like the specific heat,
\begin{equation}  \label{cvoverscreened}
{{C_v}\over{T}} \propto \left( {{T}\over{T_K}}
\right)^{{n-2}\over{n+2}}~.
\end{equation}
For $n=2$, the power law behavior on the right hand side is replaced by a
$\ln T$ term.

Another way of thinking about the problem is in the spirit of Wilson's
renormalization group: Consider the problem of two channels
interacting with a $S=1/2$ impurity. The conduction states can be
expressed as linear combinations of concentric orbitals of conduction
electrons around the impurity. These successive  orbitals peak further
and further  away from the impurity. Consider first the exchange
coupling of the orbitals in each of the two channels peaking at the
impurity site. Each of them has $S=1/2$. Only one linear combination
of the two channels, call it red, can couple, while the other (blue)
does not. So after the singlet with the impurity is formed, we are
left with a $S=1/2$, color blue problem. We must now consider the
interaction of this effective impurity with the next orbital and so
on. It is obvious that to any order we will be left with a spin 1/2
problem in a color. Conformal field theory methods first showed that
the ground state is left with $1/2 \ln 2 $ impurity. A nice
application of the bosonization method \cite{Emery-1992}
identifies the red and blue
above as linear combinations of the fermions in the two channels so
that one is purely real, the other purely imaginary. The emergence of
new types of particles --- the Majorana fermions in this case ---
often happens at Singular Fermi-liquids.

As a detailed calculation confirms \cite{affleck}, the
$J_1=J_2$ fixed point is unstable, and the flow is like sketched in
Fig. \ref{j1j2flow}. This means that the $J_1=J_2$ fixed point is a
Quantum Critical Point: in the $T-J_1/J_2$ phase diagram, there is a
critical point at $T=0, J_1/J_2=1$. Moreover, it confirms that
asymmetry in the couplings is a relevant perturbation, so that the SFL
behavior is unstable to any introduction of differences between the
couplings to the different channels. The crossover temperature
$T_{\times} $ below which two-channel behavior is replaced by the
approach to the Kondo fixed point is  \cite{affleck}
\begin{equation}
T_{\times} = {\mathcal O}\left( T_K(\bar{J})
\left[(J_1-J_2)/\bar{J}\right]^2\right)~,
\end{equation}
where $\bar{J}= (J_1+J_2)/2$.
\begin{figure}
\begin{center}   
  \epsfig{figure=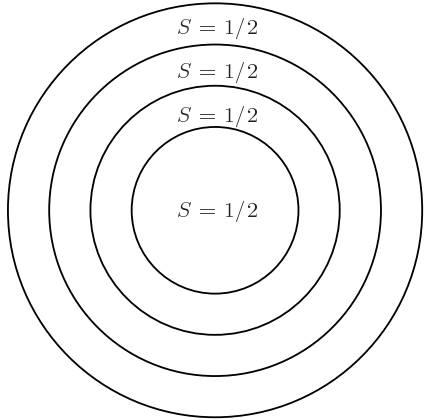,width=0.4\linewidth}
\end{center}
\caption[]{Effective shells in Energy space in Wilson's method. The
  first shell integrates over a fraction $\lambda$ of the top of the
  band, the next shell $\lambda$ of the rest, and so on. In the
  two-channel $S=1/2$ problem, a $S=1/2$ effective impurity is left at
  every stage of interpretation.} \label{shells}
\end{figure}

\begin{figure}
\begin{center}   
  \epsfig{figure=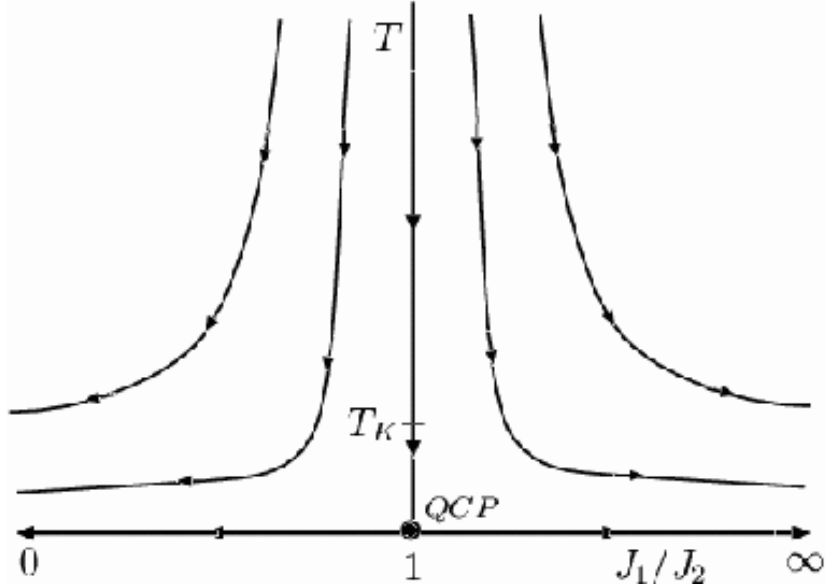,width=0.8\linewidth}\label{fig0}
\end{center}
\caption[]{Flow in temperature-anisotropy plane for the two-channel
  Kondo problem with coupling constant $J_1$ and $J_2$.} \label{j1j2flow}
\end{figure}

The overscreened Kondo problem  illustrates
that the SFL behavior is associated with the occurrence of degeneracy:
the critical point requires degeneracy of the two orthogonal channels.

\begin{figure}

\begin{center}   
  \epsfig{figure=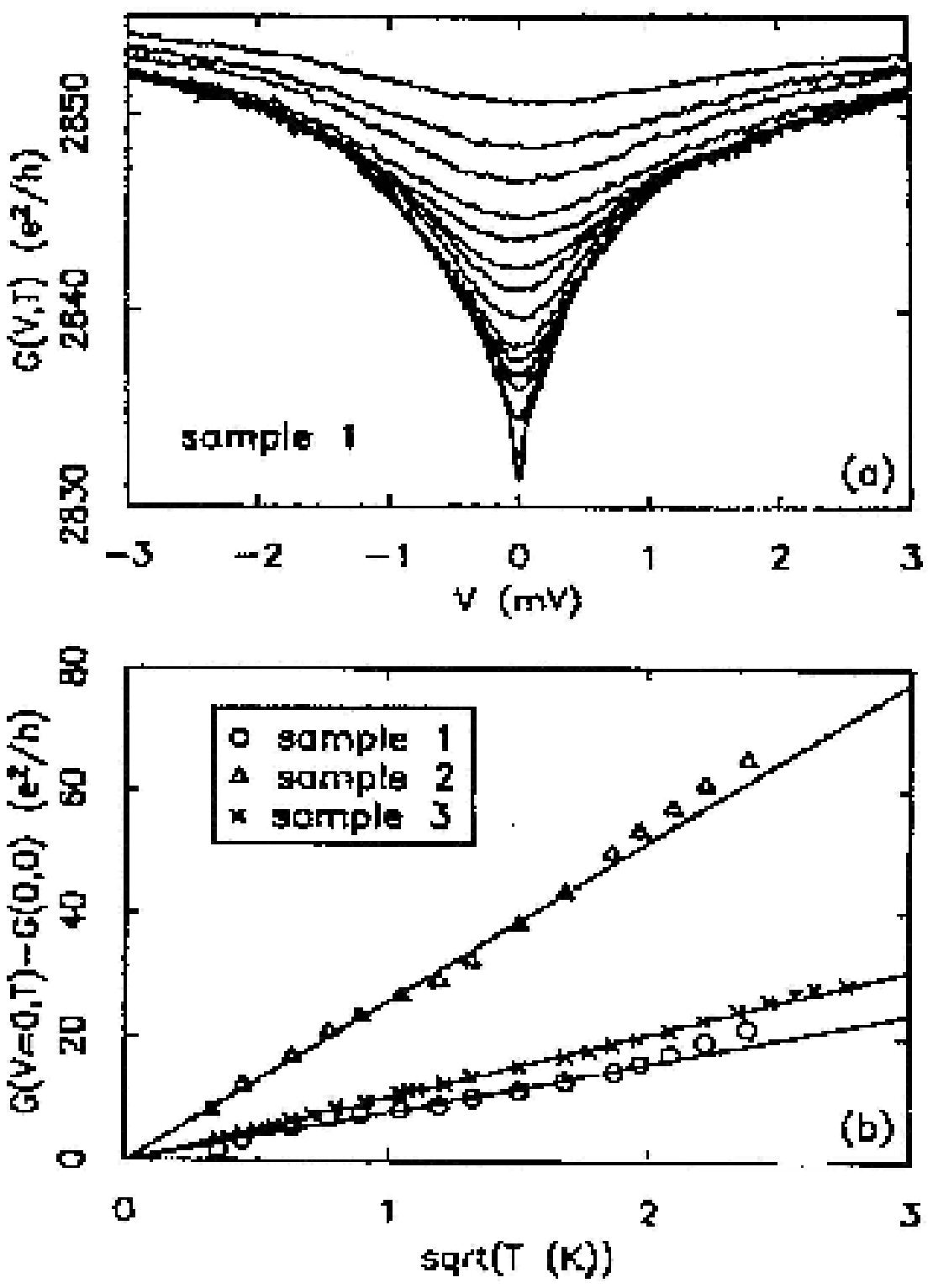,angle=0.5,width=0.62\linewidth}
\end{center}
\caption[]{{\em (a)} Differential conductance $G(V,T)$ as a function
of voltage $V$ in measurements on metal point contacts by Ralph {\em
et al.} \cite{ralph}, for various temperatures ranging from 0.4 K
(bottom curve) to 5.6 K (upper curve). Note the $\sqrt{V}$ type
behavior developing as the temperature decreases. {\em (b)} The zero bias
($V=0$) conductance as a function of temperature for three different
samples shows a $G(0,T)-G(0,0)\sim \sqrt{T}$ behavior. The scaling behavior as
a function of voltage and temperature is consistent with two-channel
Kondo behavior \cite{ralph}.} \label{tunnelingplot}
\end{figure}

An interesting application of the two-channel Kondo problem
is obtained by considering the spin label
to play the role of the channel index, while the Kondo coupling is in
the orbital angular momentum or crystal field states for impurities at
symmetry sites in crystals \cite{kim,cox,multichannelreview}.
 Another possibility that has been considered is that of
scattering of conduction electrons of two-level tunneling centers with
different angular momentum states \cite{zarand}.
In this case, the tunneling model
translates into a  model with $x$ and $z$ coupling only, but this
model flows towards a Kondo-type model with equal $x$ and $y$ spin
coupling.
For both types of proposed applications, one has to worry
about the breaking of the symmetry, and about the question of how dilute the system
has to be for a Kondo-type model to be realistic. Interesting results
in the tunneling conductance of two metals through a narrow
constriction, shown in Fig. \ref{tunnelingplot}
appear to bear resemblance to the properties expected of a degenerate
two-channel Kondo effect \cite{ralph,cox3}, but this interpretation is
not undisputed \cite{altshulercomment,ralphreply}.
Applications to impurities in heavy fermions will
be briefly discussed in section \ref{specialcompl}.

\subsection{The Two-Kondo-Impurities Problem}\label{twokondosection}

In a metal with a  finite concentration of magnetic impurities, an important
question is what the (weak) interaction between the
magnetic impurities does to the Kondo physics --- that the effect
might be substantial is already clear from the fact that the Kondo
effect is seen in logarithmic corrections about the high-temperature
local moment fixed point while the RKKY interaction between the
moments mediated by the conduction electrons occur as a power law
correction\footnote{Stated technically, the Kondo Hamiltonian is a marginal
operator while the RKKY operator is a relevant operator about the
local moment fixed point: In a perturbation calculation the interaction produces
corrections of ${\mathcal O}(1/T)$ compared to a $\ln(T)$ correction of the Kondo effect.}.
 Stated simply $T_K \sim {{1}\over{N(0)}} \exp(
-{{1}\over{N(0) J}} )$ while the RKKY interaction $I\sim
J^2 N(0)$.  The existence
of mixed-valence and heavy
fermion metals makes this much more than an academic question.
 The question of the competition between these two effects,
and in particular whether long range magnetic order can arise, was
first posed by Varma \cite{varma2} and by Doniach
\cite{doniach},  who gave the obvious answer that
RKKY interactions will be ineffective only when the Kondo temperature
below which the local spin at each
impurity is zero is much larger than the RKKY interaction I. Considering
that $ J N(0)$ is usually $\gg 1$, this is unlikely for $S=1/2$
problems, but for large $S$, as encountered typically in rare earths
and actinides, it is possible in some cases.  But the vast majority of
rare earths  and actinide compounds show magnetic order and no Heavy
fermion magnetic behavior.

 The two-Kondo-impurity problem  serves as a first step to understanding
some of the physics of heavy fermions. We will however only consider
the $S=1/2$ problem and work with unrealistic $ J N(0)$ so that the
competition between the Kondo effect and RKKY is possible.

In this subsection we will summarize the results \cite{jones,Jones-1988,affleck2,affleck3,Sire-1993}
 for the {\em two
Kondo impurity} problem; like the models we discussed above, this one
also has a quantum critical point at which SFL behavior is
found. However as for other impurity problems, an unrealistic symmetry
must be assumed for a QCP and attendant SFL behavior.

The two-Kondo-impurity Hamiltonian is defined as
\begin{equation}
\label{2kondo}
{\mathcal H} =  t \sum_{k,\sigma}
c^{\dagger}_{k\sigma} c_{k \sigma} + J\; \left[ {\bf S}_1
\cdot  {\bf \sigma}(r_1)  +  {\bf S}_2 \cdot {\bf \sigma}(r_2) \right]~.
\end{equation}
In this form, the problem has a symmetry with respect to the midpoint
between the two impurity sites $r_1$ and $r_2$, and hence one can
define even ({\em e}) and odd ({\em o}) parity states relative to this
point. In the approximation that the $k$-dependence of the couplings
is neglected \cite{jones} the two Kondo impurity
Hamiltonian can then be transformed to
\begin{eqnarray}
{\mathcal H} =  {\mathcal H}_{non-interacting} &+&  \sum_{k k'} \left(
{\bf S}_1 + {\bf S}_2 \right) \cdot \left[ J_e c^\dagger_{k' e} {\bf \sigma}
c_{ke} + J_o c^\dagger_{k'o} {\bf \sigma} c_{ko} \right] \nonumber \\
 &+&  \sum_{k k'} J_m \left(
{\bf S}_1 - {\bf S}_2 \right) \cdot \left[ c^\dagger_{k' e} {\bf \sigma}
c_{ko} +  c^\dagger_{k'o} {\bf \sigma} c_{ke} \right]~. \label{2kondo2}
\end{eqnarray}
The coupling between the spin and orbital channels generates an effective RKKY
interaction
\begin{equation}
{\mathcal H}_{RKKY} = I_0(J_e, J_o, J_m) ~{\bf S}_1 \cdot {\bf S}_2
\label{Hrkky}
\end{equation}
 between the two impurity spins, with $I_0 = 2 \ln 2
(J_e^2+J_o^2-2J_m^2) $ for $t=1$.

The main results of a numerical Wilson-type renormalization group
treatment of this model are the following:

{\em (i)} For ferromagnetic coupling $I_0>0$ or for a small
antiferromagnetic coupling $I_0 > -2.24 T_K$, where $T_K$ is the Kondo
temperature of the single-impurity problem, one finds that there is a
Kondo effect with
\begin{equation} \label{rkkycor}
\langle {\bf S}_1 \cdot {\bf S}_2 \rangle \neq 0~,
\end{equation}
{\em unless} $I_0$ is very small, $|I_0 / T_K| \ll 1$. Since for
uncorrelated impurity spins $\langle {\bf S}_1 \cdot {\bf S}_2 \rangle
=0$, (\ref{rkkycor}) expresses that although in the RG language the RKKY
interaction is an irrelevant perturbation, it is quite
important in calculating physical properties due to large
``corrections to scaling''. Another  feature of
the solution in this regime is the fact that the phase shift is
$\pi/2$ in {\em both} channels. This means that at the fixed point
the even-parity
channel and the odd-parity channel have {\em independent} Kondo effects,
each one having one electron pushed below the chemical potential in
the Kondo resonance. As discussed below, this is due to particle-hole
 symmetry assumed in
the model --- without it, only the sum of the phase shifts in the two-channels is fixed.

{\em (ii)} There is no Kondo effect for $I_0 < -2.24~T_K$. In this
case, the coupling between the impurities is so strong that the
impurities form a singlet among themselves and decouple from  the conduction
electrons. There is no phase shift at the fixed
point. Also, in this case the total spin $S_{tot} =0$, but the
impurity spins become only singlet like, $\langle {\bf S}_1 \cdot
{\bf S}_2 \rangle \approx -3/4$ for very strong coupling, $I_0 \ll
-2.24~T_K$. So again there are important ``corrections to scaling''.

{\em (iii)} The point $I_0 = -2.24~T_K$ is a true critical point, at which the staggered
susceptibility $\langle ({\bf S}_1-{\bf S}_2)^2 \rangle /T $
diverges. Moreover, at this point the specific heat has a logarithmic
correction to the linear $T$ dependence,  $C_v \sim T\ln T$, while the
impurity spin correlation function $\langle {\bf S}_1\cdot {\bf S}_2
\rangle$ becomes equal to $-1/4$ at this value.

Although the approximate Hamiltonian (\ref{2kondo2}) has a true
quantum critical point with associated SFL behavior, we stress that
the analysis shows that this critical point is destroyed by any
$c^\dagger_{ke}c_{ko}$ coupling. A coupling of this type is not
particle-hole symmetric; since the approximate Hamiltonian
(\ref{2kondo2}) is particle-hole symmetric, these terms are not
generated under the renormalization group flow for
(\ref{2kondo2}). The physical two-Kondo-impurity
problem (\ref{2kondo}), on the other hand, is {\em not} particle-hole
symmetric. Therefore, the physical two
Kondo impurity problem does {\em not} have a true quantum critical
point --- in other words, when the two-Kondo-impurity Hamiltonian
(\ref{2kondo}) is approximated by (\ref{2kondo2}) by ignoring $k$-{\em
independent} interactions, relevant terms which destroy the quantum
critical point of the latter Hamiltonian are also dropped. This has
been verified in explicit analysis keeping the symmetry breaking terms
\cite{Sire-1993}.

An illuminating way to understand the result for the two-impurity
Kondo problem, is to note that the Hamiltonian can be written in the
following form:
\begin{equation}
{\mathcal H} = \left[  \hspace*{2mm} \begin{array}{cc}
\left[ \begin{array}{cc} {\mathcal H}^{even}_{S=0} & 0 \\ 0 &
{\mathcal H}^{odd}_{S=0} \end{array}   \right]
&
 \left[ \begin{array}{cc} 0 & {\mathcal H}_{mix}  \\
{\mathcal H}_{mix} & 0 \end{array}  \right]
\\ \\
\left[ \begin{array}{cc} 0 & {\mathcal H}_{mix}  \\
{\mathcal H}_{mix} & 0 \end{array} \right]
&
\left[ \begin{array}{cc} {\mathcal H}^{even}_{S=1} & 0 \\ 0 &
{\mathcal H}^{odd}_{S=1} \end{array} \right]
\end{array} \hspace*{2mm}
\right] ~,\label{2kondorepr}
\end{equation}
where ${\bf S}={\bf S}_1+{\bf S}_2$ is the total impurity spin.

In this representation, the Hamiltonian ${\mathcal H}_{mix} $ couples
the $S=0$ and $S=1$ state, and the following interpretation naturally
emerges:  for large antiferromagnetic
values of $I_0$, ${\mathcal H}_{S=0}$ is lower in energy than
${\mathcal H}_{S=1}$, while for large ferromagnetic coupling $I_0$,
the converse is true. The two-Kondo impurity coupling can thus be
viewed as one in which by changing $I_0$, we can tune the relative
importance of the upper left block and the lower right block of the
Hamiltonian. In general, the two types of states are mixed by
${\mathcal H}_{mix}$, but at the fixed point ${\mathcal H}^*_{mix}
\rightarrow 0$. This implies that there is a critical value of $I_0/T_K$ where
the $S=0$ and $S=1$ states are degenerate, and where SFL behavior
occurs. At this critical value, the impurity spin is a linear
combination of a singlet and triplet state with $\langle
{\bf S}_1\cdot {\bf S}_2 \rangle =-1/4$
(i.e. a value in between the singlet value
-3/4 and the triplet value 1/4) and the
singular low-energy fluctuations give
rise to the anomalous specific
heat behavior.

Within this scenario, the fact that the susceptibility $\chi$ is
divergent at the critical point signals that a term ${\bf H}\cdot
({\bf S}_1-{\bf S}_2) $ lifts  the  spin degeneracy.  Moreover, the
leading irrelevant operators about the  fixed point are all
divergent at the critical point --- of course, this just reflects the
breakdown of the Fermi-liquid description.

 The reason for ${\mathcal H}_{mix}
\rightarrow 0$ is as follows. It arises from the last term in
(\ref{2kondo2}) which is not  particle-hole symmetric because under
even-odd interchange both the spin term and the fermion
electron terms change sign. At the Kondo fixed
point, the leading operators must all be biquadratic in fermions. An
${\mathcal H}_{mix}$ in that case would be of the form
$c^\dagger_{k\sigma e} c_{k'\sigma o}$ and and such a term by itself
would break particle hole
symmetry.

In the two-Kondo impurity problem, one encounters again the
essentials of degeneracy for quantum critical points and the need
for (unphysical) constraints to preserve the singularity. Once again
new types of quantum numbers can be invoked in the excitations about
the QCP.

From the point of view of understanding actual phenomena for problems
with a moderate concentration of impurities or in reference to
heavy fermion compounds, the importance of the solution to the
two-Kondo impurity problem is the large correction to scaling found in
the Wilson-type solution away from the special symmetries required to
have a QCP. These survive quite generally and must be
taken into account in constructing low-energy effective Hamiltonians
in physical situations.

\section{SFL behavior for interacting fermions in one dimension}\label{1dphysics}

A variety of elegant mathematical techniques, including exact
solutions in certain nontrivial limits,  have been employed to
analyze the
problem of interacting electrons in one dimension. We  discuss the
 essential features of
these models here and their SFL properties. We also discuss whether the methods and the
results can be extended to higher dimensions.

 Some
cases where the models solved are experimentally realized include the
edge states of Quantum Hall liquids  and quasi-one
dimensional organic and inorganic compounds \cite{emery1,emery2}. In the
latter case, the asymptotic low energy properties are, however, unlikely
to be those of the 1D models because of the inevitable coupling to the
other dimensions which proves to be a relevant
perturbation. Nonetheless,
data on carbon nanotubes \cite{dekker} discussed in
 section \ref{nanotubessection} show  clear evidence of
one-dimensional interacting electron physics. Several
one-dimensional
 spin chains problems can also be transformed into problems
 of one-dimensional fermions \cite{sachdev}.

Our aim in this chapter is to highlight the physics leading to the
Singular Fermi Liquid behavior and to present the most important
results obtained.  Detailed reviews of the technical steps in the
various solutions as well as numerical calculations may be found in
\cite{gogolin,solyom}. As pointed out already in section \ref{routesto},
logarithmic singularities appear in the second order perturbative
calculation of the vertices in the 1D problem (both in the
particle-hole and particle-particle channels).  We present the $T=0$
phase diagram as obtained by summing those singularities by
perturbative RG. This will be followed by a discussion of the results
of the exact solution of the Tomonaga-Luttinger and the more general
model for special values of the coupling constants (along the so
called ``Luther-Emery'' line \cite{lutheremery}).

One special feature of 1D physics is that the low energy excitations
can be described by either fermions or bosons. The bosonic description
of the Tomonaga-Luttinger model is especially attractive and will be
presented below.

A distinctive feature of one dimensional physics is that
single-particle as well as multiple-particle correlation
functions are expressible in terms of independent charge
and spin excitations, which, in general, propagate with
different velocities\footnote{Even a Fermi-liquid
displays distinct energy scales for charge and spin
(particle-hole) fluctuations because of the difference in the Landau
parameters in the spin-symmetric and spin-antisymmetric channels.
The phrase spin-charge separation should therfore be reserved for
situations, as in one-dimension, where the single-particle
excitations separate into objects
carrying charge alone and which carry spin alone.}  This feature
has been shown to arise due to extra conservation laws in one
dimensions \cite{castellani,metznerdic}. As we shall discuss,
 an extension of charge-spin separation
to higher dimensions is unlikely because there are no such
conservation laws.

The one dimensional singularities may be also seen as a manifestation of the
orthogonality catastrophe \cite{Anderson-1967} that we discussed
in section \ref{orthogonalitysection}. We shall see that this feature disappears
in higher dimension due to the effects of recoil.

\subsection{The One Dimensional Electron Gas}

The Hamiltonian describing the low energy dynamics reads
\begin{equation}
{\mathcal H}= {\mathcal H}_{0}+ {\mathcal H}_{forward} + {\mathcal
  H}_{backward} +{\mathcal H}_{Umklapp}~.  \label{centralzohar}
\end{equation}
In (\ref{centralzohar}) ${\mathcal H}_0$ is the free electron Hamiltonian
\begin{equation}
{\mathcal H}_{0} = \sum_{k} \epsilon_{k} c^{\dagger}_{k} c_{k}~.
\end{equation}
In one dimension the particles move in two branches, either to the left or the right.
Define the charge density
operators $\rho_r$ and the spin density operators $S_r$
for the two branches, $r=\pm $,
\begin{equation}
\rho_{r}=\Sigma_{\sigma=\pm 1} \psi_{r,\sigma}^{\dagger} \psi_{r,\sigma}~,~~~~~  ~ S^{z}_{r} = \frac{1}{2} \Sigma_{\sigma,\sigma^{\prime}} \psi^{\dagger}_{r,\sigma} \tau^{z}_{\sigma,\sigma^{\prime}} \psi_{r,\sigma^{\prime}}~,
\end{equation}
where $\tau^{z}$ is a Pauli-matrix.
 The Fourier components of the particle density operators
read
\begin{equation}
\rho_{r,\sigma}(q) = \sum_{k} c_{r,\sigma,k+q}^{\dagger} c_{r,\sigma,k} = \rho^{\dagger}_{r,\sigma}(-q)~.
\end{equation}
These bilinears may be explicitly shown to obey Bose commutation relastions.
If we linearize $\epsilon_k$ about the two Fermi-points, the energy of a particle-hole pair
created by $\rho_{r,\sigma}(q)$,
\begin{equation}
\epsilon_{r,k+q}- \epsilon_{r,k} = rv_{F}q~,
\end{equation}
is independent of $k$ in one dimension (provided
the band cutoff is taken to infinity). States created by
$\rho_{r,\sigma}(q)$ are
linear combinations of individual electron-hole excitations all of
which have the same energy and are therefore eigenstates of ${\mathcal
H}_{0}$.
It follows that for $q>0$, $\rho_{r=+,\sigma}(q)$
[$\rho_{r=-,\sigma}(q)$] is a raising [lowering] operator.
Consequently ${\mathcal H}_{0}$ may be expressed in terms of the
density operators:
\begin{equation}
{\mathcal H}_{0}= \frac{2 \pi v_{F}}{L} \sum_{r=\pm } \sum_{q>0} \rho_{r,\sigma}(rq)
\rho_{r,\sigma}(-rq)~.
\end{equation}

Upon separating the densities on a given branch into charge and spin pieces,
\begin{equation}
\rho_{r \sigma}(x) = \frac{1}{2} [ \rho_{r}(x)+ \sigma S_{r}^{z}(x)]~,
\end{equation}
the {\em{free Hamiltonian}} may be expressed as a
sum  in the spin and charge degrees of
freedom:
\begin{equation}
{\mathcal H}_{0}= \sum_{r} {\cal{H}}_{0}[\rho_{r}] + \sum_{r} {\cal{H}}_{0}[S_{r}^{z}].
\end{equation}
 It follows that in the noninteracting problem spin and charge have
identical dynamics and
propagate in unison. Once interactions are introduced, the electron
will ``{\em fractionalize}'' and spin and charge dynamics will, in general,
differ.

We now discuss the various interaction terms in the Hamiltonian.
The piece describing the forward scattering events
$(k_{F},\sigma;-k_{F},\sigma^{\prime})$ $ \rightarrow$ $ (k_{F},\sigma;-k_{F},\sigma^{\prime})$
and $(k_{F},\sigma;k_{F},\sigma^{\prime})$
$ \rightarrow (k_{F},\sigma;k_{F},\sigma^{\prime})$ reads
\begin{equation}
{\mathcal H}_{forward} =  {\mathcal H}_{2}+ {\mathcal H}_{4}~,
\end{equation}
with
\begin{eqnarray}
{\mathcal H}_{2} & = & \frac{1}{L} \sum_{q}
\sum_{\sigma \sigma^{\prime}} g_{2}^{\sigma \sigma^{\prime}}  \rho_{+, \sigma} (q)
\rho_{-,\sigma^{\prime}} (-q) ~,
\\ H_{4} & = &  \frac{1}{2 L} \sum_{q} \sum_{\sigma \sigma^{\prime}} g_{4}^{\sigma \sigma^{\prime}} [ \rho_{+,\sigma}(q)
\rho_{+,\sigma^{\prime}}(-q) + \rho_{-,\sigma}(q)
\rho_{-,\sigma^{\prime}}(-q)]~.
\end{eqnarray}
 The operators
$\rho_{r,\sigma}$ involve a creation and annihilation operator  on the
same branch. We also  define operators $\rho^r_\sigma$ which involve
operators on opposite branches,
\begin{equation}
\rho_{\sigma}^{r}= \sum_{k} c^{\dagger}_{r,\sigma,k} c_{-r,\sigma,k+q}~.
\end{equation}
In terms of these
the Hamiltonian describing  backscattering
interactions (the scattering event $(+k_{F},\sigma;-k_{F},\sigma^{\prime})
\rightarrow (-k_{F},\sigma;k_{F},\sigma^{\prime})$ and its reverse)
becomes
\begin{equation}
{\mathcal H}_{backwards}= {\mathcal H}_{1}= g_{1} \sum_{q} \sum_{\sigma \sigma^{\prime}} g_{1}^{\sigma \sigma^{\prime}} \rho_{\sigma}^{+}(q) \rho_{\sigma^{\prime}}^{-}(-q)~.
\end{equation}
and the Umklapp term reads
\begin{equation}
{\mathcal H}_{3} = \frac{1}{2L} \sum_{q} \sum_{\sigma \sigma^{\prime}} g_{3}^{\sigma \sigma^{\prime}} [\rho_{\sigma}^{+}(q) \rho_{\sigma^{\prime}}^{+} (-q) + \rho^{-}_{\sigma}(q) \rho^{-}_{\sigma^{\prime}} (-q)]~.
\end{equation}

\begin{figure}
\begin{center}   
  \epsfig{figure=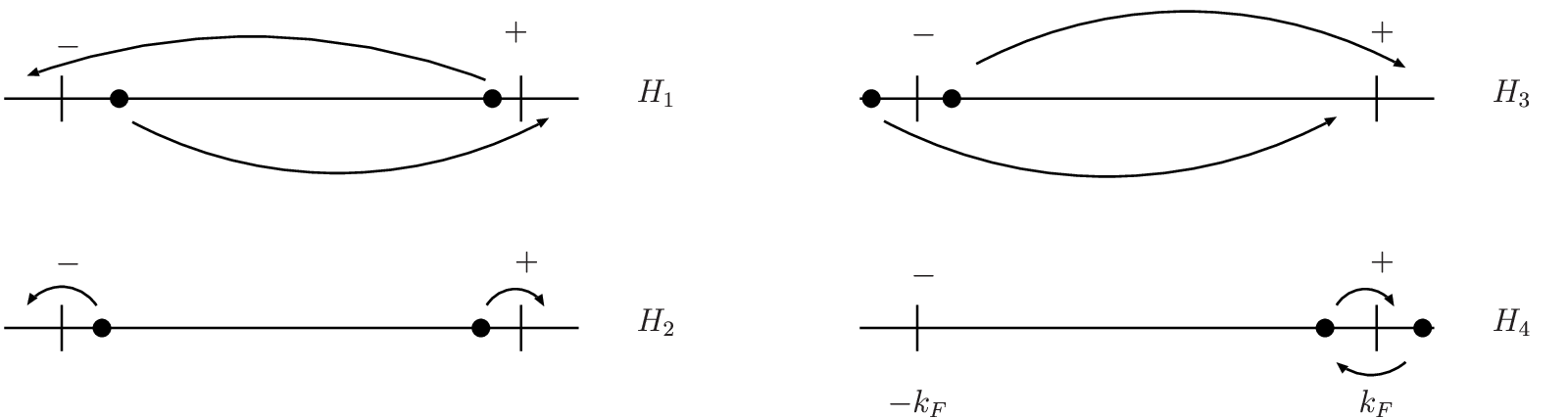,width=0.9\linewidth}
\end{center}
\caption[]{Pictorial representation of the low-energy interaction
  terms in the one-dimensional problem. After
\cite{metznercastellanidicastro}. The ``+'' and ``-'' points
are a shorthand for the two Fermi points $k=k_{F}$ and $(-k_{F})$
respectively.} \label{zoharfig1}
\end{figure}

In all these expressions, the coupling constants may be spin
dependent:
\begin{equation}
g_{i}^{\sigma \sigma^{\prime}} = g_{i ||}
\delta_{\sigma \sigma^{\prime}} + g_{i \perp}
\delta_{\sigma -\sigma^{\prime}} ~.
\label{spin_g}
\end{equation}

As the terms $H_{1 ||}$ and $H_{2 ||}$ describe the same process
we may set $g_{1 ||}=0$ with no loss of
generality. Umklapp
processes are important only
when $4 k_{F}$ is a reciprocal lattice
vector so that all scattering particles
may be near the Fermi points.
If the one dimensional electronic
system is sufficiently incommensurate
then these processes may be neglected.
 
The condition for spin rotation invariance $[{\mathcal H},\vec{S}]=0$ reads
\begin{equation}
g_{2 \perp} - g_{1 \perp} = g_{2 ||}- g_{1 ||}~.
\end{equation}

\begin{figure}
\begin{center}   
  \epsfig{figure=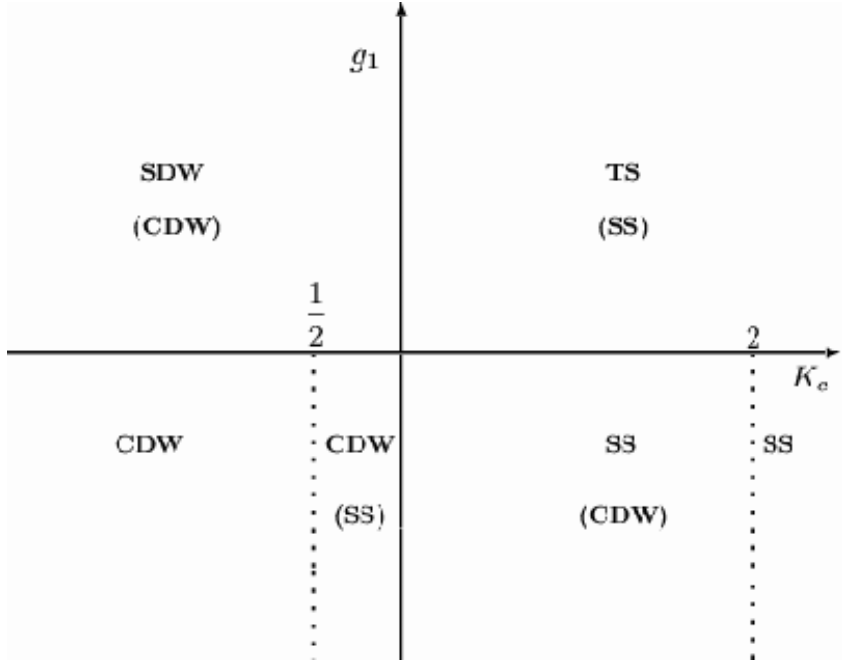,width=0.8\linewidth}
\end{center}
\caption[]{Phase diagram for one-dimensional interacting fermions in
the $(K_{c},g_{1})$ plane with $g_{1||}=0$ in  Eqn. (\ref{spin_g}) and $K_c$
given by Eq.\ref{kcsexpressions}~ .
$SDW$ and $(CDW)$ in the upper left quadrant indicates
that both the spin density  wave susceptibility and the charge density
wave susceptibility diverge as $T\to 0$, but that the spin density
wave susceptibility diverges a factor $\ln^2 T$ faster than the charge
density wave susceptibility --- see Eq. (\ref{onedchieqs}). Other sectors are
labeled accordingly. From \cite{schulzlectures}.} \label{zoharfig11}\label{phasediagram1d}
\end{figure}

A guide to the behavior of the  Hamiltonian
(\ref{centralzohar}) is obtained from the perturbative Renormalization Group
flow equations. In one dimension, there are no truly
ordered phases, but at $T=0$ there is algebraic long range order. One
may thus determine a ``phase diagram'' according to which
susceptibilities diverge as $T\rightarrow 0$: the one associated with
singlet superconductivity ($SS$), triplet superconductivity
($TS$), charge density wave ($CDW$) state at $2 k_{F}$,
and spin density wave ($SDW$) state at $2k_{F}$. The expressions for
these susceptibilities are given in Eq. (\ref{onedchieqs}) below, and
the resulting ``phase diagram'' is shown  in Fig. \ref{zoharfig11}.

\subsection{The Tomonaga-Luttinger Model}\label{tmmodel}

With forward scattering alone in (\ref{centralzohar}) and
after linearizing the kinetic energy
we obtain the Tomonaga-Luttinger (T-L)model. In terms of
fermion field operators $\psi_{r,\sigma}(x)$ and
$\psi_{r,\sigma}^{\dagger}(x)$, and density and spin
operators $\rho_r(x)$ and $S_{r}^{z}(x)$ in real space, the T-L
model is
\begin{eqnarray}
& {\mathcal H}^{T-L}& = \int dx~ \left[  - v_{F} \sum_{r,\sigma= \pm } r \psi_{r,\sigma}^{\dagger} i \partial_{x}
\psi_{r,\sigma} + \frac{1}{2} \sum_{r=\pm } g_{2,c} \rho_{r}(x)
\rho_{-r}(x)\right]\nonumber \\& & \hspace{-6mm} +\left. \sum_{r=\pm}  g_{4,c}
\rho_{r}(x) \rho_{r}(x)  + 2 \sum_{r= \pm } g_{2,s} S_{r}^{z}(x)
S_{-r}^{z}(x) + g_{4,s} S_{r}^{z}(x) S_{r}^{z}(x) \right].
\end{eqnarray}
where
\begin{equation}
g_{i}^{c}=\frac{g_{i ||}+ g_{i \perp}}{2} ~,~~~~~~ g_{i}^{s} = \frac{g_{i ||} - g_{i \perp}}{2}~.
\end{equation}

Note that the $g_{2,s}$ term is the only term
which break $SU(2)$ spin symmetry.
The T-L model is exactly solvable. After
all, as previously
noted, the (Dirac-like) kinetic energy Hamiltonian
${\mathcal H}_{0}$ is also quadratic in the density operators.
So the Hamiltonian is readily diagonalized by a Bogoliubov
transformation whereupon the Hamiltonian
becomes a sum of two {\em independent} (harmonic) parts describing noninteracting
charge and spin density waves: the charge and spin
density waves are the collective eigenmodes
of the system.

The simplest way to describe the TL  model
and to explicitly track down these collective
modes is via the bosonization of the electronic
degrees of freedom.  The reader should
be warned that many different conventions
abound in the literature. The bosonic
representation of the fermionic
fields proceeds by writing \cite{gogolin,nagaosa2,sachdev}
\begin{equation}
\psi_{r,\sigma}(x) = \lim_{a \rightarrow 0} \frac{\exp[ir(k_{F}x+ \Phi_{r \sigma}(x))]}{\sqrt{2 \pi a}} F_{r \sigma}~,
\end{equation}
where $a$ is a short distance
regulator.
$\Phi_{r \sigma}(x)$ satisfies
\begin{equation}
[\Phi_{r
\sigma}(x),\Phi^{\dagger}_{r^{\prime},\sigma^{\prime}}(x^{\prime})]
= - i \pi \delta_{r, r^{\prime}} \delta_{\sigma, \sigma^{\prime}}
\mbox{sign} (x-x^{\prime})~.
\end{equation}

The ``Klein factors'' $F_{r \sigma}$ are chosen such that
the proper fermionic anticommutation relations
are reproduced. The exponential envelope $\exp[i \Phi_{r \sigma}(x)]$
represents the slow bosonic collective degrees of freedom
which dress the rapidly oscillating part
$F_{r \sigma} \exp[i k_{F}x]$
 describing the energetic particle
excitations near the Fermi points.

$\Phi$ may be written in terms of the
bosonic fields $\phi_{c,s}$ and their conjugate momenta
$\partial_{x}\theta_{c,s}$
\begin{equation}
\Phi_{r,\sigma} = \sqrt{\pi/2}[(\theta_{c}-r \phi_{c})+\sigma(\theta_{s}-r \phi_{s})]~.
\end{equation}

In terms of the new variables, the charge and
spin densities read
\begin{equation}
\rho(x) = \sum_{r} \rho_{r}(x) = \sqrt{\frac{2}{\pi}} \partial_{x} \phi_{c} ~,~~~~~~ S^{z}(x) = \sum_{r} S_{r}^{z}(x) = \sqrt{\frac{1}{2 \pi}} \partial_{x} \phi_{s}~.
\end{equation}

In the $(\theta_{c,s},\phi_{c,s})$ representation
the Tomonaga-Luttinger Hamiltonian
becomes a sum of two decoupled oscillators describing
the gapless charge and spin density wave eigenmodes
\begin{equation}
{\mathcal H}^{T-L}= \int dx ~\sum_{\nu=c,s}
\frac{v_{\nu}}{2}\left[K_{\nu}(\partial_{x} \theta_{\nu})^{2}+
\frac{(\partial_{x} \phi_{\nu})^{2}}{K_{\nu}}\right]
\equiv {\cal{H}}_{s}^{T-L}+ {\cal{H}}_{c}^{T-L}.
\end{equation}

The velocities of the collective charge and spin modes
are easily read of by analogy to a harmonic string:
\begin{equation}
v_{c,s} = \sqrt{(v_{F}+ \frac{g_{4}^{c,s}}{\pi})^{2} -
(\frac{g_{2}^{c,s}}{\pi})^{2}}~.\label{vcs}
\end{equation}
Likewise, the moduli determining the power-law decay
of the correlations are
\begin{equation}
K_{c,s} = \sqrt{\frac{\pi v_{F} + g_{4}^{c,s} - g_{2}^{c,s}}
{\pi v_{F} + g_{4}^{c,s} + g_{2}^{c,s}}}~.\label{kcsexpressions}
\end{equation}
In section \ref{spinchargeseparationsection} we shall show how the above
expressions for the spin and charge density wave velocities follow
simply from the conservation of left and right moving particles in
the T-L model.
 
As previously noted, the
charge and spin velocities are degenerate
in the noninteracting model.
When interactions are introduced, the charge
and spin velocities ($v_{c}$ and $v_{s}$)
as well as the energy to create
spin and charge excitations
($(v_{s}/K_{s}$ and $v_{c}/K_{c}$
respectively) become different.
The charge constant $K_{c}$ is less than 1 for repulsive interactions,
which elevates the energy of the charge excitations,
while $K_{c}$ is greater than 1 for
attractive interactions.

\subsection{One-particle Spectral functions}\label{onepspf}

\begin{figure}
\begin{center}   
  \epsfig{figure=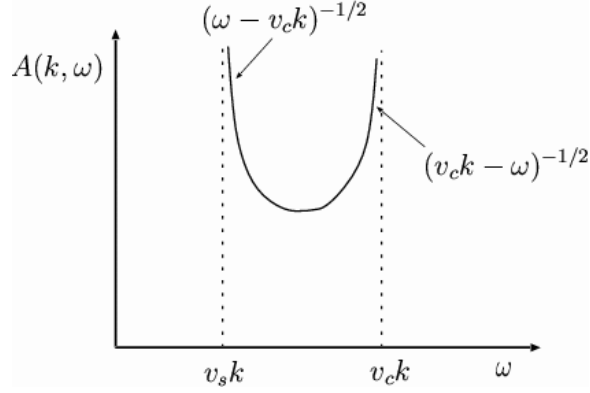,width=0.55\linewidth}
\end{center}
\caption[]{
The zero temperature spectral function
$A (k,\omega) = \mbox{Im} \{ G_{r=+1}^{<}(k,\omega) \}$
 as a function of $\omega$ for the case
($g_{2}=0,~ g_{4} \neq 0$) --- the ``one-branch Luttinger
liquid'' in which
the spin and charge velocities differ
but for which the the correlation exponents keep their
canonical $K_{c}=1$ according to (\ref{kcsexpressions}). In the
figure $v_{c}>v_{s}$ and $k>0$
are assumed. Note the inverse square root singularities.
 This is  a consequence of
 $K_{c}=1$ which makes $\gamma_{c}=0$. After Voit \cite{voit}.} \label{voitfig2}
\end{figure}

\begin{figure}
\begin{center}   
\epsfig{figure=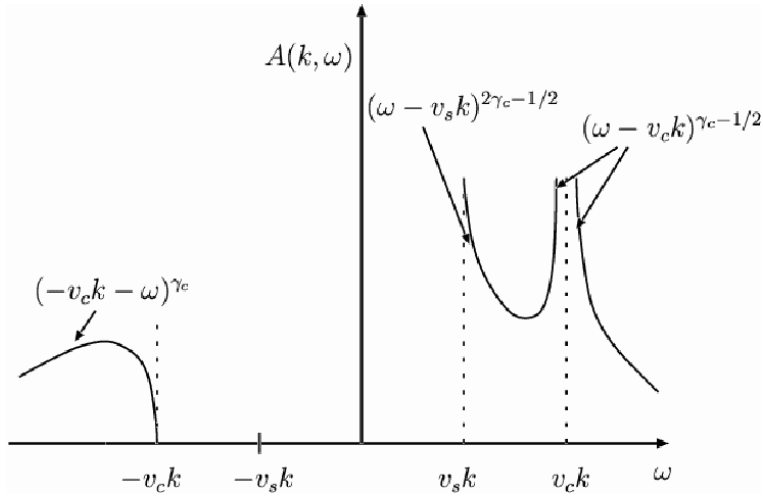,width=0.72\linewidth}
\end{center}
\caption[]{
The generic $( g_{2} \neq 0, g_{4} \neq 0$) zero temperature
spectral function. Note the broader range of non-trivial
singularities near $v_{s}k$ and $\pm v_{c} k$. Here both the
effect of spin-charge velocity difference
and the emergence of non-trivial exponents
is visible. After Voit \cite{voit}} \label{voitfig3}
\end{figure}

\begin{figure}
\begin{center}
  \epsfig{figure=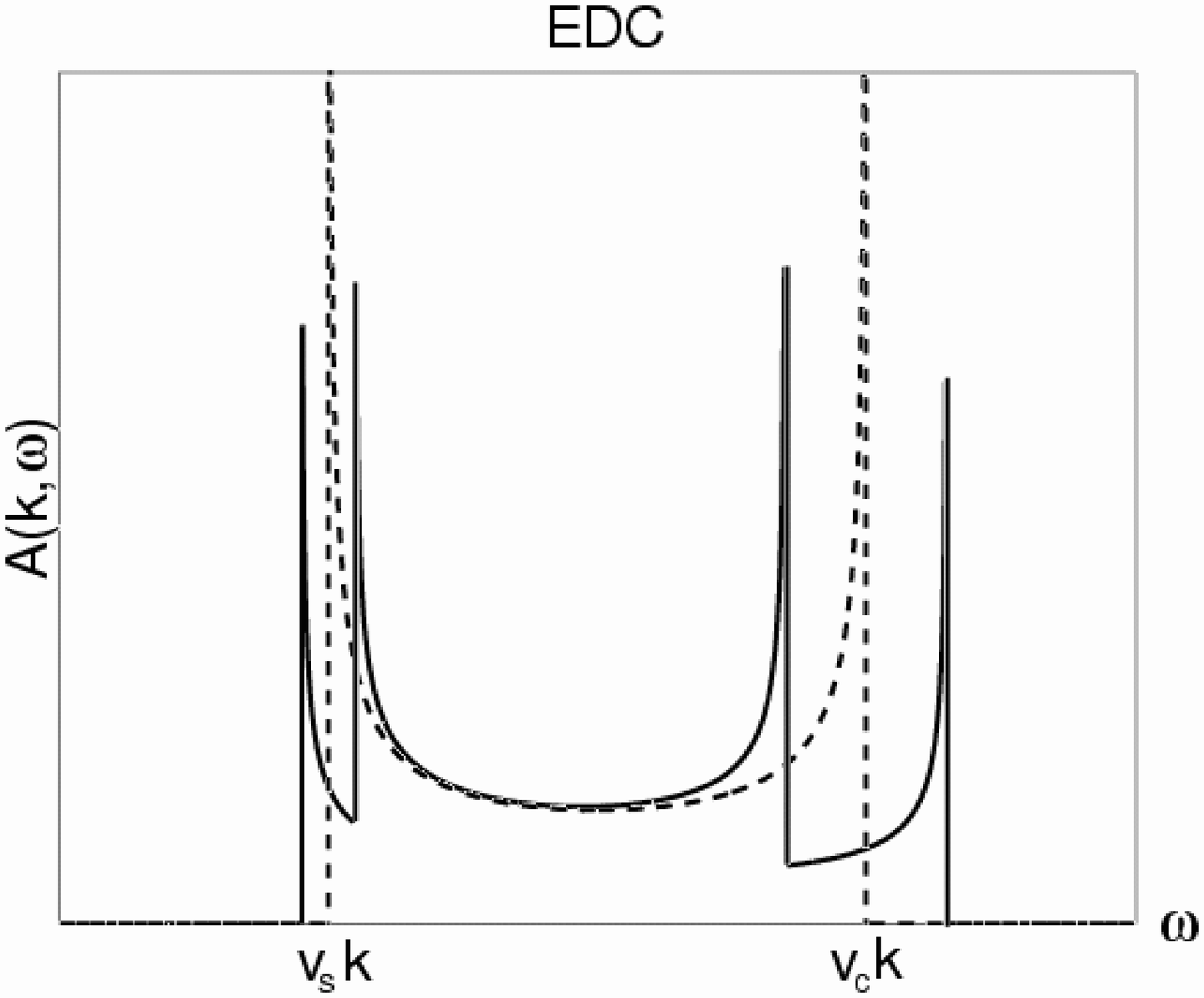,width=0.5\linewidth}
\end{center}
\caption[]{The energy distribution curve (the spectrum $A(k,\omega)$
at fixed $k$) as a function of $\omega$ in the presence of a magnetic
field. The dashed
line is the zero-field result of Fig. \ref{voitfig2}. The magnitude
of the Zeeman splitting is enhanced
with respect to $(v_{c}-v_{s})k$ for
clarity. From Rabello and Si \cite{rabello}.} \label{energycurves}
\end{figure}

The  zero temperature
spectral functions
of the T-L model are
\begin{eqnarray}
A(k,\omega) \approx (\omega - v_{c}(k-k_{F}))^{2 \gamma_{c} - 1/2}
|\omega- v_{c}(k-k_{F})|^{\gamma_{c} - 1/2} ~~~(v_{c} > v_{s}) \nonumber
\\ A(k,\omega) \approx(\omega - v_{s}(k-k_{F}))^{\gamma_{c}-1/2} |\omega-v_{s}(k-k_{F})|^{2 \gamma_{c}-1/2} ~~~(v_{c}<v_{s}).\label{akomegaexpressions}
\end{eqnarray}
These spectral functions are sketched in Fig. \ref{voitfig2} for the
case $g_2=0$ and in Fig. \ref{voitfig3} for the general case.
Notice that  the  quasiparticle pole in $A(k,\omega)$ in a Landau
 Fermi-liquid is smeared out to produce a {\em branch
cut}  extending from the spin mode excitation
energy to the charge mode
excitation energy.  These branch cuts
split into two in an applied magnetic field, see  Fig. \ref{energycurves}.
These results are
exact for small $\omega$ and small
$|k-k_F|$. Another manifestation of
the SFL behavior is the behavior of
the momentum distribution function
\begin{equation}
n_{k} \sim n_{k_{F}} -  const \times \mbox{sign} (k-k_{F})
|k-k_{F}|^{2 \gamma_{c}}~,
\end{equation}
where
\begin{equation}
\gamma_{c,s}  = \frac{1}{8} \left( K_{c,s} + K_{c,s}^{-1} -2\right)
~.\label{gammaeq}
\end{equation}

In contrast to a Fermi liquid, the expression for $n_k$
does not exhibit a step-like discontinuity
at the Fermi points.  Note that the exponent
$ 2 \gamma_{c}$ is {\it non-universal}
(as usual, an outcome of a line
of critical points).

The single particle density of states
\begin{equation}
N(\omega) \approx |\omega|^{2 \gamma_{c}}~,\label{1ddensityofstates}
\end{equation}
 vanishes at the Fermi surface.

\subsection{Thermodynamics}

The contributions of the independent charge and
spin modes appears independently
in most physical quantities.

The specific heat coefficient is found to be
\begin{equation}
\gamma/\gamma_{0}= \frac{v_{F}}{2}\left(\frac{1}{v_{c}}+\frac{1}{v_{s}}\right).
\end{equation}
where $\gamma=\gamma_{0}$ for the non-interacting
system.

The spin susceptibility
and the compressibility
are readily computed
\begin{equation}
\chi_{0} = v_{F}/v_{s}~,~~ ~~~~~ \kappa/\kappa_{0} = v_{F} K_{c}/v_{c}
\end{equation}
where $\chi_{0}$ and $\kappa_{0}$ are the susceptibility
and compressibility of the noninteracting gas.
The Wilson ratio, already encountered in our discussion of the Kondo
problem in section \ref{flkondo},
\begin{equation}
R_{W} = \frac{\chi/\chi_{0}}{\gamma/\gamma_{0}} = \frac{2 v_{c}}{v_{c}+v_{s}}~,
\end{equation}
deviates from its Fermi Liquid value of
unity by an amount dependent on the
relative separation between the
spin and charge velocities.

\subsection{Correlation Functions}\label{correlationfunctionssection}

Since the Hamiltonian is separable in charge and
spin and as $\psi$ is a product of independent charge
and spin degree of freedom all real
space correlation functions
are products of independent charge and
spin factors.
 We will only focus on the most important correlation functions in the
illustrative examples below,  and refer for a summary of the various exact
expressions to \cite{orgad}.

 The most important feature of the large distance behavior
of the charge and spin correlators is their algebraic
decay at zero temperature:
\begin{eqnarray}
\langle \rho(x)\rho(x^{\prime}) \rangle& \simeq& \frac{K_{c}}{(\pi (x-x^{\prime}))^{2}} + B_{1,c}
\frac{\cos(2 k_{F}(x-x^{\prime}))}{|x-x^{\prime}|^{1+K_{c}}} \ln^{-3/2}|x-x^{\prime}| \nonumber
\\ & & \hspace{2cm}  B_{2,c} \frac{\cos(4 k_{F}(x-x^{\prime}))}{|x-x^{\prime}|^{4K_{c}}} + ...
\\ \langle \vec{S}(x) \cdot \vec{S}(x^{\prime}) \rangle&  \simeq &
\frac{1}{(\pi (x-x^{\prime}))^{2}} + B_{1,s} \frac{\cos(2 k_{F} (x-x^{\prime}))}{|x-x^{\prime}|^{1+K_{c}}} \ln^{1/2}|x-x^{\prime}| + ...\nonumber
\end{eqnarray}
at asymptotically long distances and $K_{s}=1$. For
not too repulsive interactions,
so that $K_c <1$,
the $2 k_{F}$ fluctuations are dominant. We have previously
seen that such a CDW/SDW instability may arise
due to the special $2k_{F}$ nesting wavevector
in one dimension.  The  $4k_{F}$ modulations
are due to Umklapp scattering.
The amplitudes $\{B_{i,c} \}$ and $\{B_{i,s}\}$ are
non-universal while the exponents are determined
by the stiffness of the free charge and
spin fields.

While the above expressions are for $K_s=1$, in the general case
$K_{s} \neq 1$
the spin correlator decays asymptotically with the
exponent $(K_{s}+K_{c})$.

\begin{figure}
\begin{center}
  \epsfig{figure=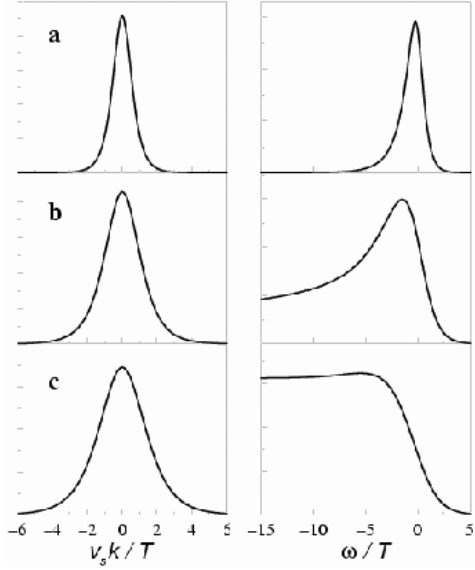,width=0.45\linewidth}
\end{center}
\caption[]{
{\em Left panel:} Momentum distribution curves at $\omega=0$ (i.e., the
spectrum at fixed $\omega=0$ as a function of $k$)
for a spin rotationally invariant Tomonaga-Luttinger
liquid, plotted as  a function of $v_{s}k/T$. {\em Right panels:}  Energy
distribution curves at $k=0$ (the spectrum at fixed $k=0$) as a
function of  $\omega/T$. In both panels,
 $v_{c}/v_{s}=3$ and  $\gamma_{c}=0$ in {\em (a)},
$\gamma_{c}=0.25$ in {\em (b)},
and $\gamma_{c}=0.5$ in {\em (c)}. From Orgad \cite{orgad}.} \label{orgadfig1}
\end{figure}

At nonzero temperatures, it is found that
 the Fourier
transforms of these correlation functions
scale as
\begin{eqnarray}
\chi_{CDW} & \approx & T^{K_{c}-1} |\ln T|^{-3/2}~,~~ ~~~ \chi_{SDW}
\approx T^{K_{c}-1}|\ln T|^{1/2} ~,\nonumber\\
\chi_{SS} & \approx & T^{K_{c}-1} |\ln T|^{-3/2}~,~~~~~ ~~~ \chi_{TS}
\approx T^{K_{c}-1}|\ln T|^{1/2} ~.
\label{onedchieqs}
\end{eqnarray}
An example of the spectral density as a function at finite
temperatures is shown in Fig. \ref{orgadfig1}.

The results above lead in principle to clear  experimental signatures.
$X$-rays, which couple to the charge density
waves, should peak at low temperatures
with intensities given by
\begin{equation}
I_{2 k_{F}} \sim T^{K_{c}},~~~I_{4 k_{F}} \sim T^{2 K_{c}-1}~.
\end{equation}
The NMR probe couples to the spin degrees
of freedom and the theoretically computed nuclear
relaxation time
scales as
\begin{equation}
T_{1} \sim T^{-K_{c}}~.
\end{equation}

\subsection{The Luther-Emery Model}

The Luther Emery model
extends the Tomonaga-Luttinger Hamiltonian
by including the backward scattering interactions
parametrized by ${\mathcal H}_1$, which scatter
from $(+k_{F},\sigma ;-k_{F},\sigma')$ to $(-k_{F},\sigma ;k_{F},\sigma')$
and vice versa. The Umklapp
processes, important (at $T=0$) only
when $4 k_{F}$ is a reciprocal lattice, continue to be discarded.

The backward scattering term
\begin{equation}
{\mathcal H}_{1}= \int dx ~ g_{1} \sum_{r= \pm 1} \psi_{r,\sigma=+1}^{\dagger} \psi_{-r,\sigma=-1}^{\dagger} \psi_{r,\sigma=-1}\psi_{-r,\sigma=1}
\end{equation}
written in terms of the bosonic variables introduces
a nontrivial sine-Gordon like interaction:
\begin{equation}
{\cal{H}}_{s}= \int dx \frac{v_{s}}{2} \left[ K_{s} (\partial_{x} \theta_{s})^{2} + \frac{
(\partial_{x} \phi_{s})^{2}}{K_{s}} \right]+ \frac{2 g_{1}}{(2 \pi a)^{2}} \cos(\sqrt{8 \pi} \phi_{s})\label{sinegordon}
\end{equation}
with rescaled values of the spin and charge velocities
and stiffness constants.
When $g_{1}>0$ (repulsive interactions), $g_{1}$ is renormalized to zero
in the long wavelength limit. Since along the RG flow trajectories
$K_{s}-1 \approx g_{1}/(\pi v_{s})$, this means that $K_s$
renormalizes to 1. The physics corresponding to
this case is in the Tomonaga-Luttinger
model that we just discussed.

When the backscattering interactions are
attractive, $g_{1}<0$, spin excitations are no longer gapless, and
a spin gap of magnitude
\begin{equation}
\Delta_{s} \sim \frac{v_{s}}{a} \left[\frac{g_{1}}{2 \pi^{2} v_{s}}\right]^{1/(2-2K_{s})}
\end{equation}
is dynamically generated. The attractive backscattering
leads to the formation of bound particle-hole pairs
which form a CDW. The spin correlation length is then finite,
\begin{equation}
\xi_{s}= \frac{v_{s}}{\Delta_{s}}~.
\end{equation}
Gaps in the charge spectrum develop in the presence of Umklapp scattering.
In the spin gapped phase the Hamiltonian can be conveniently
expressed in terms of new  {\em refermionized} spin
fields $\Psi_{r}(x)$.
\begin{equation}
\Psi_{r} = F_{r} \exp[-i \sqrt{\pi/2} (\theta_{s}-2 r \phi_{s})]~.
\end{equation}

Luther and Emery observed that at the point
$K_{s}= 1/2$ the Hamiltonian in terms of
these new spin fields
becomes that of noninteracting free fermions
having a mass gap $\Delta_{s} = g_{1}/(2 \pi a)$.
\begin{equation}
{\cal{H}}_{s}= \int dx \sum_{r=\pm 1} [-i v_{s} r \Psi_{r}^{\dagger} \partial_{x} \Psi_{r} + \Delta_{s} \Psi_{r}^{\dagger} \Psi_{-r}]~,
\end{equation}
leading to the spin excitation spectrum
\begin{equation}
E_{s} = \sqrt{\Delta_{s}^{2}+v_{s}^{2}(k-k_{F})^{2}}~.
\end{equation}

\subsection{Spin charge separation}\label{spinchargeseparationsection}

As in many other physical problems, the availability of
an exact solution to the one-dimensional
electron gas problem is intimately linked
to the existence of additional conservation
laws or symmetries. One may attack the Luttinger Liquid
problem by looking for its symmetries.

The $U(1)_{L} \otimes U(1)_{R}$ symmetry
 in the absence of Umklapp scattering may be
exhibited by considering the effect
of the separate left and right
rotations by angles $\Gamma_{L,R}$ on the fermion
variables:
\begin{equation}
\psi_{L \sigma}(x,t) \rightarrow e^{i \Gamma_{L}} \psi_{L \sigma}(x,t)
~,~~~~~~ \psi_{R \sigma}(x,t) \rightarrow e^{i \Gamma_{R}} \psi_{R \sigma}(x,t)~.
\end{equation}
All of the currents are trivially invariant
under this transformation as the $\psi^{\dagger}$
fields transform with opposite phases. Physically
this corresponds to the conservation of the number
and net spin of left and right moving particles.

As discussed by  Metzner and
 Di Castro \cite{metznerdic}, these
separate conservation laws for the left and right
moving charge and spin currents lead to Ward identities
which enable the computation of the single
particle correlation functions.

In the absence of Umklapp scattering, spin and
charge are conserved about each
individual Fermi point. The net total charge density
$\rho \equiv \rho_+ + \rho_-$ and charge density asymmetry
$\tilde{\rho} \equiv \rho_+-\rho_-$ in the Tomonaga-Luttinger
Hamiltonian
satisfy the continuity
equations
\begin{equation}
\partial_{\tau} \rho = [{\mathcal H}, \rho] = - q {j} ~,~~~~~~
\partial_{\tau} \tilde{\rho} = [{\mathcal H}, \tilde{\rho}] = - q \tilde{j}~,
\end{equation}
where
\begin{equation}
j(q) = u_{c}[\rho_{+}-\rho_{-}] ~,~~~~~~\tilde{j} = \tilde{u}_{c}[\rho_{+}+\rho_{-}]~,
\end{equation}
and where the velocities are given by
\begin{equation}
u_{c} = v_{F} + \frac{g_{4}^{c}-g_{2}^{c}}{\pi} ~,~~~~~~ \tilde{u}_{c} = v_{F} + \frac{g_{4}^{c}+g_{2}^{c}}{\pi}~.
\end{equation}
These results follow straightforwardly from the form of ${\mathcal
H}^{T-L}$ in combination with the fact that the only nonzero
commutator is
\begin{equation}
[\rho_{r,\sigma}(q),\rho_{r',\sigma'}(-q')] = \delta_{qq'} \delta_{rr'}
\delta_{\sigma \sigma \prime} \left( {{q L}\over{2\pi}}\right) ~.
\end{equation}

Let us illustrate simply
how many of the results
derived via bosonization
may also be directly
computed by employing these conservation laws.
The existence of gapless charge modes is a direct consequence
of the right-left charge conservation laws.
The two first-order continuity equations given
above lead to
\begin{equation}
[\partial_{\tau}^{2} + u_{c} \tilde{u}_{c} q^{2}] \rho = 0~,
\end{equation}
from which we can read off a linear charge dispersion
mode
\begin{equation}
\omega = v_{c} |q|
\end{equation}
with velocity $v_{c} = \sqrt{u_{c} \tilde{u}_{c}}$, in agreement with
the earlier result (\ref{vcs}). Thus,
 collective charge excitations
propagate with a velocity $v_{c}$.
A similar relation may be found for the
spin velocity $v_s$ which  in general
is different from $v_{c}$.
This {\it spin-charge
separation} also becomes clear from the
explicit form of the expectation values of the charge
and spin densities:
\begin{eqnarray}
\langle 0| \psi_{r}(x_{0})  \rho_r(x,t) \psi^{\dagger}_{r}(x_{0}) |0 \rangle
& = & \delta(x-x_{0}-rv_{c}t)~, \nonumber
\\ \langle 0| \psi_{r}(x_{0})  S_r^{z}(x,t) \psi^{\dagger}_{r}(x_{0}) |0 \rangle
& = & \delta(x-x_{0}-r v_{s}t)~,
\end{eqnarray}
where $|0 \rangle$ denotes the ground state.

The separate
Right--Left conservation laws cease to hold if (back-scattering) impurities
are present. Accordingly, as shown by Giamarchi and Schulz, \cite{giaschulz}
spin-charge
separation then no longer holds.

 \subsection{Spin-charge Separation in more than one-dimension?}\label{spinchargesepinmore}
 
Spin-Charge separation in one dimension
 requires extra conservation laws. Can something analogous happen in
  more than one-dimension?  No extra conservation laws are discernible in the
  generic Hamiltonians
  in two-dimension, although such Hamiltonians can doubtless be constructed.
 Are there conditions in which generic Hamiltonians become dynamically equivalent to
 such special Hamiltonians (because the unwelcome operators are "irrelevant").
 No definite answers to these questions are known. In  section (5.2)and later in this section
 we shall briefly review some interesting attempts towards spin-charge separation
 in higher dimensions. First we present
  qualitative arguments pointing out the difficulty in this quest.

There is a  simple caricature
given by Schulz \cite{schulzlectures} for qualitatively
visualizing charge-spin separation for a special one-dimensional case:
 the $U \rightarrow \infty$
Hubbard model.  This model is characterized  (at half-filling)
by algebraic decay of Spin-Density
correlations, which at short distances appear as almost antiferromagnetic alignments of
spins. Let us track the
motion of a hole introduced into an antiferromagnetically
ordered chain. The hole is subject
to only the lattice kinetic term which enables
it to move by swapping with a nearby spin.

An initial configuration will be
\begin{equation}
...\Downarrow \Uparrow \Downarrow \Uparrow \Downarrow O \Downarrow \Uparrow \Downarrow \Uparrow \Downarrow ...
\end{equation}

After one move the configuration is
\begin{equation}
... \Downarrow \Uparrow \Downarrow \Uparrow O \Downarrow \Downarrow \Uparrow
\Downarrow \Uparrow \Downarrow ...
\end{equation}
After two additional moves to the left
the configuration reads
\begin{equation}
... \Downarrow \Uparrow O \Downarrow \Uparrow \Downarrow \Downarrow
\Uparrow \Downarrow \Uparrow \Downarrow ...
\end{equation}
Thus the initial hole surrounded by two spins of the same
polarization has broken into a charge excitation
(``holon'' or ``chargon'' --- a hole surrounded by
antiferromagnetically aligned spins)
and a spin excitation (``spinon'')
composed of two consecutive parallel
spins in an antiferromagnetic environment.
 The statistics
of the localized spinons
and holons in this model
must be such that their
product is fermionic.

The feasibility of well defined
spin and charge excitations
hinges on the commuting nature
of the right and left
kinetic (hopping)
operators $T_{Right}, T_{Left}$
which are the inverse each other.
Any general term of the
form
\begin{eqnarray}
(T_{Left})^{n_{1}^{Left}} (T_{Right})^{n_{1}^{Right}}
(T_{Left})^{n_{2}^{Left}} (T_{Right})^{n_{2}^{Right}}... &  & \nonumber
\\  = (T_{Left}T_{Right})^{N_{R}} T_{Left}^{N_{L}-N_{R}}&  = & I \times T_{Left}^{N_{L}-N_{R}}
\end{eqnarray}
where $N_{R,L} = \sum_{i} n_{i}^{R,L}$.
We have just shown that terms of the
form $T_{L}^{n_{L}}$ lead to a cartoon
of the sort
depicted above which
gives rise to spin-charge separation
and therefore our result holds for the general
perturbative term. The  proof of spin-charge separation
for the one dimensional electron gas
rests on the existence of separate conservation
laws for the left and right moving fronts, as a result of the fact
that the operators $T_{Right}$ and $T_{Left}$ commute.

Such a simple ``proof'' cannot be extended
to higher dimensions. In higher dimensions
this suggestive illustration for spin-charge
separation is
made impossible by the noncommuting
(frustrating) nature of the permutation operators
$T_{Up},T_{Down},T_{Right},T_{Left}$,
... Moreover, even if the exchange operators commuted
we would be left with terms of the
form $T_{Left}^{n_{Left}} T_{Up}^{n_{Up}}$
which when acting on the single
hole state will no longer give rise to
states that may be seen as a direct product of
localized holon and spinon like entities.

Let us simply illustrate this by applying a sequence
of various exchanges on the planar state $| \psi \rangle$:
\begin{equation}
\begin{tabular}{cccccccc}
- & +& -& +& -& +& -& +
\\ +&  -& +& -& +& -& +& -
\\  -&  +& -& +& -& +& -& +
\\  + & -& +& 0 & + & -& +& -
\\  - & +& -& +& -& +& -&  +
\end{tabular}~~,\nonumber
\end{equation}
where $+$ and $-$ denote up and down spins, respectively.
Applying \\
$T_{Down}T_{Right}^{2} T_{Up} T_{Left}^{2}$
we arrive at $| \psi^{\prime} \rangle$
\begin{equation}
\begin{tabular}{cccccccc}
 - &  + & - & +&  -&  +&  - & +
\\  + & -& + & -& +& -& +& -
\\  - & - &  +& +& -& +& -& +
\\  + & + & - & 0 &+& -& +& -
\\  - & +& -& +& -& +& -& +
\end{tabular}~~,\nonumber
\end{equation}
a state which obviously differs
from $T_{Down}T_{up} T_{Right}^{2} T_{Left}^{2} |\psi \rangle =
|\psi \rangle$.  Unlike the one dimensional case,
damage is not kept under check. Note the extended
domain wall neighboring the hole,
enclosing a $2 \times 2$ region of spins of the
incorrect registry. Note also that the hole is now
surrounded by a pair of antiferromagnetically
aligned spins along one axis
and ferromagnetically aligned
spins along the other.
A path closing on itself
does not lead to the fusion of the
a ``holon'' and ``spinon'' like entities
back into a simple hole. As the hole
continues to further explore
both dimensions damage is continuously
compounded. The state
$T_{Down}^{2} T_{Right} T_{Down} T_{Right} T_{Up}^{3} T_{Left}^{2}
|\psi \rangle = |\tilde{\psi} \rangle$  contains a string
of eight spins of incorrect orientation surrounded by a
domain wall whose perimeter is $ 16$ lattice units long:
\begin{equation}
\begin{tabular}{cccccccc}
 - &  -& +& +& -& +& -& +
\\  + &  +& -& +& +& -& +& -
\\   - & -& -& +& -& +& -& +
\\  + & + &  - & 0 &+& - & +& - \nonumber
\\  - & +& -& +& -& +& -& + \nonumber
\end{tabular}\nonumber
\end{equation}

As seen, the moving
electron leaves a string
of bad magnetic bonds in
its wake. The energy penalty
of such a string is
linear in its extent. It is therefore
expected that this (magnetic string)
potential leads, in more than
one dimension, to a confining force
amongst the spin and charge
degrees of freedom. As this caricature for the {\em single hole}
makes clear, the notion of localized
``spinons'' and ``holons'' is unlikely to
hold water for the $U \rightarrow
\infty$ Hubbard model in more than one
dimension. The well-defined SFL solutions for special models with
nested Fermi-surface in two-dimensions should however be noted \cite{fjaer}.

A certain form of spin-charge
separation in two dimensions may be sought in the very special
{\em hole aggregates} (or stripes) that have been detected
in some of the cuprates \cite{Tranquada} and the nickelates
\cite{cheong}. Here, holes arrange themselves
along lines which concurrently act as antiferromagnetic domain
walls (i.e. behave like holons) in the background spin texture.
Charge and spin literally separate and occupy different regions of
space. In effect the two-dimensional material breaks up into
one-dimensional lines with weak inter-connections\footnote{This
observation has led to a line of thought which is of some interest
in the context of the issues discussed here. If one focusses on
the quantum mechanics of a single line of holes by formulating it as a
quantum mechanical lattice string model \cite{eskes}, the string
traces out a two-dimensional world sheet in space-time. Quantum
mechanical particles in one dimension, on the other hand, trace out
world lines in space-time. It is claimed that one can recover most of the power law
correlation functions of one-dimensional interacting fermions from the
classical statistical mechanics of fluctuating lines, and along these
lines approach stripe formation as some form of spin-charge
separation in two dimensions \cite{zaanen3}.}. Related behavior
is also found  in numerical work \cite{Dagotto-2000} in
 the so called, $t^{\prime}$-$t$-$J$ model\footnote{In  this paper it
was further
observed that  the kinetic
motion of single holes may  scramble the background
spin environment in such a way that, on average,
the holes may become surrounded by antiferromagnetically
ordered spins on all sides (i.e. both along the horizontal
and along vertical axis) --- this is claimed to be a higher dimensional
generalization of the holons encountered
thus far.}. An important question for such models is
the extent to which the interconnections
between stripes are ``irrelevant'' --- i.e.
the coupled-chains problem, which we briefly
allude to in section \ref{coupledchainssec}.

\subsection{Recoil and the Orthogonality Catastrophe in 1d and higher}\label{1dorthog}

Here we show how the SFL behavior in one-dimension is intimately tied
to the issue of
orthogonality we discussed in sections \ref{landauwavefren} and \ref{orthogonalitysection}.
 This line of thinking is emphasized by
Anderson \cite{Anderson-1967,andersonbook} who has also argued that this line of reasoning
gives SFL behavior in two dimensions for arbitrary small interactions.

We consider the effect of interactions through the explicit
 computation of our old friend from section \ref{landauwavefren}, the
quasiparticle weight
\begin{equation}
Z_{k} = \langle  \psi_{k}^{N+1}| c_{k}^{\dagger} | \psi^{N} \rangle.
\end{equation}
As we have seen, this indeed vanishes in all
canonical one-dimensional models.
Consider the  model \cite{castella} of
a Hamiltonian describing $N$ fermions interacting with
an injected particle via a delta function potential
\begin{equation}
H = -\frac{1}{2m} \sum_{i=1}^{N} \frac{\partial^{2}}{\partial x_{i}^{2}}
- \frac{1}{2 m} \frac{\partial^{2}}{\partial x_{0}^{2}} + U \sum_{i=1}^{N}
\delta(x_{i}-x_{0}).
\end{equation}
The calculation for the small $s$-wave phase shifts for all
single particle states $\{\phi_{i}( x_{i})\}$ is
relatively straightforward.
The quasiparticle
weight $Z$ reduces to an overlap integral between
two $(N+1)$-particle Slater determinants, and one finds
\begin{equation}
Z \sim N^{-2 (\delta_{F}/\pi)^{2}}
\end{equation}
with
\begin{equation}
\delta_{F} = - \tan^{-1}[U k_{F}/2].
\end{equation}
In the thermodynamic limit $Z = 0$
and no quasiparticles exist. As we see,
the scattering phase shifts must conspire to
give rise to anomalous behavior
(exponents) for the electronic
correlation function in such a
way that they lead to a vanishing
density of states at the Fermi
level. We have already given explicit
expressions for the anomalous exponent(s)
under the presence of general
scattering terms.

As indicated in sections \ref{landausfermiliquid} and
\ref{localfermil}, an identically
vanishing overlap integral
between two (N+1) particle states
could be a natural outcome
of the emergence of additional
Quantum numbers labeling
orthogonal states. These states could
correspond to different topological excitation
sectors (e.g. solitons). Each quantum
number corresponds to some conserved quantity
in the system. In the one-dimensional
electron gas this may be derived as we saw as a consequence
of separate conservation law for
left and right movers.


An illustrative example of how singular Fermi liquid behavior due to
orthogonality of the wavefunction is
robust in one dimension but easily destroyed  in higher dimensions, is
provided by the $X$-ray edge singularity problem, already
discussed in section \ref{xrayedge}. As sketched in Fig. \ref{xrayfig},
we consider the transition of an electron from a deep core level to
the conduction band through absorption of a photon. This problem is
essentially the same as that of  optical absorption in degenerate
semiconductors, and from this point of view it is natural to analyze,
following Nozi\`eres \cite{nozieresrecoil},
the effect of dispersion in the hole band, the analogue of the deep
level state. For optical absorption in a semiconductor, the transition
conserves momentum; hence in the absence of final state interactions, the
threshold absorption is  associated with
the transition indicated with the arrowed line in
Fig. \ref{xrayfig}, and absorption starts discontinuously above the threshold energy
$\omega_D$ proveded the hole mass is infinite. For finite hole mass the
threshold gets rounded on the scale of the dispersion of the hole band. But in one dimension,
the edge
 singularity {\em does survive} because
low-energy electron-hole excitations in one dimension have
momenta only near 0 and near $2k_F$; see Fig. \ref{spectrum1d},
 electron-hole pairs can not carry away arbitrary momenta.
This  is seen in the following
calculation  \cite{nozieresrecoil}.

Assume a simple
featureless final state potential $V$, and consider first the case
without recoil.  The relevant quantity to
calculate is the transient propagator for the scatterer,
\begin{equation}\label{Gxray}
G(t) = \langle 0 | d e^{i {\mathcal H } t} d^\dagger | 0 \rangle ~,
\end{equation}
as the spectrum is the Fourier transform of $G(t)$. In (\ref{Gxray}),
the potential $V$ is turned on at time 0 and turned off at time
$t$. In a linked cluster expansion, we may write $G(t)= e^{C(t)}$,
where $C(t)$ is the contribution of a single closed loop. In lowest
order perturbation theory, $C(t)$ becomes
\begin{equation}\label{Cxray}
C(t)= \int_0^t d\tau \int_0^t d\tau' V^2 g(0,\tau - \tau')
g(\tau'-\tau)~,
\end{equation}
where $g(0,\tau)$ is the free conduction electron propagator at the
origin. For large times,  one has $g(0,\tau) \approx -i N(0)/ \tau$, and
when this is used  in (\ref{Cxray}) one immediately find that for
large times
\begin{equation} \label{CtoG}
C(t) = V^2 N(0)^2 \ln t ~, ~~~~ \Rightarrow ~~~~G(t) \propto 1/ t^n~,
\end{equation}
where
\begin{equation}
n= V^2 N(0)^2 = \delta^2 / \pi^2 +1
\end{equation}
is the phase shift exponent due to the orthogonality effect, compare
Eq. (\ref{Aequation}). A power law decay of $G(t)\sim 1/t^n$ at long times
corresponds to power law dependence $\sim (\omega-\omega_D)^{n-1}=(\omega-\omega_D)^{\delta^2/\pi^2}$
just above the absorption edge.

If we now take into account the recoil effect;  the dispersion
of the lower band implies that the hole in this band can hop from site
to site. The propagator $G(t)$ is then obtained as  a  sum over all
trajectories $R(\tau)$ of the scattering hole which begin and end at
$R=0$. For a given history, we can extend the above analysis to lowest
order by replacing the propagator $g(0,\tau) $ by $g(\rho,\tau)$,
where $\rho(\tau)=R(\tau)-R(\tau')$. For positive time difference, we
can then write
\begin{equation}
g(\rho, \tau) = \sum_{k > k_F} e^{i\epsilon_k \tau - i {\bf k} \cdot
\rho}~.
\end{equation}
For small hopping rates and large times, the integration over the
modulus is dominated
by the energy term, and this yields a term proportional to $-1/\tau$
as in the recoilless case. The trajectory of the hole enters through
the  average $ \overline{\exp(-i {\bf k} \cdot \rho)}^{FS}$ over the
Fermi sphere. A simple calculation yields
\begin{equation}
 \overline{\exp(-i {\bf k} \cdot \rho)}^{FS} = \left\{ \begin{array}{ll}
\cos(k_F \rho)\hspace{3mm} & d=1 \\ J_0(k_F\rho) & d=2\\ {{\sin
(k_F\rho)}\over{k_F\rho}} & d=3 \end{array}  \right.  ~~.
\end{equation}
In order to calculate the large time behavior of the Green's function,
we  finally have to average the square of this result over the
distribution function of the trajectories $\rho$ for large
times. Using the large-$\rho$ behavior of the expressions
found above, one then easily finds that for large times
\begin{equation}
g^2(\rho, \tau) \left< \left( \overline{\exp(-i {\bf k} \cdot
      \rho)}^{FS} \right)^2  \right>_{\rho} =  \left\{ \begin{array}{ll}
\frac{1}{2} N^2(0)/\tau^2  & d=1 \\ \propto \ln \rho_{typ}(\tau)/(\rho_{typ}(\tau) \tau^2) \hspace{2mm} & d=2\\ \propto 1 /(\rho^2_{typ}(\tau) \tau^2) & d=3 \end{array}  \right.  ,
\end{equation}
where $\rho_{typ}$ is the typical distance the hole trajectory moves
away from the origin.
In one dimension, we see that $g^2 $ still falls of as $1/\tau^2$ and
hence in analogy to (\ref{CtoG}) that $G(t)$  has power law long time
behavior: in the presence of recoil, an edge singularity persisits
but the exponent $n$ is now
only half that in  the absence of dispersion of the lower
state, due to the averaging over $\rho$.

Since $\rho_{typ}(\tau)$ grows diffusively as $\tau^{\half}$ for large
$\tau$, the integrand in the expression (\ref{Cxray}) for $C(t)$
converges faster than $1/t^2$, and hence $C(t)$ converges to a finite
limit for large times. The singular $X$-ray edge effect is washed out
in two and three dimensions due to the recoil.

If the above argument is extended to arbitrary dimension by
analytically continuing the  angular average over the Fermi surface to
continuous dimensions, one finds that the orthogonality and
concomitant singular behavior is destroyed for any dimension
$d>1$. Nevertheless, the subdominant behavior of the integrals will
contain noninteger powers of time, and this gives rise to subdominant
nonanalytic terms in the spectrum for noninteger $d$.  This behavior
is completely in agreement with an analysis of the dimensional
crossover from Luttinger liquid behavior to Fermi liquid behavior as a
function of dimension \cite{castellani}.

\subsection{Coupled One-dimensional Chains}\label{coupledchainssec}

The two-coupled chain problem has been thoroughly considered
\cite{fabrizio,finkelsteinlarkin,schulzchains}
\cite{balentsfisher,noack} following
earlier one-loop calculations \cite{varma0} on a related model. The two-chain or ladder
is especially interesting both theoretically and experimentally.
In general interchain coupling is a relevant parameter, changing the behavior
qualitatively. In the model of coupled Luttinger chains, the weight of the
massless bosons characteristic of one dimension goes down with
the number of chains.

\subsection{Experimental observations of one-dimensional Luttinger liquid behavior}\label{nanotubessection}
There has, of course,  been a long-standing interest to observe the
fascinating one-dimensional Luttinger liquid type  SFL behavior
experimentally, but the possibility of clear signatures has arisen
only in the last few years. The clearest way to probe for Luttinger liquid
behavior is to measure the tunneling into the one-dimensional system:
Associated with the power law behavior
(\ref{1ddensityofstates}) in one dimension, one has a power law
behavior  for the single electron tunneling
amplitude into the wire. For fixed voltage, this leads
to a differential conductance
$dI/dV\sim V^{\alpha}$, with the exponent $\alpha$
determined by the charge stiffness $K_{c}$, the geometry,
and the band structure.
Hence, from the measurement of the tunneling
as a function of temperature or voltage,  $\alpha$ and thus $K_c$
can be extracted.
Recent experiments on resonant tunneling \cite{auslaender}
of small islands embedded in one-dimensional
quantum wires in semi-conductors, grown with a so-called cleaved edge
overgrowth method,  do indeed yield a power law temperature
behavior of the conductance which is consistent with Luttinger liquid
behavior, but the value of the exponent is substantially different
from the one expected theoretically.

 It has recently also
been realized that nature has been kind
enough to give us an almost ideal one-dimensional wire to study
one-dimensional electron physics:  the wave functions of carbon nanotubes turn
out to be coherent over very large distances \cite{dekker}. Though
the circumference of the nanotubes is rather large,  due to the band
structure of the graphite-like
structure the conduction in nanotubes can be described in terms
of two gapless one-dimensional bands.
Moreover, it was realized by Kane {\em et
al.} \cite{kane} that due to the special geometry the backscattering
in nanotubes is strongly suppressed, so that they are very good
realizations of the Tomonaga-Luttinger model of  section
\ref{tmmodel}, with an interaction constant $K_c$ which is  determined
by the Coulomb energy on a cylinder. Their calculation based on this idea
gives a value  $K_c \approx 0.2$.

\begin{figure}
\begin{center}   
  \epsfig{figure=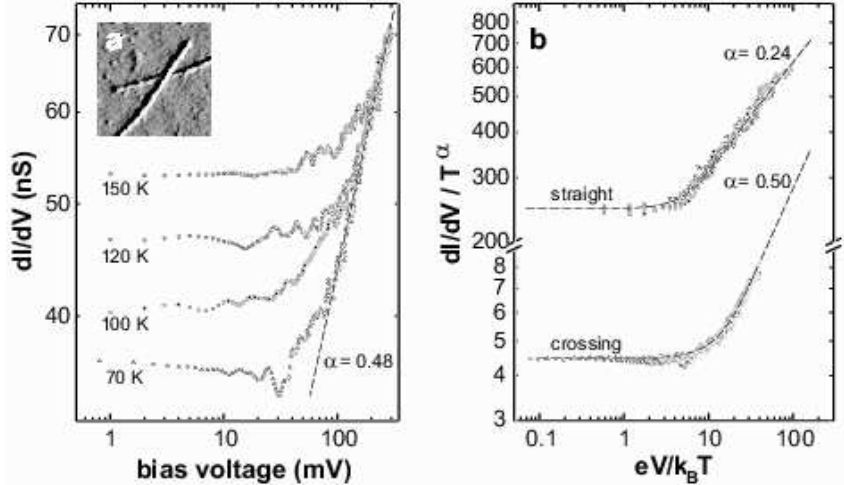,width=0.80\linewidth}
\end{center}
\caption[]{Differential conduction $dI/dV$ measured by Postma {\em et
al.} \cite{postma} for carbon nanotubes. At low voltages or
temperatures,
Coulomb blockade effects dominate, but at higher temperatures or bias
voltages, one probes the one-dimensional SFL behavior. Panel {\em (a)}
shows the differential conductance as a function of bias voltage for
various temperatures (note that these temperatures are relatively
high, reflecting the fact that the electronic energy scales of the
nanotubes is high). The effective exponent $\alpha$ for the large $V$
behavior is 0.48; since these data are for tunneling between two
nanotubes, $\alpha=2\gamma_c$, so $\gamma_c\approx 0.24$ and
$K_c\approx 0.27$. The predicted value is $K_c\approx 0.2$ \cite{kane}.
The data in panel {\em (b)} show the differential
conductance as a function of temperature at fixed bias for two
nanotubes which cross, as well as for a single nanotube with a
bend.}\label{nanotubes}
\end{figure}

Fig. \ref{nanotubes} shows recent nanotube data \cite{postma} which beautifully confirm
the predictions by Kane {\em et al.} \cite{kane}: the differential
conductance $dI/dV$ is found to vary as $V^{\alpha}$ as a function of
the bias voltage at low but fixed temperatures, or as $T^{\alpha}$ as a
function of temperature at fixed bias. For tunneling into
the bulk of a carbon nanotube, the relevant density of states
\begin{equation}
N_{bulk}(\omega) \sim \omega^{\alpha_{bulk}} \mbox{ with }
\alpha_{bulk} = \gamma_{c}
\end{equation}
The fact that the exponent here is only half the value
attained in the simple Luttinger Liquid --- see
Eq. (\ref{1ddensityofstates}) ---
is due to the fact that  there are {\em two} Luttinger Liquid-like bands
present in the carbon nanotubes\footnote{The electrostatic charging
energy depends only on the symmetrized band mode, which in bosonic
variables can be 
written as
$\theta_{c,+} = (\theta_{c,band=1}+\theta_{c,band=2})/\sqrt{2}$. The
essential reason that exponents can change depending on the number of
bands is that the normalization factor $1/\sqrt{2}$ in this bosonic
variable enters in the exponent when the electron variables are
written in terms of the bosonic modes, as discussed in section
\ref{tmmodel}.  A simple way to illustrate the halving of the
exponents  in the context of the
various results we have discussed is by considering the difference
between a the spinfull Luttinger liquid that we have discussed and the
spinless Luttinger liquid (a system only having charge degrees of
freedom): the spinfull model has a density of states exponent which is
half of the spinless one. A calculation proceeds along the following
lines: for the spinfull case, the relevant Greens function is
$G_{R \sigma=+1}(x,t) = \langle e^{i (\Phi_{cR}(x,t) +
\Phi_{sR}(x,t))/\sqrt{2})}
e^{-i(\Phi_{cR}(0,0)+\Phi_{sR}(0,0))/\sqrt{2})}\rangle$. Note the factor
$1/\sqrt{2}$ in the exponent, coming from the projection onto the proper
bosonic variables. This expection value can be written as
$G_{R \sigma=+1}(x,t) = \langle e^{i \Phi_{cR}(x,t)/\sqrt{2}} e^{-i
\Phi_{cR}(0,0)/\sqrt{2}} \rangle $ $ \times $ $\langle e^{i \Phi_{sR}(x,t)/\sqrt{2}} e^{-i \Phi_{sR}(0,0)/\sqrt{2}}
\rangle $, which  with the aid of the results
of section \ref{onepspf} becomes
$G_{R \sigma=+1}(x,t) =   \frac{1}{|x-v_{c}t|^{1/2}}
\frac{1}{|x^{2}-v_{c}^{2}t^{2}|^{\gamma_{c}}} $ $ \times $ $
\frac{1}{|x-v_{s}t|^{1/2}} \frac{1}{|x^{2}-v_{s}^{2}t^{2}|^{\gamma_{s}}}$.
From this result, one immediately obtains the density of states
$N(\omega) = \int dt ~G(x=0,t)~ e^{i \omega t}.$ A
simple integration then yields $
N(\omega)=  [v_{c}^{-2 \gamma_{c}} v_{s}^{-2 \gamma_{s}}]
\int dt~
t^{-1}~ t^{-2(\gamma_{c}+\gamma_{s})} ~
e^{i \omega t}
$ $\sim $ $ \omega^{2 (\gamma_{c} + \gamma_{s})}.$ Since $K_s$
renormalizes to 1 for repulsive interactions [see the remark just
after (\ref{sinegordon})], one usually has
$\gamma_{s}=0$ and so in this case the density of states exponent is
simply $2\gamma_s$. This is precisely (\ref{1ddensityofstates}). Now
consider what one gets for the spinless Luttinger liquid. In this case
neither the bosonic spin modes {\em nor the projection factor} $
1/\sqrt{2}$ are  present in the above expression for
$G$. This results in a spatial decay with an exponent $2\gamma_c$
instead of $\gamma_c$, and hence an exponent $4\gamma_c$ in the
density of states. In other words, for $\gamma_s=0$ the density of
states exponent in the spinful case is half of what it is in the
spinless case. The same mechanism is at work in the nanotubes.}.  Only one
linear combination of the two associated charge
modes attains a nontrivial stiffness
$K_{c} \neq 1$ \cite{kane}.

By contrast, at the tips of the nanotubes, surplus
electronic charge can propagate in only one direction
and, as a consequence, the tunneling is more restricted,
\begin{equation}
N_{tip}(\omega) \sim \omega^{\alpha_{tip}}~~~ \mbox{ with }~ \alpha_{tip}=
(K_{c}-1)/4~.
\end{equation}
By Fermi's golden rule, the relevant exponents
for (tip-tip) or (bulk-bulk) tunneling
are $\alpha_{t-t} = (K_{c}-1)/2$ and
$\alpha_{b-b} = 2 \gamma_{c}$ respectively.
Bulk-bulk tunneling is achieved by arranging
the nanotubes according to the crossing geometry
depicted in the inset of Fig. \ref{nanotubes} above.
By extracting the value of the charge stiffness
$K_{c}$ from each of the independent measurements
of $\alpha_{t-t},\alpha_{b-b}$ for the two different
geometries, a single consistent stiffness $K_{c} \approx 0.27$
was found \cite{postma}, in good agreement with
theoretical prediction. This is a beautiful
realization of Luttinger Liquid physics!


Other, older, canonical realizations of Luttinger liquids
include the Quantum Hall edge states. These represent
a chiral spinless Luttinger liquid. Here, low lying
energy states can only prevail at the edge of the sample,
and, concurrently, disperse linearly about the Fermi energy.
Edge states can attain macroscopic linear (perimeter)
extent, and the tunneling experiments between such states \cite{webb,grayson}
have observed several features predicted theoretically  \cite{wen,mpafisher}.
 We refer the interested reader to the review articles by
Schulz {\em et al.} \cite{schulzlectures} and  Fisher and Glazman \cite{fisherglazman}
for further details.

\section{Singular Fermi-liquid behavior due to gauge fields}\label{sflfromgauge}

\subsection{SFL behavior due to coupling to the electromagnetic field}\label{gaugesec}

Almost thirty years ago Holstein, Pincus and Norton
 \cite{holstein} (see also Reizer \cite{reizer}) showed that the coupling
of electrons to the
electromagnetic fields  gives rise to SFL behavior.  Since the typical
temperatures where the effects become important are of the order of
$10^{-15}$ $K$, the
effect is not important  in practice. But the theory is of considerable general
interest.

If we work in the Coulomb gauge in which $\nabla \cdot {\bf A}=0$ for
the electromagnetic ${\bf A}$ field, then the transverse propagator
$D_{ij}^0$ in free space is given by
\begin{equation}
D_{ij}^0({\bf k}, \omega) = \langle A_i A_j \rangle ({\bf k}, \omega)
= {{\delta _{ij} -\hat{k}_i \hat{k}_j}\over{c^2 k^2 -\omega^2
-i\epsilon}}~,
\label{freeAprop}
\end{equation}
where $\epsilon$ is an infinitesimal positive number. The interaction
of the electrons with the electromagnetic field is described by the
coupling term
\begin{equation}
{\bf j}\cdot{\bf A} + \rho A^2\label{coupling}~,
\end{equation}
where ${\bf j}$ is the electron current operator and $\rho$ the
density operator.

\begin{figure}
\begin{center}   
  \epsfig{figure=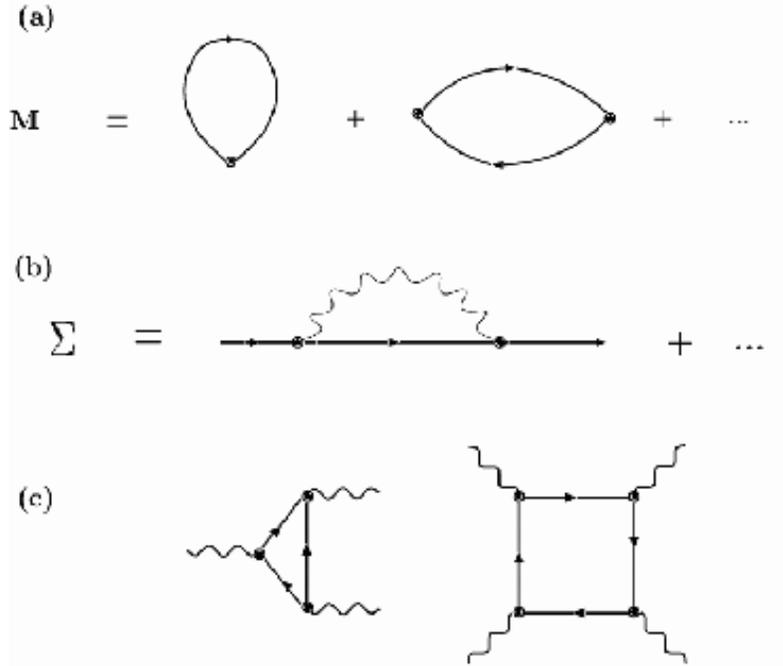,width=0.75\linewidth}
\end{center}
\caption[]{{\em (a)} The diagrams for M; {\em (b)} The diagram for the
 self-energy from the  in the discussion of
the SFL effect arising from the coupling of the electrons to the
electromagnetic gauge fields; {\em (c)} Anharmonic interaction of
fluctuations such as  {\em (c)} are non-singular in the SFL problem of
coupling of the electrons in metals to electromagnetic fields.}
\label{Mdiagrams}
\end{figure}

Quite generally in the Coulomb gauge, one finds from perturbation
theory, or phenomenologically from the Maxwell 
equations, that the electromagnetic propagator in a metal
can be written as
\begin{equation}
D_{ij}^{-1}= \left( D^0_{ij} \right)^{-1} - M \left(\delta_{ij}-\hat{k}_i\hat{k}_j \right)^{-1} ~.
\end{equation}
The perturbative diagrammatic expansion of $M$ is indicated in
Fig. \ref{Mdiagrams}. The first term leads to a term proportional to the
density while the second term is the first correction due to
particle-hole excitations; in formulas, these terms yield
\begin{equation}
M(k,\omega) = {{4 \pi \alpha}\over{m}} \left( n   + {{1}\over{m}} \int
{{{\rm d}^d p }\over{(2 \pi)^d}}  {{ [p^2 - ({\bf p}\cdot {\hat{\bf
k}})^2] [f_{p-k/2} -f_{p+k/2}] }\over {\omega
-(\epsilon_{p-k/2}-\epsilon_{p+k/2})}} \right). \label{Meq}
\end{equation}
Here $n$ is the electron density, and $\alpha=1/137$ is the fine
structure constant. For $\omega/k \rightarrow 0$ the two terms
combined yield
\begin{equation}
\label{Msmallomega}
M(k,\omega) \approx {{3 \pi \alpha n}\over{m}} \left[ 4 \left(
{{\omega}\over{k v_F}}\right)^2 + 2 i {{\omega}\over{kv_F}} \right]~.
\end{equation}
At small frequencies, this yields for the propagator $D$
\begin{equation}
\label{Deq}
D_{ij}({\bf k}, \omega) \approx {{\delta_{ij} - \hat{k}_i \hat{k}_j} \over {i 6 \pi \alpha n {{\omega}\over{nk v_F}} + c^2 k^2 -i \omega}}~,
\end{equation}
which corresponds to an overdamped mode with dispersion $\omega \sim
k^3$.

Before  discussing how  such a dispersion gives   SFL
behavior in three dimensions, it is instructive to point out that
although (\ref{Deq}) was obtained perturbatively, Maxwell's equations
ensure that the field propagator must generally be of this form at low
frequencies and momenta. Indeed,  for a metal we can write the current
${\bf j}$ as ${\bf j}= \sigma(k,\omega) {\bf E}$; if we combine this with the
Maxwell equation $\nabla \times {\bf H} = {\bf j} + \partial {\bf E}/
\partial t$ we easily find that the general form of the propagator is
\begin{equation} \label{Dgeneral}
D_{ij}({\bf k}, \omega) = {{\delta_{ij}- \hat{k}_i\hat{k}_j}\over
{4i \pi \mu \omega \sigma({\bf k},\omega) + c^2 k^2}} ~.
\end{equation}
Here $\mu $ is the diamagnetic permeability.
For pure metals, the low-frequency limit is determined by the
anomalous skin effect \cite{skineffect} and  $\sigma({\bf k}, 0) \sim k^{-1}
$. According to the  expression (\ref{Dgeneral}), this $1/k$
behavior then implies that the dispersion at small frequencies
generally goes as $\omega \sim k^3$ for a pure metal. For dirty
metals, the fact that $\sigma(k,0)$ approaches a finite limit
$\sigma_0$ at small wavenumbers, so for dirty metals there is
according to (\ref{Dgeneral}) a crossover to a behavior $\omega\sim
k^2$ at small wavenumbers.

Gauge invariance of the theory requires that the photon cannot
 acquire a mass (a finite energy
in the limit $k\rightarrow0$) in
the interaction process with the electrons,
 and hence the form of Eq.\ (\ref{Deq}) remains
 unchanged. Thus the anharmonic corrections to the photon propagator
 due to processes such as shown in  Fig. \ref{Mdiagrams}{\em (c)}
 do not change the form of  Eq.\ (\ref{Deq}).
The self-energy of the electrons due to photon exchange,
Fig. \ref{Mdiagrams}{\em (b)}, may now be calculated with confidence
given the small coupling constant in lowest order. The leading
contribution in $d=3$ is
\begin{equation}
\label{sigmaem}
\Sigma(k_F, \omega) \sim \alpha (\omega \ln \omega + i \omega
\mbox{sgn} \omega )
\end{equation}
 The momentum
dependence, on the other hand, is {\em nonsingular} as a function of
$k-k_F$.

The nonanalytic behavior of the self-energy as a function  of
frequency implies that the  resistivity
of a pure metal in $d=3$ is proportional to $T^{5/3}$ at
 low temperatures\footnote{This follows from the fact that
 in the quasi-elastic approximation, the
transport relaxation rate $\tau_{tr}^{-1}$ is related to the
 single-particle relaxation
rate $\tau^{-1}(\theta)$ due to scattering through an angle
$\theta$ near the Fermi-surface
\begin{equation}
 \tau_{tr}^{-1} =\int d\Omega (1-cos(\theta))\tau^{-1}(\theta).  \nonumber
 \end{equation}
For small
$T$, $\omega \sim k^3 \sim T$, and hence the characterestic angle of scattering,
 $\theta \sim (k/k_F)\sim (T/E_F)^{1/3}$ is small. Upon expanding
$(1-\cos \theta)\simeq \theta^2/2$, one finds that the effective
transport scattering rate goes as $\int d^3k \;k^2 f(k^3/T) \sim T^{5/3}$.}.

The simplicity and strength of the above example lies in the fact that
the theory has no uncontrolled
approximations, and the gauge-invariance of the photon field
 dictates the low-energy low-momentum behavior of the photon
propagator $D_{ij}({\bf k}, \omega)$. Moreover,  vertex corrections
are not important because the Migdal theorem is valid
 \cite{polchinski} when the frequency of fluctuations is very small
compared to their momenta, as in $\omega\sim q^3$.

However, the SFL behavior as a result of the coupling to the
electromagnetic field is not relevant in practice. This can
most easily be argued as follows. For a Fermi gas, the entropy per
particle is
\begin{equation}
\label{entropygas}
{{S}\over{N}} = {{ \pi^2 m k_B^2 T} \over {\hbar^2 k_F^2}} ~,
\end{equation}
while for the entropy for the electrons interacting with the electromagnetic
field one finds from the above results  \cite{holstein}
\begin{equation}
{{S}\over{N}} \approx {{2 \pi^2 \alpha \mu}\over{3}} {{T}\over{c k_F}}
\ln \left( {{\omega_0}\over{T}} \right) ~,
\end{equation}
with
\begin{equation}
\omega_0 = {{c \epsilon_F}\over{\alpha \mu v_F}} ~.
\end{equation}
Upon comparing these two results, one concludes that the SFL effects
start to become important for temperatures
\begin{equation}
T \stackrel{<}{\sim} \omega_0 e^{-{{3 m^* c k_B^2}\over{2 \alpha \mu \hbar^2 k_F}}}~.
\end{equation}
Since the numerical factor in the exponent is typically of order $10^5
\mu^{-1}$ for ordinary metals, the temperature range one finds from
this is of order $10^{-15}$ Kelvin for values of $\mu$ of order unity, according to this
estimate.  Note however that for pure ferromagnetic metals $\mu $ can
be as large as $10^4$. Possibly, in some ferromagnets, the effects can
become real.

\subsection{Generalized gauge theories}\label{generalgaugesec}

The example of coupling to the electromagnetic field
identifies one possible theoretical route to SFL behavior, but as we
have seen the crossover temperatures that one estimates for this scenario
are too small for observable physical properties. The smallness of the
estimated crossover
temperature is essentially due to the fact that the coupling to the
electromagnetic field is determined by the product $\alpha v_F/c$, where
$\alpha=1/137$ and typically $v_F/c = {\mathcal O}( 10^{-2})$. Motivated by this
observation, many researchers have
 been led to explore the possibility of getting  SFL
effects from coupling to different, more general gauge bosons which
might be generated dynamically in strongly correlated fermions.
For a recent review with references  to the literature, see
\cite{leegauge,nagaosa1,nagaosa2}. The
hope is that if one
could consistently find such a theory in which the small factor
$\alpha v_F /c$ arising in the electromagnetic theory is replaced by a
term of order unity, realistic crossover temperatures might arise.
Much of the motivation in this direction comes from Anderson's proposal
\cite{andersonbook}
of spin-charge separation and resonating valence bonds
in the high-temperature superconductivity problem which we discuss in
section \ref{hightcsection}.

The essence of approaches along these lines is most easily illustrated
by considering electrons on a lattice in the case in which strong on-site
(Hubbard-type) repulsions forbid
two electrons to occupy the same site. Then each site is either
occupied by an electron with an up or down spin, or by a hole. If we
introduce fictitious fermionic creation and annihilation operators $f^\dagger$
and $f$ for the electrons and fictitious bosonic hole creation and annihilation
operators $b^\dagger$ and $b$ for the holes, we can express the
constraint that there can only be one electron or one hole on each
lattice site by
\begin{equation}
\sum_{\alpha=1}^2 f^\dagger_{i\alpha} f_{i\alpha} + b^\dagger_i b_i =1
~~~~~\mbox{for each } i
~.\label{occupationconstr}
\end{equation}
With this convention, the real electron field $\psi_{i\alpha}$ can be
written as a product of these fermion and boson operators
\begin{equation}\label{splitpsi}
\psi_{i\alpha} = f_{i\alpha} b^\dagger_i ~.
\end{equation}
This expresses the fact that given the constraint
(\ref{occupationconstr}), a real fermion annihilation at a site
creates a hole.
This way of writing the electron field may be motivated by the physics of
the one-dimensional Hubbard model: there a local excitation may indeed
be expressed in terms of  a charged spinless holon and an uncharged
spinon. In general, in
a transformation to boson and fermion operators as in (\ref{splitpsi})
there is some freedom with which operator we associate the charge and
with which one the spin.

Whenever we split a single electron operator up into two, as in
(\ref{splitpsi}), then there is a gauge invariance, as the product is
unchanged by the transformation
\begin{equation}
f_{i\alpha} (t) \longrightarrow e^{i\lambda(t)} f_{i\alpha}(t)~,~~~~~
b_{i} (t) \longrightarrow e^{i\lambda(t)} b_{i}(t)~.
\label{invariance}
\end{equation}
We can promote this invariance to a dynamical gauge symmetry by
introducing a gauge field ${\bf a}$ and writing the Hamiltonian in the continuum
limit as
\begin{eqnarray}
{\mathcal H} & = & \int d{\bf r} \sum_{\sigma}~ f^\dagger_\sigma
\left( -{{i \nabla+ {\bf a}}\over{2 m_f}} \right)^2 f_\sigma + (\phi
-\mu_f) f^\dagger_\sigma f + \nonumber \\
 & & \hspace*{2cm}  + \int d{\bf r}~ b^\dagger \left( -{{i
\nabla+{\bf a}}\over{2 m_b}}  \right)^2 b + (\phi -\mu_b) b^\dagger_\sigma b \label{gaugeham}~.
\end{eqnarray}
Here $\phi$ is a Lagrange multiplier field which is introduced to
implement the constraint (\ref{occupationconstr}). Note that the $\bf a$
field enters in much the same way in the Hamiltonian
as the electromagnetic field usually does --- indeed, a change in the
individual
$f$'s and $b$'s by a space dependent phase factor as in
(\ref{invariance}) can be reabsorbed into a change of ${\bf a}$.  Note
also the presence of the chemical potentials $\mu_f$ and $\mu_b$ to
enforce that for a deviation of the (average) density $n$ from one
per site, the density of holes is $1-n$ and of fermions is
$n$.

The next step in the theory is to find the fluctuation propagator for
the ${\bf a}$ field, as a function of $\mu_f $ and $\mu_b$. For finite $\mu_f$
and negative $\mu_b$ (bosons uncondensed), the fluctuation propagator
is similar to that of the previous section but with $(\alpha v_F/c)$
replaced by a term of ${\mathcal O} (1)$. Spinon and holon
self-energies can now be calculated and composition laws
\cite{ioffelarkin}
 are derived
to relate physical correlation functions to correlation functions of
spinons and holons.

   Unfortunately, this very attractive route has turned out to be less viable than had
been hoped. It is not clear whether the difficulties are purely
technical; they are certainly formidable. The essential reason is that while photons have
no mass and are not conserved,
and hence can not Bose condense, in a gauge theory obtained by
introducing additional bosons, the bosons generally {\em can and will}
Bose
condense because they do have a chemical potential. Bose condensation leads to a
 mass term in the propagator for the
gauge fields. The singularities in the fermions due to the gauge fields then disappear.
It  is the analogue of the fact that
superconductivity leads to the Meissner effect --- there the emergence
of the superconducting field breaks gauge invariance and leads to the
expulsion of the magnetic field from the superconductor.
The latter effect can also be thought of as being due to the
generation of a mass term for the gauge fields.

Several variants of these ideas have been proposed with and without
attempts to suppress the unphysical condensation through
fluctuations \cite{moessner}.
 The trouble is that such fluctuations
tend to bind the
spinons and holons  and the happy situation in one dimension where
they exist independently --- being protected by (extra) conservation
laws, see section \ref{1dorthog} --- is hard to realize. As usual, it
appears that introduction of new quantum numbers requires new symmetries.

Interesting variants using the idea of spin-charge separation have
 recently appeared \cite{senthilfisher,moessner}.
 
\section{Quantum Critical Points in fermionic systems}\label{qcpsection}

As mentioned in the introduction, quantum critical behavior is
associated with the existence of a $T=0$ phase transition; of course,
in practice one can only study experimentally the behavior at nonzero
temperatures, but in this sense the situation is no different from
ordinary critical phenomena: one never accesses the critical point
itself, but observes the critical scaling of various experimentally
accessible quantities in its neighborhood.

In practice, the most common situation in which one observes a quantum
critical point is the one sketched in Fig.  \ref{figphasediagram}, in
which there is a low temperature ordered state --- a ferromagnetic
state, antiferromagnetic state or charge density wave ordered state,
for instance --- whose transition temperature to the disordered state
or some other ordered state goes to zero upon varying some parameter.
In this case, the quantum critical point is then {\em also} the end
point of a $T=0$ ordered state. However, sometimes the ``ordered''
state really only exists {\em at} $T=0$; for example in
metal-insulator transitions and in quantum Hall effect transitions
\cite{sondhi,prange}. A well-known example of this case in spin models is
in two-dimensional antiferromagnetic quantum Heisenberg models with
"quantumness' as a parameter $g$ \cite{chn}, which
do not order at any finite temperature, but which show genuine ordered
phases at $T=0$ below some value $g<g_c$.

Although the question concerning the origin of the
behavior of high temperature superconductors is not settled yet, there
are strong indications, discussed in the next section, that much of
their behavior is governed by the proximity to a quantum critical
point.

One of the first formulations of what we now refer to as quantum
critical behavior was due to Moriya \cite{moriya1,moriya2} and Ramakrishnan
\cite{ramakrishnan}, who did an RPA calculation for a model of
itinerant fermions with a Stoner-type instability to a ferromagnetic
state. In modern language, their approach amounts to a $1/N$
expansion. Various other important contributions
were made \cite{bealmonod,doniacheng}.  The now standard more modern
formulation, which we will
follow, is due to Hertz \cite{hertz}. A nice introduction can be found
in the article by Sondhi {\em et al.}  \cite{sondhi}, and for a
detailed expose we refer to the book by Sachdev \cite{sachdev}. See
also the review by Continento \cite{continentino2}.

\subsection{Quantum critical points in ferromagnets, antiferromagnets,
  and charge density waves}

\begin{figure}
\begin{center}   
  \epsfig{figure=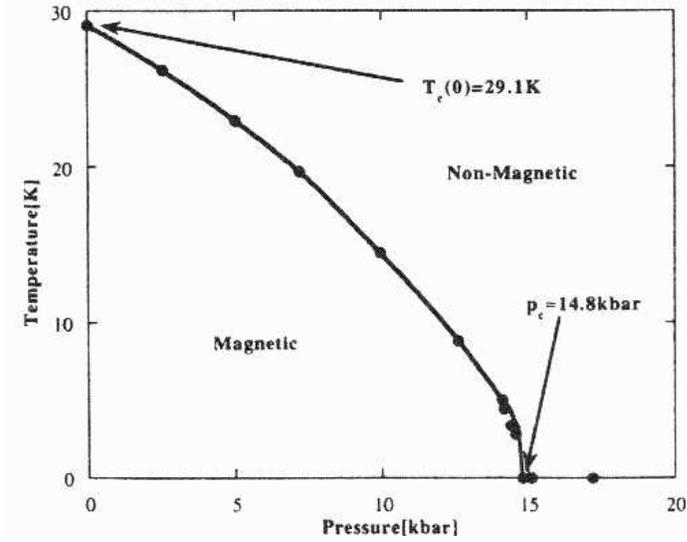,width=0.65\linewidth}
\end{center}
\caption[]{Magnetic phase diagram as a function of pressure of $MnSi$
\cite{lonzarich,lonzarich3,lonzarich7}. From \cite{lonzarich7}.}
\label{datamnsi1}
\end{figure}

\begin{figure}
\begin{center}   
  \epsfig{figure=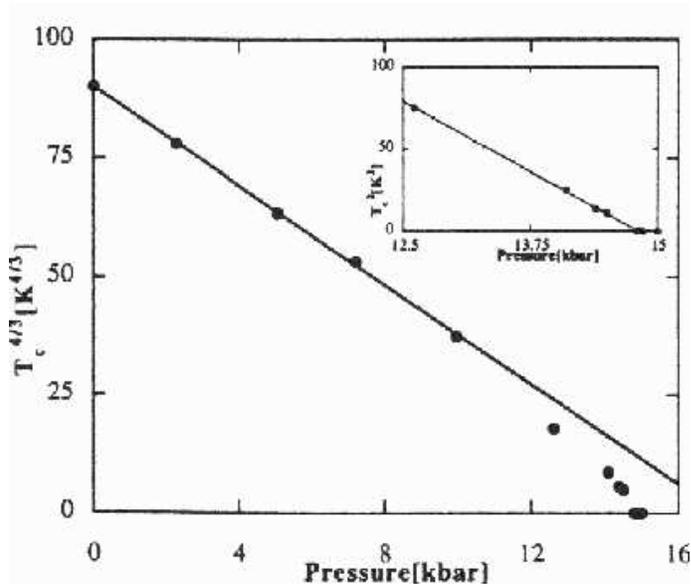,width=0.65\linewidth}
\end{center}
\caption[]{ Power law dependence of the
  Curie-temperature as a function of pressure for $MnSi$. From \cite{lonzarich7}.}
\label{datamnsi2}
\end{figure}

\begin{figure}
\begin{center}   
  \epsfig{figure=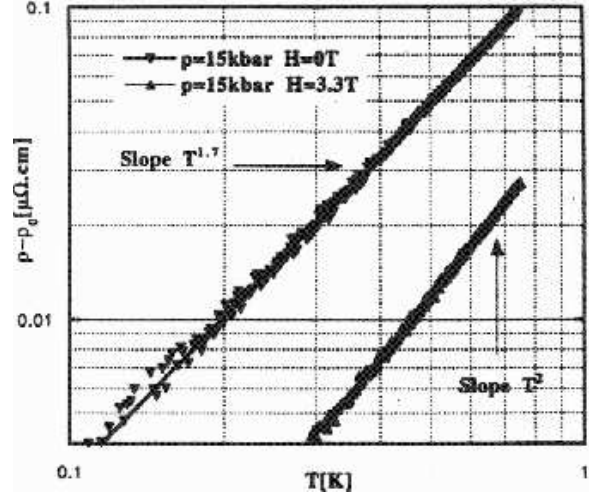,width=0.55\linewidth}
\end{center}
\caption[]{$T$-dependence of $(\rho(T)-\rho_{T=0})$
near the critical pressure $p_{c}$
with and without an external
magnetic field. From \cite{lonzarich7}.}
\label{datamnsi3}
\end{figure}

\begin{figure}
\begin{center}   
  \epsfig{figure=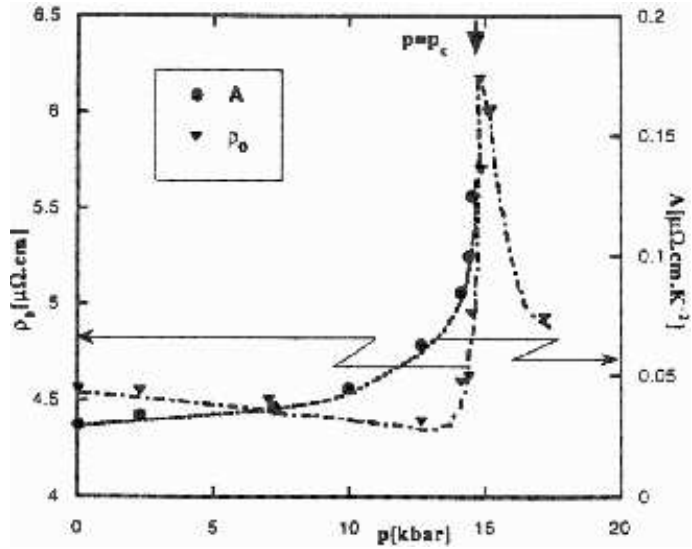,width=0.65\linewidth}
\end{center}
\caption[]{ Evolution of $\rho_{0}$ and A under pressure, when the
temperature dependence of the resistivity is fitted to  $(\rho =
\rho_{0}+ A T^{2})$. From \cite{lonzarich7}. }
\label{datamnsi4}
\end{figure}

A clear example of quantum critical behavior, and actually one for
which one can compare with theoretical predictions, is summarized in
Figs. \ref{datamnsi1}-\ref{datamnsi4}. The figures show various data from \cite{lonzarich7} 
 on the
magnetic compound $MnSi$ \cite{lonzarich,lonzarich3,lonzarich4,lonzarich5,lonzarich7}. Fig. \ref{datamnsi1} shows that for low pressures and temperatures, this compound
exhibits a
magnetic phase whose transition temperature $T_c$ vanishes
as the pressure is increased up to $p_c=14.8 ~kbar$. This
value of the pressure then identifies the quantum critical
point. Fig. \ref{datamnsi2} shows that when the same data are plotted as
$T_c^{4/3}$ versus pressure, the data fall nicely on a straight line
except close to the critical pressure. This observed scaling of
$T_c^{4/3}$ with $p-p_c$ away from $p_c$ is  in accord with the behavior
predicted by the theory discussed below. Actually  the transition is weakly
 first order near $p_c$;
so very detailed verification of the theory is not possible.

Fig. \ref{datamnsi3}  shows data for the temperature dependence
of $\rho - \rho_0$ near $p_c$, where $\rho_0$ is the residual low temperature
resistivity. In the presence of a field of 3 Tesla, one observes the
usual $\rho-\rho_0\sim T^2$ Fermi-liquid scaling, but at zero field
 the results are consistent with$\rho-\rho_0\sim T^{5/3}$ behavior predicted by the
theory. But if we write the low temperature resistivity behavior
as
\begin{equation}
\rho=\rho_0 + A T^\theta \label{rhoATformula}
\end{equation}
then both the residual resistivity $\rho_0$ and the amplitude $A$ are
found to show a sharp peak at $p_c$ as a function of pressure --- see
Fig. \ref{datamnsi4}. This behavior is not understood nor is the fact that
the exponent $\theta$ does not appear to regain the Fermi-liquid value
of 2 for significant values of $p > p_c$ (at $H=0$) and in a temperature regime
where the theory would put the material in the quantum-disordered
Fermi-liquid regime.

\subsection{Quantum critical scaling}\label{qcs}

Before discussing other experimental examples of  quantum critical
behavior it is expedient to summarize some of the essential quantum
critical scaling ideas.

As is well known, at a finite temperature transition, the critical
behavior is classical and we can use classical statistical
mechanics to calculate the correlation functions. This is so because due to critical
slowing down the characteristic time scale $\tau$ diverges with the
correlation length,
\begin{equation}
\tau \sim \xi^z~.
\end{equation}
Near a critical point the correlation length $\xi$ diverges as
\begin{equation}
\xi \sim |T-T_c|^{-\nu}~.
\end{equation}
The combination of these two results shows that
critical slowing down implies that near any finite temperature
critical point the characteristic frequency scale $\omega_c$ goes to
zero as
\begin{equation}
\omega_c \sim |T-T_c|^{\nu z}~.
\end{equation}
Therefore near any phase transition  $\omega_c \ll T_c$, and as a result
the phase transition is governed by classical statistical physics: the
Matsubara frequencies are closely spaced relative to the temperature,
the thermal occupation of bosonic modes is large and hence classical, etc.

In
classical statistical mechanics, the dynamics
is a {\em slave} to the statics; usually, the dynamical behavior is
adequately described by time-dependent Landau-Ginzburg type of
equations or Langevin equations which are obtained by building in the
appropriate conservation laws and equilibrium scaling
behavior \cite{halperinandhohenberg}.  At a quantum critical point,
 the
dynamics must be determined  a priori
from the quantum-mechanical equations of motion.

The general scaling behavior near a $T=0$ quantum critical point can
however be discussed with the formalism of dynamical scaling \cite{halperinandhohenberg,ma},
 just as
near classical critical points. Consider for
example the susceptibility for the case of $MnSi$ that we considered
above. The scaling ansatz for the singular part of the susceptibility
\begin{equation}
\chi(k, \omega, p) = \langle M M \rangle (k, \omega,p)
\end{equation}
implies that near the critical point where the correlation length and
time scale diverge, the zero-temperature susceptibility $\chi$
is a universal scaling
function of the scaled momentum and frequency,
\begin{equation}
\chi (k,\omega,p) = \xi^{-d_M} {\Upsilon} (k\xi, \omega \xi_t )~, \label{chieq}
\end{equation}
where now
\begin{equation}
\xi \sim |p-p_c|^{-\nu}~,~~~~~ \xi_t \equiv \tau =
\xi^z~. \label{nuandz}
\end{equation}
This is just like the classical scaling with $T-T_c$ replaced by
$p-p_c$. The reason for writing $\xi_t$ instead of $\tau$ is that in
quantum statistical calculations, the ``time-wise'' direction becomes
like an additional dimension, so that $\xi_t$ plays the role of a
correlation ``length'' in this direction. However, the time-direction
has both a long-time cutoff given by $1/k_B T$ and a short-time cutoff
given by the high energy cutoff in the problem --- exchange energy or
Fermi-energy, whichever is smaller in the ferromagnetic problem. The
short-time cutoff has its analog in the spatial scale. The long-time
cutoff, which determines the crossover from classical to quantum
behavior, plays a crucial role in the properties discussed
below.
The  crucial point is that when $z\neq 1$, there is
anisotropic scaling between the space and time-wise direction, and as
we shall discuss below, this implies that as far as the critical
behavior is concerned, the effective dimension of
the problem is $d+z$, not $d+1$.

The exponent $d_M$ in (\ref{chieq}) reflects that a correlation
function like $\chi$ has some physical dimension which often is inevitably related to
the spatial dimension. The dependence of
critical properties on spatial dimensions
must be expressible purely in terms of the divergent correlation
length $\xi$.
 Often $d_M$ is
fixed by dimensional considerations (in the language of field theory,
it is then given by the ``engineering dimension'' of the field), but
this may not be true in general. It must be so,
however, if $\chi$ is a correlation function of a conserved
quantity\footnote{E.g., if we consider the free energy {\em per unit
volume} at a classical transition, the energy scale is set by $k_BT$,
and $d_M=d$; likewise, if we consider the surface tension of an
interface, whose physical dimension is energy per unit surface  area,
$d_M= d-1$.}.

\begin{figure}
\begin{center}   
  \epsfig{figure=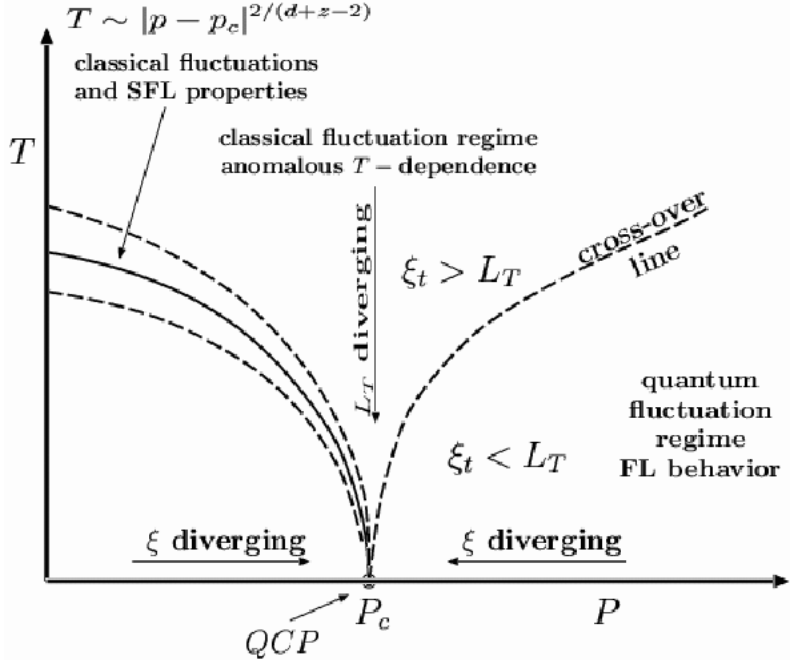,width=0.75\linewidth}
\end{center}
\caption[]{Generic phase diagram and cross-overs for quantum critical
  points with the various regimes indicated. $\xi $ is the correlation
  length at $T=0$; $L_T$ is the ``thermal correlation length''.}
\label{qcpscaling}
\end{figure}

Let us now address the finite temperature scaling, taking again the
case of MnSi as an example.  The various regimes in the $T$-$p$
diagram discussed below
are indicated in Fig. \ref{qcpscaling}. To distinguish these regimes
it is necessary to define an additional
 quantity, the
thermal length
\begin{equation}
L_T \equiv {{\hbar}\over{k_B T}}~.
\label{LT}
\end{equation}
 $L_T$ corresponds dimensionally to a timescale. It marks the crossover
between  phenomena at long time scales which can be treated
essentially classically from those on a shorter scale which are inherently
quantum mechanical. Whenever $\xi_t < L_T$,  the correlation
length and time are  finite and  quantum
mechanics begins to dominate. This is the regime on the right in
the figure. Fermi-liquid behavior is expected in this
regime. However, if one approaches the critical point $(T\!=\!0,~  p\!=\!p_c)$
from above along the vertical line, then $\L_T$ diverges but so does
$\xi_t$. Moreover, $\xi_t$ diverges faster than $L_T$ since $z$ is
usually larger $>1$. This means that
the characteristic fluctuation energy and temperature are similar. So the behavior is quasiclassical
 throughout each
 correlated region down to zero
 temperature (In a path integral formulation  \cite{sondhi,sachdev} one
considers the model on a infinite strip whose width is finite in the
timewise direction and equal to $L_T$. Hence for $\xi_t>L_T$ the model
is fully correlated across the strip in this direction). This regime is
therefore characterized by anomalous $T$
dependence in the physical quantities up to some ultra-violet cutoff.
It is important to stress that this so-called ``quantum critical
scaling behavior''\footnote{The term is somewhat problematic;
it refers to the quasiclassical fluctuation regime around
a quantum critical point.} is expected in the observable properties
 in the whole  region above $p_c$.
So SFL can be observed over a whole range of parameter $p$
and temperatures between
the left and right cross-over lines.

If one approaches the line of phase transitions to the ordered phase,
which is marked with a solid line in the figure, one has a region
with SFL properties dominated by classical
fluctuations  close to the
transition. Millis   \cite{millis} has corrected Hertz's results
 \cite{hertz} on this point, and has
found that the critical temperature  of the
phase transition
scales as $T_{c} \sim |p-p_c|^{z/(d+z-2)}$.  Estimates of the classical
critical region are also given. Results along similar lines
may also be found in
\cite{continentino1,continentino2}.

If we include both the temperature and the parameter $p$, the scaling
ansatz for the imaginary part of $\chi$ becomes
\begin{equation}
\chi''(k,\omega,p,T)~\sim ~\xi^{-d_M} { \Upsilon}_1 \left( k \xi , \omega \xi^z,
{{\omega}\over{T}} \right)~. \label{upsilon1}
\end{equation}
This can be rewritten in other forms depending on which experiment is
being analyzed. For example, the above form is especially suitable for
analysis as a function of  $(p-p_c)$. For analyzing data as a function
of temperature, we may instead rewrite
\begin{equation}
\chi''(k,\omega,p,T)
= L_T^{-d_M /z} { \Upsilon}_2 \left( k
L_T^{1/z}, \omega L_T, {{L_T}\over{\xi_t}} \right)~,
\end{equation}
and for analyzing data as a function of frequency
\begin{equation}
\chi''(k,\omega,p,T)= T^{-d_M/z} { \Upsilon}_3 \left( {{k}\over{T^{1/z}}} ,
{{\omega}\over{T}}, {{1}\over{T \xi_t}} \right) ~.
\end{equation}

Moreover, the scaling of the free energy ${\mathcal F}$ can be
obtained from the
argument that it is of the order of the thermal energy $k_B T$ per
correlated volume $\xi^d$. Moreover, since $L_T$ acts as a finite
cutoff for $\xi_t$ in the timewise direction, we then get the scaling
\begin{equation}
{\mathcal F} \sim T \xi^{-d} \sim T (\xi_t)^{-d/z} \sim T^{1 + d/z}~.
\end{equation}
By differentiating twice, this also gives immediately the specific
heat behavior at low temperatures. In writing the above scaling forms,
 we have assumed that no ``dangerously irrelevant
variables'' exist, as these could change $\omega/T$ scaling to
$\omega/T^{\Delta}$ scaling\footnote{``Dangerously irrelevant
variables'' are irrelevant variables which come in as prefactors of
scaling behavior of quantities like the Free Energy
\cite{irrelevant}. Within the
Renormalization Group scenario, the hyperscaling relation $d \nu
=2-\alpha$ is violated above the upper critical dimension because of
the presence of dangerously irrelevant variables. Presumably,
dangerously irrelevant variables are more important
than usually  at QCP's, since the effective fluctuation
dimension is above the upper critical dimension for $d=3 $ and $z>1$. Some examples
are discussed in  \cite{sachdev}.}.

In order to get the critical exponents and the cross-over scales, one
has to turn to a
microscopic theory. The theory for this particular case of the
quantum critical point in $MnSi$ is essentially a Random Phase
Approximation and proceeds along the following lines: {\em (i)} One
starts with a model of interacting fermions; {\em (ii)} An
ordering field $M(k,\omega)$ is introduced; {\em (iii)} One assumes
the fermions can be eliminated near the critical point to get a free
energy in terms of $M$ of the form
\begin{eqnarray}
{\mathcal F} & = & \int {\rm d}\omega \int {\rm d}^{d}k \chi^{-1}(k,\omega)
|M(k,\omega)|^2 + \nonumber \\
& & \hspace*{1.2cm} + \int {\rm d} \{ \omega \} \int {\rm d} \{ k \}
~V
M(k_1,\omega_1) M (k_2, \omega_2) M k_3
, \omega_3) M (k_4, \omega_4)\nonumber \\
& & \hspace*{2.2cm} \times ~ \delta(\omega_1 + \omega_2 + \omega_3 +
\omega_4 ) \delta(k_1+k_2+k_3+k_4) + \cdots  \label{glexpr}
\end{eqnarray}
Note that this is essentially an extension of the usual
Landau-Ginzburg-Wilson free energy to the frequency domain. Indeed,
from here on one can follow the usual analysis of critical phenomena,
treating the frequency $\omega$ on an equal footing with the momentum
$k$.

The important result of such an analysis is that the effective
dimension as far as the critical behavior is $d+z$, {\em not} $d+1$ as
one might naively expect. Since $z\ge 1$ in all known examples, the
fact that the effective dimension is bigger than $d+1$ reflects the
fact that the correlation ``length'' $\xi_t$ in the timewise direction grows
as $ \xi^z$, i.e. faster than the spatial correlation
length\footnote{This has important consequences for a scaling analysis
of numerical data, aimed at determining the critical behavior. For it
implies that the finite size scaling has to be done {\em
anisotropically}, with the anisotropy depending on the exponent $z$
which itself is one of the exponents to be determined from the
analysis}. Moreover, the fact that $z\ge 1$ implies that the
effective dimension of a $d=3$ dimensional problem is always larger or
equal than four. Since the upper critical dimension above which mean
field behavior is observed equals four for most
critical phenomena, one thus arrives immediately at the important
conclusion that most quantum critical points
in three dimensions should exhibit classical
fluctuations with mean field scaling exponents! It
also implies that the critical behavior can typically be seen over a
large parameter or temperature range --- the question of the width of
the critical region, which normally is determined by the Ginzburg
criterion, does not arise. On the other hand, questions concerning the
existence of dangerously irrelevant variables, due for example to the
scaling of the parameters $V$ in Eq. (\ref{glexpr}), do arise.

In order to judge the validity and generality of these results, it is
important to keep in mind that they are derived {\em assuming} that
the coefficients of the $M^2$, $M^4$ terms are analytic functions of
$k$, $\omega$ and the pressure $p$, etc. This is completely in line
with the usual assumption of analyticity of the bare coupling
parameters in a renormalization group approach. This assumption may
well be violated --- in fact none of the impurity models discussed
earlier can be treated along these lines: the fermions can not be
integrated there, and if one attempts to apply the above procedure,
one finds singular contributions to the bare coupling
parameters. Later on we shall discuss a three-dimensional
experimental example where this assumption appears to be
invalid. Secondly, it is inherently an expansion
about the nonmagnetic state, which can not apply in the ordered
phase: In the ordered phase with nonzero magnetization, $M\neq 0$,
there is a gap for some momenta in the fermionic spectra. This gap can
not be removed perturbatively.

In a  {\em ferromagnet}, the ground state susceptibility on the disordered
 side is given by
\begin{equation}
\chi^{-1} (k, \omega) = \left[ (p-p_c) + k^2 + {{ i \omega}\over{k
v_F}} \right]  ~. \label{fmchi}
\end{equation}
In the first two terms we recognize the usual mean field type behavior
with a correlation length that diverges as $\xi \sim (p-p_c)^{1/2}$,
hence the critical exponent $\nu=1/2$. The last term, which describes
Landau damping of the spin wave modes, is very special here as it arises
from  fluctuations of magnetization, a quantity which is
conserved, i.e., commutes with the Hamiltonian. Therefore the
characteristic damping rate must approach zero as $k\rightarrow 0$.  Since at
the critical point (\ref{fmchi}) leads to a damping $\omega \sim k^3$,
the critical exponent $z=3$. According to the theory described above,
the critical behavior at the quantum critical point $(T\!=\!0,\; p\!=\!p_c)$ is
therefore of the mean-field type for any physical dimension $d>1$, since the
effective dimension $d+z > 4$ for $d>1$.

The scattering of electrons off the long wavelength spin waves  is dominated by
small angle  scattering, and it is easy to calculate the resulting
dominant behavior of the self-energy of the electrons. Near the critical
point the behavior of
$\chi$ is very similar to  the electromagnetic problem we
discussed in section \ref{sflfromgauge}. Analogously,
one also finds SFL behavior here: in $d=3$, $\Sigma(k_F, \omega) \sim
\omega \ln \omega + i |\omega |$ while in $d=2$ one obtains
$\Sigma(k_F,\omega) \sim \omega^{2/3} + i|\omega|^{2/3}$. Furthermore, for
the resistivity one finds in three dimensions $\rho \sim T^{5/3}$ ---
this is consistent with the behavior found in MnSi in the absence of a
field, see Fig. \ref{datamnsi3}. Moreover, as we mentioned earlier,
 according to  the theory, near the critical point $T_c$ should
vanish as $|p-p_c|^{z/(d+z-2)}$; with $d=z=3$ this yields $T_c \sim
|p-p_c|^{3/4}$. As we saw in Fig. \ref{datamnsi2}, this is
the scaling observed over a large range of pressures, except very near
$p_c$. $ZrZn_2$ \cite{grosche} is an example in which
the Ferromagnetic transition is shifted
to $T=0$ under pressure continuously. The properties are again
consistent with the
simple theory outlined. There is however trouble on the horizon
\cite{lonzarich2}. The asymptotic temperature dependence
for $p>p_c$ is not proportional to $ T^2$, as expected. We will return
to this point
in sections \ref{specialcompl}.

For {\em antiferromagnets} or {\em charge density waves} the critical
exponents are different. In these cases, the order parameter is not
conserved, and the inverse susceptibility in these cases is of the form
\begin{equation}
\label{chiaf}
\chi^{-1}(k,\omega) = \left[ (p-p_c) + (k-k_0)^2  + {{i
\omega}\over{\Gamma}} \right] ~,
\end{equation}
where $k_0$ is the wavenumber of the antiferromagnetic or charge
density wave order. From this expression we immediately read off the
mean field exponents
$z=2$ and $\nu=1/2$. Since the effective dimension $d+z$ is above the upper
critical dimension for $d=3$, the mean field behavior is robust in
three dimensions. In $d=2$, on the other hand, the effective dimension
$d+z=4$ is {\em equal to} the upper critical dimension, and hence one
expects logarithmic corrections to the mean field
behavior. Indeed, in
two dimensions one finds for the self-energy
\cite{hlubina} $\Sigma(\hat{\bk}_F,\omega) \sim
\omega \ln \omega +i|\omega| $ (only)for those $\hat{\bk}_F$ from which
spanning vectors to other regions of the Fermi-surface separated by
$\bk_0$ can be found;  the
resistivity  goes as $\rho(T)\sim
T^2\ln T$ in this case. However, if several bands cross the
Fermi-surface, as often happens in heavy fermions, Umklapp-type
scattering may enforce the same temperature dependence in the resistivity
as in the single particle self-energy, except at some very low crossover
temperature. The physical reason for this dependence despite the fact
that the soft modes are at large momentums (and therefore vertex
corrections do not change the temperature dependence of
 transport relaxation rates) is that the set of $\hat{\bk}_F$ usually covers a
small portion of the Fermi-surface.

\begin{figure}
\begin{center}   
  \epsfig{figure=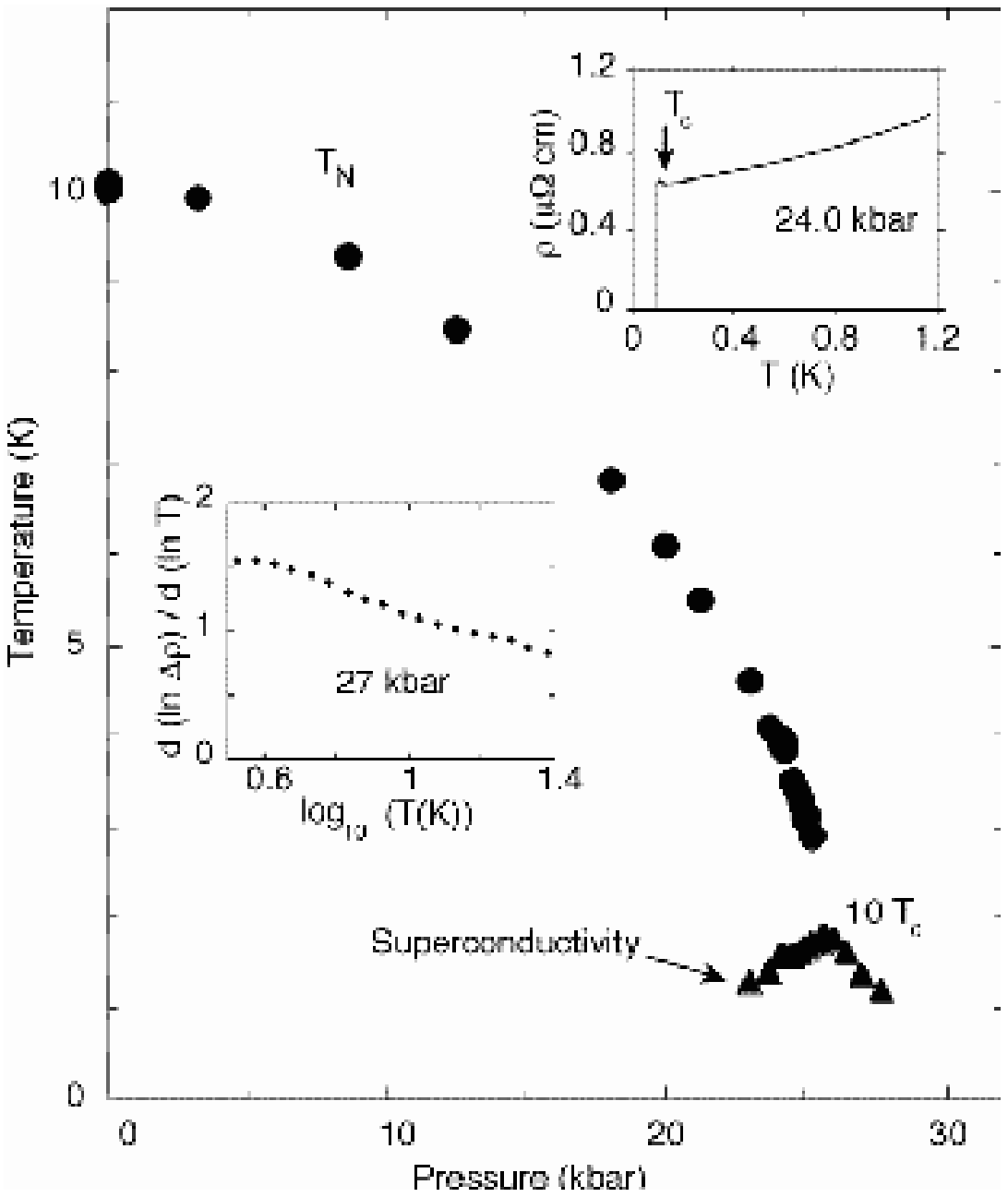,width=0.45\linewidth}
  \epsfig{figure=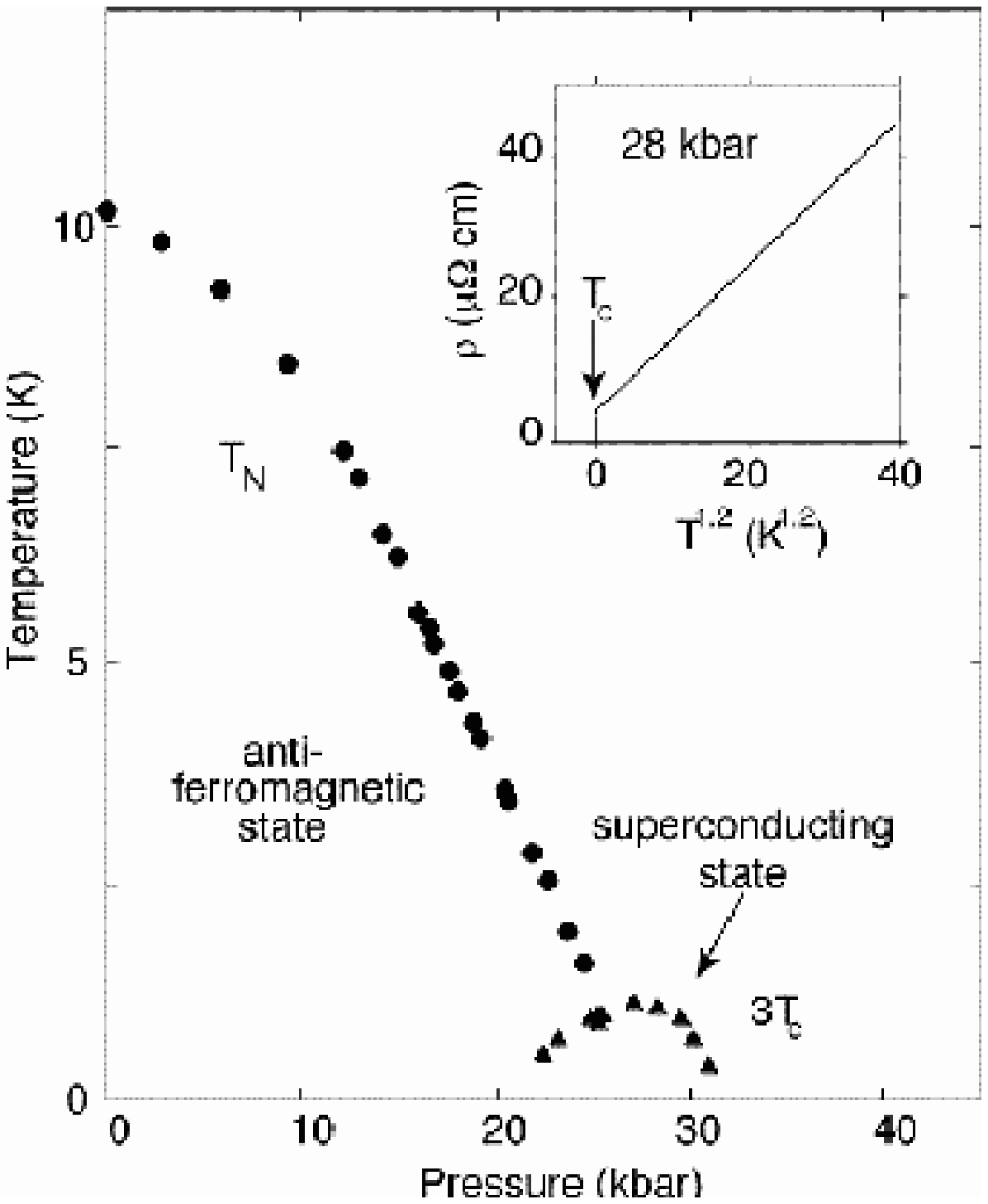,width=0.45\linewidth}
\end{center}
\caption[]{{\em Left panel:}
Temperature-pressure phase diagram of high-purity single-crystal
$CeIn_{3}$. A sharp drop in the resistivity consistent with the onset of
superconductivity below $T_{c}$  is observed in a narrow
window near $p_c$ the pressure at which the Néel temperature $T_{N}$
tends to absolute zero. Upper inset: this transition is complete
even below $p_{c}$ itself. Lower inset: just
above $p_{c}$, where there is no Néel transition, a
plot of the temperature dependence of d($\ln \Delta \rho$ )/d($\ln$ T)
is best able to demonstrate that the normal state resistivity varies as
$T^{1.6 \pm 0.2}$ below about 3 K.
$\Delta \rho$ is the difference between
the normal state resistivity and its residual
value (which is calculated by extrapolating the normal-state
resistivity to absolute zero). For clarity, the values
of $T_c$ have been scaled by a factor of ten.
{\em Right panel:}
Temperature-pressure phase diagram of high-purity single-crystal
$CePd_{2}Si_{2}$. Superconductivity appears below $T_{c}$ in
a narrow window where the Néel temperature $T_{N}$ tends to
absolute zero. Inset shows the normal state a-axis resistivity above
the superconducting transition varies as $T^{1.2 \pm 0.1}$ over
nearly two decades in temperature.
The upper critical field $B_{c2}$ at the maximum value of $T_{c}$
varies near $T_{c}$ at a rate of approximately $-6 T/K$.
For clarity, the values of $T_{c}$ have been
scaled by a factor of three,
and the origin of the inset has
been set at 5 K below
absolute zero.
 Both plots are from Mathur {\em et al.} \cite{lonzarich5}. }\label{cein}
\end{figure}

Fig. \ref{cein} shows the phase diagram of two compounds that order
antiferromagnetically at low temperatures. The first one, $CeIn_3$
is a three-dimensional antiferromagnet.
 A superconducting phase intervenes at very low  temperatures
(note the different scale on
which the transition to the superconducting phase is drawn), covering
 the region around the
quantum critical point at a pressure of about 26 kbar. At this
pressure, the normal state resistivity is found to vary as $\rho(T)
\sim T^{1.6 \pm 0.2}$ which is consistent with the theoretical
prediction that at a quantum critical point dominated by
antiferromagnetic fluctuations the resistivity should go
as
$\rho \sim T^{1.5}$. The right panel in Fig. \ref{cein} shows the phase diagram
and resistivity data of the three dimensional antiferromagnet
$CePd_2Si_2$; the data in this case are best fitted by $\rho \sim
T^{1.2}$; this is consistent with the theoretical prediction $\rho\sim
T^{1.25}$ which  results if one has a  $(k-k_0)^4$   dispersion around the AFM vector $k_0$
in one
direction  and the usual $(k-k_0)^2$ dispersion in the other two. No
independent evidence for such dispersion is however available yet.
In both of these cases, part of the region of superconductivity, in a
region bounded by a line
emanating from the QCP and going on to the antiferromagnetic to the
normal state line, is expected to be antiferromagnetic as well.

These two compounds are also of interest because the phase diagram bears
a  resemblance to the phase diagram of the high- $T_c$
copper-oxide based superconductors in which the conduction
electron
density is the parameter varied --- see. Fig. \ref{genericphasediagram}.
Unlike the heavy fermion compounds
where the ordered phase is antiferromagnetic, the  order in copper-oxides
near the QCP is not AFM. Its nature is in fact unknown. In the
heavy fermion compounds
superconductivity promoted by antiferromagnetic fluctuations
is expected to be of the $d$-wave variety \cite{miyake} as it is in
the high- $T_c$ copper-oxide compounds.
 
\subsection{Experimental Examples of SFL due to Quantum-Criticality: Open
  Theoretical Problems}\label{experimentalexamples}

We have discussed the observed quantum critical behavior in some
system which is largely consistent with the simple RPA-like
theoretical predictions. There are however quite a few experimentally observed
signatures of  singularities near QCPs, especially in heavy fermion compounds,
 which are not
understood theoretically by the simple RPA theory of the previous subsection.
In this section we present some prominent
examples of these.

The experimental observations fall in  two general categories, in
both of which the low-temperature resistivity does not obey the power
laws expected of Fermi-liquids: Compounds in which resistivity
decreases from its limit at $T=0$ and those in which it increases. In
both cases the $C_v/T$ is singular for $T\rightarrow 0$. It is reasonable
to associate the former with the behavior due to impurities and the
latter with the QCP properties of the pure system.  But as we
discussed in section \ref{twokondosection} the QCP due to impurities requires
tuning to special symmetries unlikely to be realized experimentally.
As we will discuss, the effect of impurities without any special symmetries
but coupling to the order parameter is
expected to be quite different near the QCP of
the pure system compared to far from it. Under some circumstances, it
is expected to be singular and may
dominate the observations.

\begin{figure}
\begin{center}   
  \epsfig{figure=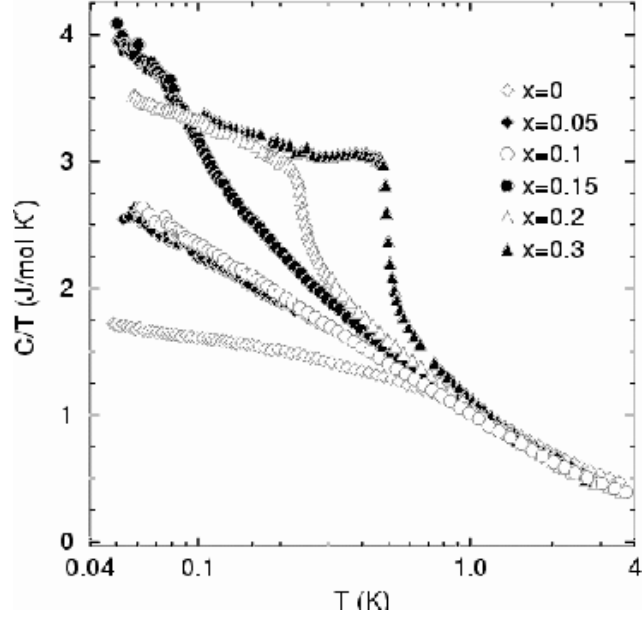,width=0.6\linewidth}
\end{center}
\caption[]{The specific heat $C/T$ of $CeCu_{6-x}Au_{x}$
versus $\log T$. From L\"ohneysen {\em et al.} \cite{lohneysen3,lohneysen2b,lohneysen4}.}  \label{cecuauspech}
\end{figure}

\begin{figure}
\begin{center}   
\epsfig{figure=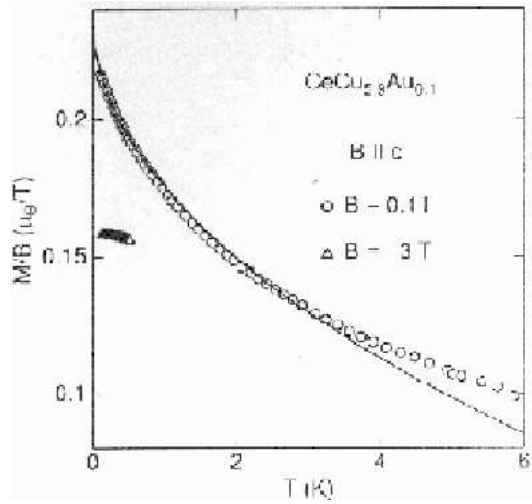,width=0.5\linewidth}
\end{center}

\caption[]{Susceptibility data for $CeCuAu$.
 From L\"ohneysen {\em et al.} \cite{lohneysen5}.}  \label{cecuaususcep}
\end{figure}

\begin{figure}
\begin{center}   
\epsfig{figure=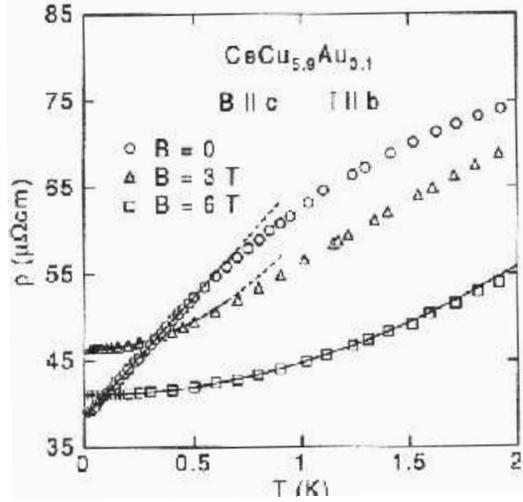,angle=0.9,width=0.5\linewidth}
\end{center}

\caption[]{Resistivity data for $CeCuAu$.
 From L\"ohneysen {\em et al.} \cite{lohneysen5}.}  \label{cecuauresist}
\end{figure}

We start with experiments in the second category.  Figs. \ref{cecuauspech}-\ref{cecuauresist}
shows several data sets for the heavy fermion compound
$CeCu_{6-x}Au_x$ for various amounts of gold. This compound shows a
low-temperature paramagnetic phase for $x < 0.1$ and a low temperature
antiferromagnetic phase for $x > 0.1$. The specific heat data of Fig. \ref{cecuauspech}
and the resistivity data of Fig. \ref{cecuauresist} show that while without
$Au$, i.e. in the paramagnetic regime, the behavior is that of a
 Heavy Fermi-liquid metal, the alloy near the quantum critical composition
 $CeCu_{5.9}Au_{0.1}$ exhibits a specific heat with an
anomalous
\begin{equation}
C_v\sim T \ln T~.
\end{equation}
 At the same composition, the resistivity shows a linear
temperature dependence, and the susceptibility data
 in Fig. \ref{cecuaususcep}  show a $\sqrt{T}$ cusp.  The
anomalous behavior is replaced by Fermi-liquid properties both by a magnetic field and by
increasing the substitution of copper by gold.  The compound $YbRh_2S_2$ seems
to have similar properties \cite{trovarelli}.  Related properties have
also been found in $U_2Pt_2In$ \cite{estrela1,estrela2} and in
$UPt_{3-x}Pd_x$ \cite{devisser}, $UBe_{13}$,$CeCu_2Si_2$, $CeNi_2Ge_2$
\cite{steglich}.  The SFL properties observed at the
Mott-Insulator-Metal Transition in $BaVS_3$ \cite{forro} are also of
related interest.

\begin{figure}
\begin{center}   
 \epsfig{figure=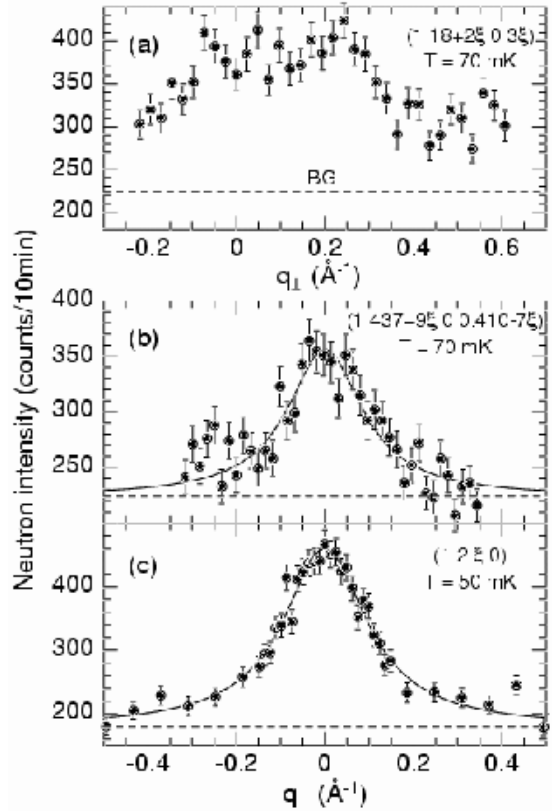,width=0.52\linewidth}
\end{center}
\caption[]{Neutron scattering  data for $CeCu_{5.9}Au_{0.1}$,  a
compound which is close to a QCP.  The figure shows $q$-scans along
three different crystallographic directions, from top to bottom in
the $a$, $b$ and $c$ directions for $\hbar \omega=0.1meV$. The figures
shows that there is only a weak $q$-dependence along the rods
($q_\perp$), while transverse scans ($q_{||}$) shows well defined
peaks with nearly the same line width. From Stockert {\em et al.}
\cite{stockert}.}  \label{cecuauscattering}
\end{figure}

None of the quantum critical properties of the $CeCuAu$ compounds
is consistent with any of the models that we have discussed.
Information on the magnetic fluctuation spectra for
$CeCu_{5.9}Au_{0.1}$ is available through
neutron scattering experiments \cite{schroeder,stockert}.  The data
shown in Fig. \ref{cecuauscattering}
show rod-like peaks, indicating that the spin fluctuations are almost
two-dimensional at this composition. The neutron scattering data
can be fitted by an expression for the
spin susceptibility of the form
\begin{equation}
\chi^{-1}(k, \omega) = C \left[ f(\delta k ) + (-i \omega + a
  T)^\alpha \right]
\end{equation}
with  a function $f$ which is consistent with an effectively
two-dimensional scattering,
\begin{equation}
f(\delta k ) = b (\delta k_\perp )^2 + c (\delta k_{\parallel})^4~.
\end{equation}
$\delta k_{\parallel}$ and $ \delta k_\perp$ are the deviations from the AFM
Bragg vector parallel and perpendicular to the c-axis in these tetragonal
crystals.
Further
\begin{equation}
a\approx 1 ~,~~~~~ \alpha = 0.74 \pm 0.1~.
\end{equation}
At present, there is no room for this anomalous exponent $\alpha$
within known theoretical frameworks. However, if one accepts this
particular form of $\chi$ as giving an adequate fit, the observed specific heat
follows: at the critical composition, we expect the
temperature scaling relation
\begin{equation}
{\mathcal F} \sim T \xi_{\perp}^{-(d-1)} \xi_{\parallel}^{-1} \sim
T  T^{(d-1)\alpha/2} T^{\alpha/4} \sim
 T^{1 +
   (d-1/2)\alpha /2}~,
\end{equation}
which immediately gives $C_V \sim d^2{\mathcal F}/dT^2 \sim T $ for $\alpha
= 4/5$ with a plausible logarithmic multiplying factor.

The observed resistivity does not follow directly from the measured $\chi$;
a further assumption is needed. The assumption that works is that fermions couple to
the fluctuations locally, as in an effective Hamiltonian
$\sim c^{\dagger}_{i,\sigma'}c_{i,\sigma}S_{i,\sigma,\sigma'}$. Then if the
measured fluctuation spectra are that of some localized
spins $S_i$, the single-particle self-energy is that due to exchange of bosons with
propagator  proportional to
\begin{equation}
 \sum_k \chi({\bf k},\omega) \sim \ln(\omega) +i \; \mbox{sgn}\;(\omega).
 \end{equation}
This ensures that the single-particle relaxation rate as well as the transport relaxation
rate is proportional to $T$. A major theoretical problem is why the non-local
 or "recoil" terms in the interaction of itinerant fermions are irrelevant; i.e
 why is the effective Hamiltonian not
 \begin{equation}
 \Sigma_{k,q}c^{\dagger}_{k+q,\sigma}c_{k,\sigma'}
(S_q+S_{-q}^{\dagger})_{\sigma',\sigma}. \nonumber
 \end{equation}
 Why has momentum conservation been legislated away?

The singularity of $\chi$ also raises the question whether the
anharmonic processes,  Fig. \ref{Mdiagrams}{\em
(c)}, which are benign and allow the
 elimination of  fermions in the RPA theory, give singular
contributions to $\chi$.
 Also can fermions really be eliminated in calculating the
 critical behavior.

\begin{figure}
\begin{center}
\epsfig{figure=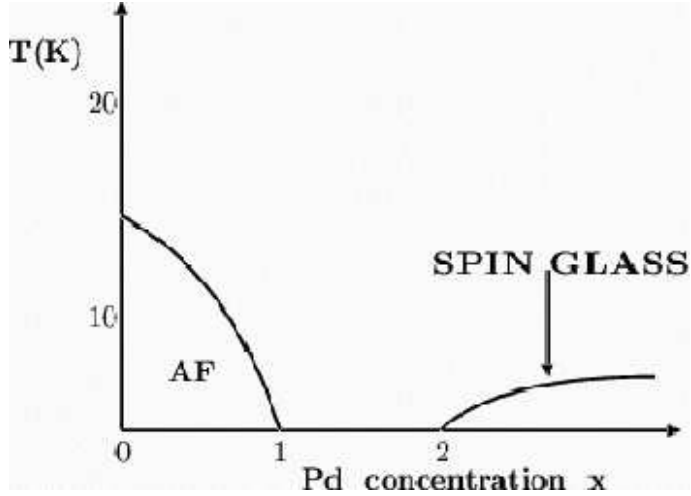,width=0.65\linewidth}
\end{center}
\caption[]{The phase diagram of $UCu_{5-x}Pd_{x}$.
At low dopings and temperatures
the system is in an antiferromgnetic
phase. In the undoped sample ($x=0$): $T_{N}$= 15 K
with a magnetization $\mu \simeq 1 \mu_{B}$.
For doping $x=1$ and 1.5 the specific
heat $C_{v}/T \sim T$ (for $x=1$) while
displaying weak logarithmic characteristics
for $x = 1.5$. Similarly, the susceptibility
$\chi(T) \sim \ln T$ and $\chi \sim T^{-0.25}$
(for $x=1,~ 1.5$ respectively). Courtesy of M. C. Aronson.}\label{aronsondata}
\end{figure}

As an example in the second category, we show in Fig.
\ref{aronsondata} the phase diagram of and some of the resistivity
data of the heavy fermion compound $U Cu_{5-x}Pd_x$.  There are
several other compounds in this category also; for a review we refer
to \cite{maple,mclaughlin}. For a
theoretical discussion of the scaling properties of some of this class
of problems see \cite{andraka}.

\begin{figure}
\begin{center}
\epsfig{figure=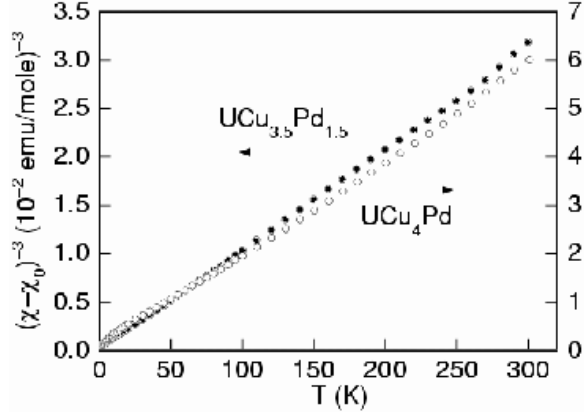,angle=90,width=0.55\linewidth}
\end{center}
\caption[]{The temperature dependence of the static susceptibility of
$\chi_0(T)$ for both $UCu_4Pd$ and $UCu_{3.5}Pd_{1.5}$, showing that
for both compounds $\xi_0(T)$ has a low temperature divergence as
$T^{-1/3}$.
 The measuring
field is 1 Tesla. From Aronson {\em et al.}
\cite{aronson1,aronson2}. }\label{figucupd}
\end{figure}

For $x < 1$, there is an (antiferromagnetic) ordered state at low
temperatures, while for $x>2$, a spin-glass phase appears. At first
sight, one would therefore expect possible SFL behavior {\em only}
near the critical composition $x=1$ and near $x=2$. The remarkable
observation however is that over a whole range of intermediate
compositions, one
observes anomalous behavior of the type \cite{aronson1,aronson2}
\begin{equation}
\rho = \rho_0 - B T^{1/3}~,~~~~~ \chi \sim T^{-1/3}
\end{equation}
The data for $\chi$ that show this  power law behavior for $UCu_{3.5}Pd_{1.5}$ and $UCu_1Pd$ are shown in Fig. \ref{figucupd}.
The fact that this is genuine scaling behavior is independently
confirmed \cite{aronson1,aronson2}
 from the fact that the frequency dependent susceptibility, measured by
 neutron scattering,
shows a very good collapse of the data with the scaling assumption
\begin{equation}
\chi'' (\omega, T ) = T^{-1/3} \Upsilon \left( {{\omega}\over{T}} \right)~.
\end{equation}
In passing, we note that this as well as the result for
$\chi$ in $CeCuAu$ are examples of an anomalous dimension,
as the engineering dimension of the susceptibility $\chi$ is
$1/energy$ --- see the remark made just after Eq. (\ref{nuandz}).

Again, none of this behavior finds a clear explanation in any of the
well-studied models. One is tempted to use the critical points of
impurity models (see for example \cite{cox} and references therein), but runs
into the difficulty of having to tune to special
symmetries. The ideas of critical points of metallic spin-glasses (see
for example \cite{sachdevread} and  also \cite{sengupta}),
although theoretically appealing, are also not obviously
applicable over such a wide range of composition.  It must be
mentioned however that NMR evidence does show clear evidence for the
inhomogeneity in the singular part of the magnetic fluctuations in
several heavy fermion compounds \cite{bernal}. This has inspired
models of varying sophistication (see \cite{miranda} and references
therein, and also \cite{castroneto}) in which
the Kondo-temperature itself has a inhomogeneous distribution. It is
possible to fit the properties with reasonable distributions but
there is room for a deeper examination of the theoretical issues
related to competition of disorder, Kondo effects and
magnetic-interaction between the magnetic moments.

\subsection{Special complications in heavy fermion physics}\label{specialcompl}

In heavy fermion compounds, there is often  an additional complication
that besets treating a QCP as a simple
antiferromagnetic transition coupled to itinerant electrons. Often
such materials exhibit magnetic order of the $f$-electrons (with
magnetic moments of the ${\mathcal O}(1 \mu_B)$  per $f$-electron).
Thus, such materials have local moments in the ordered phase; so
the disappearance of the (anti)ferromagnetic order at a quantum
critical point is accompanied by a metal-insulator transition of the
$f$-electrons. This means that the volume of the Fermi-surface changes
in the transition.

\begin{figure}
\begin{center}
\epsfig{figure=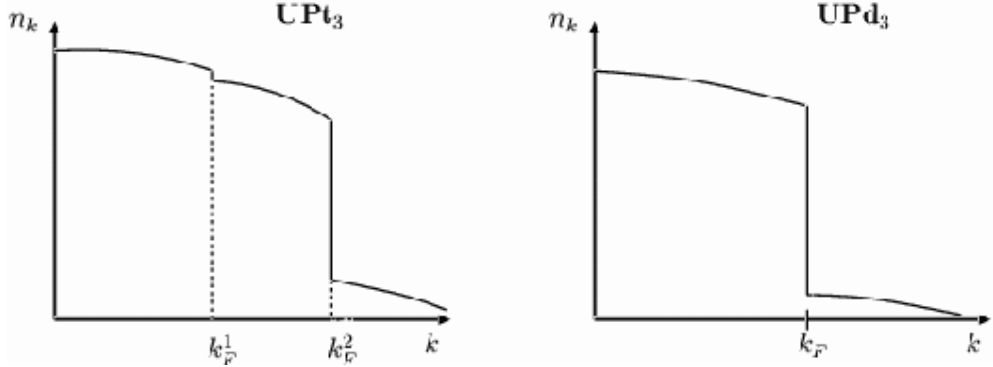,width=0.95\linewidth}
\end{center}
\caption[]{Schematic occupation number function $n_k$ for $UPt_3$ and for $UPd_3$.}\label{uptpd}
\end{figure}

We may illustrate the above scenario by comparing $U Pt_3$, a Heavy
fermion compound with effective mass of the order of 100, with
$UPd_3$, an ``ordinary'' metal with effective mass of ${\mathcal O}(1)$
 in which the f-electrons are localized. A schematic summary of the
 momentum occupation $n_k$
for the two cases is shown in Fig. \ref{uptpd}: in the former $n_k$
is shown with two discontinuities, one small ${\mathcal O}(10^{-2})$
representing the large renormalisation in
the effective mass of ``$f$-electrons'' while the other is close to 1 representing
the modest renormalisation of $s$ and $d$-electrons. The other case,
representative of $UPd_3$ has just one Fermi surface with a
 jump in
$n_k$ of close to 1. The Fermi surface in the former encloses the
number of electrons equal to the sum of the $f$ and $s$-$d$ electrons while the latter includes
only the $s$-$d$ electrons. This is consistent with de Haas-van Alphen measurements as
well as the band structures calculations of the two compounds; but the band-structure
calculation must be done with the f-electrons assumed as itinerant in the former and
part of the localized core in the latter. The magnetic transitions in heavy fermion
compounds (with ordered moment of ${\mathcal O}(1\mu_B)$)  occurs through the conversion
of itinerant $f$-electrons to localized electrons. So the Fermi surface on the two sides
of the transition must switch between the two schematic representations in
Fig. \ref{uptpd}.  The problem couples the
"metal-insulator transition" of the $f$-electrons to the magnetic
fluctuations --- those of itinerant electrons on one side and of
interacting local moments on the other. The fluctuations of the metal-insulator transition
and the Fermi surfaces is an important part of the problem. Some theoretical work with
these ideas in mind is available \cite{sirabello}. Another possible approach is to
generate an effective Hamiltonian for the heavy fermion lattice
from a pairwise sum of the effective Hamiltonian
deduced from the two-Kondo impurity problems discussed in section \ref{twokondosection}
and study its instabilities. The two impurity problem contains the rudiments
 of some of the essential physics.

In connection with the data in $CeCu_{6-x} Au_x$, we have discussed
two important puzzles: the
non-trivial exponent ${\em d_M/z}$ measured by $\chi(k,\omega)$,
and the coupling of fermions to the local fluctuations
alone for transport properties. In the other category (impurity-dominated)
the first puzzle reoccurs; the second puzzle may be explained more
easily since the
measured fluctuation spectrum $\chi(k,\omega)$ is in fact $k-$independent.
Both puzzles reoccur in the SFL phenomena in the cuprate
compounds to be discussed in section \ref{hightcsection}.

 \subsection{Effects of impurities on Quantum Critical
Points}\label{impuritysec}
 
 As is well known, randomness can have an important effect on
 classical phase transitions. Two classes of quenched disorder are
distinguished: First,
impurities coupling quadratically to the order parameter \cite{harris}
or, equivalently, impurities which may be used to define a local transition temperature
$T_c(r)$; the second class concerns impurities coupling linearly to the order
parameter \cite {imryma}.
The so-called Harris
 criterion, for the  first class, tells us that the disorder is relevant,
i.e. changes the exponents or turns the transition to a crossover,
 if the specific heat exponent $\alpha$ of the pure system is positive
or, equivalently, if
\begin{equation}
d\nu-2 <0~. \label{harriscriterion}
\end{equation}
For application to QCP phenomena the value of $\nu$ to be used is
different in the
quantum fluctuation regime and the quasi-classical
regime\footnote{Actually, the Harris criterion is derived
 in the form  (\ref{harriscriterion}) for $\nu$ and $d$, {\em not} in terms of
the exponent $\alpha$. This is particularly important at QCP's, since
as we discussed in section \ref{qcs} for QCP's one is  often {\em above} the
upper critical dimension where the hyperscaling relation $d \nu
-2=\alpha$ breaks down.} For the latter
 $\nu$ should be defined by the correlation
length $\xi\sim(T-T_c)^{-\nu_1}$ for a fixed $(p-p_c)$ while deep in the former,
near $T=0$ it should be defined by  $\xi\sim (p-p_c)^{-\nu_2}$.
Accordingly the effect of disorder depends on the direction one
 approaches the
QCP.
Similarly the celebrated Imry-Ma argument \cite{imryma} for linearly-coupled disorder
can be generalized  to QCP's.

In problems of fermions, additional effects of disorder arise because the
vertices coupling the impurity to the fermions can be renormalized
due to the singularity in the fluctuation of the pure system \cite{varma10}.
Not too much work has been done along these lines. A simple example is the effect
of magnetic impurities near a ferromagnetic transition \cite{larkinmelnikov}.
The growth of the
magnetic correlation length converts a single-channel Kondo effect to a multi-channel
Kondo effect with a regime in which the singularities discussed in
section \ref{multichannelsection} for the
degenerate multi-channel Kondo effect may be realized without tuning
any parameters \cite{maebashi}. This may be relevant to
the deviations from the predictions of the
pure case discussed here in the properties near the QCP in $MnSi$.
Extensions of these ideas to antiferromagnetic and other QCP's would
be quite worthwhile.

\section{The High-$T_c$ Problem in the Copper-Oxide Based Compounds} \label{hightcsection}

About $10^5$ scientific papers have appeared in the
field of high-$T_c$ superconductors since their discovery
in 1987. For reviews, see the Proceedings \cite{mms} of the latest in a
series of Tri-annual Conferences or \cite{ginsberg}.
  Although no consensus on the theory of the
phenomena has been arrived at, the intensive
investigation has resulted in a body of consistent
experimental information.  Here we emphasize only those
properties which are common to all members of the
copper-oxide family and in which Singular Fermi-liquid
properties appear to play the governing role.

The high-$T_c$ materials are complicated, and many
fundamental condensed matter physics phenomena play a role in some
or other part of their phase diagram.  As we shall
see, the normal state near the composition of the
highest $T_c$ shows convincing evidence of being
 a weak form of a SFL, a Marginal Fermi Liquid. Since the vertices
 coupling fermions to the fluctuations for transport
 in the normal state and those for Cooper pairing through exchange of
 fluctuations can be derived from each other,
the physics of SFL and the mechanism for superconductivity in the
cuprates are intimately related.

\begin{figure}
\begin{center}   
{\em (a)}
\epsfig{figure=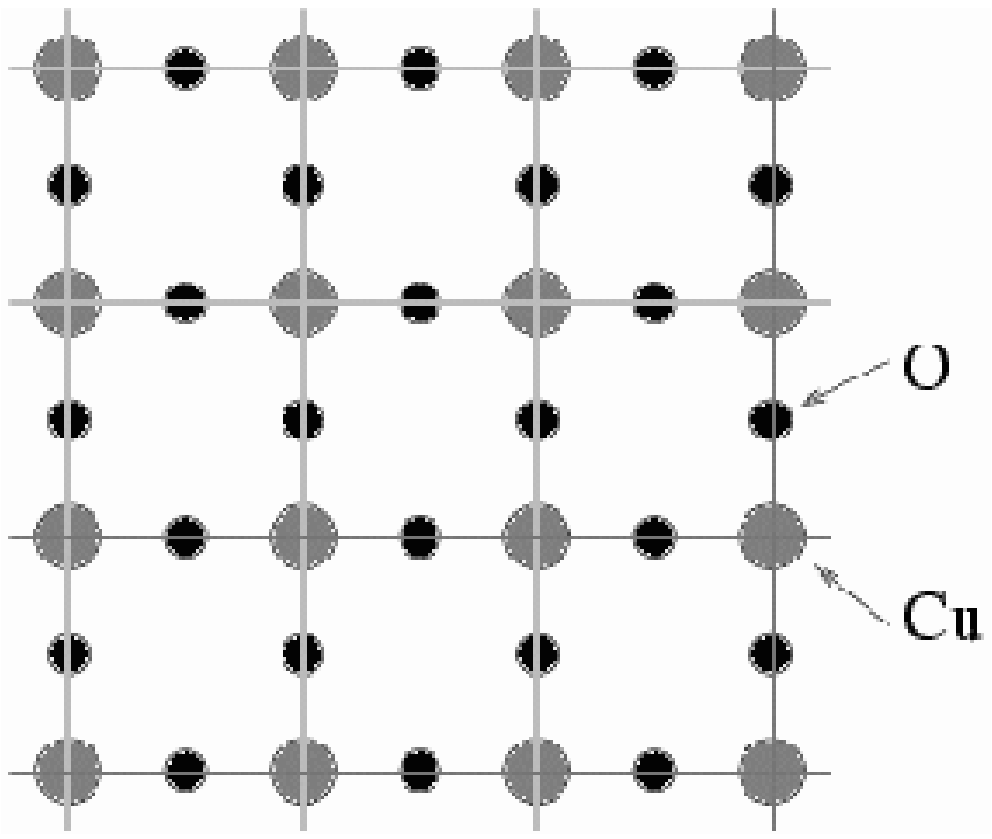,width=0.45\linewidth} \hspace*{1cm}
  \epsfig{figure=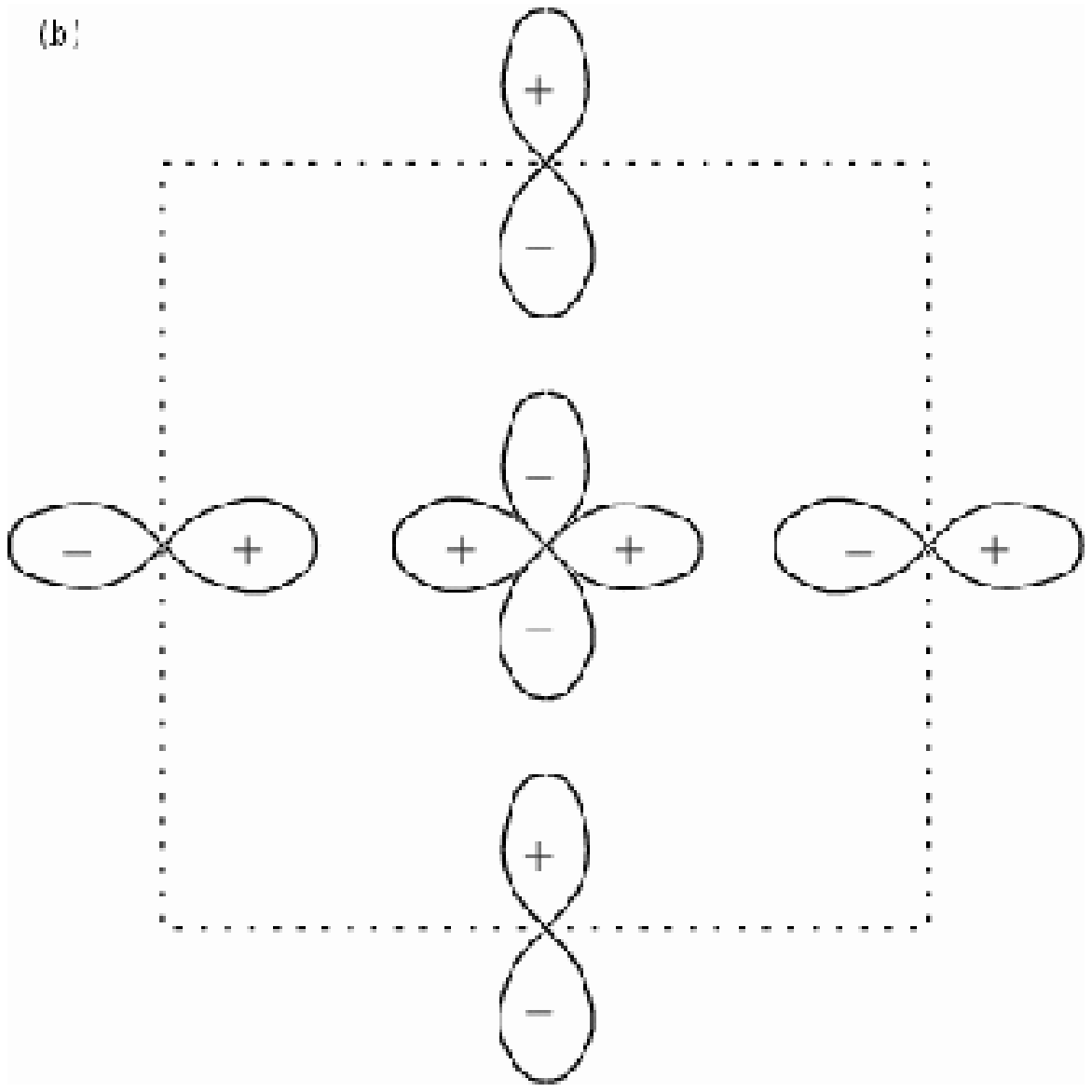,width=0.35\linewidth}
\end{center}
\caption[]{{\em (a)}
Schematic structure of the copper-oxide $ab$-planes in
$La_{2}CuO_{4}$. $Ba$ or $Sr$ substitution
for $La$ in the parent compound $La_{2}CuO_4$ introduces holes in the
$CuO_{2}$ planes. The structure of
other high T$_{c}$ materials differs only in ways which do not affect
the central issues, e.g., it is oxygen doping in
$YBa_{2}Cu_{3}O_{6+x}$ that
provides planar holes. The magnetic moments of the planar
copper atoms are ordered antiferromagnetically in the
ground state of the undoped compounds. From \cite{jacklic}. {\em (b)}
The ``orbital unit cell'' of the $Cu$-$O$ compounds in the
$ab$ plane. The minimal orbital set contains a $d_{x^{2}-y^{2}}$
orbital of $Cu$ and $p_{x}$ and a $p_{y}$ orbital of oxygen
per unit cell.}
\label{cuoplanes}
\end{figure}

\begin{figure}
\begin{center}   
   \epsfig{figure=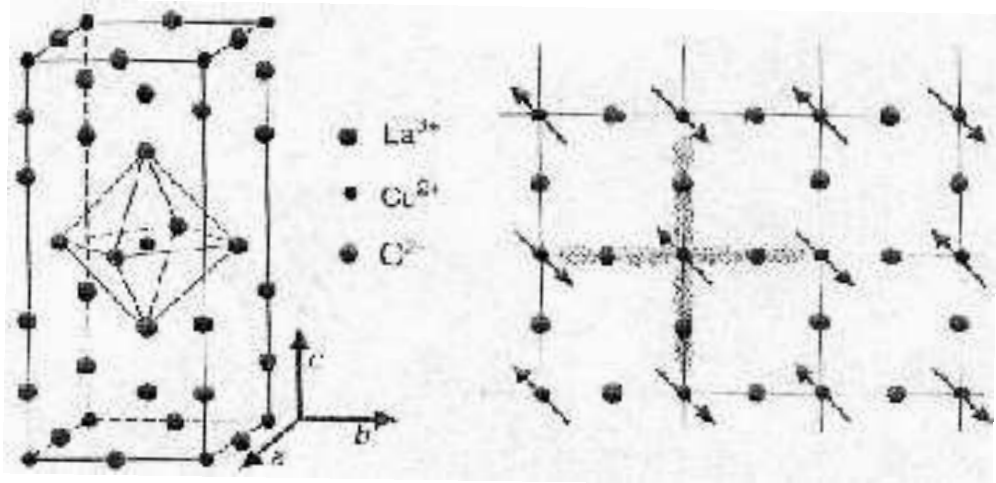,angle=-0.9,width=0.95\linewidth}
\end{center}
\caption[]{ {\em{a}})
The crystal structure of $La_{2}CuO_{4}$. From \cite{orenstein}.
Electronic couplings along the c direction are weak. {\em{b}}) Schematic of the
$CuO_{2}$ plane. The arrows indicate the alignment of spins in
the antiferromagnetic ground state of $La_{2}CuO_{4}$.
Speckeled shading indicates oxygen $p_{\sigma}$ orbitals;
coupling through these leads to superexchange in the insulating state.}
\label{lacuofigures}
\end{figure}

\subsection{Some basic features of the high-$T_c$ materials}\label{basicfeaturessection}

A wide variety of $Cu$-$O$ containing compounds with different
chemical formulae and different structures belong in
the high-$T_c$ family.  The common chemical and
structural features are that they all contain two-dimensional
stacks of $Cu$$O_2$ planes which are negatively charged with
neutralizing ions and other structures in between the
planes.  The minimal information about the structure in the $Cu-O$
planes and the important electronic orbitals on Cu and on Oxygen
is shown in Fig. \ref{cuoplanes}. The structure of one of the simpler compound
$La_{2-x}Sr_xCuO_4$ is shown in Fig. \ref{lacuofigures}{\em (a)}
with the
$Cu O_2$ plane  shown again in Fig. \ref{lacuofigures}{\em (b)}.
 For $x = 0$ the
$Cu O_2$ plane has a negative charge of 2 per unit cell
which is nominally ascribed to the
$Cu^{2+} (O^{2-})_2$ ionic configuration. Since $O^{2-}$ has a filled
shell while $Cu^{2+}$ has a hole in the 3d shell, the $Cu$-$O$ planes
 have a half-filled
band according to the non-interacting or one-electron model.
 But at $x = 0$ the compound is an antiferromagnetic
insulator with $S=1/2$ at the Copper sites.
  This is well known to be characteristic of a
Mott-type insulator in which the electron-electron interactions
determine the ground state. Actually
 \cite{zsa,varma4,varma5}
copper-oxide compounds at $x=0$
belong to the Charge-Transfer  sub-category of
Mott- insulators. But at $x=0$ the ground state and low-energy
properties of all Mott-insulators  are {\em qualitatively} the same.
By substituting divalent $Sr$ for the trivalent $La$ in
the above example, ``holes'' are introduced
in the copper-oxide planes with density $x$ per unit cell.

\begin{figure}
\begin{center}   
  \epsfig{figure=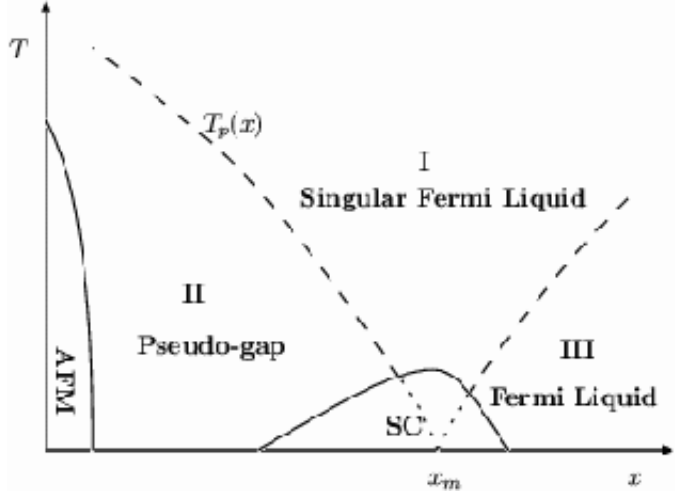,width=0.65\linewidth}
\end{center}
\caption[]{Generic phase diagram of the cuprates for hole doping. The
portion labeled by AFM is the antiferromagnetic phase, and
the dome marked by SC is the superconducting phase.
Crossovers to other characteristic properties are marked and
discussed in the text. Not
shown is a low-temperature ``insulating phase'' in region II due to
disorder. }\label{genericphasediagram}\label{batloggfigure}
\end{figure}

Fig. \ref{genericphasediagram} is the
generic phase
diagram of the $Cu$-$O$ compounds in the $T$-$ x$ plane.  In the few
compounds with electron doping which have been synthesized
properties vary with doping density in a similar way.

 Antiferromagnetism
disappears
for $x$ typically less than 0.05 to be replaced by a superconducting
ground state
starting at somewhat larger $x$. The superconducting transition temperature
is peaked for $x$ typically between 0.15 and 0.20 and disappears for
$x$ typically less than 0.25. We will define $x_m$ to be the
density for maximum $T_c$. Conventionally, copper-oxides with
$x <  x_m$ are referred to as underdoped, with $x=x_m$ as optimally doped
and with $x>x_m$ as overdoped.
Superconductivity is of the ``$d$-wave" singlet
symmetry.

The
superconducting region in the $T$-$x$ plane is surrounded by three distinct
regions:  a region marked (III) with properties characteristic
of a Landau Fermi-liquid, a region marked (I) in which
(marginally) SFL properties are observed and a region
marked (II) which is often called the
{\it pseudo gap} region whose correlations in the ground
state still remain a matter of conjecture.
  The
{\it topology} of Fig. \ref{genericphasediagram} around the superconducting
region is that expected around a QCP.  Indeed it resembles
the phase diagram of some heavy fermion superconductors
(see e.g. Fig. \ref{cein}) except that region II has no long-range
antiferromagnetic order --- the best experimental information
is that, generically, spin rotational invariance as well as (lattice)
translational invariance remains unbroken in the passage
from (I) to (II) in the $Cu$-$O$ compounds.

\begin{figure}
\begin{center}   
  \epsfig{figure=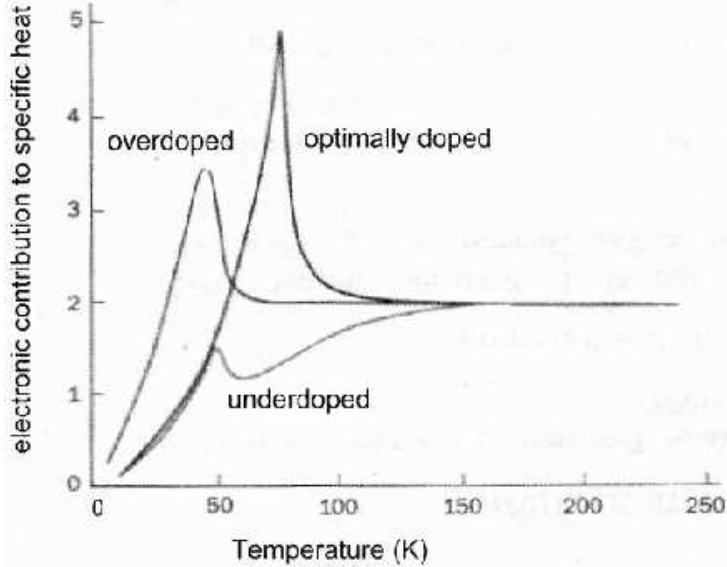,width=0.7\linewidth}
\end{center}
\caption[]{ The electronic contribution to the
specific heat as a function of temperature
for underdoped, optimally doped and overdoped samples of Y$_{0.8}$Ca$_{0.2}$Ba$_{2}$Cu$_{3}$O$_{7-x}$.
For optimally doped and overdoped
samples the heat capacity remains constant as the temperature
is lowered, then shows the characteristic features at the superconducting transition
temperature $T_{c}$ and rapidly
approaches zero in the superconducting state. For underdoped samples, however,
the heat capacity starts to fall well above $T_{c}$ as the temperature
is reduced, and there is only a small peak at $T_{c}$ indicating much smaller condensation
energy than the optimal and overdoped compounds \cite{loram}. From \cite{batloggref}.
 }\label{physworld2}
\end{figure}

\begin{figure}
\begin{center}   
  \epsfig{figure=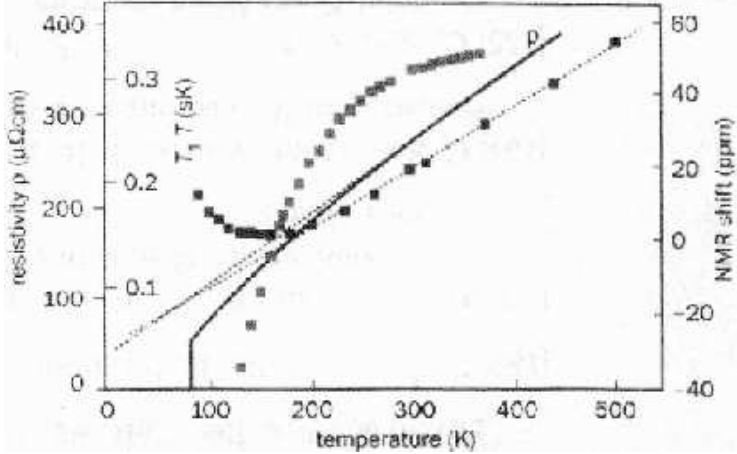,width=0.7\linewidth}
\end{center}
\caption[]{Signatures of the pseudogap in various transport properties
for the underdoped compound $YBa_{2}Cu_{4}O_{8}$.
At high temperatures the resistivity (solid line)
decreases linearly with temperature. In the
pseudogap region it drops faster with
temperature before falling to zero
at the superconducting transition
temperature (about 85K). Similarly the NMR
relaxation time displays characteristics of the optimum doped compounds
above about 200 K
(squares on dashed line) but deviates strongly from it in the pseudogap region.
The NMR shift  (top squares) also deviates from the
temperature-independent  behavior (not shown)
below the inset of the pseudogap. Note
that the pseudogap expresses itself as a sharper change with temperature
in the transport properties compared to the equilibrium properties
- specific heat and magnetic susceptibility.
 \cite{buechner,yasuoka,alloul}. From \cite{batloggref}.
 }\label{physworld3}
\end{figure}

The quantity
$\gamma (T) \equiv C_v(T)/T$ and the magnetic susceptibility
$\chi(T)$, which are temperature independent  for a
Landau-Fermi-liquid begin to decline rapidly \cite{loram} in the
passage from region I
to region II, which we will call $T_p(x)$, but without
 any singular feature.
However, the transport properties --- resistivity, Nuclear relaxation rate (NMR),
etc. ---  show  sharper change in their temperature dependence at $T_p(x)$.
The generic deduced electronic contribution to the specific heat for
overdoped, optimally-doped and underdoped compounds is shown in Fig. \ref{physworld2}.
 The generic behavior for an
 underdoped compound for the resisitivity, Nuclear relaxation rate and
Knight shift --- proportional to the uniform susceptibility --- is shown in
Fig. \ref{physworld3}.
ARPES (Angle Resolved Photoemission) measurements show a diminution of
the electronic density of
states starting at about $T_p(x)$ with a four-fold symmetry: no change
along the $(\pi, \pi)$ directions and maximum change along the $(\pi,0)$
directions. The magnitude of the anisotropic ``pseudogap'' is several times
$T_p(x)$.

It is important to note that given the observed change in the single-particle spectra,
the measured specific heat and the magnetic susceptibility in the pseudogap
 region are consistent
with the simple calculation using
the single-particle density of states alone. Nothing fancier
is demanded by the data, at least at
in its present state. Moreover, the transport properties as well as
 the thermodynamic properties at different $x$ can be collapsed to
  scaling functions with the same $T_p(x)$ as a parameter \cite{wuyts}.

\subsection{Marginal Fermi Liquid behavior of the normal state}

\begin{figure}
\vspace*{0.3cm}
\begin{center}   
  \epsfig{figure=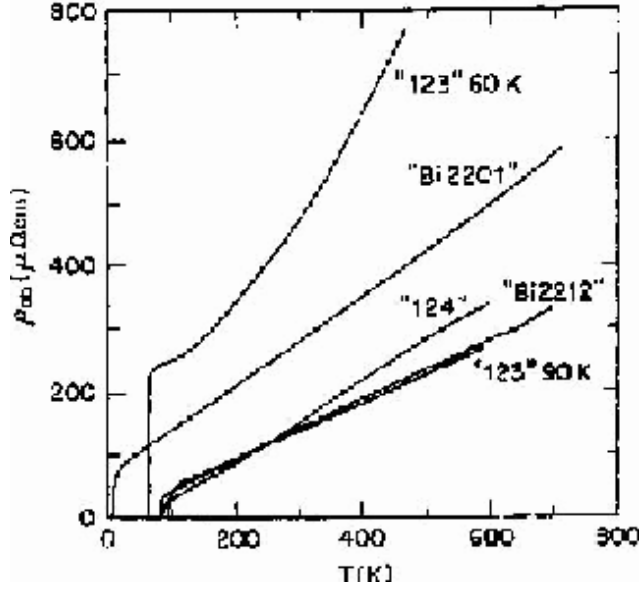,width=0.6\linewidth}
\vspace*{2mm}
\end{center}
\caption[]{Resistivity as a function of temperature for various high temperature superconductors.
From \cite{resistivitydata}.}\label{hightcresistivity}
\end{figure}

Every measured transport property in Region I is unlike
those of Landau Fermi-liquids.  The most commonly measured
of these is the dc resistivity shown for many different
compounds at the ``optimum'' composition in Fig. \ref{hightcresistivity}
including one with $T_c \approx 10K$.  The resistivity
is linear from near $T_c$ to the decomposition
temperature of the compound.   As shown in Fig. \ref{physworld3}, in
the ``under-doped''
region the departure from linearity commences at a
temperature $\sim T_p (x)$ marked in Fig. \ref{genericphasediagram}.
Similarly,
the cross-over into region (III) shown in Fig. \ref{genericphasediagram} is
accompanied by Fermi-liquid like properties.
Wherever measurements are available, every other measured
transport property shows similar changes.

The different measured transport properties study the
response of the compounds over quite different momentum
and energy regions.  For example, the Raman scattering studies
the long wavelength density and current response at long
wavelength but over a range of frequencies from low
${\mathcal O} (1 cm^{-1})$ to high, ${\mathcal O}(10^4 cm^{-1})$.  On the
other hand nuclear relaxation rate $T_1^{-1}$ depends on
the magnetic fluctuations at very low frequencies but
integrates over all momenta, so that the short
wave-length fluctuations dominate.  In 1989 it was proposed
 \cite{varma6,varma7} that a single hypothesis about the particle-hole
excitation spectra  captures most of the diverse
transport anomalies.The hypothesis is that the density as well as magnetic
fluctuation spectrum has an absorptive part with the
following property:
\begin{equation}
\chi ''(q,\omega ) \left\{  \matrix{ =\  -\chi _{o}''~\omega /T~,~~\  \mbox{for}\
\omega \ll T ~,\cr  =\  -\chi _{o}''sgn(\omega)  ~,~~~~~  \mbox{for}\  \omega _{c}>\omega \gg
T~.} \right.  \label{mflchi}
\end{equation}
Here $\chi _{o}''$ is the order of the bare single-particle
density of states $N (0)$ and $\omega_c$ is an
upper cut-off.  The fluctuation spectrum is assumed to have
only a weak momentum dependence, except at very long
wavelength, where a $q^2$ dependence is required for
fluctuations of conserved quantities like density or
spin (in the absence of spin-orbit interactions).  A form which
implements these requirements for the
conserved quantities with a rather arbitrary crossover function
to get the different regimes of $\omega/T$ is
\begin{equation}
\chi ''({\bf q},\omega )\  \sim \  {-xq^{2}\over \omega (\omega ^{2}+\pi ^{2}x^{2})}
,~~~\  \mbox{for}\  v_{F}q\ll \sqrt {\omega x} , \label{mflchi2}
\end{equation}
where $x = \omega$ for $\omega /T \ll T$ and
$x= T$ for $\omega / T \gg 1$.

Using the Kramers-Kronig relations, one deduces that
the real part of the corresponding correlation functions
have a $\log (x)$ divergence at all momenta except the
conserved quantities where the divergence does not extend
to $v_F q \geq x$.  Thus compressibility and magnetic
susceptibility are finite. (Besides (\ref{mflchi}), (\ref{mflchi2}) analytic fluctuation
spectra of the Fermi-liquid form is of course also present.)

 The spectral function (\ref{mflchi})
is unlike those for  a Landau Fermi-liquid discussed in section \ref{landausfermiliquid},
which always have a scale --- the Fermi-energy,  Debye-frequency, or spin-wave
energy, etc. --- which is obtained from parameters of the Hamiltonian.
Such parameters have been replaced by $T$. As we have discussed in
section \ref{qcpsection}, this  scale-invariance of
(\ref{mflchi}) is characteristic of fluctuations in the quasi-classical
regime of a QCP. Eq. (\ref{mflchi}) characterizes the fluctuations
around the QCP: comparing with Eq. (\ref{upsilon1}),
 the
exponent $ d_M/z = 0 $ and $1/z = 0 $.
These are equivalent to the statement
that the momentum dependence is negligibly important compared
to the frequency dependence. This is a very unusual requirements for a QCP
in an itinerant problem: the spatial correlation length plays no role in determining
the frequency dependence of the critical properties.

The experimental results for the various transport properties
for compositions near the optimum are consistent in detail
with Eq. (\ref{mflchi}), supplemented with the elastic scattering rate
due to impurities (see later).  We refer the reader to the original literature
for the details.  The temperature independence and the frequency
independence in Raman scattering intensity up to $\omega$
of ${\mathcal O} (1 eV)$ directly follows from (\ref{mflchi2}).
 Eq. (\ref{mflchi2}) also
gives a temperature independent contribution to the nuclear
relaxation rate $T_1^{-1}$ as is observed for $Cu$ nuclei.
The transport scattering rates have the same temperature dependence as
the single-particle scattering rate.
The observed anomalous optical conductivity can be directly
obtained by using the continuity equation together with
Eq. (\ref{mflchi2}), or by first calculating the single-particle
scattering rate and the transport scattering rate.  The
single-particle scattering rate is independently measured
in ARPES  experiments and provides
the most detailed test of the assumed hypothesis.

\begin{figure}
\begin{center}   
  \epsfig{figure=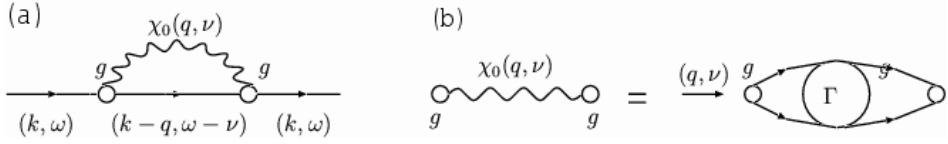,width=0.9\linewidth}
\end{center}
\caption[]{{\em (a) } Diagram for the singular contribution to the one-particle
  self-energy with the fluctuating $\chi(q,\nu)$. $g$'s are the
  vertices which in microscopic theory \cite{mflmodel} is shown to have
  important momentum dependence, but which gives negligible momentum
  dependence to the self-energy. {\em (b)} For $\chi_0(q,\nu)$ which
  is momentum independent, a total vertex $\Gamma$ may sometimes be
  usefully defined, which has the same frequency dependence as
  Eq. (\ref{mflsigmaeq}) and which is also $q$-independent. The lines
  are the exact single-particle Green's functions. }\label{mflscattering}
\end{figure}

To calculate the single-particle scattering rate, assume to
begin with a constant coupling matrix element $g$ for the
scattering of particles by the singular fluctuations.  We
shall return to this point in the section on microscopic theory.
Then provided there is no singular contribution to the
self-energy from particle-particle fluctuations, the graph
in Fig. \ref{mflscattering} represents the singular self-energy exactly.
It is important to note that for this to be true, Eq. (\ref{mflchi2})
is to be regarded as the exact (not irreducible) propagator for
particle-hole fluctuations; it should not be iterated.

\begin{figure}
\begin{center}   
  \epsfig{figure=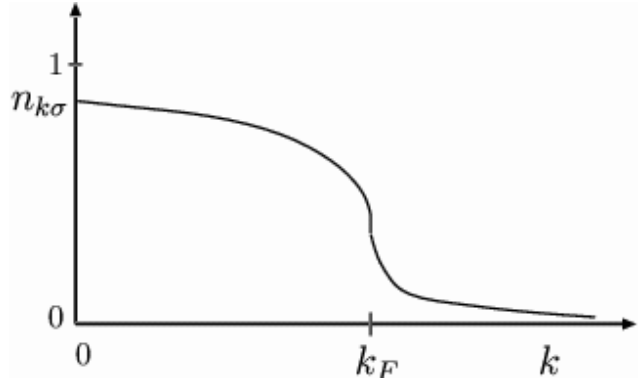,width=0.6\linewidth}
\end{center}
\caption[]{The $T=0$ distribution of bare particles in a marginal
  Fermi liquid. No discontinuity exists at $k_F$ but the derivative of
  the distribution is discontinuous.}\label{mfloccupation}
\end{figure}

$\Sigma ({\bf q}, \omega )$ can now be evaluated straight
forwardly to find a singular contribution,
\begin{equation}
\Sigma (\omega ,q)\  \approx \  g^{2}(\chi _{o}'')^{2}\  \left(\omega
  \ln {x\over \omega _{c}}
-i\  {\pi \over 2}
\  x\right) \label{mflsigmaeq}
\end{equation}
for $x \ll \omega_c$ and $v_F |(q -k_F )| \gg \omega_c$.
The noteworthy points about (\ref{mflsigmaeq}) are:

\noindent(1) The single particle scattering rate
is proportional to $ x$ rather than to $ x^2$ as in Landau Fermi-liquids.\\
\noindent(2) The momentum independence of the
single-particle scattering rate.\\
\noindent(3) The quasi-particle renormalization amplitude
\begin{equation}
Z=\  \left(1-\lambda \ln \  {x\over \omega _{c}}
\right)^{-1}
\end{equation}
scales to zero logarithmically as $x \rightarrow 0$.
Hence the name Marginal Fermi-liquid.\\
\noindent(4) The single-particle Green's function
\begin{equation}
G(\omega ,q)\  =\  {1\over \omega -(\epsilon _{q}-\mu )-\Sigma
  (q,\omega )} \label{mflgreens}
\end{equation}
has a branch cut rather than a pole.  It may be written as
\begin{equation}
{Z(x)\over \omega -(\tilde \epsilon _{q}-\tilde \mu )-i/\tilde \tau }~,
\end{equation}
where $\tilde \epsilon_{q}$ is the renormalized single-particle
energy:
\begin{equation}
\tilde \epsilon _{q}-\tilde \mu \  =Z(\epsilon _{q}-\mu )\  \approx \
Z {\bf v}_{F}\cdot ({\bf q}-{\bf k}_{F})
\end{equation}
for small $|q - k_F |$.  Also
$\tilde \tau^{-1}(x) = Z \; \mbox{Im} \Sigma (\omega )$, and
the effective Fermi-velocity $\tilde v_F = Z v_F$ has
a frequency and temperature dependent correction.  \\
\noindent(5) The
single-particle occupation number has no discontinuity at
the Fermi-surface, but its derivative does, see Fig. \ref{mfloccupation}.
So the Fermi-surface remains a well-defined concept both
in energy and in momentum space.

The predictions of
(\ref{mflgreens}) have been tested in detail in ARPES measurements
only recently.  ARPES measures the spectral function
\begin{equation}
A(q,\omega )=-{1\over \pi }
\  {\Sigma ''(q,\omega )\over [\omega -(\epsilon _{q}-\mu )-\Sigma '(q,\omega
)]^{2}+[\Sigma ''(q,\omega )]^{2}} ~. \label{mflspectrum}
\end{equation}

\begin{figure}
\begin{center}   
\epsfig{figure=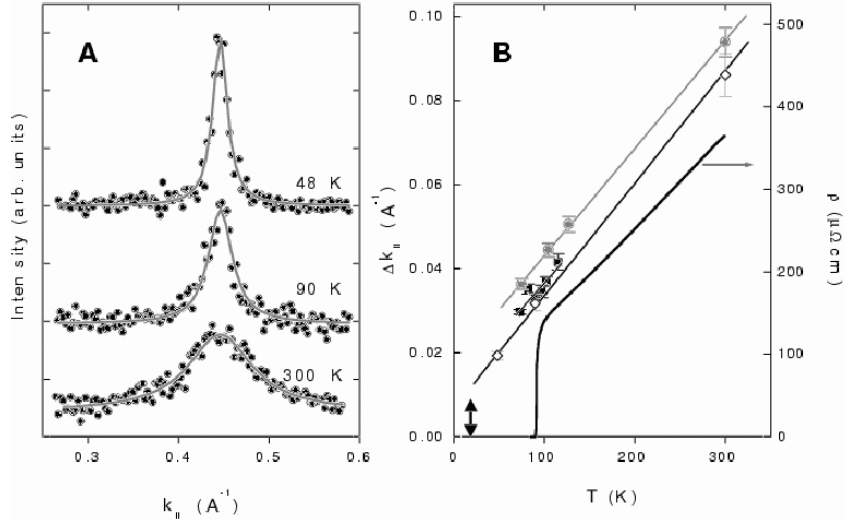,width=0.80\linewidth}
\end{center}

\caption[]{{\em (a)} Momentum distribution curves for different
  temperatures. The solid lines are Lorentzian fits. {\em (b)}
  Momentum widths of MDCs for three samples (circles, squares, and
  diamonds). The thin lines are $T$-linear fits
which show consistency with Eqn.(\ref{mflspectrum}) and
the MFL hypothesis. The resistivity
  (solid black line) is also shown. The double-headed arrow shows the
  momentum resolution of the experiment. From
   Johnson {\em et al.} \cite{johnson}.} \label{mfllineshape1}
\end{figure}

\begin{figure}
\begin{center}   
 \epsfig{figure=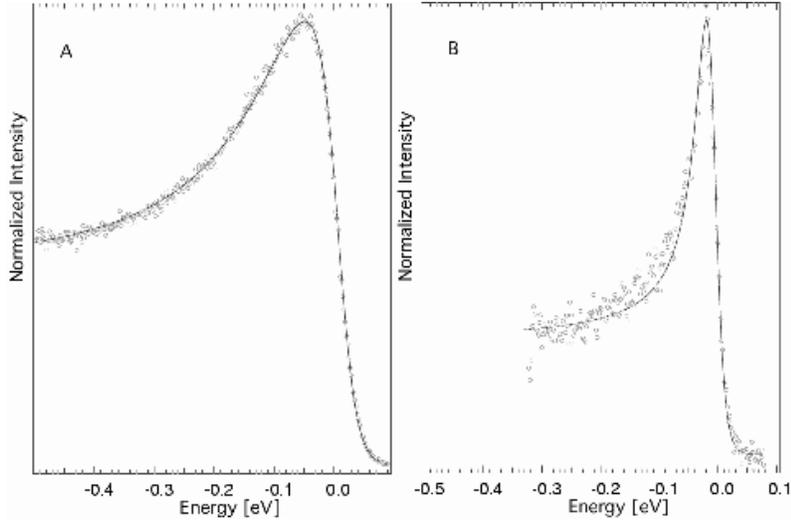,width=0.75\linewidth}
\end{center}
\caption[]{ Fits of the MFL self-energy
  $\Gamma+ \lambda\hbar \omega$ to the experimental data, according to
(\ref{arpestosigmafit}). Estimated
uncertainties of $\pm15$\% in $\Gamma$ and
  $\pm 25$\% in $\lambda$. {\em (a)} A scan along the (1,0) direction,
  $\Gamma=0.12$, $\lambda=0.27$;
{\em (b)} A scan along the (1,1) direction,
  $\Gamma=0.035$, $\lambda=0.35$.  From Kaminsky and co-workers,
  \cite{kaminsky}.}\label{mfllineshape3}\label{mfllineshape2}
\end{figure}

In ARPES experiments, the energy distribution curve at fixed
momentum (EDC) and the momentum distribution curve at
fixed energy (MDC) can both be measured.  It follows from
Eq. (\ref{mflspectrum}) that if $\Sigma$ is momentum independent perpendicular
to the Fermi-surface, then an MDC scanned along
${\bf k}_{\perp}$ for $\omega \approx \mu$ should have a
Lorentzian shape plotted against
$(\bk - \bk_F)_{\perp}$ with a width proportional to
$\Sigma ''(\omega )/v_{F}(\hat \bk)$.
For a Marginal Fermi Liquid (MFL), this width should be proportional to $x$.
The agreement of the measured lineshape to a
Lorentzian and the variation of the width with temperature
are shown in Figs. \ref{mfllineshape1}.
    Fig. \ref{mfllineshape2}
shows the
momentum width measured in different directions
$\hat \bk$.  The Fermi-velocity $v (\hat \bk )$ is
measured through the EDC with the conclusion that it is
independent of $\hat \bk$ to within the experimental errors.
The data shows that the temperature dependence is consistent
with linearity with a coefficient independent of
$\hat \bk$ although the error bars are huge near the
$(\pi , 0)$ direction.  Besides the MFL contribution,
there is also a {\it temperature independent} contribution
to the width which is strongly angle-dependent, to
which we will soon turn.  The ambiguity of the temperature
(and frequency) dependence near the $(\pi , 0)$ direction
is removed by the EDC measurements.  In Fig. \ref{mfllineshape2}
EDC at the Fermi-surface in the $( \pi , \pi )$
direction and the $( \pi, 0)$  directions are shown
together with a fit to the MFL spectral function with a
constant contribution added to $\Sigma ''$.
EDC's have the additional problem of an energy-independent
experimental background.  This has also been added in
the fit.  In both directions $\Sigma ''$ has a contribution
proportional to $ \omega$ with the same coefficient within the
experimental uncertainty.
In summary, the ARPES experiments give that
\begin{equation}
\Sigma ''(k,\omega ;T)\  \cong \  \Gamma (\hat k_{F})\  +\Sigma _{MFL}''(\omega
,T) . \label{arpestosigmafit}
\end{equation}
The $(\omega , T)$-independent contribution
$\Gamma (\hat{\bk}_F )$ can only be understood as due to impurity
scattering \cite{abrahams2}. Its dependence on $\hat{\bk}_F$ can be understood
by the assumption that in well-prepared samples, the
impurities lie between the $Cu$-$O$ planes and therefore
only lead to small angle scattering of electrons in the
plane.  The contribution $\Gamma (\hat \bk_F)$ at
$\hat \bk_F$ then depends on the forward scattering matrix element
and the local density of states
at $\hat \bk_F$ which  increase  from
the $( \pi , \pi )$ direction to the
$(\pi ,  0)$ direction.    This small angle contribution
has several very important consequences: {\em (i)} relative
insensitivity of residual resistivity to disorder, {\em (ii)} relative insensitivity
of $d$-wave superconductivity transition temperature
 to the elastic part of the single-particle
scattering rate \cite{kee},
and {\em (iii)} most significantly relative insensitivity to the anomalous
Hall effect and magneto-resistance. Such anomalous magneto-transport properties
follow from a proper solution of the Boltzmann equation including
both small-angle elastic scattering and angle-independent MFL inelastic
scattering \cite{varma8}.

\begin{figure}
\begin{center}   
 \epsfig{figure=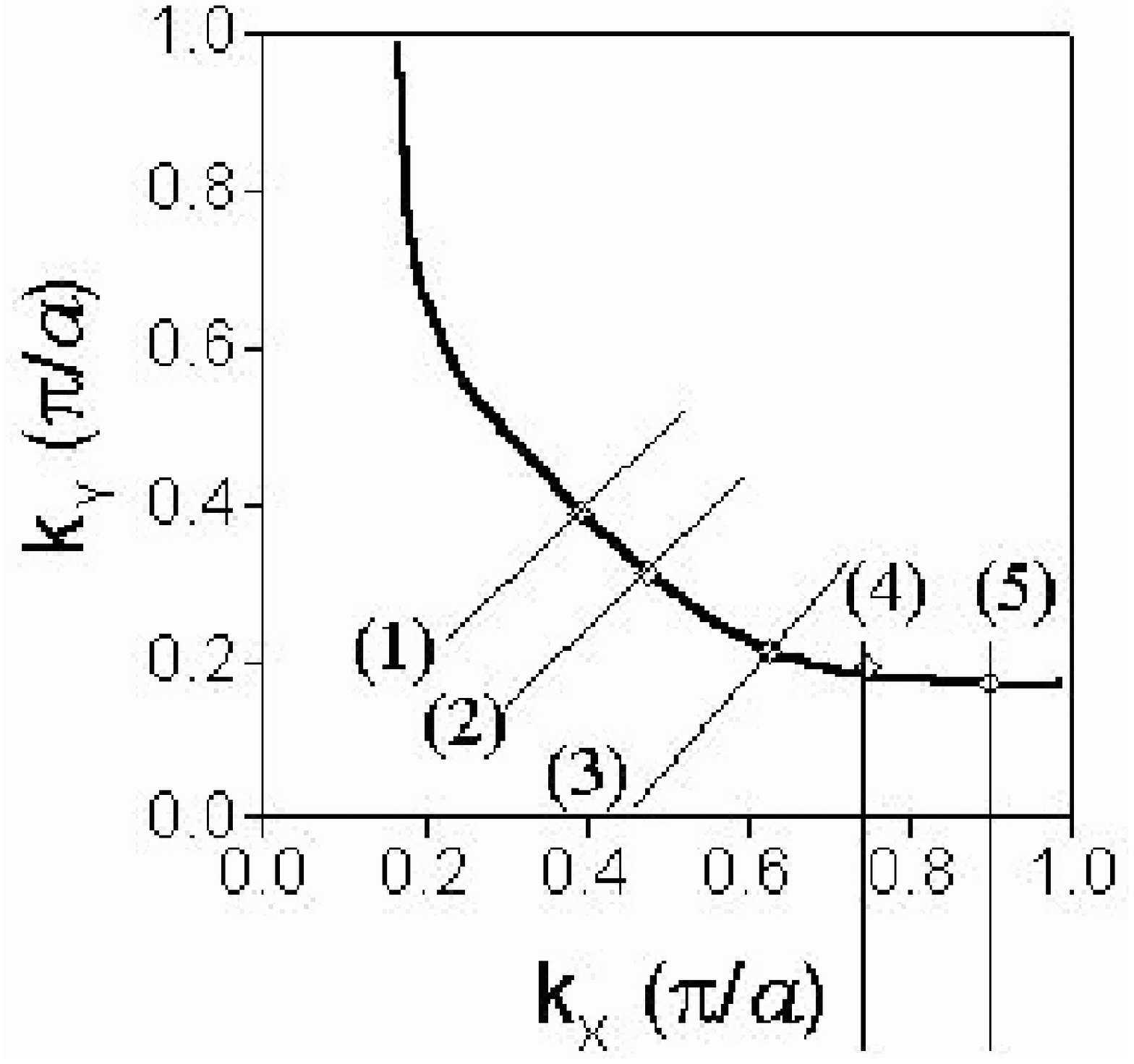,width=0.45\linewidth}\hspace*{4mm}
 \epsfig{figure=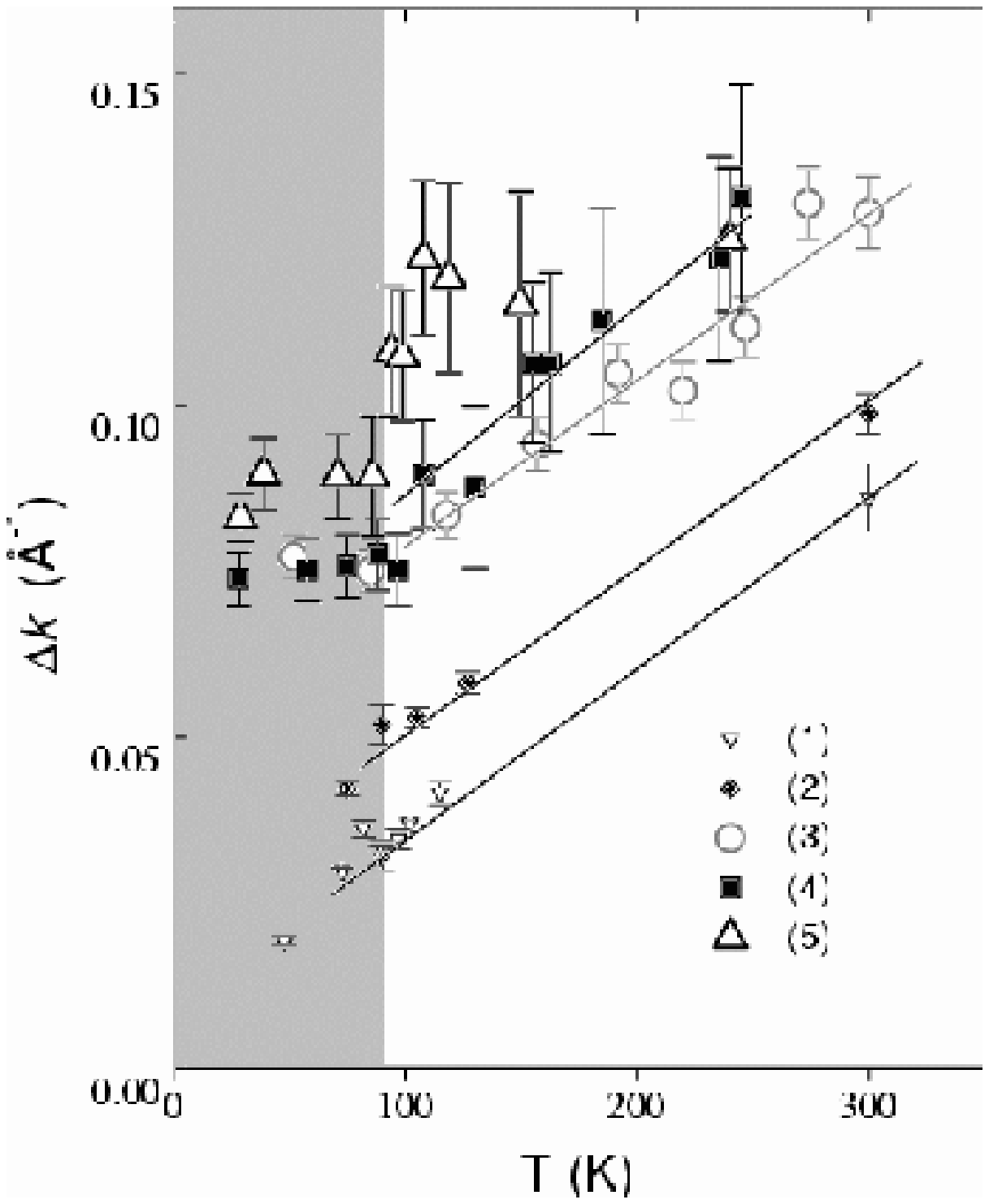,width=0.4\linewidth}
\end{center}
\vspace*{1cm}

\caption[]{{\em Left panel:} Fermi surface of the optimally doped Bi2212,
measure in the superconducting state. Indicated are the lines (1) to
(5) on which the temperature dependence is measured in the
experiments by Valla {\em et al.}  \cite{valla}. Typical spectra are
shown for line (2) in the normal (b) and superconducting state (c), as
well as for line (4), in the normal (d) and superconducting state
(c). {\em Right panel:} momentum widths as a function of
temperature for different
positions on the Fermi surface, obtained by fitting the Momentum
Distribution Curves with Lorentzian lineshapes in the same
experiments. Widths are measured at
the Fermi level and at the leading edge, in the normal and in
superconducting (grey region) state, respectively.}\label{mfllineshape4}
\end{figure}

The momentum independence of the inelastic part of
$\Sigma ''$ is crucial to the SFL properties of the
cuprates.
Because the inelastic scattering to all angles on the
Fermi-surface is the same, i.e. $s$-wave scattering,
the vertex corrections to transport of vector quantities like
particle current and energy current are zero.  It follows
that the momentum transport scattering rate measured
in resistivity or optical conductivity and the energy
transport rate measured in thermal conductivity have
the same $( \omega , T)$ dependence as the single-particle
scattering rate $1 / \tau ( \omega , T)$.

Recently far-infrared conductivity measurements \cite{corson}
have
been analyzed and shown to be consistent with
$1 / \tau ( \omega , T)$ deduced from MFL including the
logarithmic corrections.

As already discussed no singular correction to the magnetic susceptibility
is to be expected on the basis of (\ref{mflchi}). But the specific heat
should have a logarithmic correction so that
\begin{equation}
\gamma(T) = \gamma_0\left[ 1+\lambda \ln(\omega_c/T)\right]~.
\end{equation}
Such a logarithmic correction has not yet been deciphered in the data
presumably because the electronic specific heat in the normal state
is less than ${\mathcal O} (10^{-2})$ of the total measured specific heat and
must be extracted by a subtraction procedure which is not sufficiently accurate.

\subsection{General Requirements on a Microscopic Theory}

The MFL self-energy, Eq. (\ref{mflsigmaeq}), has been verified in such
detail in its $( \omega , T, {\bf q} )$ dependence
that it is hard to see how any theory of $CuO$ compounds
can be relevant to the experiments without reproducing
 it (or a very close approximation to it) in Region I of the phase diagram of
Fig. \ref{genericphasediagram}.  Such a scale-invariant self-energy is characteristic
of the quasi-classical regime of a QCP and indeed the
topological features of the phase diagram are
consistent with their being a QCP at $x_c$ near the
composition for the highest $T_c$ (Alternatively, a QCP in the
overdoped region where $T_c$ vanishes is predicted in some approaches,
like in \cite{senthilfisher}).  To date,  no way has been found
to obtain Eq. (\ref{mflchi2}) except through the scale-invariant form of
fluctuations which is momentum independent
$({\it z} \approx \infty )$ as in Eq. (\ref{mflchi}).

A consistent and applicable microscopic theory of the copper-oxides
must show a QCP with fluctuations of the form (\ref{mflchi2}).
This is a very unusual requirement for a QCP in a homogeneous extended
system for at least two reasons. First, the fluctuations must have a negligible
 ${\bf q}$-dependence compared to the frequency dependence, i.e. ${
z}\approx \infty$
and second, the singularity in the spectrum should just have
logarithmic form; i.e. there should be no exponentiation of the
logarithm giving rise to power laws. Such singularities
do arise, as we discussed in section \ref{localfermil}, in models of
isolated impurities
under certain conditions but they disappear when the impurities are
coupled; recoil kills the singularities. The requirement of negligible
$q/\omega$ dependence runs contrary to the idea of critical slowing
 down in the fluctuation
regime of the usual transitions, in which the frequency dependence
 of the
fluctuations becomes strongly peaked at zero-frequency
because the spatial correlation
length diverges.

Another crucial thing to note is that
any known QCP (in more than one dimension) is the end of a line
 of continuous transitions
at $T = 0$.  Region II (at least at $T = 0$) must
then have a broken symmetry (This includes part of
Region III, which is also superconducting). Experiments
appear to exclude any broken translational symmetry
or spin-rotational symmetry for this region\footnote{A new lattice
symmetry due to lattice distortions  or antiferromagnetism, if
significant, would change the fermi-surface because the size of the
Brillouin zone would decrease. This would be visible both in ARPES
measurements as well as in hall effect measurements.} although as discussed
in section \ref{basicfeaturessection},
a sharp change in
transport properties is observed along with a four-fold
symmetric diminution of the ARPES intensity for low
energies at $T \approx T_{p} (x)$. If there is indeed
a broken symmetry, it is of a very elusive kind;
experiments have not yet found it.

\begin{figure}
\begin{center}   
 \epsfig{figure=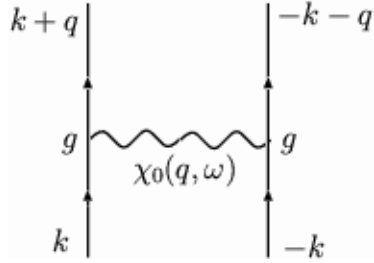,width=0.35\linewidth}
\end{center}
\caption[]{The Cooper pair-vertex and the normal state self-energy,
  Fig. \ref{mflscattering}, are intimately related. }\label{mflcooper}
\end{figure}

A related  question is how a momentum-independent
$\chi ({\bf q}, \omega , T)$ can be the fluctuation
spectrum of a transition which leads to an anisotropic
state as in Region II. Furthermore,  how can such
a spectrum lead to an anisotropic superconducting
state because it is unavoidable that $\chi ({\bf q}, \omega ,T)$
of Eq. (\ref{mflchi2})
which determines the inelastic properties in Region I
also be responsible for the superconductive instability.
After all, the process leading to the normal self-energy,
Fig. \ref{mflscattering},  the superconductive self-energy,
and the Cooper pair vertex, Fig. \ref{mflcooper}, are  intimately related.
Given Im$ \chi ({\bf q}, \omega , T)$ the effective interaction
in the particle-particle channel is
\begin{equation}
V_{pair}({\bf k},{\bf k}\pm {\bf q})\  =\  g^{2\  }Re\  \chi \
({\bf q},\omega )~. \label{vparieq}
\end{equation}
Re$ \chi (q; \omega )$ is negative for all  $q$ and
for all $-\omega _{c}\leq \omega \leq \omega _{c}$. So we do have a mechanism for
 superconductive pairing in the Cu-O problem given by the
normal state properties just as the normal state self-energy and transport
properties of say $Pb$ tell us about the mechanism for superconductivity in $Pb$.
In fact given that the normal state properties give that the upper cut-of frequency
is of ${\mathcal O}(10^3)$K and the coupling constant $\lambda\sim g^2N(0)$ is of
${\mathcal O}(1)$, the correct scale of $T_c$ is obtained. The important puzzle is,
 how can
 this mechanism produce $d$-wave pairing given that $\chi({\bf
q},\omega)$ is momentum independent. How can one obtain momentum-independent
 inelastic self-energy in the normal state and a $d$-wave superconducting
 order parameter from the same fluctuations?

We summarize in the next section a microscopic theory which
attempts to meet these requirements and answer some of the questions raised.

\subsection{Microscopic Theory} \label{microscopicsection}

There is no consensus on even the minimum necessary model Hamiltonian
to describe the essential properties in the phase-diagram, Fig.
\ref{genericphasediagram}, of the $CuO$ compounds. It is generally
agreed that, since other transition metal compounds do not share the
properties of $CuO$ compounds, a model Hamiltonian with
some rather special features is called for. Two such features are \\
{\em (A)} They are two-dimensional with an insulating antiferromagnetic ground state and
spin $S=1/2$ per unit cell at
half filling. Although not unique this feature is rare. If this is the
determining feature, a two-dimensional Hubbard model is adequate.
\cite{andersonbook}. Even
this model is not soluble in $d=2$.\\
{\em (B)} $CuO$ are the extreme limit of charge transfer compounds
\cite{zsa} in which charge fluctuations in the metallic state occur
almost equally on Oxygen and Copper. Then the longer-range ionic
interactions, which in magnitude are comparable to the on-site
interactions have a crucial role to play in the low-energy dynamics
in the metallic state
through excitonic effects. A model with both $Cu$ and $O$ orbitals,
hopping between them, and the excitonic interactions besides the
on-site repulsions is then required \cite{varma4,varma5}. This is of
course even harder to deal with than the Hubbard model.

Numerous attempts have been made using one or the other such models to
obtain SFL behavior. We briefly discuss the motivations for the pursuit of
model ${\em (A)}$ before summarizing in a little more detail the only attempt to
obtain the phenomenlogical fluctuation spectrum of Eq.
(\ref{mflchi}), and which relies on a model of type
{\em (B)}.

\subsubsection{The Doped Hubbard Model}
The investigations of the copper-oxide problem from this point of view asks some
valid and deep questions \cite{andersonbook}.
How does a low concentration of holes move through the spin configurations in a
two-dimensional model
with restriction of zero double occupancy? We have sketched in section \ref{spinchargesepinmore}
the difficulties of connecting to the same problem in one dimension when the ground state at
zero doping is an antiferromagnet. In fact
analytic \cite{schmittrink} and numerical \cite{manousakis} answers to
the question for a
single hole show the spectral weight of a heavy particle with an incoherent part
composed of multiple spin-wave polaronic cloud. Simply extrapolated (a dangerous thing to do),
a Fermi-liquid is expected. The larger zero-point fluctuations of the $S=1/2$ model, compared
to a large spin model only change the relative weight of the coherent and the incoherent parts.
But more subtle possibilities must be considered. The antiferromagnetic ground state
of a Heisenberg $S=1/2$ model (or the undoped non-degenerate Hubbard model) in two-dimensions
is close in energy to a singlet ground state. A possible description of such a state
is as linear combination on a basis of  singlet-bonds of pairs of
spins. As noted earlier, such itinerant
bond states have been termed Resonating Valence Bonds
(by analogy to the ground state of Benzene like molecules).
The massive degeneracy of the singlet bond-basis raises
interesting possibilities. If the quantum fluctuations of spins were (significantly)
larger than allowed by
$S=1/2$, such states would indeed be the ground state, as they are in the one-dimensional model
or two-dimensional models with additional frustrating interactions
\cite{affleckliebetal}. It
is possible that by doping with  holes in the $S=1/2$ Heisenberg
model, the additional quantum fluctuations induce a
ground state and low-lying excitations which utilize the massive degeneracy of RVB states.
Especially intriguing is the fact that resonating valence ground state may be looked on
as the projection of the BCS ground state to a fixed number of particles \cite{andersonbook}.
 Furthermore in the normal state
this line of reasoning is
likely to lead to a SFL.

A specific proposal incorporating the RVB idea \cite{kivelson-vison} relies on
the ground state of the half-filled model to be localised dimers. Then
defects in this state due to deviation from half-filling
can plausibly support excitations which
are charged spinless bosons. Further work on this
idea may be found in \cite{sachdev,moessner}.
Related ideas were put forth in \cite{dzyalo}.
 
These are very attractive set of ideas and no proof
exists that they are disallowed.
We have already considered an implementation of these ideas in
section 5.2 on generalized gauge theories. As discussed, controlled
calculations using these ideas are hard to come by.
Moreover, what theoretical results do
exist do not correspond in a persuasive way to the experimental results
on the copper-oxide materials.

One should take special note here of the idea of Anyon
superconductivity
which besides
being a lovely theoretical idea,
is founded on the solution to a well defined model, and has clear
experimental predictions.
Laughlin and collaborators \cite{laughlin,laughlin1,laughlin2}
found a specific model with long-range four spin interactions
for which his quantum hall wavefunction
is the ground state. Therefore Time-reversal and Parity are
spontaneously broken in this state. This state is superconducting.
The predicted Time-reversal broken properties
have not been observed experimentally \cite{spielman}.

An alternative idea from the microscopic characterization of these
materials as doped
Hubbard models is that a dilute concentration of holes in the
Hubbard model is likely to
phase separate or form ordered one-dimensional
charge-density wave/spin-density wave structures
\cite{zaanen,zaanen2,emerykivelson}. There exists both empirical
\cite{Tranquada} and
computational \cite{whitescalapino} support for
this idea at least for a very dilute concentration of holes.
For concentrations close to
optimum compositions these structures appear in experiment to exist
only at high energies with short correlation lengths and times and
small amplitudes
in the majority of copper-oxygen compounds.
Their relation to SFL properties is again
not theoretically or empirically persuasive.

\subsubsection{The excitonic Interactions Model}

This relies on a model of type ${\em (B)}$. A brief sketch of the calculations
leading to a QCP and a MFL spectrum is given here.
We refer the reader to \cite{varma4,varma6,varma8}
for details.

\begin{figure}
\begin{center}   
  \epsfig{figure=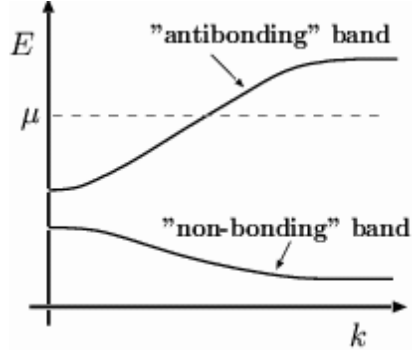,width=0.4\linewidth}
\end{center}
\caption[]{Three bands result from the orbitals shown in
  Fig. \ref{cuoplanes}{\em (b)} in a one-electron calculation; two of these
are shown. The
  chemical potential lies in the ``anti-bonding'' band and is varied
  by the doping concentration. The other band shown is crucial for the
  theory using excitonic effects as in \cite{mflmodel}.
  }\label{cuobands}
\end{figure}

At half-filling the ground state and the low-lying excitations of
such models are identical to the Hubbard model. But important
differences can arise in
the metallic state. Consider the one-electron structure of such models.
The $O$-$O$ hopping in the lattice structure with
$d_{x^2-y^2}$ orbitals in $Cu$ and $p_x$, $p_y$ orbitals on $O$, as
shown in Fig. \ref{cuoplanes}, produces a weakly dispersing
``non-bonding'' band while the $Cu$-$O$ hopping produces ``bonding''
and ``anti-bonding'' bands --- see Fig. \ref{cuobands}. We need
consider only the filled non-bonding band and the partially filled
anti-bonding bands shown in Fig. \ref{cuobands}.

\begin{figure}
\begin{center}   
  \epsfig{figure=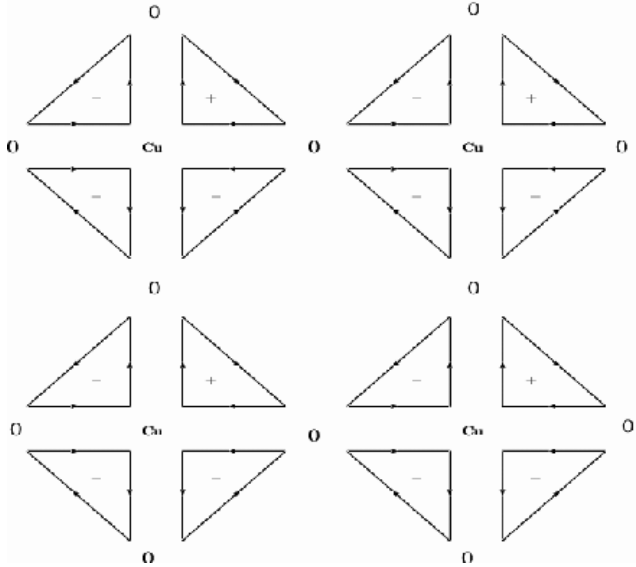,width=0.6\linewidth}
\end{center}
\caption[]{The current pattern in the time-reversal breaking phase
  predicted for Region II of the phase diagram.}\label{cuocurrent}
\end{figure}

In the mean-field approximation such an electronic structure together
with the on-site interaction and the ionic interactions is unstable to
an unusual phase provided the latter summed over the nearest neighbors
is of the order or larger than the bandwidth. In this phase
translational symmetry is preserved but time-reversal symmetry and the
four-fold rotation symmetry about the $Cu$ sites is broken. The ground
state has a current pattern, sketched in Fig. \ref{cuocurrent}, in
which each unit cell breaks up into four plaquettes with currents in
the direction shown. The variation of the transition temperature with
doping $x$ is similar to the line $T_p(x)$ in Fig. \ref{qcpscaling},
so that there is a QCP at $x=x_c$.
Experiments have been proposed to look for the elusive broken symmetry
sketched in Fig. \ref{cuocurrent} \cite{varma9}.

The long-range interactions in the model also favor other time-reversal
breaking phases which also break translational invariance. This is known
 from calculations on ladder models \cite{orignac}. Such states
have also been proposed for copper-oxide compounds
\cite{kotliar,houghton,chakravarty2}.

Our primary interest here is how the mechanism of transition to such a
phase produces the particular SFL fluctuation spectrum (\ref{mflchi2})
in Region I of the phase diagram. The driving mechanism for the
transition is the excitonic singularity, due to the scattering between
the states of the partially filled conduction band $c$ and the valence
band $v$ of Fig. \ref{cuobands}. This scattering is of course what we
considered in sections \ref{xrayedge} and \ref{1dorthog} for the
problem of $X$-ray edge singularities for the case that the interband
interaction $V$ in Eq. (\ref{xrayedgehamiltonian}) is small and the
valence band is dispersionless (i.e., the no recoil case). Actually,
the problem is exactly soluble for the no-recoil case even for large
$V$ \cite{combescot}.  For large enough $V$ the energy to create the
exciton, $\omega_{ex}$, is less than the $v$-$c$ splitting $\Delta$.
The excitonic lineshape is essentially the one sketched in Fig.
\ref{xrayfig}{\em (b)} and given by Eq. (\ref{Aequation}) for $\omega
> \omega_{ex}$, but $\delta $ is now the phase shift modulo $\pi$
which is the value required to pull a bound state below $\Delta$. The
excitonic instability arises when $\omega_{ex} \to 0$.

\begin{figure}
\begin{center}   
  \epsfig{figure=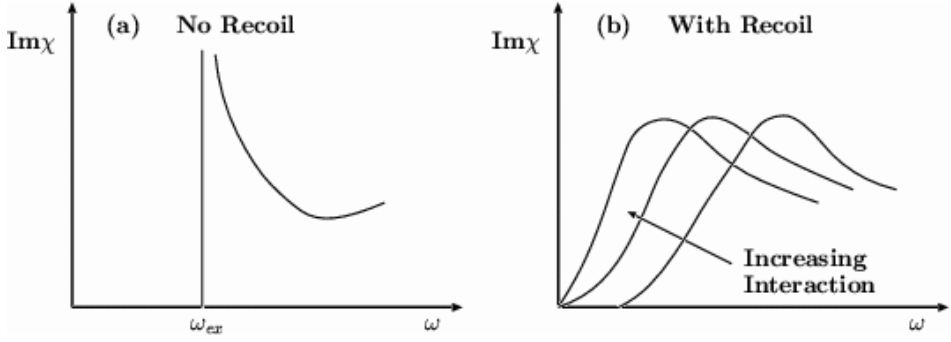,width=0.9\linewidth}
\end{center}
\caption[]{Sketch of the development of the particle-hole spectra in the microscopic model for the cuprates. }\label{phspectradev}
\end{figure}

The effect of a finite mass or recoil on the excitonic spectra is to
smoothen the edge singularity on the scale of the valence band
dispersion between $k=0$ and $k=k_F$. The phase shift $\delta$ or the
interaction energy $V$ no longer determines the low-energy shape of
the resonance. $V$ does determine its location. The low-energy
fluctuation spectra is determined by the following argument: Let us
concentrate on $q=0$ which is obviously where Im$\chi_{ex}({\bf q},\omega)$
is largest. The absorptive part of a particle-hole spectra must be odd
in frequency:
\begin{equation}
\mbox{Im}~\chi({\bf q},\omega) = -\mbox{Im}~\chi({\bf q},-\omega)~.
\end{equation}

As $V$ increases, Im$\chi_{ex}(0,\omega) $ must shift its weight
towards zero-frequency as shown in Fig. \ref{phspectradev}. Let us
continue to denote by $|\omega_{ex}|$ the characteristic energy of the
maximum in Im$\chi$. For $|\omega |$ small compared to
$|\omega_{ex}|$, Im$\chi(0,\omega)\sim \omega$ while for $|\omega |$
large compared to $|\omega_{ex}|$, it is very slowly varying up to a
cut-off $\omega_c$ on the scale of the Fermi-energy.  Then by
Kramers-Kronig transform the leading term in
\begin{equation}
\mbox{Re}\chi(0,\omega) \sim \ln \left(\omega_c/\max(\omega_x,\omega)\right)~.
\end{equation}

For any finite $|\omega_{ex}|$, Re$\chi$ is finite and there is no
instability. Only for $|\omega_{ex}|\to 0$, i.e. Im$\chi(0,\omega)
\to \mbox{sgn}\omega$, Re$\chi(0,\omega)$ is singular $\sim \ln
|\omega|$ and there is an instability. Thus in an excitonic
instability of a Fermi-sea with a dispersive valence band, the
zero-temperature spectrum has the form $\chi(\omega,0) \sim \ln |
\omega | + i \;\mbox{sgn}\; \omega $ at the instability. Given a parameter
$p$ such that the instability occurs only at $p_c$, i.e. $\omega_{ex}
(p \to p_c) \to 0$, the generalization for finite temperature $T$ and
momentum $q$ and $p\neq p_c$ is
\begin{eqnarray}
\nonumber \chi({\bf q},\omega) & = &
 \left[ \left( {{i
\omega}\over{\mbox{max} (\omega , T , \omega_{ex}(p)) }} + \ln
{{\omega_c}\over{\mbox{max} (\omega , T , \omega_{ex}(p)) }}
\right)^{-1} \right. \\
 & & \hspace{1.5cm} \left. + a^2 q^2 + (p_c(T) -p)
\right]^{-1}~. \label{chichmodel}
\end{eqnarray}
Here $\omega_c$ is the frequency cut-off of ${\mathcal O}(\Delta)$.
Since the binding energy is ${\mathcal O} $(1eV), the size of the
exciton, $a$, is of the order of the lattice constant. The
${\bf q}$'
 dependence of (\ref{chichmodel}) is negligible compared to the
frequency dependence. The exponent ${ z} $  is effectively infinite. At $p\approx
p_c$ to logarithmic accuracy, the above expression (\ref{chichmodel})
is identical to the phenomenological hypothesis (\ref{mflchi2}).

\begin{figure}
\begin{center}   
  \epsfig{figure=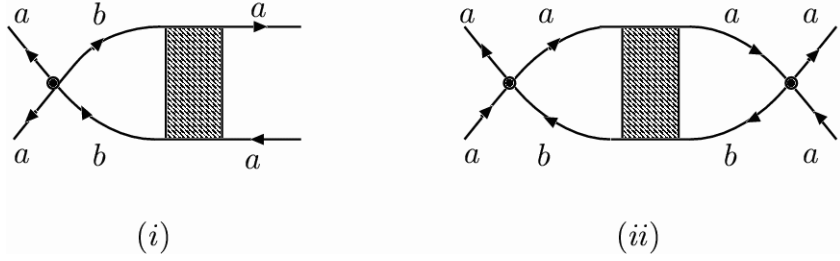,width=0.8\linewidth}
\end{center}
\caption[]{Singularity of interaction $\Gamma_{aaaa}$ between states
`$a$' near the chemical potential generated by the excitonic
singularity between the states `$a$' of the conduction band and states
`$b$' of the valence band. The excitonic singularity is indicated by
the shaded block.}\label{chmodelinteract}
\end{figure}

The effective low-energy interaction for states near the chemical
potential, which is sketched in Fig. \ref{chmodelinteract}, is
singular when the excitonic resonance is at low frequency. Here is an
example of the mechanism mentioned under {\em (v)} in section
\ref{routesto}
where the irreducible interaction obtained by integrating over
non-perturbed high-energy states is singular. This is of course only
possible when the interactions represented by the shaded block in Fig.
\ref{chmodelinteract} are large enough.

In relation to some of the questions raised about the phenemenology at
the end of the last subsection, the momentum dependence of the vertex
coupling the low energy fermions to the fluctuations in. Figs.
\ref{mflscattering} and \ref{mflcooper} has been evaluated
\cite{varma1999}. It is non-local with a form depending on the wave
functions of the conduction and valence bands and the leading result
is
\begin{equation}\label{gvertex}
g(k,k')\sim \left( \sin(k_xa/2) \sin(k'_xa/2)-\sin(k_ya/2) \sin(k'_ya/2) \right).
\end{equation}
Note that at $(k-k')=0$ this is proportional to
$[\cos(k_xa)-\cos(k_ya)]$. This is intimately
 related  to the $d$-wave current distribution in the broken-symmetry phase
 predicted for region (II), shown in Fig. \ref{cuocurrent}.
Eq. (\ref {gvertex}) is such that that when
diagram \ref{mflscattering} is evaluated, the
intermediate state momentum integration makes the self-energy depend
very weakly on the incoming momentum.  But when the pairing kernel of
Fig. \ref{mflcooper} is evaluated, its momentum dependent and exhibits
attraction in the $d$-wave channel.

Similarly, as has been shown \cite{varma1999},
the vertex of Eq. (\ref{gvertex}) leads to an
anisotropic state with properties of the pseudo-gap state of region II
below a temperature $T_p(x)$.  The principal theoretical problem remaining with
this point of view is that a transition of the Ising class occurs at
$T_p(x)$at least in mean-field theory. This would
 be accompanied by a feature in the specific heat
 unlike the observations.\footnote{One might appeal to disorder
  to round off the transition, but this appears implausible quantitatively. More likely,
  the nature of the transition is strongly affected by the Fluctuation spectra
  of the form of Eq. (\ref{mflchi}) and is unlikely to be of the Ising class.}

The microscopic theory reviewed above reproduces the principal
features of the phase diagram Fig. \ref{genericphasediagram} of the
copper-oxide
superconductors, and of the SFL properties. It also gives a mechanism
for high temperature superconductivity of the right symmetry. Further
confidence in the applicability of the theory to the cuprates will rest on
the observation of the predicted current pattern of Fig. \ref{cuocurrent} in
region II of the phase diagram \footnote{As already mentioned
in section \ref{routesto}, ferromagnetism in some compounds has an excitonic
origin. The dynamics near such a transition should also exhibit
features of the edge-singularity as modified by recoil.}.

\section{The Metallic State in Two-Dimensions }\label{mitrans}

The distinction between metals and insulators and the metal-insulator
transition has been a central problem in condensed matter physics for
seven decades.  Despite the accumulated theoretical
and empirical wisdom acquired over all these years, the experimental
observation made in 1995 of a metal-insulator transition in
two-dimensions \cite{midiscovery} was a major surprise and is a subject
of great
current controversy. The theoretical work in the 1980's
\cite{altshuler4,finkelstein,finkelstein2,finkelstein3,lee,altshuler} on
 disordered interacting electrons pointed to a major
unsolved theoretical problem in two dimensions. Infrared singularities
were discovered
in the scattering amplitudes which scaled to strong coupling where the
theory is uncontrolled (The situation is similar to that after the
singularities in the one- or two-loop approximations
in the Kondo problem were discovered revealing a fascinating  problem
without providing the properties of the asymptotic low temperature state).
 However the problem was not pursued
and the field lapsed till the new experiments came along.

The 1980's theoretical work shows that this problem naturally
belongs as a
subject in our study of Singular Fermi-Liquids.
We will first summarize the principal theoretical ideas relevant to the
problem before the 1995 experiments.  We then briefly summarize the principal
results of these and subsequent experiments. Reviews of the
experiments have appeared in \cite{aks,altshuler2,altshuler5}. There
are two types of
theoretical problems raised: the nature of the metallic state and the
mechanism of the metal-insulator transition. We will address the former and
only briefly touch on the latter.

\subsection{The two-dimensional Electron Gas}\label{2degas}

\begin{figure}
\begin{center}   
  \epsfig{figure=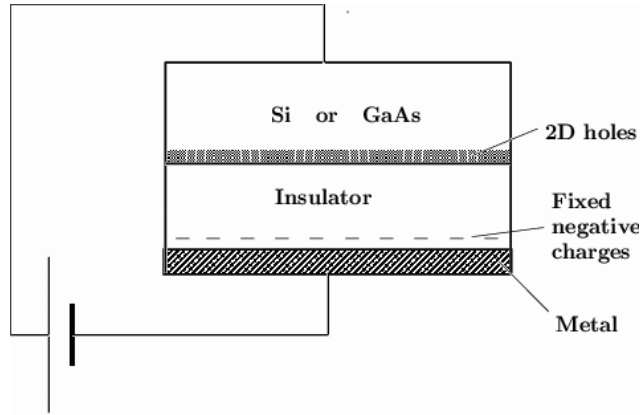,width=0.6\linewidth}
\end{center}
\caption[]{Sketch of a MOSFET. Holes (or electrons) are trapped at the
interface of the semi-conductor and the insulator due to the band gap
difference between them, the dipole layer and the applied electric
field. Two-dimensional electrons (holes) may also be found by layered
structures (heterostructures) of semi-conductors with different band
gaps such as $GaAs$ and $AlGaAs$.}\label{2dexpfig}
\end{figure}

We consider an electron gas with a uniform positive background with
no complications arising from the lattice structure --- this is how the
many-electron problem was originally formulated: the Jellium model.
This situation is indeed realized experimentally in MOSFETS
(and heterostructures) in which an insulator is typically sandwiched
between a metallic plate and a semiconductor --- see
Fig. \ref{2dexpfig}.
By applying an electric field a two-dimensional charge layer
 accumulates on the
surface of the semiconductor adjacent to the insulator, whose density
can simply be changed by varying the field strength  (For details
see \cite{ando}).. Similar
geometries have been used to observe the Quantum Hall Effects and the
metal-insulator transition by varying the density\footnote{Recently, novel ``high temperature
superconductivity''  was observed
in such a FET geometry. Gate induced doping in such
a Field Effect Transistor was employed to introduce
significant hole densities into $C_{60}$
which became superconducting at $T_{c}= 52K$!
\cite{Scho-2000}.}.
 
Typically we will be interested in  phenomena when the average
inter-electron distance is
${\mathcal O} (10 ) - {\mathcal O} (10^2 )\;nm$.  The thickness of the insulating layer
is typically more than $100 \;nm$,
so that the positive (capactive) charge on the insulator provides
a uniform background to a first approximation. In $Si$ samles, surface
roughness is the
principle source of disorder at high densities, while at low densities
(in the regime where the transition takes place)
 ionized impurity scattering dominates
due to the fact that there is much less screening. In $GaAs$, remote
impurity scattering dominates, and this scattering is mostly
small-angle. This is the main reason the mobility in these samples is large.

Neglecting disorder,
the problem is characterized by $r_s$, defined as the ratio of the
potential energy to the kinetic energy:
\begin{equation}
r_s = \frac{m \: e^2}{4 \pi \epsilon \hbar^2 \sqrt{\pi n}} \label{rs1}~.
\end{equation}
Here $n$ is the electron density, $m$ the band mass, and $\epsilon$
the background static dielectric constant.  We can also write
\begin{equation}
\pi \: r^2_s \: a^2_o = \frac{1}{n}~,\label{rs2}
\end{equation}
which expresses that $r_s$ is the radius of the circle whose area is
equal to the area per conduction electron, measured in units of the
Bohr radius $a_0$.  For a two-valley band structure,
as on the $110$ surface of $Si$, the kinetic energy is reduced
 and $r_s$ is twice that
defined by Eqs. (\ref{rs1}) and (\ref{rs2}).

For $r_s \ll 1$, (the dense electron limit) the kinetic energy dominates
and metallic behavior is expected.  For $r_s \gg 1$, the potential
energy dominates and a crystalline state (Wigner crystal) is
expected.  The best current
numerical estimates put the transition to the crystalline state
at $r_{sc} \approx 37$ \cite{tanatar}.
The entropy at the transition is tiny,
indicating that the radial distribution function for the liquid state
at low densities is similar to that of the crystal for distances up to
a few times $r_sa_0$.

It is important to note for our subsequent study that magnetism
is always lurking close by. Reliable numerical calculations show that the
magnetic state in the Wigner crystal near the critical density is
determined by multiple-particle exchanges \cite{ceperley}.  On
the metallic side the energy of the ferromagnetic state is only a
few percent above the unpolarized metallic or crystalline states
for $r_s \approx r_{sc}$ \cite{tanatar}.  Disorder is expected to make
the
metal-insulator state continuous.  On the insulating side at
$T \rightarrow 0$, the disordered Wigner crystal is expected to
be glassy and have low-energy properties of a Coulomb glass \cite{efros}.
 On
the metallic side fluctuations in the local density of electrons
might be expected
to lead to locally polarized magnetic states or possibly to some unusual
frustrated magnetic states \cite{chakravarty}.
The perturbative calculations with disorder and interactions, already
alluded to \cite{finkelstein,finkelstein2,finkelstein3,castellani4}
also hint
at formation of magnetic moments in the metallic state.
It is the
interplay of such magnetic fluctuations with itinerancy which
is one of the principal
theoretical problem in understanding the metallic state.

\subsection{Noninteracting Disordered Electrons:  Scaling
Theory of Localization}

Detailed reviews on the material in this section may be found in
\cite{thouless,lee}
\cite{altshuler,imry}.

The concept of localization of non-interacting electrons
for strong enough disorder was invented in 1958 by
Anderson \cite{anderson58}.
In one dimension all electronic states are localized for arbitrarily
small disorder while in three dimension a critical value of
disorder is required. That two dimensions is the marginal
dimension in the problem was discovered through the
scaling theory of localization.

The conceptual foundations for the scaling theory of localization
were laid by Thouless and co-workers \cite{thoulessphysrep,thouless}
and by Abrahams {\em et al.} \cite{abrahams}, and were
 developed formally by Wegner \cite{wegner}.
 Abrahams {\em et al.} \cite{abrahams} also made predictions which
could be tested experimentally.  Thouless noted first of all
that the {\it conductance} $G$ of a hypercube of volume
$L^D$ in any dimension $d$ is dimensionless
when expressed in units of $(e^2 / h)$,  thus defining a
scale-independent quantity
\begin{equation}
g = G / \left( e^2 / h \right)   .
\end{equation}
Next, he argued  that $g$ for a box of linear size $2L$
may be obtained from the properties  of a box of size L and the
connection between two of them.
The conductance of a box of size $L$ itself increases with the
transition amplitude  $t$ between energy levels in the two boxes
and decreases
with the characteristic width of the distribution of the energy
levels in the boxes $\Delta W(L)$ due to the disorder,
\begin{equation}
g (L) \approx f\left( \frac{\Delta t(L)}{\Delta W (L)}\right) ~.
\end{equation}
For weak-Gaussian disorder, the bandwidth may be expected to be
proportional to the square-root of the number of impurities in
the box, so
$\Delta W (L) \sim L^{d/2}$. The transition amplitude $ t$ is obtained
by the hopping
between near-neighbors near the surface of the boxes of size $L$.
It is therefore proportional to the surface area $L^{d-1}$. Thus\footnote{This line of
reasoning of course breaks down when we include electron-electron interactions}
\begin{equation}
g (L) = f\left( L^{(d-2)/2} \right)  \:.\label{gscaling}
\end{equation}
Now, in three dimensions the conductivity should approach a constant
for large $L$ (Ohm's law!), and hence the conductance should scale as $L$. This
implies that the scaling function $f(x)$  should go for large $L$ as
$f(x)\sim x^2$. Note that while for $d>2$ the $g$ therefore increases
with increasing $L$ while for $d<2$ the large $L$ behavior is
determined by the small argument behavior of the scaling function, while
 $d = 2$ is the marginal dimension.

In a very influential paper,
Abrahams {\em et al. } \cite{abrahams} analyzed the
$\beta$-function of the RG flow
\begin{equation}
\beta (g)\equiv d (\ln g )/ d (\ln L ) ~,
\end{equation}
and showed by a perturbative calculation in $1/g$ that
\begin{equation}
\beta (g) = \left( d - 2 \right) -
\frac{1}{\pi^2} \: \frac{1}{g}~,\label{gpert}
\end{equation}
where the first part comes  from Eq. (\ref{gscaling}) with $f(x)\sim x^2$.

\begin{figure}
\begin{center}   
  \epsfig{figure=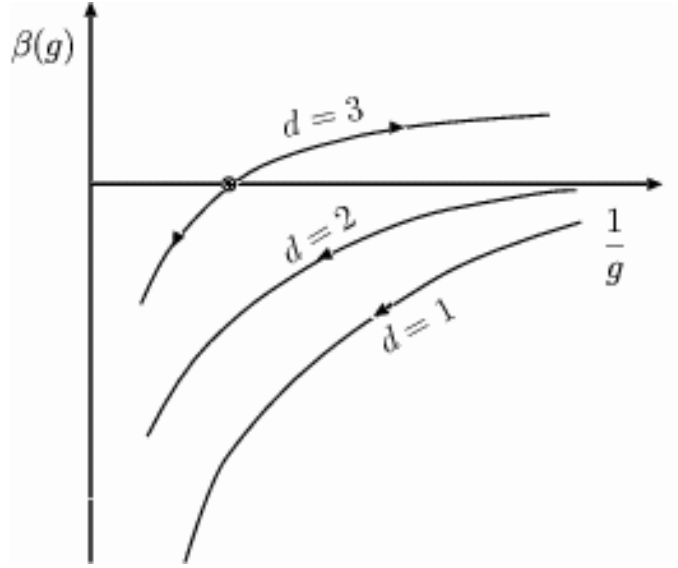,width=0.65\linewidth}
\end{center}
\caption[]{The scaling function for non-interacting electrons with
disorder deduced by Abrahams {\em et al.} \cite{abrahams}.  }\label{gscalingplot}
\end{figure}

For small enough $g$ (i.e. for large disorder) we expect
exponential localization
$g (L) \sim e^{-L / \zeta}$, where $\zeta$ is the localization
length, so that $\beta(g) \sim (-L/\zeta)$.  The smooth connection between the
perturbative result (\ref{gpert}) for large $g$ and the exponentially
localized solution at small $g$ is shown in Fig. \ref{gscalingplot}.
While for $d = 3$ (or any $d>2$), a critical disorder $g_c$ is required for
localization, in $d = 2$ states are asymptotically localized for any disorder
for non-interacting fermions.  The characteristic value of the
localization length in $d = 2$ is estimated from the perturbative solution:
\begin{equation}
g (L) = g_o - \frac{1}{\pi^2} \: \ln
\left( \frac{L}{\ell} \right)~, \label{gLeq}
\end{equation}
where $g_o$ is the dimensionless conductance at
$L \approx \ell$.  In conventional Boltzmann transport
theory $g_o = ( e^2 / 2 \pi \hbar ) k_F \ell$.
$\zeta$ is of the order of  the value of $L$ at which the correction term
is of order $g_0$, so that
\begin{equation}
\zeta \approx \ell \exp \: \left( \frac{\pi}{2} k_F \ell \right)~.
\label{weak-loc}
\end{equation}
At $T\rightarrow 0$, the sample size of a sample with
$k_F\ell \gg 1 $ has to be very large indeed for weak localization to be observable.

The theory described above must be modified at finite
temperatures due to inelastic scattering.  If the inelastic
scattering rate is much less than the elastic scattering rate,
$\tau^{-1}_{in} \ll \tau^{-1}$,
localization effects are cut off at a
length scale $L_{Th} (T)$, the Thouless length scale:
\begin{equation}
L_{Th} = \left( D \tau_{in} \right)^{1/2}
\end{equation}
where $D = (v^2_F \tau / d )$ is the (Boltzmann)
diffusion constant.  However as noted by Altshuler,
Aronov and Khmelnitski \cite{altshuler3,altshuler7}, the correct scale for the
cut-off is
$\tau^{-1}_\phi$, the phase breaking rate.  In
an individual collision the energy change $\Delta E$ may
be such that the phase  changes only by a very small amount,
$ \tau_{in} \Delta E \ll 2 \pi$.
The phase breaking time is then longer and is shown to be
given by
$\tau_\phi \sim (\Delta E \: \tau_{in} )^{-2/3} \: \tau_{in}$.
The $T = 0$ theory with
\begin{equation}
L_\phi = \left( D \tau_\phi \right)^{1/2} \label{Lpsi}
\end{equation}
replacing $L$ then gives the finite temperature scaling behavior to
which experiments may be compared.

The characteristic temperature $T_x$ at which weak-localization
effects become prominent may be estimated in a manner similar to
(\ref{gLeq}),
\begin{equation}
 T_x\tau_\phi(T_x) = \exp \: \left( -\pi k_F \ell \right)~. \label{Tpsi}
\end{equation}
This expression puts useful bounds on the temperatures required
to observe weak localization.

\begin{figure}
\begin{center}   
  \epsfig{figure=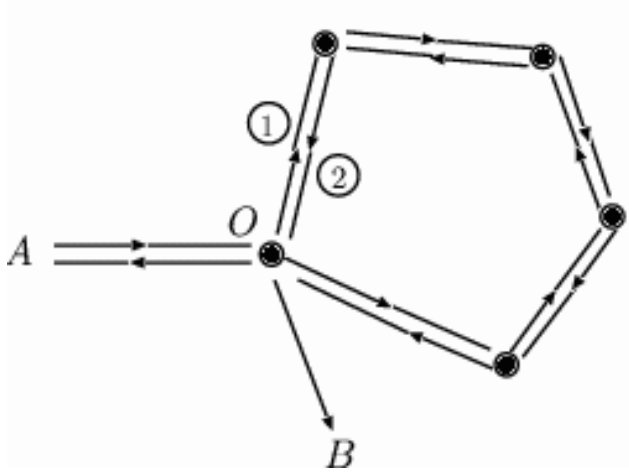,width=0.6\linewidth}
\end{center}
\caption[]{Interfering (time-reversed) parts in elastic scattering off
a fixed set of impurities. The probability for the particle to arrive
at B is reduced because of the enhance probability for the particle to
arrive back at A, as a result of interference.}\label{timerevfig}
\end{figure}

Eq. (\ref{gpert}) is derived microscopically by considering
repeated backward scattering between impurities.  It can
also be derived by considering quantum interference between
different paths to go from one point A to another B \cite{bergmann}.  The
total probability $\Omega$ for this process is
\begin{equation}
\Omega = \left| \sum_i a_i \right|^2 =
\sum_i \left| a_i \right|^2 + \sum_{i \neq j} a_i^* a_j ~,   \label{Omegaeq}
\end{equation}
where $a_i$ is the amplitude of the $i$-th path.  The second term
in Eq. (\ref{Omegaeq}) is non-zero only for classical trajectories which cross,
for example at the point O in Fig. \ref{timerevfig}.
The probability of finding a particle at the
point $O$ is increased from $2 | a_i |^2$ to
\begin{equation}
| a_1 |^2 + | a_2 |^2 + 2 \: Re \; a^*_1 a_2 = 4 |a_1 |^2~,
\end{equation}
because the two paths are mutually time-reversed.
Increasing this probability of course leads to a
decrease in the probability of the particle to arrive
at $B$, and hence to a decrease in the  conductivity.

This argument makes clear why the interfering paths
must be shorter than the phase relaxation rate due
to inelastic processes and why magnetic impurities or a
magnetic-field which introduces phase-shift between two
otherwise time-reversed paths suppress weak-localization.
In two-dimensions \cite{altshuler}
\begin{equation}
\sigma (H, T) - \sigma (O, T) =
\frac{e^2}{2 \pi^2 \hbar}
\left[ \psi \left( \frac{1}{2} +  \frac{1}{x} \right) + \ln x
\right] ~, \label{altshuleraronov}
\end{equation}
where $\psi$ is the digamma function and
\begin{equation}
x = 4 L^2_\phi \: e H/ \hbar c \equiv
\left( L_\phi / L_H \right)^2 \:.
\end{equation}
The quantity in brackets in (\ref{altshuleraronov})is equal to $x^2/24$
for $x \rightarrow 0$ and to $\ln(x/4)-\gamma$ for $x \rightarrow \infty$.

Spin-orbit scattering preserves time-reversal symmetry but
spin is no longer a good quantum number.  The spins are
rotated in opposite directions in the two self-intersecting
paths of Fig. (\ref{timerevfig}) if the impurities are spin-orbit scatters
\cite{altshuler,lee}.
This has been shown to lead to an average overlap of
the spin-function of
$ - \frac{1}{2}$ (because a rotation by $2 \pi$ of wave
function of a spin $1/2$ particle leads to a wave
function of opposite sign).  The correction to the
$\beta$-function of Eq. (\ref{gpert}) due to this effect is
\begin{equation}
\frac{1}{2 \pi^2} \: \frac{1}{g}~.
\end{equation}
This effect tends to  {\em enhance} the conductivity.

\subsection{Interactions in Disordered Electrons}\label{interdis}

\begin{figure}
\begin{center}   
  \epsfig{figure=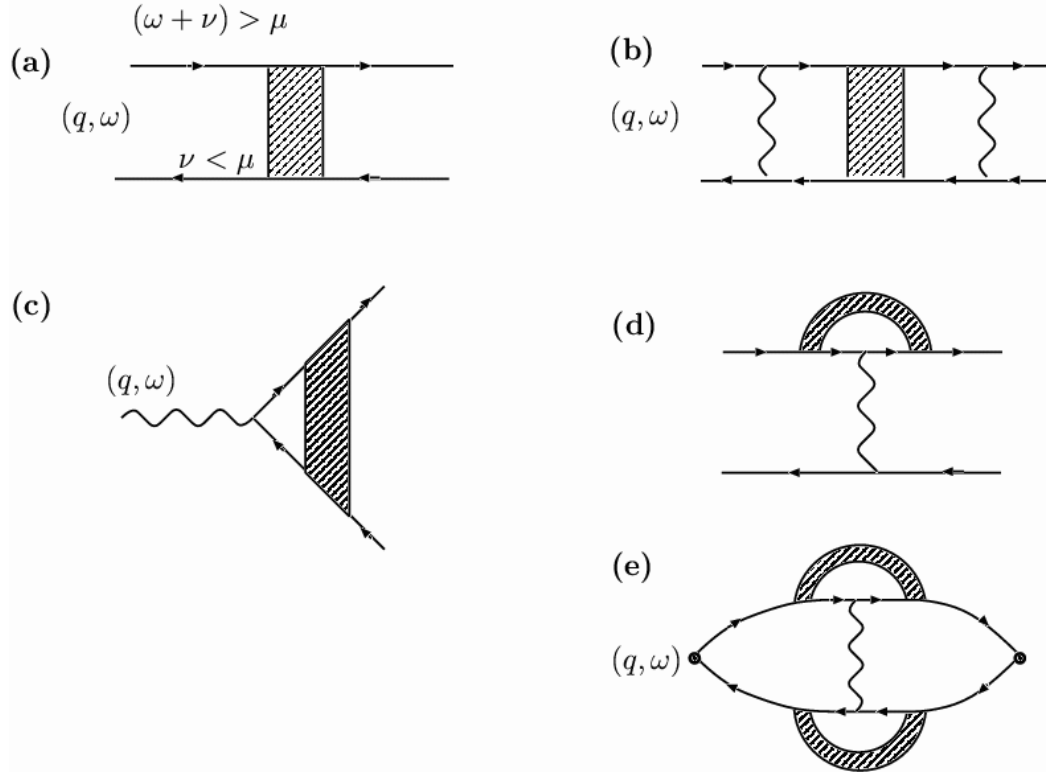,width=0.99\linewidth}
\end{center}
\caption[]{Elementary processes important in the problem of
two-dimensional disordered interacting electrons and referred to in
the text. {\em (a)} Representation of the diffusion propagator due to
impurity scattering vertices and corresponding self-energy. The
particle lines and hole lines should be on opposite sides of the
chemical potential. {\em (b)} Singular second-order interactions. {\em
(c)} Singular vertex in the density channel (and in the spin-density
channel for the spin-conserving problem). {\em (d)} Singular
(irreducible) first-order interactions. {\em (e)} Elementary singular
polarization propagator. }\label{diagramsmichap}
\end{figure}

Fermi-liquid theory for interacting
electrons survives in three dimensions in the presence of
a dilute concentration of impurities \cite{betbeder}.
Some noteworthy differences from the pure case are:
\begin{enumerate}
\item
 Because of the lack of momentum conservation, the
concept of a Fermi-surface in momentum space is lost
but it is preserved in energy space, i.e., a discontinuity in
particle occupation as a function of energy occurs
at the chemical potential. Momentum of particles may be
defined after impurity averaging. General techniques for calculating
impurity-averaged quantities are well developed; see for example
\cite{agd,betbeder}. Here and subsequently in
this chapter the self-energies, vertices etc. refer to their form
 after impurity averaging.
 \item
In the presence of impurities, the density-density
correlation (and spin-density correlation, if spin is conserved)
at low frequencies and small momentum must
have a diffusive form (This is required by particle-number conservation
and the continuity equation)
\begin{equation}
\pi ( q, \omega ) = \kappa \:
\frac{D q^2}{i \omega + D q^2}~,~~~~~~
q \ll \ell^{-1} ~\mbox{and}~ \omega \ll \tau^{-1}~.  \label{diffusiveform}
\end{equation}
Here $\kappa = dn / d \mu$ is the compressibility and $D$
is the diffusion constant.  For non-interacting electrons
$D =\frac{1}{3} v^2_F \tau$. Interactions
renormalize $D$ and $\kappa$ \cite{betbeder}.
In the diagrammatic representation used below, the diffusive propagator
is shown by a cross-hatched line connecting a particle and a hole line
as in Fig. \ref{diagramsmichap}.

\item
Because of statement {\it 1}, the {\it impurity-averaged} single-particle
spectral function at a fixed $\bf {k}$ is spread out
over an energy $\Gamma$, so that for frequencies
within a range $\Gamma$ of the
chemical potential, it has both a hole-part
(for $\omega < \mu$) and a particle part (for
$\omega > \mu $).  This is an important technical
point in microscopic calculations.
\item
The Ward-identities relating the coupling of vertices to external
perturbations change for the
coupling to unconserved
quantities  (For the pure case they are given in section \ref{routesto}).  For example
no Ward-identity can be derived for the vertex
needed for the conductivity calculation, i.e.
$Lim_{\omega \rightarrow 0} \: Lim_{q \rightarrow 0} \: \Lambda^{impure}_\alpha$,
because current is not conserved.

\begin{equation}
Lim_{q \rightarrow 0} \: Lim_{\omega \rightarrow 0}
\: \Lambda^{impure}_\alpha =
\frac{k_\alpha}{m} -
\frac{\partial \Sigma (k, \omega )}{\partial k_\alpha}
\end{equation}
holds because after impurity averaging momentum is conserved.
But microscopic calculations show that at least when
Fermi-Liquid theory is valid (Cf. section \ref{landauconsist})
\begin{equation}
Lim_{\omega \rightarrow 0} \: Lim_{q \rightarrow 0}
\: \Lambda^{impure}_\alpha =
Lim_{q \rightarrow 0} \: Lim_{\omega \rightarrow 0}
\: \Lambda^{impure}_\alpha~. \label{vertexcond}
\end{equation}
Indeed, if this were not so, one would not get a
finite d.c. conductivity at $T = 0$ for a disordered
metal in $d=3$. An argument for this is as follows: Normally, we
calculate the
conductivity by first taking the limit $q\to 0$ and then the limit
$\omega \to 0$, as on the left side of (\ref{vertexcond}). In
practice, however, even when we apply a homogeneous
field to a system, the electrons in a disordered medium experience a
field which varies on the scale of the distance between the
impurities, and so the physically relevant limit is the one on the
right hand side of (\ref{vertexcond}), where the limit $\omega \to 0$
is taken first. But the validity of (\ref{vertexcond})
appears not to extend to the case of Singular Fermi Liquids, at least for the
present case where the singularities are $q$-dependent. This is one of the
important difficulties in developing a consistent theory for disordered
interacting electrons in $d=2$.

\end{enumerate}

 The diffusive form of
the density correlation function and spin-density correlation
 is the genesis of the singularities due to interactions in
two-dimensions.
For example the elementary effective vertex in Fig. \ref{diagramsmichap}
due to a bare frequency-independent short-range
interactions in two-dimensions is
\begin{equation} \label{diffvert}
v^{2}\  \intop _{o}^{\ell ^{-1}}\  d q ~ q \left({1\over
i\omega +Dq^{2}}
\right)\approx \  v^{2}\  N(0)~\ln (\omega \tau )\  \:  .
\end{equation}

The singularity arises because
$\pi ( \omega , q ) = f ( \omega /D q^2 )$.
Recall that for pure electrons
$\pi ( \omega , q ) = f ( \omega /v_F q )$
leading to a logarithmic singularity for the second-order
vertex in one-dimension and regular behavior in higher dimensions.
 Similarly
$\pi ( \omega , q ) =  f ( \omega / q^3 )$ leads
to a logarithmic singularity in the second-order vertex in three-dimensions,
as we saw in  section \ref{gaugesec} on SFL's due to Gauge
interactions.

Note that in Eq. (\ref{diffvert}) and other singular integrals in the
problem have ultra-violet cutoffs at $q\approx\ell^{-1}$ and $\omega\approx
\tau^{-1}$ since the diffusive form is not applicable at shorter length scales
or time scales. It also follows that Boltzmann transport theory is valid
at temperatures larger than $\tau^{-1}$.

\begin{figure}
\begin{center}   
  \epsfig{figure=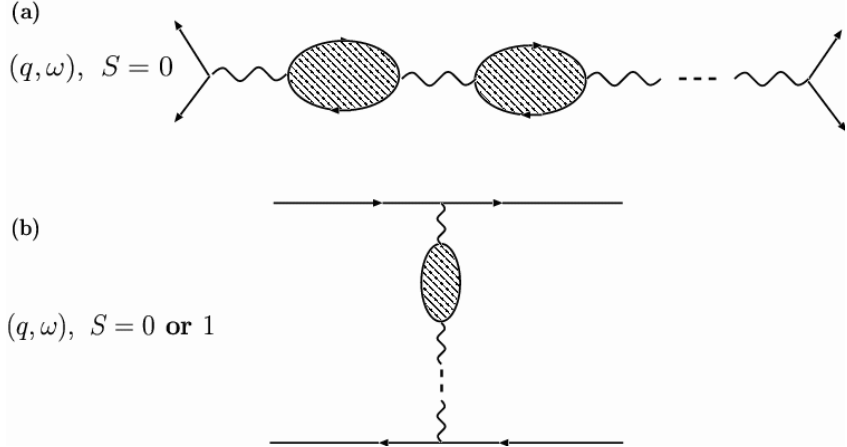,width=0.8\linewidth}
\end{center}
\caption[]{Effective interactions can be split up into singlet and
triplet channels. In the singlet-only channel {\em (a)}, the
density-density interaction is screened by the Coulomb interaction and
is universal at long wavelengths. In the triplet channel and in the
singlet channel for large momentum, the screened density-density
interaction appears only in the cross channel
and is therefore non-universal.}\label{diagramsmichap3}
\end{figure}

Actually even the first order interaction dressed by diffusion
fluctuations
 is singular. Consider first the diffusion correction to
the vertex shown in Fig. \ref{vertexcoupl}
\begin{equation}
{\Lambda \over \Lambda _{o}}
\  =\  {1\over \tau }
\  \left(i\omega \  +\  Dq^{2}\right)^{-1} \:,
\end{equation}
provided $\epsilon < 0 , \epsilon - \omega > 0$ or
vice-versa.
The restriction is a manifestation of point {\em (3)}
and arises because in the
diffusion process, only intermediate states with one
line above (particle) and the other below (hole)
the chemical potential contribute as they alone
define the physical density. This leads to the first order
{\em irreducible} interaction and the polarization graph shown
in Fig. \ref{diagramsmichap3} to
be logarithmically singular.

For the small $q$ of
interest for singular properties, one need consider
interactions only in the $s$-wave
channel.  One then has two interaction parameters, one in
the singlet channel and the other in the triplet channel.

\begin{figure}
\begin{center}   
  \epsfig{figure=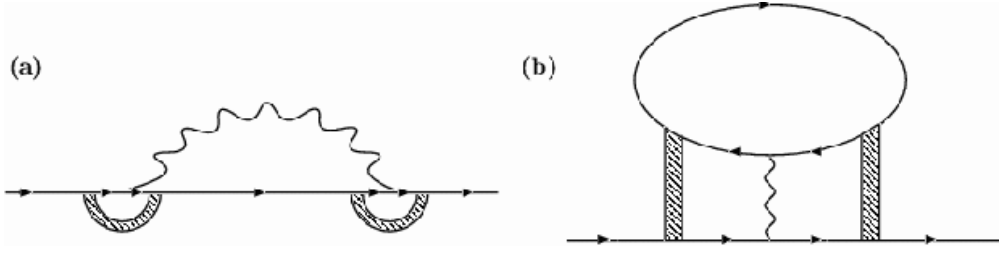,width=0.95\linewidth}
\end{center}
\caption[]{Simplest processes contributing to the singular
self-energy. {\em (a)} Exchange process. {\em (b)} Hartree process. }\label{diagramsmichap2}
\end{figure}

Consider the problem with Coulomb interactions.  Then the
effective interaction in the singlet channel sums the
polarization bubbles connected by Coulomb interactions.
Using (\ref{diffusiveform}) for the polarization bubble, it is shown
\cite{altshuler} that
for small momentum transfer the interaction in the spin singlet $(S=0)$
channel, Fig. \ref{diagramsmichap3}, becomes
\begin{equation}
V_{\mbox {\footnotesize singlet}} = 2 \kappa~.
\end{equation}
In the non-interacting limit $\kappa = N(0)$,
independent of density.  Consider now the
ladder-type interactions illustrated in Fig. \ref{diagramsmichap2}.
These involve both the singlet and the triplet interactions.
The momentum carried by the interaction lines is however to
be integrated over.  Therefore the triplet-interactions do
not have a universal behavior, unlike the singlet interactions.

Altshuler, Aronov and collaborators \cite{altshuler}
calculated the logarithmic corrections
to first-order in the interactions for various physical
quantities.  To these one can add the contribution
already discussed due to weak-localization.  The corrections to the
single-particle density of states, the specific heat
and the conductivity over the non-interacting values are
respectively:
\begin{eqnarray}
{\delta N\over N}\label{nchange}
\  & = & {1\over 4\pi \epsilon _{F}\tau }
\  \ln |\omega \tau | \left|  \ln |{\omega \over \tau (Ds^{-2})^2}|
\right| ~,  \label{correctioneq1}\\
{\delta C\over C} \label{ceqchange}
\  & = & \  {1\over \pi \epsilon _{F}\tau }
\  \left(1-{3\over 2}
F\right)\ln |T\tau | ~,\\
{\delta \sigma \over \sigma } \label{sigmaeqchange}
\  & = & \  {1\over 4\pi ^{2}}
\ \left(2-{3\over 2} F \right)\ln |T\tau | \: . \label{correctioneq3}
\end{eqnarray}
The compressibility has no logarithmic correction. In these equations,
$s$ is the screening length and $F$ is a parameter which is of the
order of the dimensionless interaction $r_s$.

The first terms in (\ref{ceqchange}) and (\ref{sigmaeqchange}) are
due to exchange
processes and the second due to the Hartree processes.
The exchange process, of which the contribution to the self-energy
is shown in Fig. \ref{diagramsmichap}, use the interaction in the singlet
channel; hence the universal coefficient.  The second
contribution uses both triplet and the (large $q$ part of the)
singlet interactions.  In first-order of interaction the
difference in sign of the two processes is natural.  In
pure systems, the Hartree process does not appear as it involves
the $ q = 0$ interaction alone which is exactly
canceled by the positive background.  For disordered systems,
due to the fluctuation in the (ground state) density, a first
order Hartree process, Fig. \ref{diagramsmichap2} contributes.

In the presence of a magnetic field the
$S_z = \pm 1$ parts of the triplet interactions
acquire a low-energy cut-off.  Therefore the logarithmic
correction to the resistivity is suppressed leading to
negative magneto-resistance proportion to $F (H/ kT)^2$
for small $H/kT$ but
$g \mu_B H >> \tau^{-1}_{so} , \tau^{-1}_s$ where
$\tau^{-1}_{so}$ and $\tau^{-1}_s$ are spin-orbit and
spin-scattering rates respectively, for appropriate impurities.

\subsection{Finkelstein Theory}\label{finkelsec}

Finkelstein \cite{finkelstein} has used field-theoretical methods to generalize
Eqs. (\ref{correctioneq1})-(\ref{correctioneq3}) beyond the
Hartree-Fock approximation. His results have been
rederived in customary diagrammatic theory
\cite{castellani2,belitz}. The interference
processes leading to weak localization are again neglected.  The
theory may be regarded as first order in $1 / k_F \ell$.  In
effect the method consists in replacing the parameter
$F$ by a scattering amplitude $\gamma_t$ for which scaling
equations are derived.  The equivalent of the $F^s_o$ parameter is
fixed by imposing that the compressibility remains unrenormalized,
i.e., does not acquire logarithmic corrections.  A second important
quantity is a scaling variable $z$, which is analogous to the dynamical scaling
exponent $z$ which we discussed in the section \ref{qcpsection}, which gives
 the relative scaling of temperature
(or frequency) with respect to the length scale. A very unusual feature of the
theory is that  $z$ itself scales!
Scaling equations are derived for $\gamma_t$ and $z$
to leading order $1 / k_F \ell$.   As $T \to 0$, both $\gamma_t$
and $z$ diverge. The divergence in $z$
 (see the discussion in section \ref{hightcsection}) usually
means that the the momentum dependence of the fluctuations is unimportant compared
to their frequency dependence.
The divergence in $\gamma_t$ as $T \rightarrow 0$ in such a case
 has been interpreted to imply divergent spatially localized magnetic-fluctuations;
 in other words, it implies the formation of local moments
\cite{finkelstein2,castellani4}.
At the same time, the scaling equations show
  conductivity flowing to  a finite value.

\begin{figure}
\begin{center}   
  \epsfig{figure=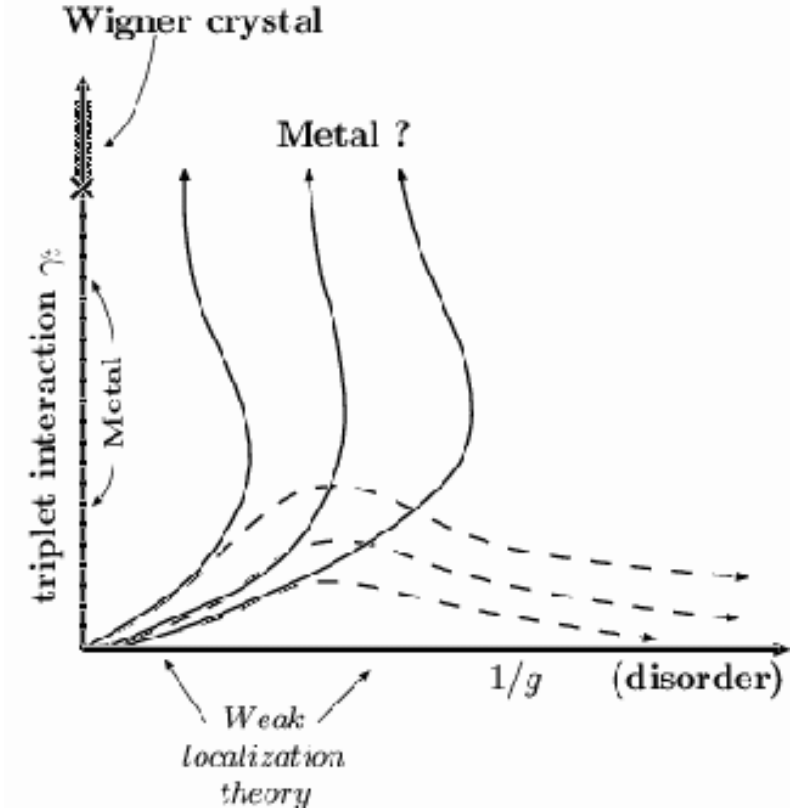,width=0.75\linewidth}
\end{center}
\caption[]{Schematic renormalization group flow for the disordered
interacting electron problem according to the Finkelstein theory.
dashed lines represent the effect on the solid lines on applying a magnetic-
field which couples to spins alone.
 }\label{finkelsteinflow}
\end{figure}

The scaling trajectories of Finkelstein's theory are shown
schematically in Fig. \ref{finkelsteinflow}.  While in the non-interacting
theory with disorder, one always has an insulator, this
theory flows always towards a metal. However the theory cannot be
trusted beyond $\gamma_t \sim 1$  beyond which it is uncontrolled.
  The theory also cannot be trusted for large disorder,
   $k_F\ell\sim {\mathcal O}(1)$ , even for small
  interactions.

It is worth emphasizing that Finkelstein's theory gives an effect of
the interactions in a direction
opposite to the leading perturbative results. The perturbative results themselves
of course are valid only for small $r_s $ while Finkelstein theory is
strictly valid only for $r_s< {\mathcal{O}}(1)$. One possibility is
that the
Finkelstein result itself is a transient
and the correct theory scales back towards an insulator.
Another possibility is that it correctly indicates (at least
for some range of $r_s$ and disorder) a strong-coupling
singular Fermi-liquid metallic fixed line. The new experiments
discussed below can be argued to point to the latter direction.

It is hard except in very simple situations (the Kondo-problem, for instance)
to obtain the approach to
 a strong-coupling fixed point analytically. In that case one may usually
 guess
 the nature of the fixed point and make an expansion about it to ascertain its
stability\footnote{Although the theory breaks down in the strong coupling
  regime, this situation is somewhat comparable to the hints that
  the weak coupling expansion gave in the early phase of the work on
  the Kondo problem: these weak coupling expansions broke down at
  temperatures comparable to the Kondo temperature, but did hint at
  the fact that the low temperature regime was a strong coupling
  regime.}.

In making such a guess, the SFL properties towards which the
Finkelstein solution flows should be kept in mind:

 {\em (i)} The conductivity flows towards a finite value in the
  theory as $T\rightarrow 0$.\\
 {\em (ii)} The density of single-particle states flows towards zero
\begin{equation}
N ( \omega ) \sim \omega^{\alpha} ~.
\end{equation}
{\em (iii)}  The magnetic susceptibility diverges at
a finite length scale (the effect of a diverging $z$)
 indicating the formation of local moments.

The last point appears to be crucial.  As may be seen from
Eq. (\ref{gammateq}) below, the growth of the triplet scattering
overrides the exchange processes which favor
the insulating state.  Indeed if the triplet divergence
is suppressed by an applied magnetic field, the theory
reverts to the perturbative form of Eq. (\ref{altshuleraronov}).  The scale
of the magnetic-field for this effect is given by the temperature. The
formation of localized regions of moments may be linked to the fact
already discussed that the ferromagnetic state is close in energy to
the paramagnetic fluid (and the crystalline states) as density is decreased.
 The experiments discussed below have a significant correspondence with this picture,
 although there are some crucial differences.

A possible strong coupling fixed point is a state  in
which the local moments form a singlet state with
a finite spin stiffness of energy of ${\mathcal O} (H_c)$.
This eliminates any perturbative instability of the triplet
channel about the fixed point.  The state is assumed to have zero density
of single-particle states at the chemical potential.
This eliminates the localization singularity as well
as well as the singularity due to the singlet channel.
This state is then perturbatively stable.
The issue of whether such a state has finite conductivity is unresolved.
  The occurrence of a new characteristic scale $H_c$
observed in the magneto-resistance experiments discussed below with
$H_c\rightarrow 0$ as the metal-insulator transition in
zero field as $n\rightarrow n_c$ is also in correspondence with these ideas.

\subsection{Compressibility, Screening length and a Mechanism for
Metal-Insulator Transition}

Suppose the metallic state in two dimensions
is described by a fixed point hinted by the Finkelstein theory and expansion
about it. Such a description must break down
 near the critical $r_s$ where a first order transition
to the Wigner transition must occur in the limit of zero disorder.
General arguments suggest that the transition for finite disorder must be
continuous \cite{imry2}.

 A suggestion for the breakdown of the Finkelstein regime
follows from the calculation of the
correction to the compressibility due to disorder \cite{sivarma}. As already
mentioned, no perturbative
singularity is found in the compressibility due
to interactions.
However, the correlation energy
contribution of the zero-point fluctuations of plasmons
 is altered due to disorder with a magnitude which depends also on $r_s$.
 The leading order contribution in powers of $(k_F\ell)^{-1}$ can be
 calculated for arbitrary $r_s$. Including this contribution the
 compressibility $\kappa$ may be written in the form
\begin{equation}
\frac{\kappa_0}{\kappa} = \frac{\kappa_0}{\kappa}_{pure}
+ 0.11 r_s^3/(\omega_0\tau) + {\mathcal O}\left((r_s^4)/(\omega_0\tau)^2\right).\label{comp}
\end{equation}
Here $\kappa_{pure}$ is the compressibility for zero disorder,
 $\kappa_0 = N(0)$, and $\omega_0$ is the
 Rydberg. In the Hartree-Fock approximation
\begin{equation}
 \frac{\kappa_0}{\kappa}_{pure} = 1- (\sqrt{2}/\pi) r_s
\end{equation}
The best available numerical calculations also give
$\kappa_0/\kappa$   varying slowly  enough
with $r_s$ that the correction term (\ref{comp}) dominates for $r_s$ of
interest near the metal-insulator
transition even for modest disorder; for example for $\omega_0\tau \approx 10$,
the disorder contribution in Eq. (\ref{comp}) is larger than the pure contribution
for $r_s\geq 10$. This has an
important bearing on the Metal-Insulator transition
because the screening length $s$ is given by
\begin{equation}
s/s_0=\kappa_0/\kappa~,
\end{equation}
where $s_0=a_0/2$.
Strictly speaking $s$ is the screening length for an external
 immobile charge and the screening of the
electron-electron interactions is modified from (\ref{comp}) due
to vertex renormalizations.
But in this case they do not change the essential results.
From Eq. (\ref{comp}) it follows that the screening length
$s(\ell) > \ell$, the mean-free path, for
\begin{equation}
r_s \geq 3(\omega_0\tau)~.\label{micondition}
\end{equation}
Suppose the condition $s(\ell)>L>\ell$ is satisfied. Here L
 again is the size of the box for which the
calculation is done, defined through $DL^{-2}\approx T$. The assumption
 of screened short-range interactions,
with which perturbative corrections
leading to results
of Eqs. (\ref{nchange}), (\ref{ceqchange}), and (\ref{sigmaeqchange}) are obtained,
 is no longer valid. In this regime
the calculations must be done with
 unscreened Coulomb interactions.
The correction proportional to
F in Eqs. (\ref{nchange}), (\ref{ceqchange}), and (\ref{sigmaeqchange}) is not
 modified but the singlet contributions
are more singular (due to the extra $q^{-1}$ in the momentum integrals).
For instance, Eq. (\ref{sigmaeqchange})  is modified to
\begin{equation}
\delta \sigma/\sigma =-(\sqrt 2/\pi^2)r_s\frac{L}{\ell}~.
\end{equation}
This implies a crossover to strong localization. It is therefore
suggested that the metallic state ceases to exist when
condition (\ref{micondition}) is satisfied.

The above line of reasoning is of particular interest because as discussed
below a sharp variation in the compressibility is indeed observed
 to accompany the transition from the metallic-like to
insulating-like state as density is decreased (as shown
in Fig. \ref{dultzplot} below).

 \begin{figure}
\begin{center}   
  \epsfig{figure=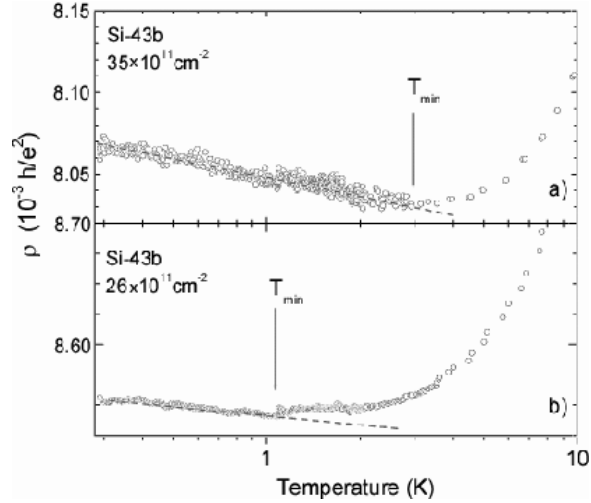,width=0.55\linewidth}
\end{center}
\caption[]{Resistivity data on a fine scale for the two highest
densities in Fig. \ref{pudalov_1plot} below, showing correspondence
with the theory of weak
localization at such high densities.  From Pudalov {\em et al.}
\cite{pudalov3}.
}\label{pudalovlogT}
\end{figure}

\subsection{Experiments}

Soon after the publication of the theory of weak
localization its predictions were seemingly verified
in experiments on $Si$-MOSFETS \cite{bishop,uren}.
The experiments measured resistivity
on not very clean samples of high density  with resistivity of
$O ( 10^{-2} h /e^2 )$.  In a limited range of temperature
the predicted logarithmic rise in resistivity with
decreasing temperature with about the right prefactor
was found \cite{bergmann}.  In view of the perturbative results
of Altshuler and Aronov \cite{altshuler}
 and the knowledge that  electron-electron interactions
 alone lead to a
Wigner insulator at low densities, one was led to the conviction that
the metallic state does not exist in two dimensions. It was expected that
samples will larger $r_s$ will simply show logarithmic corrections
to the resistivity at a higher temperature and pure samples at a
lower temperature. Not too much attention was paid to Finkelstein's results
which pointed to the more interesting possibility of
corrections in the opposite direction.

\begin{figure}
\begin{center}   
  \epsfig{figure=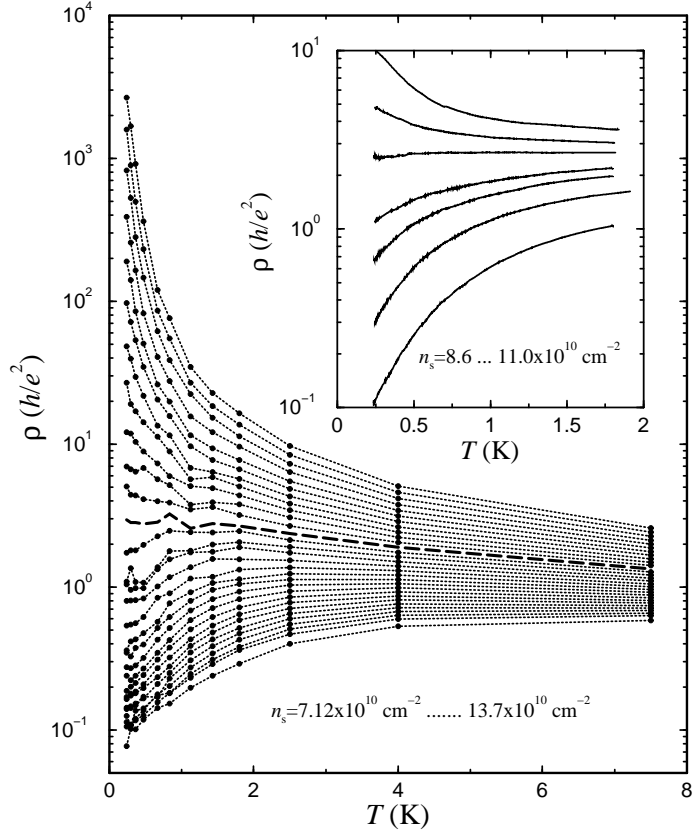,width=0.65\linewidth}
\end{center}
\caption[]{Resisitivity as a function of temperature for a wide range
of densities (and Fermi-energy) in a disordered $Si$ MOSFET. The inset
shows accurate measurements of $\rho(T)$ close to the speparatrix for
another sample. From Sarachik and Kravchenko \cite{sarachikkravchenko,aks}. }
\label{resistversusTkrav1}
\end{figure}

\begin{figure}
\begin{center}   
  \epsfig{figure=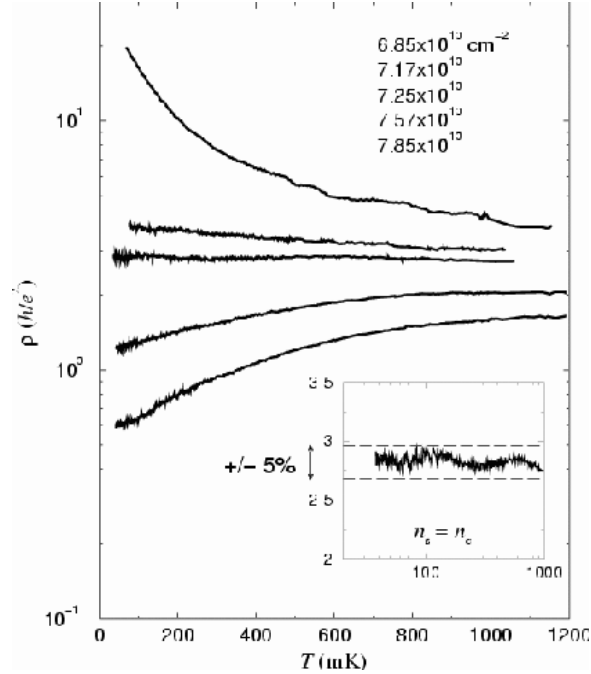,width=0.55\linewidth}
\end{center}
\caption[]{Resistivity versus temperature at five different electron
densities in the experiments of  Kravchenko and Klapwijk
\cite{kravchenko2}. The inset shows that the middle curve
($n_s=7.25\times 10^{10} cm^{-2}$) changes by less than $\pm 5 $\% in
the entire temperature range.}
\label{resistversusTkrav2}
\end{figure}

\begin{figure}
\begin{center}   
 \epsfig{figure=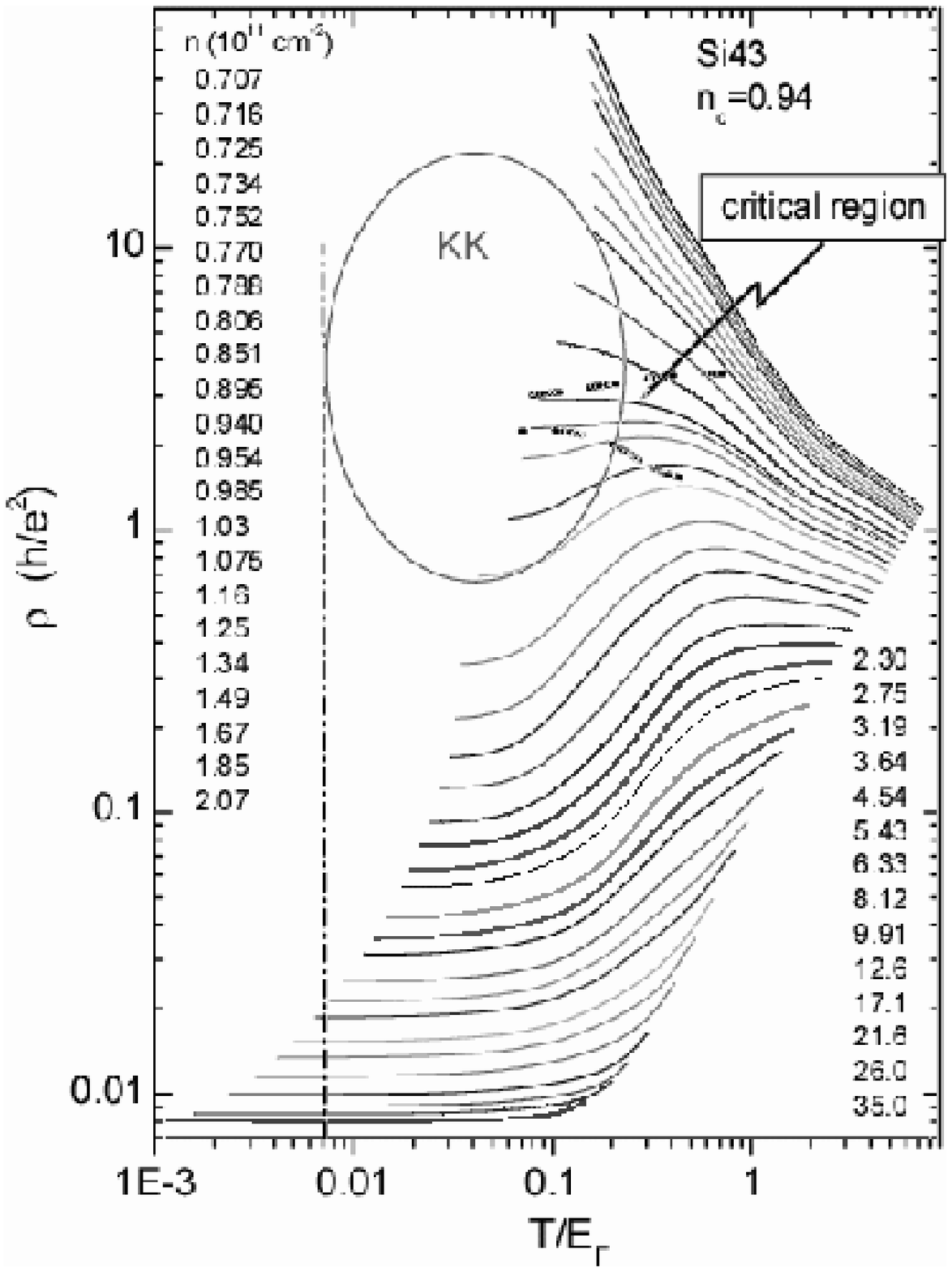,width=0.6\linewidth}
\end{center}
\caption[]{Plot of the resistivity as a function of the scaled
temperature $T/E_F$. The encircled region indicates the range of
parameters explored in Fig. \ref{resistversusTkrav2}  and in
\cite{kravchenko2}. The dash-dotted vertical line depicts the
empirical temperature $T_Q=0.007 E_F$ below which the logarithmic
temperature dependence like that of Fig. \ref{pudalovlogT} sets
in. From Prinz {\em et al.} \cite{prinz}.}
\label{resistivityscaledT}
\end{figure}

The more recent experiments on a variety of samples on a wider
range of density and of higher purity than earlier
have re-focused attention on the problem of
 disorder and interactions in two dimensions and,
by implication, in three dimensions as well.
 Several
 reviews of the experiments are available
\cite{aks,altshuler2,altshuler5}.
We will present only a few experimental data to
highlight the theoretical problems posed, and will focus on the
behavior of the data as function of temperature. The scaling of the
data as a function of the electron density $n-n_c$ or field $E$ will
not be discussed; there is a considerable body of data on nonlinear
$E$-dependence (see e.g. \cite{shashkin} and references therein) but the
significance of the data is not clear at present.

The first thing to note is that results consistent with the
 earlier data \cite{bishop,uren}
 are indeed obtained for high enough densities. Fig. \ref{pudalovlogT}
 shows the resistance versus temperature in $Si$ for $r_s \sim {\mathcal O}(1)$.
 The magnitude of the
  temperature dependence is  consistent with the predictions of weak
localization
  corrections. As we will show below in the same region of densities, the
  negative magneto-resistance predicted as correction to weak-localization,
  discussed above, is also observed.

\begin{figure}
\begin{center}   
  \epsfig{figure=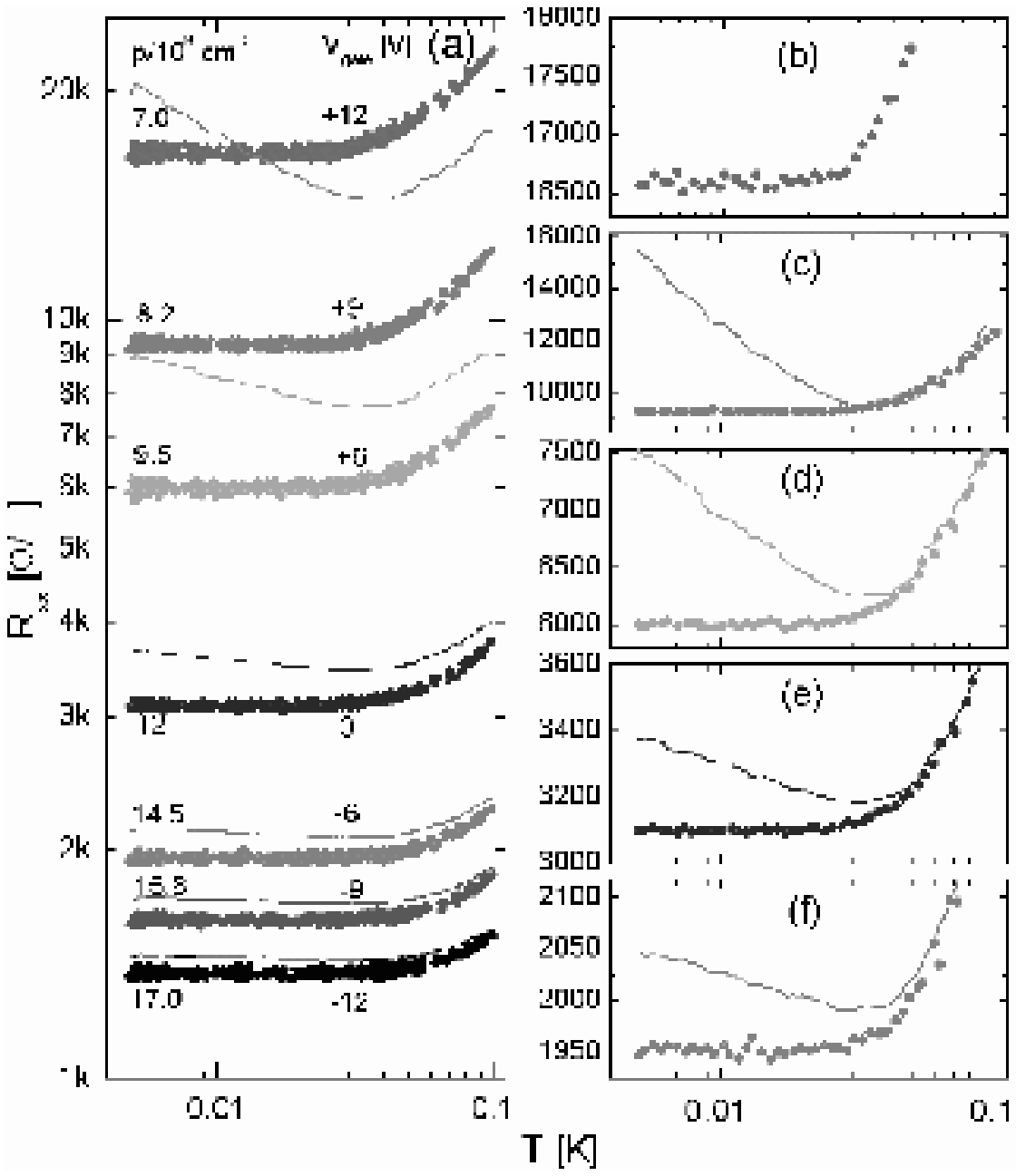,width=0.7\linewidth}
\end{center}
\caption[]{ {\em (a)}  Temperature dependence of the longitudinal resistivity
of a 2D hole gas for various gate biases and associated densities in
experiments on $GaAs$. The
solid curves are estimated weak localization predictions.
{\em (b)-(f)}: Magnified view of the data in {\em (a)} averaged over a 5\%
temperature interval. The estimated weak localization prediction has
been shifted to coincide with the data curve at $T=50 mK$. From Mills {\em et al.} \cite{mills} }\label{millsdata}
\end{figure}

Fig. \ref{resistversusTkrav1}  shows the resistivity as a function of $T$
over a  wide range of densities.  Similar data from \cite{kravchenko2}
over a large small of densities is shown in
Fig. \ref{resistversusTkrav2}, and data over a large range
of densities are plotted as a function of $T/E_F$
in Fig. \ref{resistivityscaledT}. The
resistivity clearly shows a change of  sign in the
curvature as
a function of density at low temperatures.  The resistivity at the
cross-over density as a function of temperature is shown
down to 20mK in the inset to Fig. \ref{resistversusTkrav2} and
 is consistent with temperature independence.
 The true electron temperature in
these samples is a question of some controversy \cite{altshuler2,aks},
but more recent experiments, whose data are shown in
Fig. \ref{millsdata}, have corroborated these results by studying this issue
 very carefully  down to 5mK.
 In the high density region
the resistivity does rise with decreasing temperature
logarithmically, consistent with earlier
measurements. The consistency of these data sets for two very
different types of samples therefore gives strong evidence that these
are genuine effects in both types of systems.

The data shown in Figs. \ref{resistversusTkrav1}-\ref{resistivityscaledT}
 is for $Si$-MOSFET samples. The data for $GaAs$
heterostructures, and $Si$ in other geometries is
qualitatively similar
\cite{hanein,coleridge1,coleridge2,popovic,hamilton,mills}.
 Fig. \ref{millsdata}
shows data on high quality gated GaAs quantum wells with densities on
the metallic side of the metal-insulator "transition" taken to
temperatures as low as 5mK. The resistivity is essentially temperature independent
at low temperatures; the logarithmic corrections expected from weak-localization (calculated using
the measured resistivity and the theoretically expected
$\tau_{\phi}$) is also shown.

\begin{figure}
\begin{center}   
\epsfig{figure=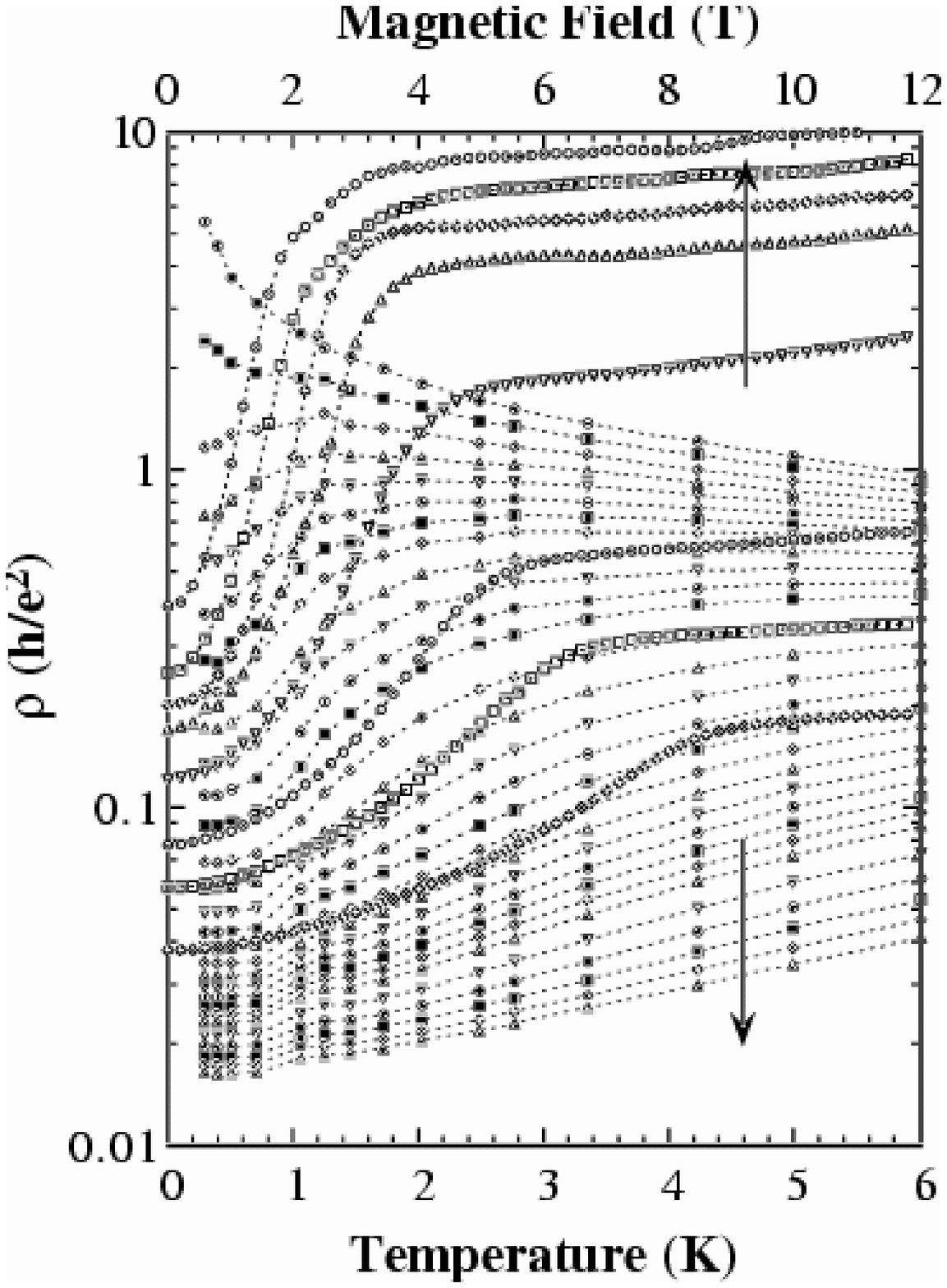,width=0.6\linewidth}
\end{center}
\caption[]{Results  for resistivity versus
 temperature and versus magnetic field applied in the plane for a few densities
 on either side of $n_c$ .
 The magnetic field is shown on the upper axis
 and the data is taken at the lowest temperature
 for some of the densities shown in the resistivity versus temperature
plots.
From Pudalov {\em et al.} \cite{pudalov}.}
\label{pudalov_1plot}
\end{figure}

 {\em Experiments in a Parallel Magnetic Field:}  A magnetic field applied
parallel to the plane
couples primarily to the spin of the electrons.
For small-fields
and for $n \gg n_c$, a positive magneto-resistance
proportional to $H^2$ is observed as expected from
perturbative calculations in the interactions.  For
fields such that $\mu_B H \approx E_F$, the electrons
are fully polarized and the resistivity saturates as expected.
The temperature dependence of the resistivity
begins to be insulating-like at low temperatures
with the cross-over temperature
increasing as $n$ decreases \cite{simonian}.  This is an
indication that the metallic state becomes unstable
as the spins are polarized. A complete set of data is shown in Fig. \ref{pudalov_1plot}
where resistivity versus temperature in a  $Si$-MOSFET
with density varying across $n_c$ is shown together with
the resistivity versus magnetic-field at the lowest temperature for some densities
on the $n>n_c$ side. It is noteworthy that
the temperature dependence of the high-field data (not shown)
appears to fall on the curve of resistivity versus temperature (at H=0) which the
high field (low temperature) data saturates asymptotically.

\begin{figure}
\begin{center}   
\epsfig{figure=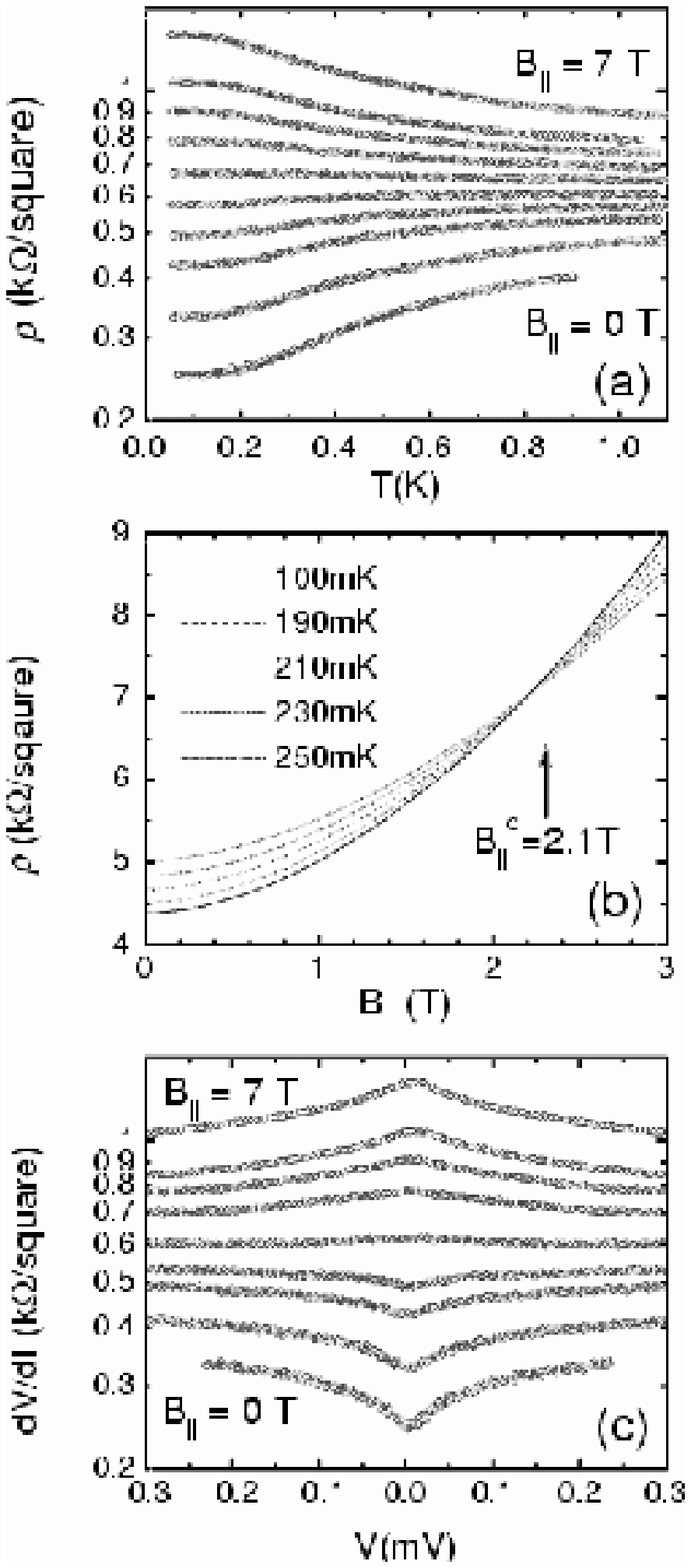,width=6.5cm}
\end{center}
\caption[]
{Plot of the magnetoresistance.
In {\em (a)} the $T$ dependence of $\rho$ in the zero field metallic phase
is shown on a semilog plot for a hole density
$3.7 \times 10^{10} cm^{-2}$ for varying
$B_{||}$ values. As $B_{||}$ increases
from zero, the strength of the
metallic behavior measured by the total
change in $\rho$ from about 1K to 50mK
weakens progressively, and for
$B_{||} \ge B_{c}$
$ d \rho/dT$  becomes negative (i.e.
the system becomes insulating).
An alternate way of demonstrating the
existence of a well defined $B_{||}^{c}$
is to plot $\rho$ against $B_{||}$
at several different temperatures.
In {\em (b)}~ $\rho$ is plotted vs $B_{||}$
at a hole density $1.5 \times 10^{10} cm^{-2}$.
$B_{||}^{c}$ is read off the crossing point
marked by the arrow. In {\em (c)}, the differential
resistivity $dV/dI$ is measured
at 50mK across the $B_{||}$ induced Metal-Insulator
transition is shown at magnetic field
strengths similar to those in {\em (a)}. From  Yoon {\em et al.}
\cite{yoon}.}
\label{plotfromyoon}
\end{figure}

  The parallel magneto-resistance  has been
examined carefully for $n$ close to but larger than
$n_c$, and is shown in Figs. \ref{plotfromyoon} for $p$-type $GaAs$
\cite{yoon}.
Qualitatively similar results are found in $n$-type $Si$
\cite{pudalov2}.
It is discovered that a critical field as a function of density
$H_c (n)$ exists such that for $H < H_c (n)$
the resistivity continues to be metallic-like
$d \rho / dT > 0$ and for $H > H_c (n)$ it is
insulating like $d \rho / dT < 0$.  The field
$H_c (n)$ tends to zero as
$n \rightarrow n_c$. The low temperature data on the high fields side is puzzling and
should be re-examined to ensure that the electron-temperature is
indeed the indicated temperature.

\begin{figure}
\begin{center}   
  \epsfig{figure=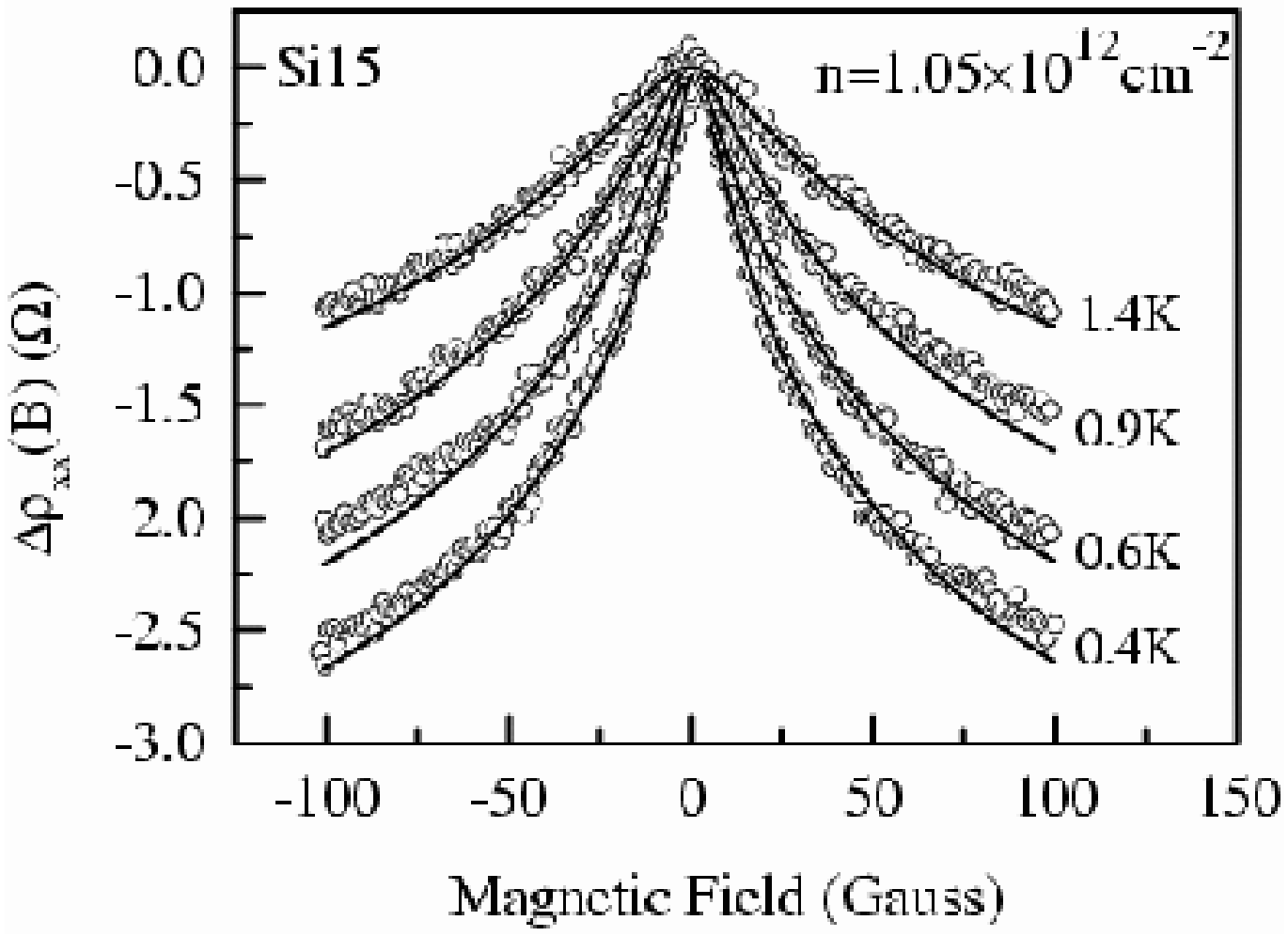,width=0.6\linewidth}
\end{center}
\caption[]{Plot of the magnetoresistance.
The change in the resistivity $\Delta \rho(B) = \rho_{xx}(B)-
\rho_{xx}(0)$ vs magnetic field $B$ at an electron density of 1.05 $\times$
10$^{12}$ cm$^{-2}$ at various temperatures. The open circles denote
the measurements and the full line is the best least square fit
according to the single electron weak localization correction
to the conductivity. From Brunthaler {\em et al.} \cite{brunthaler}.
}
\label{datalowfields}
\end{figure}

\begin{figure}
\vspace*{0.6cm}
\begin{center}   
\epsfig{figure=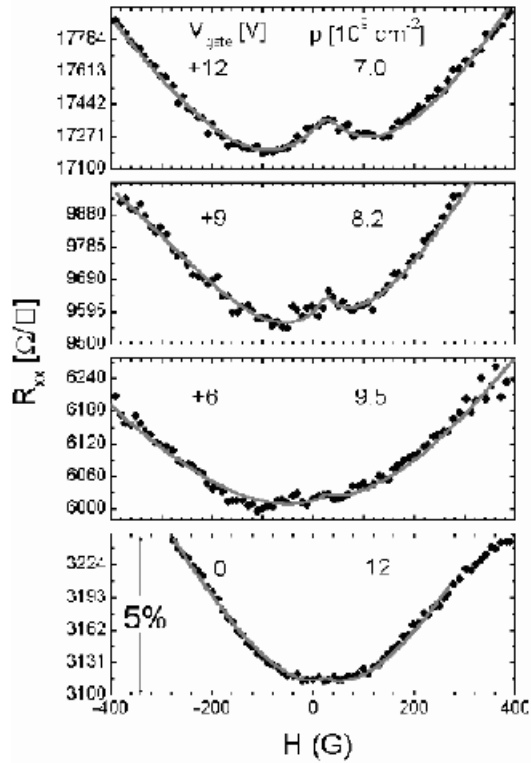,width=7cm}
\end{center}
\caption[]{Variation of the longitudinal resistance with
perpendicular magnetic field for 2D sample at $T=9 mK$ and at various
indicated densities. The weak-localization correction is estimated to be
${\mathcal O} (10^2)$
larger than the observations at these densities.
 From Mills {\em et al.} \cite{mills}.}\label{millsplot}
\end{figure}

\begin{figure}
\begin{center}   
  \epsfig{figure=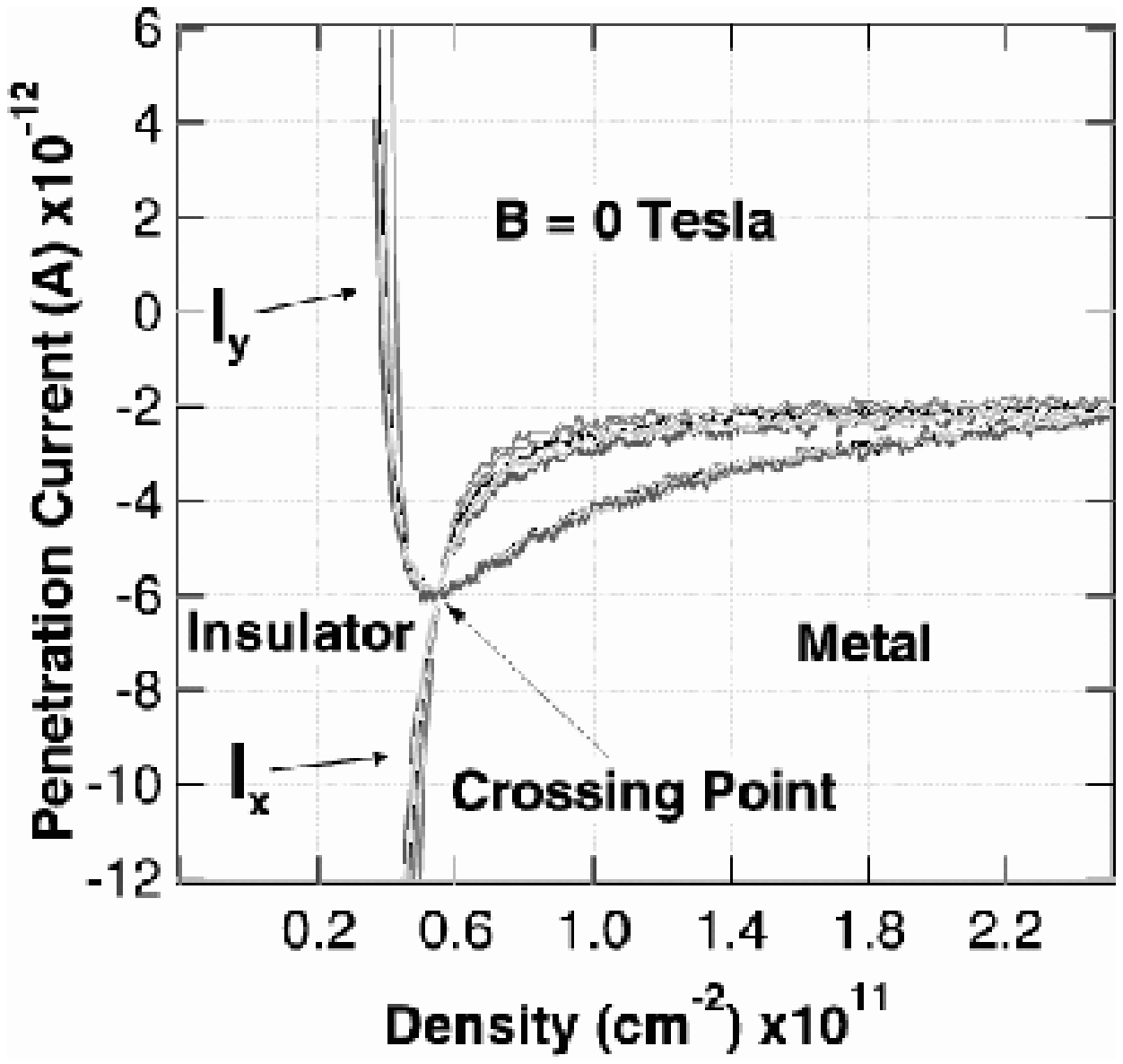,width=0.6\linewidth}
\end{center}
\caption[]{Compressibility data.  In this experiment, at low frequencies,
$I_{x}$ is directly proportional to
$R_{x}$, the dissipation of the two dimensional hole system,
while $I_{y}$ is
proportional to  the inverse compressibility.
$I_{x}$ and $I_{y}$ are shown as a function of density for five different
temperatures ranging from 0.33 K to 1.28 K at an excitation frequency of 100Hz.
The crossing point of the five dissipation channel
curves corresponds to the metal-insulator transition at $B=0$.
The minimum of the inverse compressibility occurs at the same
hole density of $5.5 \times 10^{10} cm^{-2}$. From Dultz {\em et al.} \cite{dultz}. }\label{dultzplot}
\end{figure}

{\em Experiments in a Perpendicular Field:}   The behavior
of a resistance in all but very small perpendicular
fields, is dominated by the
Quantum Hall Effect (QHE).  The connection of
the Quantum Hall transitions to the metal-insulator
transition at $n = n_c$ and $H = 0$ is an interesting
question which we will not touch on.  At low fields and for
$n\geq n_c$, outside the  QHE regime, negative
magneto-resistance predicted by weak-localization
theory are observed.  Data for $n \gg n_c$ is shown
in Fig. \ref{datalowfields} and agrees quite well with the theoretical curves
as shown; similar results for $n$ close to $n_c$
are also reported \cite{hamilton}. More
recent low temperature data in $GaAs$ heterostructures\cite{mills} is
reproduced in Fig.
\ref{millsplot} for densities $n>n_c$ but close to $n_c$.
A magnetoresistance two orders of magnitude
smaller than the weak localization theory is estimated although the width of the
negative magneto-resistance region is not in disaccord with the weak-localization
correction.

 {\em Compressibility measurements}
Compressibility $(\kappa)$ measurements \cite{dultz,ilani}
in the region around $n=n_c$ show a rapid change in $\kappa^{-1}$
from the negative value
characteristic of high $r_s$ metallic state to positive values --- see
Fig. \ref{dultzplot}. These are
very important measurements which show that a thermodynamic quantity has a
very rapid variation near $n\approx n_c$. We have already discussed that
such changes
were predicted \cite{sivarma}
 to occur through
perturbative corrections due to disorder to the energy
of interacting electrons. Some more recent measurements \cite{yacoby}
 of the local
electrostatic potential show that in the region $n\approx n_c$ large scale
density fluctuations (puddles) occur with very weak connections between them.
Completely isolated puddles evidently have $\kappa^{-1}=0$.

\subsection{Discussion of the Experiments in light of the theory of Interacting
  Disordered Electrons}\label{discussionofexpt}

In comparing the experimental results with the theory,
it is necessary to separate out the effects due to
"customary-physics" ---  for instance electron-phonon
interactions, creation of ionized impurities with
temperature \cite{altshuler6},
 change of screening from its quantum
to its classical form as a function of temperature \cite{dassarma}
change of single particle wavefunctions with a magnetic
fields \cite{klapwijk}, inter-valley scattering \cite{yaish}, etc. ---
from  the  singular effects due to
impurities and interactions.  The separation is at
present a matter of some debate.
However it seems that the following features of the
experimental data in relation to the theoretical ideas
summarized in
sections \ref{interdis}, \ref{finkelsec} are especially
noteworthy. These must be read bearing in mind our earlier discussion that
most of the interesting experiments are in a range of $r_s$
and disorder where the theoretical problems are unresolved
and only hints about the correct form of a theory are available.

$\bullet$ At $r_{s} \leq {\mathcal O}(1)$
and $k_F \ell \gg 1$ a logarithmic increase in
resistance with decreasing $T$ consistent with
weak-localization as well as with the perturbative interaction
correction is observed.  A positive magneto-resistance
consistent with the latter  is also observed.
Also observed is the correction to weak-localization due to
phase-breaking of backscattering in a perpendicular magnetic
field.  The latter yields sensible values and temperature
dependence for the phase relaxation rate given by
the theory. It appears that at high enough density,
 the weak localization theory supplemented by the
 perturbative theory of interactions is in excellent agreement with
the experiments in the range of temperatures examined.

$\bullet$ As $r_s$ is increased (and $k_F \ell$ decreased) the
logarithmic resistance is lost in the observed temperature
range, whereas weak-localization theory predicts
that the coefficient of such terms (as well as the onset
temperature for their occurrence) should increase.  For $r_s$ not
 too large the decreased logarithmic term may be
associated with the perturbative corrections
 (\ref{altshuleraronov}) due to
interactions.

$\bullet$  Upon further increasing  $r_s$, the derivative
 $d \rho / d T $ becomes positive in
the low temperature region as in a metal. The
magneto-resistance in a parallel field is positive
$\sim H^2$ as is predicted by Finkelstein [although
the variation is closer to $H^2 /T$ rather than as
$( H / T )^2$] \cite{pudalov2}.  The phase breaking correction in a
perpendicular field continues to be observed.  But
quite curiously the deduced $\tau_\phi$ is larger than
$\tau$ deduced from resistivity --- by definition a
phase-breaking rate serves as a cutoff only if
$\tau_\phi < \tau$.

$\bullet$ In the ``metallic'' regime for intermediate $r_s$, a strongly temperature
dependent contribution for $T \leq E_F$ is found which
may be fitted to the form
$\rho^\prime (n) \exp ( -E_a (n) /T)$.  The magnitude of
this term rapidly decreases as the density $n$ decreases.
No accepted explanation for this contribution has been
given. In $Si$ the change of resistivity at $n\approx 10 n_c$
due to this contribution is an order of magnitude larger than in $GaAs$.
 It has been proposed \cite{altshuler,pudalov} that this
contribution together
with the weak-localization contribution may well account for
all the data in the ``metallic'' regime since it pushes the
minimum of the resistivity below which the logarithmic
temperature dependence is visible to lower temperature than
the available data at lower densities.

This issue can be resolved by experiments at lower
temperatures. At this point, especially in view of the consistency of
the recent results of
Mills {\em et al.} --- see Figs. \ref{millsdata} and \ref{millsplot}
--- with the earlier experiments, one can say that
 it requires an unlikely conspirancy of contributions to remove the
temperature dependence over a wide range for different materials
and with different degrees of disorder.

A quite different scenario also consistent with the existing
data is that the logarithmic upturn in the resistivity
observed in high density samples is a transient that on
further decreasing the temperature disappears to be replaced
by $\frac{d \rho}{d T} \rightarrow 0$ as
$T \rightarrow 0$, at least above some characterestic density which is a
function of diorder.
We will come back to this  issue when we discuss
the possible phase-diagram.

$\bullet$ As $r_s$ approaches $r_{sc}$, $d \rho / d T$ tends
to zero (through positive values).  A separatrix
is observed with $d \rho / d T \approx 0$ over about
2 orders of magnitude in temperature for $Si$ and
over an order of magnitude in $GaAs$.
For $r_s > r_{sc}$, $d \rho / d T$ is negative befitting an insulator.
$r_{sc}$ appears to be smaller for dirtier samples but not enough
systematic data is available for drawing a functional relation.

$\bullet$ The electronic compressibility rapidly changes near the "transition"
and rapidly becomes small on the insulating side. Its
  value on the insulating side is consistent with approaching
zero in the limit of
zero temperature.
Although this is in qualitative accord with the theoretical suggestion
\cite{sivarma}, further experiments measuring
simultaneously the compressibility and the
conductivity at low temperatures are necessary to correlate the metal-insulator
 transition with the rapid variation of compressibility or the screening length.
Note that it follows from the Einstein relation $\sigma= D\kappa$ that if
$\kappa$ is finite in the metallic state ($\sigma$ finite) and zero in the
insulating state ($\sigma =0$), $\kappa$ must go to zero {\it at the transition}.
Otherwise we would have the absurd conclusion that $D \rightarrow \infty$
at the transition.

Interesting phenomenological connections between the transport properties and
formation of "puddles of electron density"
 of decreasing size as the metal-insulator "transition' is approached
have been drawn \cite{meier}.

$\bullet$ Near $r_s = r_{sc}$ the
resistivity as a function of temperature on the insulating
side appears to be a reflection of that on the metallic
side about the $d \rho / dT = 0$ line if the data is not
considered to low temperatures \cite{midiscovery}. Now with more complete data,
we know that the resistivity flattens to zero slope at low temperatures
on the "metallic" side of $n_c$ . The
most likely behavior appears to be that the resistivity approaches a finite value
at low temperature on one side of $n_c$ and an infinite value on the other. A
 one-parameter scaling ansatz \cite{dobrascaliwcz} for the problem
 with interaction and disorder
 gave
$\rho \rightarrow \infty $ for $n<n_c$ and $\rho \rightarrow 0$ for $n>n_c$
as $T \rightarrow 0$ and reflection symmetry just as at any second-order transition
with one scaling parameter\footnote{The data also led to suggestions for a superconducting
ground state on the metallic side \cite{phillips}, and to an anyonic state \cite{zhang}!}.
 Does this necessarily imply that multi-dimensional scaling is required
 near this transition? But another
  important point to bear in mind is that since
resistance does not depend on a length scale in two-dimensions, it
need not be a function,
 in particular, of the correlation length near the transition.
The resistivity is allowed to be finite on one side of a
metal-insulator transition and infinite on the other
even though the transition may be continuous and the correlation
length diverges on either side with the same exponent. The glassy nature
(Coulomb glass) of the insulating
state is also expected  to change the critical properties.

$\bullet$ The resisitivity at low temperatures for $n \ll n_c$ has been
fitted to an activated form $\propto \exp(\Delta/T)^{\alpha}$ with
$\alpha \approx 1/2$ and with $\Delta \rightarrow 0$ as
$n \rightarrow n_c$. This is characteristic of a Coulomb glass \cite{efros}.
Whether this is indeed the asymptotic low temperature
 form is not completely settled.

$\bullet$ The resistance at $n \approx n_c$
appears to vary from sample to sample but is within a factor of 3 of the
quantum of resistance. It is worth emphasizing that $n_c$ is close to the
 density expected for Wigner crystallization. With Coulomb
interactions and disorder,
 the insulating state is indeed expected to be Wigner glass.
In that case one might expect singular frequency dependent properties and hysteretic
behavior near the transition.

$\bullet$ For $r_{s}\approx r_{sc}$
the resistance in a parallel-field
is especially noteworthy.  In a parallel field
$d \rho / d T$ decreases untill at a field
$H = H_c (n)$ it  changes sign.
$H_c$ vanishes at $n_c$, the density where
$d \rho / d T = 0$ for $H_\parallel = 0$.  In this regime,
$\rho (H, T, n )$ can be scaled as \cite{yoon}
\begin{equation}
\rho \left( \frac{(n - n_c)}{T^\alpha} , \:
\frac{(H - H_c (n - n_c))}{T^\beta} \right)
\end{equation}

This means that the transition from the metallic state to the insulating state
can be driven by a magnetic field. It appears that the "metallic" state
owes its existence to low-energy magnetic fluctuations which are quenched by
a magnetic field. This is in line with Finkelstein theory and the
flow diagram of Fig. \ref{finkelsteinflow}  yet
the existence of a scale $H_c$ is not anticipated
by the calculations of Finkelstein (nor, of course, is
the mere existence of $n_c$).  As $H_\parallel$ is further
increased $d \rho / d T$ approaches the insulating
behavior characteristic of $n > n_c$ at
$H_\parallel = 0$.  At a fixed temperature the
resistivity saturates for
$g \mu_{B }H_\parallel \geq E_{F}$, i.e. for
a fully polarized band.

For small perpendicular fields, negative
magneto-resistance of the form of  (\ref{altshuleraronov})
continues to be observed at least for $Si$ for $n \gg n_c$.
In $GaAs$ this contribution at least in the range $n_c \geq n \geq 2n_c$
is negligible.

$\bullet$ The Hall coefficient $R_H$ is continuous across
the transition, obeying the kinetic theory result
$R_H \sim 1/n$.  On the ``metallic'' side this is not surprising.
On the ``insulating'' side this is reminiscent of the properties
of Wigner glasses \cite{chitra,giamarchi}

\subsection{Phase Diagram and Concluding Remarks}\label{conclremarks}

\begin{figure}
\begin{center}   
  \epsfig{figure=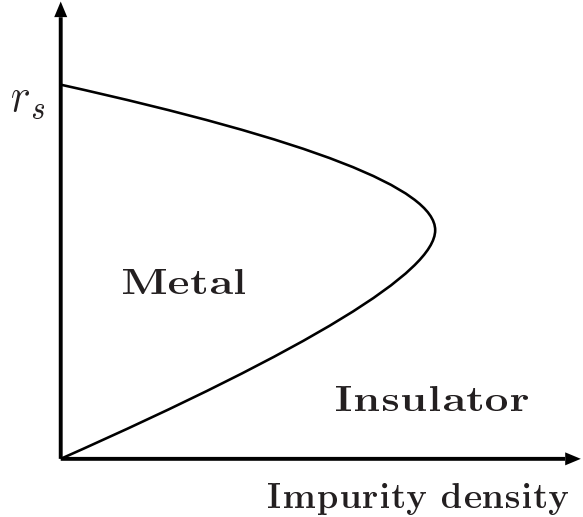,width=0.55\linewidth}
\end{center}
\caption[]{A tentative phase diagram at $T=0$ for two-dimensional disordered
  electrons with interactions.}\label{phasediagrammi}
\end{figure}

It is worthwhile to try to guess the $T = 0$ phase-diagram of
interacting disordered electrons on the basis of the data and the
available theory, inadequate though it is.  A convenient set of axes
is $r_s$ (or $n^{-1/2}$), as it parameterizes the dimensionless
interaction, and the resistance in units of $h / e^2$ as it
parametrizes the dimensionless disorder, see Fig. \ref{phasediagrammi}.

Reliable theoretical results are found only along the two axes of Fig.
\ref{phasediagrammi}.  States are localized all along the horizontal
axis.  Localized states at $T = 0$ must be organized into one or
another kind of magnetically ordered state.  On the vertical axis a
Fermi-liquid gives way via a first order transition to a Wigner
crystal which may have various magnetic phases.

In light of the experiments, the assumption made too long that the
entire region in Fig. \ref{phasediagrammi} is an insulator has to be
abandoned in all likelihood. There does appear to be a "metallic
state". With disorder a crossover to a Wigner glass must occur at
large $r_s$. It is also clear that at high densities weak-localization theory
supplemented by perturbative corrections due to interactions works quite well
in the range of temperatures examined. At moderate $r_s$ for small disorder
the Finkelstein correction appears
to take over and a ``metallic'' state takes over. The best evidence for
 this, paradoxically, is the magnetic filed
 (parallel to the plane) dependence of the resistivity which appears to
 eliminate the ``metallic'' state.
 
 Based on these considerations, the phase diagram
 Fig. \ref{phasediagrammi} is put forth.
 It is surmised that the weak localization correction
  flows to strong localization
 for sufficiently strong disorder and small enough $r_s$,
  but that it gives way to
 a metallic state at weak-disorder and larger $r_s$. What determines the boundary?
 A possible criterion is that on one side
the Finkelstein renormalization is more important and on the other
side localization due to disorder is more important.  The cross-over
to strong localization occurs at a length scale $\zeta$ given by Eq.
(\ref{weak-loc}) where the resistivity doubles.

The scaling equation for the triplet interaction parameter is
\cite{finkelstein,castellani,lee}
\begin{equation}
d\gamma _{t}/d \ln L  =\  \pi \
 k_{F}\ell \left(1+2\gamma _{t}\right)^2~, \label{gammateq}
\end{equation}
so that for small initial value $\gamma_t^o$ at $L = \ell$ one gets
$\gamma_t (L) = \gamma_t^o + ( \pi k_F \ell ) \ln (L / \gamma )$.
As $L = \zeta$, $\gamma_t \approx \gamma_t^o +1$.  The boundary
between the ``metallic'' and the ``insulating'' regions on this basis
is linear at small $r_s$ and small $g^{-1}$, as shown in
Fig. \ref{phasediagrammi}. This is highly conjectural but the existence of
the phase boundary at the point $r_s\rightarrow 0$ and $1/g\rightarrow 0$
is more robust.

This scenario can be tested in high
density, low disorder samples by measurements of resistivity at very
low temperatures. If correct in some regime of parameters near the boundary,
 the logarithmic
weak-localization correction should appear at high temperatures
and disappear at lower temperatures.

We have stressed that the ``metallic state'' in 2-dimensions is likely
to be a Singular Fermi-liquid with an interesting magnetic-ground
state.  Direct or indirect measurements of the magnetic susceptibility
through, for instance the magnetic field dependence of the
compressibility should yield very interesting results. Also
interesting would be measurements of the single-particle density of
states through tunneling measurements. Further systematic and careful measurements of
the compressibility are also required to correlate the transition from
the ``metallic'' state to the increase in
susceptibility.

The basic theoretical and experimental problem remains the
characterization of the ``metallic
state'' its low temperature entropy,magnetic susceptibility,
single-particle density of states etc. The experimental
and theoretical problems
are many but one hopes not insurmountable.

\section{Acknowledgements}

This article is an outgrowth of lectures delivered in spring 2000 by
C. M. Varma during his tenure as Lorentz Professor at the Universiteit
 Leiden. He wishes to thank the faculty and staff of the Physics
department and the deep interest shown by the attendees of the
lectures.

Special thanks are due to numerous colleagues who
provided the experimental data and who explained their ideas
and clarified countless issues.

ZN also wishes to take this opportunity
to greatly thank his former mentor, S. A. Kivelson,
for coaching in one-dimensional physics.

Finally, ZN and WvS also wish to express their thanks to Debabrata Panja,
Michael Patra, Kees Storm and especially Carlo  Beenakker from the
Instituut-Lorentz for all their help in generating, scanning, and compressing
the numerous figures.

\end{document}